\documentclass[phd, titlesmallcaps, examinerscopy, copyrightpage, foronline]{mqthesis}
\usepackage[bottom]{footmisc}
\usepackage[euler]{textgreek}
\usepackage{capt-of}
\usepackage{breqn}
\usepackage{longtable}
\usepackage{tabularx}
\usepackage{algorithm}
\usepackage{algcompatible}
\usepackage{rotating}  
\usepackage{theorem} 
\usepackage{bm}  
\usepackage{booktabs}  
\usepackage{pdfsync}  
\usepackage{graphicx}
\usepackage{latexsym}
\usepackage{fancyvrb}
\usepackage{colortbl}
\usepackage{enumerate}
\usepackage{pifont}
\usepackage{stmaryrd}
\usepackage{textcomp}
\usepackage{fncylab}
\usepackage{multirow}
\usepackage{paralist}
\usepackage{wrapfig}
\usepackage{colortbl}
\usepackage{longtable}
\usepackage[table]{xcolor}
\usepackage[protrusion=true,expansion=true]{microtype}
\usepackage{mathtools}
\usepackage{graphicx}
\usepackage{pgfplots}
\pgfplotsset{compat=1.7}
\usepackage{diagbox}
\usepackage{adjustbox}

\begin{document}

\frontmatter

\title{Learning Complex\\Users' Preferences for\\Recommender Systems}

\ifthenelse{\boolean{foronline}}{
  \author{\href{mailto:Shahpar.yakhchi@hdr.mq.edu.au}{Shahpar Yakhchi}}
  \department{Computing}
}{
  \author{Shahpar Yakhchi}
  \department{Computing}
}
\degrees{}
 \submitdate{2020}
\renewcommand{\degreetext}
{for the degree of Doctor of Philosophy}
\titlepage

\chapter{Dedication}

This work is dedicated to my lovely husband for his understanding of my commitment to my research and for always encouraging me to be the best version of myself, and to my mother, father, and little sister without whose inspiration, drive and support I might not be the person I am today.
It is also dedicated to my supervisor, Dr Amin Beheshti, for his support and his strong commitment to teaching. He believed in me, helped me become a better researcher, and shed light on the issues I faced during my PhD studies. I also thank my associate supervisor Professor Mehmet Orgun for always being supportive and sharing his knowledge with me. 

At the end, as Winston S. Churchill once said, `Success is not final; failure is not fatal: It is the courage to continue that counts'. I hope this thesis will be a starting point for my future research career. Last, but not the least, I dedicate this work to Macquarie University and the community that gave me the chance to conduct my research over the past three years.


\chapter{Acknowledgements}

I would like to express my sincere appreciation to Dr Amin Beheshti, my principal supervisor, for all his support, encouragement, and endless guidance during my PhD study. Furthermore, I would like to express my deepest appreciation to him for always helping me put myself in the right direction with his attitude for a high quality of research. It has been my great fortune to have Dr Beheshti as my supervisor at Macquarie University (MQU\footnote{https://www.mq.edu.au/}).

I would like to also thank my associate supervisor Prof Mehmet Orgun who has always supported me in various stages of my research.

I am thankful to my team members in the Data Analytics Research Lab\footnote{https://data-science-group.github.io/} at MQU for their friendship, support and helpful comments. In addition, I would like to thank the review panels and the anonymous reviewers who provided suggestions and helpful feedback on my publications.

I would also like to thank the staff members of the Department of Computing at Macquarie University for their administrative help.


Last, but not the least, I would like to dedicate this thesis to my lovely family, for their love, patience, and understanding. They allowed me to spend much of my time on this thesis. Their love, support, and encouragement
have been the foundation of my life. Without their endless love and inspiration, this thesis would have never been started nor completed.


\chapter{Dissertation Examiners}
\begin{itemize}
\item Professor Abdul Sattar, Griffith University, Australia
\item Associate Professor Aamir Cheema, Monash University, Australia
\item Associate Professor Hakim Hacid, Zayed University, UAE
    
\end{itemize}

\chapter{Publications}
This thesis is based on my research conducted during my PhD program at the Department of Computing,
Macquarie University, between 2017 and 2020. Some parts of my research have been
published at the following venues:

\begin{itemize}
\item [$\bullet$] \textbf{Shahpar Yakhchi}, Amin Beheshti,  Seyed Mohssen Ghafari, Mehmet Orgun, and Guanfeng Liu, `Towards a Deep Attention-based Sequential Recommender System', IEEE Access, vol. 8, pp. 178073--178084, 2020. (\textbf{Impact Factor: 3.75})


\item[$\bullet$]\textbf{Shahpar Yakhchi}, Amin Beheshti, Seyed Mohssen Ghafari, and Mehmet Orgun, `Enabling the Analysis of Personality Aspects in Recommender Systems', Published in the Proceedings of $26^{th}$ Pacific Asia Conference on Information Systems (PACIS), Xian, China, pp. 1--15, 2019. (\textbf{Core Rank: A})

\item[$\bullet$] \textbf{Shahpar Yakhchi}, Seyed Mohssen Ghafari, Amin Beheshti, and Mehmet Orgun, `CNR: Cross-network Recommendation Embedding User's Personality', Published in the Proceedings of Web Information Systems Engineering (WISE) Workshop QUAT'18, pp. 62--77, 2018. 
(\textbf{Core Rank: A})


\item [$\bullet$] \textbf{Shahpar Yakhchi},  Seyed Mohssen Ghafari, and Mehmet Orgun, `TAP: A Two-Level Trust and Personality-Aware Recommender System', Accepted by The 1st International Workshop on
AI-enabled Process Automation, ICSOC, December 2020. (\textbf{Core Rank:~A})

\item[$\bullet$] Amin Beheshti, \textbf{Shahpar Yakhchi}, Salman Mousaeirad, Seyed Mohssen Ghafari, Srinivasa Reddy Goluguri, and Mohammad Amin Edrisi, 'Towards Cognitive Recommender Systems',  Algorithms Journal 2020, 13(8), pp. 1--27.

\item[$\bullet$] Seyed Mohssen Ghafari, \textbf{Shahpar Yakhchi}, Amin Beheshti, and Mehmet Orgun, `Social Context-Aware Trust Prediction: Methods for Identifying Fake News', Published in the Proceedings of Web Information Systems Engineering (WISE), pp. 161--177, 2018. 
(\textbf{Core Rank: A})

\item[$\bullet$] Seyed Mohssen Ghafari, \textbf{Shahpar Yakhchi}, Amin Beheshti, and Mehmet Orgun, `SETTRUST: Social Exchange Theory Based Context-Aware Trust Prediction in Online Social Networks', Published in Web Information Systems Engineering (WISE) Workshop QUAT'18, pp. 46--61, 2018. (\textbf{Core Rank: A})

\item[$\bullet$] Amin Beheshti, Vahid Moraveji Hashemi, \textbf{Shahpar Yakhchi}, Hamid Reza Motahari-Nezhad, Seyed Mohssen Ghafari, and Jian Yang, `Personality2vec: Enabling the Analysis ofBehavioural Disorders in Social Networks', Published in the Proceedings of $13^{th}$ ACM International WSDM Conference, Houston, USA, Texas, pp. 1--4, 2020. (\textbf{Core Rank:~A*})  

\item[$\bullet$] Amin Beheshti, Vahid Hashemi Moraveji, and \textbf{Shahpar Yakhchi}, `Towards Context-Aware Social Behavioral Analytics', Published by the 17th International Conference on Advances in Mobile Computing \& Multimedia (MoMM2019), Germany, December, 2019. \textbf{(Core Rank: B)}

\item[$\bullet$] Seyed Mohssen Ghafari, Amin Beheshti, Aditya Joshi, Cecile Paris, \textbf{Shahpar Yakhchi}, Alireza Jolfaei, Mehmet Orgun, Quan Z. Sheng, and Jia Wu, `Modeling Personality Effect in Trust Prediction', Published by Journal of Data Intelligence 2021.

\item[$\bullet$] Seyed Mohssen Ghafari, Amin Beheshti, Aditya Joshi, Cecile Paris, \textbf {Shahpar Yakhchi}, Alireza Jolfaei, and Mehmet Orgun, `A Dynamic Deep Trust Prediction Approach for Online Social Networks', Accepted by the  $18^{th}$ International Conference on Advances in Mobile Computing and Multimedia (MoMM'20). (\textbf{Core Rank: B})

\item[$\bullet$] Seyed Mohssen Ghafari, Aditya Joshi, Amin Beheshti, Cecile Paris, \textbf{Shahpar Yakhchi}, and Mehmet Orgun, `DCAT: A Deep Context-Aware Trust Prediction Approach for Online Social Networks', Published in the Proceedings of $17^{th}$ International Conference on Advances in Mobile Computing and Multimedia (MoMM'19), Munich, Germany, December, pp. 1--8, 2019. (\textbf{Core Rank: B})

\item[$\bullet$] Seyed Mohssen Ghafari, Amin Beheshti, Aditya Joshi, Cecile Paris, \textbf{Shahpar Yakhchi}, and Mehmet Orgun, `Intelligent Trust Prediction: Methods for Identifying Fake News', Accepted by the Cyber Defence Next Generation Technology and science Conference (CDNG), Brisbane, Australia, 2020.

\item [$\bullet$] Seyed Mohssen Ghafari, Amin Beheshti, Aditya Joshi, Cecile Paris, Adnan Mahmood, \textbf{Shahpar Yakhchi}, and Mehmet Orgun; `A Survey on Trust Prediction in Online Social Networks', IEEE Access, vol. 8, pp. 144292--144309, 2020. (\textbf{Impact Factor: 3.75})

\end{itemize}

\chapter{Abstract}

Recommender systems (RSs) have emerged as very useful tools to help customers with their decision-making process, find items of their interest, and alleviate the information overload problem. There are two different lines of approaches in RSs: (1) general recommenders with the main goal of discovering long-term users’ preferences, and (2) sequential recommenders with the main focus of capturing short-term users’ preferences in a session of user-item interaction (here, a session refers to a record of purchasing multiple items in one shopping event). While considering short-term users’ preferences may satisfy their current needs and interests, long-term users’ preferences provide users with the items that they may interact with, eventually.

In this thesis, we first focus on improving the performance of general RSs. Most of the existing general RSs tend to exploit the users’ rating patterns on common items to detect similar users. The data sparsity problem (i.e. the lack of available information) is one of the major challenges for the current general RSs, and they may fail to have any recommendations when there are no common items of interest among users. We call this problem `data sparsity with no feedback on common items' (DSW-n-FCI). To overcome this problem, we propose a personality-based RS in which similar users are identified based on the similarity of their personality traits.

Next, we focus on one of the major difficulties that sequential recommenders are confronted with, which is how to model a noisy session. Current studies may assume that all the adjacent items in a session are highly dependent, which may not be practical in real-world scenarios because of the uncertainty of the customers' shopping behaviour. A user-item interaction session may contain some irrelevant items which in turn may lead to false dependencies. Furthermore, long-term users’ preferences may be ignored by most of the existing sequential recommenders. To address this issue, we propose an attention-based framework to discriminately learn the dependencies among items in both long-term and short-term users' preferences.

Finally, sequential recommenders assume that each user-item interaction in a session is independent, which may be a very simplistic assumption. In real-world cases, people may have a particular purpose for buying successive items in a session. Unfortunately, a user's behaviour pattern is not completely exploited by most of the current sequential recommenders, and they neglect the distinction between users' purposes and their preferences in both users' long-term and short-term item sets. We propose an approach using a purpose-specific attention unit (PSAU) to attend to important items differently depending on the user purposes and preferences for the next-item recommendation task. All of the proposed models in this thesis were tested on real-world datasets, and the experimental results demonstrated the effectiveness of these approaches as compared to the state-of-the-art RSs.

\tableofcontents
\listoffigures
\listoftables

\mainmatter

\chapter{Introduction}
\label{chap:introduction}
\section{Background}
\subsection{An Overview on Recommender Systems}
In the last two decades, we have witnessed the emerging growth of generated information by people's daily activities (e.g. browsing, clicking, listening to music, and purchasing items). This may lead to an information overload problem. Therefore, there is a need for a filtering tool which can deal with this huge amount of data and overcome the information overload problem. Recommender systems (RSs) have emerged as a filtering tool that can help people with their daily living activities from eating and selecting clothes to housing and travelling. Research on RSs has its roots in information retrieval~\cite{DBLP:books/aw/Salton89,iStory}, forecasting theories~\cite{DBLP:conf/sofsem/Pelikan99}, user option modelling in business~\cite{DBLP:journals/ior/LilienR76,Adaptiverule}, and cognitive science~\cite{DBLP:journals/cogsci/Rich79}. In the mid-1990s, RSs appeared as an independent research domain~\cite{DBLP:journals/tkde/AdomaviciusT05}. In general, classic RSs try to understand the users’ general tastes from the users' and items' historical interactions (e.g. clicks, purchases, and likes). Therefore,  RSs collect the users' preferences on a set of items (e.g. movies, songs, books, applications, websites, products, and services) explicitly (basically
by analysing the provided users’ ratings) or implicitly (typically
by capturing the users’ activities, such as clicked on items, browsed websites, heard songs, and watched movies). RSs are used to provide suggestions in environments such
as content-production industries and e-commerce platforms, e.g. Netflix~\footnote{https://www.netflix.com/}, YouTube~\footnote{https://www.youtube.com/}, and LinkedIn~\footnote{https://www.linkedin.com/}, comparing member interests and creating recommendation items corresponding to the categorised topics or other members. For example, 80\% of the movies watched on Netflix~\cite{DBLP:journals/tmis/Gomez-UribeH16}, 60\% of the video clicks on YouTube~\cite{DBLP:conf/imc/ZhouKG10}, and more than 50\% of the hiring in LinkedIn can be attributed to recommendations~\cite{DBLP:conf/kdd/Posse12}.

\begin{figure*}[t!]
\includegraphics [width=0.9\textwidth]{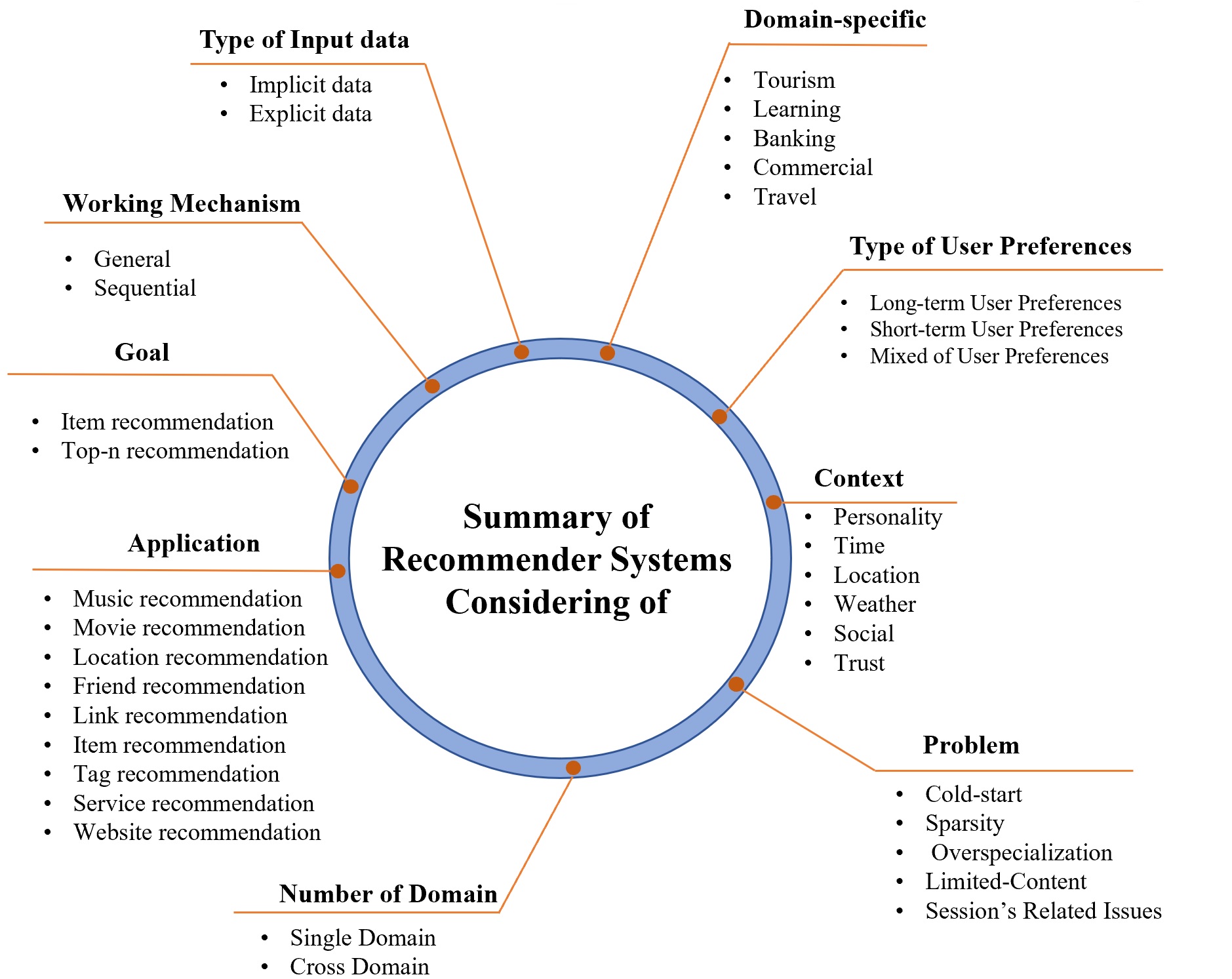}
\centering
  \caption{General summary of recommender systems' categories.}
  \label{ch1:figureRSCategory}
  \end{figure*}

An RS is known as one of the most valuable applications of machine learning and artificial intelligence and has a considerable impact on every aspect of our daily lives~\cite{Ricci2011,iProcess}. To show the position of RSs in today's cutting-edge technologies, we present Figure~\ref{ch1:figureRSCategory}, which is a summary of RSs' division from different aspects. Obviously, in this figure, if we contextually look into RSs~\cite{DBLP:reference/sp/AdomaviciusT15}, there are a wide range of context-aware RSs, including personality-aware~\cite{DBLP:conf/recsys/Recio-GarciaJSD09}, time-aware~\cite{DBLP:journals/umuai/CamposDC14}, location-aware~\cite{6228105}, weather-aware~\cite{DBLP:conf/aiia/BraunhoferEGRS13}, social-aware~\cite{DBLP:journals/comcom/YangGLS14}, and trust-aware~\cite{DBLP:journals/eswa/BediS12} RSs. In this figure, there are eight major categories with different challenges and solutions. Each of these eight classes of approaches has its distinct characteristics and is designed to address a different set of problems. 


For instance, if we take a domain (i.e. domain refers to the types of items that share similar characteristics, such as movies or books~\cite{article}) as a context (i.e. context is the information about the condition of an entity~\cite{Zheng/ShortPaper/2014}), traditional RSs can be roughly classified into two main categories: single-domain recommenders, which recommend items belonging to a single domain, and cross-domain recommenders, which aim to exploit knowledge from a source domain to perform or improve recommendations in a target domain. 
The main focus of this thesis is on the single-domain RSs and we leave the cross-domain recommenders for a future work.

Basically, on  the basis of how the users' preferences are modelled, we can classify approaches in a single domain RS into two main classes of general and sequential recommenders. The main goal of general recommenders is to discover the users’ long-term preferences by exploiting their past interactions with items. Sequential recommenders try to  understand the sequential users' behaviours and capture short-term users' preferences by modelling the item dependency. We further discuss these two lines of recommenders and review the existing studies in detail in Chapter~\ref{ch:chapter2RW}.

General recommenders can be divided into three main categories according to their input, goal, and mechanism: i) collaborative filtering models (CF): they look at like-minded people to predict the users' preferences, ii) content-based approaches: use the past preferred items' descriptions for the recommendation process, and iii) hybrid methods: they combine some of the previous methods in a unified manner. The core principle behind CF-based techniques is that among all the users of online social networks (OSNs), those who have had similar tastes in the past are more likely to share similar interests in the future. The CF-based approaches can be grouped into two different categories, namely memory-based and model-based approaches. Matrix factorisation (MF) is one of the most popular methods in model-based CF methods which learns users’ preferences by decomposing the original user-item interaction matrix into the inner product of the two low-dimensional matrices. MF as a dimensionality reduction technique factorises the user-item interaction matrix in the latent features of the ratings~\cite{DBLP:books/daglib/0033056,CDCRSLR,curationwww}.

Content-based approaches mostly rely on analysing item contents and user profiles to recommend items that match with the users’ interests. One of the most common methods in the content-based class of approaches is term frequency/inverse document frequency (TF-IDF), which is a well-known measure in information retrieval~\cite{DBLP:books/aw/Salton89}. Each of the above content-based and CF techniques has its own advantages and disadvantages. Hence, hybrid approaches have appeared as a unified model which help to avoid certain limitations of content-based and collaborative systems~\cite{DBLP:conf/aaai/BasuHC98}, \cite{DBLP:journals/air/Pazzani99}. 

In RSs, there are several successful real-world applications such as Netflix which suggests movies and TV series and Last.fm~\footnote{https://www.last.fm/} that offers songs and music albums. Usually, most of the aforementioned traditional RSs treat a recommendation task as a matrix completion problem, in which a sparse user-item rating matrix is given as an input. The main job of these RSs is to predict the value of the missing rating, which by itself can demonstrate the general users' preferences (i.e. long-term users' preferences). While traditional approaches, which are generally well studied, are good at modelling users' general tastes, they usually ignore short-term users' preferences, which may result in missing interest drifts of individual users over time~\cite{DBLP:conf/ismir/MooreCTJ13}. 

In contrast to the traditional RSs, sequential RSs (SRSs) are emerging as a new line of RSs which differ from traditional RSs in terms of their input, output, computational tasks, and abstract problem characterisation~\cite{DBLP:journals/corr/abs-1802-08452}. In SRSs,
a set of purchased items in one shopping transaction, which can be considered a session, is given to the recommendation engine as an input. Differently, SRSs take a sequence of user-item interactions to predict a set of next user-item interactions, which is the SRSs' output. In general, the main purpose of SRSs is to model the sequential dependencies among user-item interactions sequences and discover the hidden information embedded from one sequence to another sequence. SRSs have attracted researchers' attention to different applications, such as point of interest (POI) recommendation~\cite{DBLP:conf/ijcai/ChengYLK13},~\cite{DBLP:conf/ijcai/FengLZCCY15}, music recommendation ~\cite{Hariri},~\cite{DBLP:conf/recsys/WuLCHLCH13}, and browsing
recommendation~\cite{DBLP:conf/bigdataconf/ZhangZLCZM15}.

\begin{figure*}[b!]
\includegraphics [width=0.9\textwidth, scale=1]{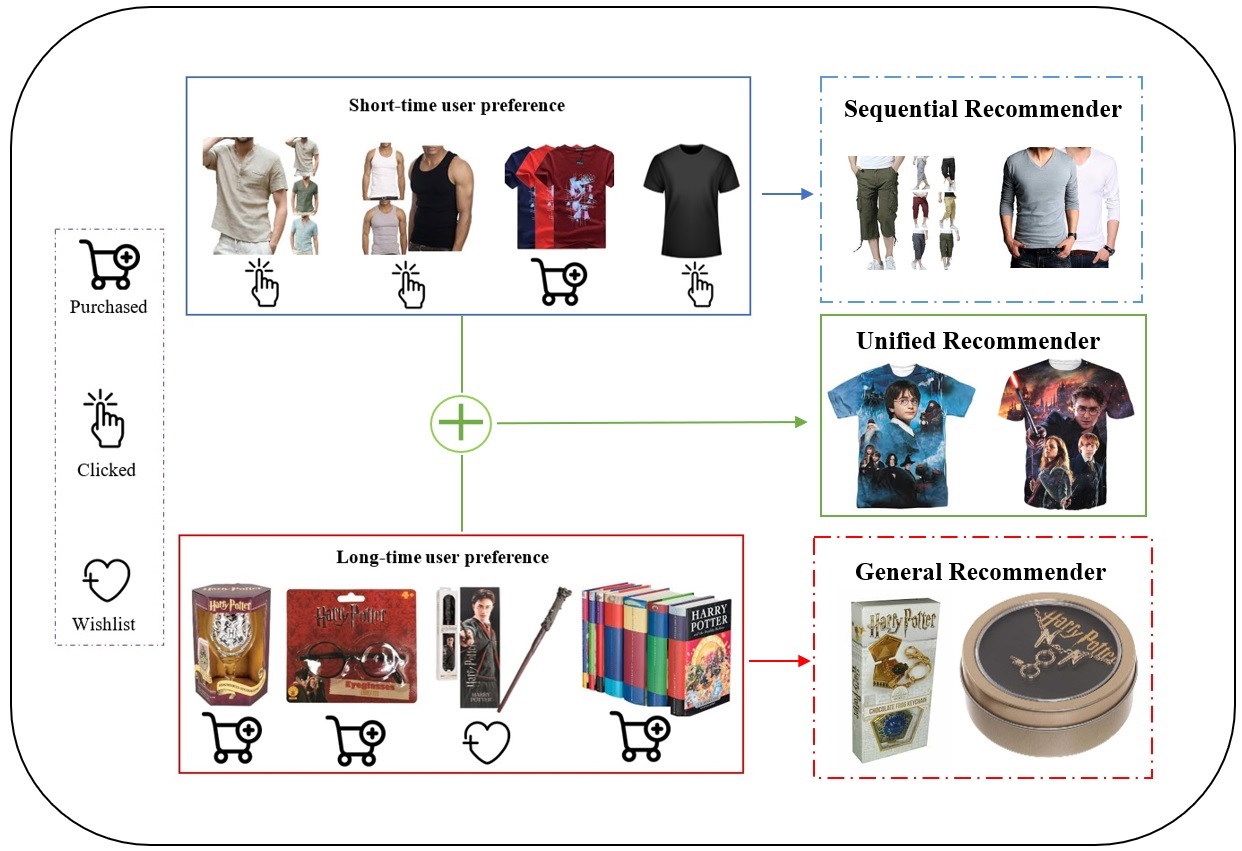}
\centering
  \caption{Real example of a user's shopping history; the blue frame shows a short-term user's interest, while a long-term user's preference is illustrated in a red frame.}
  \label{ch1:intropic}
\end{figure*}
\noindent Both types of the mentioned RSs have achieved considerable success and have been applied in a large variety of areas, such as item recommendations, music recommendations, movie recommendations,  POI recommendations, news recommendations, tag recommendations, friend recommendations, article recommendations, and link recommendations. Moreover, there are more popular real-world applications in different domains. For instance, Amazon is an example of product recommendation ~\cite{DBLP:journals/internet/LindenSY03}, Facebook~\footnote{https://www.facebook.com/} and Twitter~\footnote{https://twitter.com/} are dominant social networking applications, and YouTube is a good example of a multimedia sharing platform.

In general, a good RS should be able to build an accurate user's profile by tracing her/his activities and behaviours. If an RS can fully understand the users' needs and preferences, then it can provide satisfying suggestions, accordingly. For instance, Figure~\ref{ch1:intropic} illustrates a typical online shopping scenario. As is clear from a user's historical interactions in the lower part (red part) of this figure, the user may be a fan of `Harry Potter', as she/he has purchased or visited different `Harry Potter'-related products, such as eyeglasses, magic wands, wine glasses, and books. On the basis of this observation, we can infer that this user has been generally interested in `Harry Potter'-related items for a long time (i.e. user's general interest). This refers to the long-term user preferences (i.e. general interest), which are mostly captured by a general recommender. In contrast, a user's current actions are shown in the upper part (blue part) of Figure~\ref{ch1:intropic}. This user is looking for a t-shirt according to her/his successive clicks history in her/his previous transaction. This may refer to the short-term user preferences (i.e. sequential behaviours), which are the most important concern of a sequential recommender. While a general recommender may recommend another `Harry Potter'-related product, such as a necklace or a keychain, other t-shirts or shorts will be recommended to the user by a sequential recommender. Both types of user preferences have their own benefits; therefore if we only consider the long-term user preferences, the users' current interests may not be taken into account. In contrast, considering only the short-term user preferences may not be very interesting to the user as the user's general interest is ignored. Therefore, we can conclude that combining the two types of user preferences can exploit different  types of user behaviours and making better suggestions for a user.

\subsection{Values and Significance of General Recommender Systems}
General recommenders  have been widely studied in the last decade and the key issue related to them is to discover users' general tastes by capturing user-item interactions, which can be acquired explicitly (e.g. ratings and comments) or implicitly (e.g. clicks and purchasing history)~\cite{DBLP:conf/icdm/HuKV08}, \cite{DBLP:conf/uai/RendleFGS09}. In order to fully model users' behaviours and preferences, there is an essential need to trace each activity of the users. From Figure~\ref{ch1:intropic}, we can infer that most of the user's activities in the long-term are related to `Harry Potter'. First, she/he has purchased a `Harry Potter'-designed wine glass, then a pair of eyeglasses of `Harry Potter' was bought, and next a magic wand was added to a wishlist; and finally, the user bought a set of `Harry Potter' books. The user's historical transaction in the long-term can determine the user's general taste. As is shown in Figure~\ref{ch1:intropic}, the user is generally interested in `Harry Potter'-associated products. Hence, this user is more willing to be recommended another set of `Harry Potter' collection, and thus, an RS which can do this job, can satisfy the user's general interest.   

On the basis of the above observation, we understand that adopting a user's general interest can be a pillar of the user's decision-making process. It is an essential task for an RS to consider this important factor, as it may influence the user's decisions at present~\cite{DBLP:conf/kdd/LiZLHMC18}. The reason behind the importance of considering the user's general taste is that it may be static with no or slow change over time. For instance, take the above-mentioned shopping basket as an example. The user is a fan of `Harry Potter', and this may not change for a long time. This means that the user may always be interested in purchasing items related to `Harry Potter'.


To address the above issues, general RSs are proposed. The main concern of the approaches in general recommenders is to fully exploit the user's general taste, which is usually stable. The basic unit of a general recommender is a user-item interaction, which can be stored in a utility matrix, known as a rating matrix. The matrix factorisation method and its variants are known as dominant techniques for general recommenders in the existing literature~\cite{DBLP:conf/icml/RennieS05}, \cite{DBLP:journals/computer/KorenBV09}. Basically, approaches in general recommenders treat a recommendation problem as a matrix completion task, as the original user-item interactions matrix is very sparse. The missing values in this rating matrix need to be predicted by a general recommender. This type of a RS has been well studied, and various models have been proposed to deal with the mentioned issues. However, as users' behaviours are inherently variable, a general recommender can be enriched by various additional factors such as context~\cite{Adomavicius2015}, personality~\cite{Yakhchi}, \cite{DBLP:conf/ecweb/Fernandez-TobiasC14}, \cite{DBLP:conf/wise/YakhchiGB18}, \cite{DBLP:conf/recsys/NunesH12}, \cite{DBLP:reference/sp/TkalcicC15} and additional knowledge~\cite{DBLP:conf/apweb/HeFZZ15,iSheets,MAKG,iCOP,CrowdCorrect,Galaxy,TPM}. A comprehensive review of the existing techniques with different assumptions and ideas which try to model the users' general tastes is given in Chapter~\ref{ch:chapter2RW}.   
\subsection{Values and Significance of Sequential Recommender Systems}
In contrast to the general recommenders, SRSs are another line of RSs with the aim of modelling users' sequential behaviour. Recently, sequential recommenders have attracted considerable attention because of their superiority in capturing short-term users' preferences. Despite the general RSs, with a basic unit of rating matrix, the basic unit of a sequential recommender is a sequence of the user and item interaction logs (a session), which are mostly available in practical applications~\cite{9123874}, \cite{,DBLP:journals/algorithms/BeheshtiYMGGE20}. A session is a timestamped list of past user actions, which can store multiple records of purchased items in one shopping event~\cite{DBLP:journals/corr/abs-1902-04864}. Eventually, users' clickstream data which are recorded into a session are considered as an input of a sequential recommender. In general recommenders, users' IDs are available to the system, where each row is dedicated to the one known user to save her/his interactions with the items. In sequential recommenders, there is no need for the users to be known by the system. Take the e-commerce domain as an example; in this domain, users can freely perform actions while they are anonymous. Users in sequential recommenders do not need to log into the system to place different order types, such as viewing items, purchasing items, adding items to cart, and adding items to wishlist. In addition, based on the application domains, further detailed information can be extracted, such as the descriptions of items during  shopping for discounted products, or any other extra user and item features. Finally, the user and item interaction sessions and their attached available information are passed to the SRSs as the inputs, and the ordered lists of items are recommended as the outputs.

Consider the real-world example in Figure~\ref{ch1:intropic}. From this figure, we infer that this user would like to buy a t-shirt with some design on it, as the user is looking for many short-sleeve shirts and a set of t-shirts is added to her/his personal cart. Hence, a sequential recommender may suggest a pair of shorts or another type of shirt to the user, as many users search for a pair of shorts after buying a t-shirt. Sequential recommenders capture the current needs and interests of a user and then make a prediction accordingly to satisfy the short-term users' preference. To do so, there are varieties of models and techniques which are widely adopted in the literature and can be roughly categorised into two main classes: (a) model-free approaches which were introduced between the late of 1990s and the early 2010s, and (b) model-based methods which have been used from the early 2010s to the present~\cite{DBLP:journals/corr/abs-1902-04864}. The model-free classes of approaches use data mining techniques, such as pattern mining~\cite{DBLP:journals/datamine/LinAR02}, association rule discovery~\cite{DBLP:conf/snpd/Zhang07c}~\cite{DBLP:conf/ifip12/GhafariT16}~\cite{DBLP:conf/ksem/YakhchiGTF17}~\cite{DBLP:journals/widm/GhafariT19}, and sequence mining~\cite{10.1007/978-3-642-54927-4_62}, to make a recommendation.  While the model-free methods are straightforward and easy to develop, the model-based approaches are implemented to take advantage of the strength of machine learning techniques in modelling complex data. Markov chain models~\cite{DBLP:conf/www/RendleFS10} and recurrent neural networks (RNNs)~\cite{DBLP:conf/sigir/YuLWWT16} are two dominant paradigms in SRSs' domain. We further discuss the existing works in the SRSs domain in detail in Chapter~\ref{ch:chapter2RW}. 

\subsection{Values and Significance of Unified Recommender Systems}
Both the aforementioned paradigms have strengths and weaknesses. While general recommenders are good at modelling the long-term users' preferences, the current users' interests are ignored in these approaches. Although the main focus of SRSs is to model sequential dependencies by exploiting an item-to-item co-occurrence matrix,  the users' general tastes are not taken into consideration. Therefore, general recommenders, by considering the long-term users' preference, may better predict items that a user will be willing to consume eventually, whereas sequential recommenders show their success in recommending items that a user may interact with next, by modelling the short-term users' preferences. 

Different from both the aforementioned frameworks, quite a few works have attempted to take advantage of the ideas of both general and sequential recommenders in one unified model. Note that in Figure~\ref{ch1:intropic}, a blue dashed line denotes the items which are recommended by a sequential recommender, whereas those recommended by a general recommender are depicted by a red dashed line. If we take a closer look at this figure, we can infer that `Harry Potter' designed t-shirts which are surrounded by a green dashed line chart in the middle part of this figure  may better satisfy the users' overall interests. Unified recommenders with the idea of combining  both types of users' preferences have attracted more attention because of several reasons. First, as illustrated in Figure~\ref{ch1:intropic}, the user is browsing for a t-shirt and is generally interested in `Harry Potter'-related products. Therefore, she/he will be happy if a `Harry Potter'-designed t-shirt will be recommended to her/him. Second, a unified recommender can create an opportunity for a user to experience a novel item, which can increase the diversity rate of recommendations. Third, understanding users' behaviour and analysing their preferences and needs make a recommender system smarter and more personalised. 

Based on the above-mentioned observations, two major types of unified RSs are proposed by the approaches in the literature. First, some approaches make the key assumption of using one type of recommenders, either general or sequential, to assist the other recommender. For example, HRM proposed by Wang et al.~\cite{Wang2015} can achieve a better performance improvement of a sequential recommender by incorporating the users’ general taste. In addition, the adoption of different aggregation operations enables the HRM to model complicated interactions. Second, certain approaches attempt  to combine the two general and sequential paradigms to benefit from the advantages of both the models. Recurrent collaborative filtering (RCF)~\cite{DBLP:conf/ijcai/DongZZW18}, for instance, combines RNN and MF models within a single model to obtain a unified recommender~\cite{DBLP:journals/csur/ZhangYST19}. 

For the sequential part, the basic deep neural network models such as RNNs and their variations such as Long Short-Term Memory unit (LSTM)~\cite{DBLP:journals/neco/HochreiterS97}, \cite{DBLP:journals/cee/NiuXAPBA20} and Gate Recurrent Unit (GRU)~\cite{DBLP:journals/datamine/HidasiT16} are mainly  adopted because of their success in capturing sequential dependencies. Different from RNNs, Convolutional Neural Networks (CNNs)~\cite{DBLP:conf/wsdm/WuABSJ17} and Graph Neural Networks (GNNs)~\cite{DBLP:conf/aaai/WuT0WXT19} have also been applied in the sequential part of a unified recommender in order to tackle the possible problems of RNNs in dealing with long sequences. In addition to the aforementioned models, a combination of some kind of basic deep neural networks (e.g. RNN and CNN)~\cite{DBLP:journals/corr/HidasiKBT15} or some advanced models, such as attention mechanisms~\cite{DBLP:conf/ijcai/YingZZLXXX018}, \cite{DBLP:conf/aaai/WangHCHL018}, memory networks~\cite{DBLP:conf/wsdm/ChenXZT0QZ18}, and mixture models~\cite{DBLP:journals/corr/abs-1809-07426}, has been proposed to build more powerful RSs. While it is important to consider both types of users' preferences, there are few limited works in which both types of preferences have been modelled and more investigations are required to fill this gap. We  further discuss unified recommenders in Chapter~\ref{ch:chapter2RW}. Next, we give an introduction to the problems and the potential gaps which motivate our proposed solutions in this thesis.

\section{Research Motivations}
In this section, we first present a brief overview of the key issues, challenges, and complexities of recommender systems and then summarise the possible solutions provided in the extant literature. Then, we introduce our proposed ideas to deal with the shortcomings of the current studies.

\begin{figure*}[h!]
\includegraphics [width=\textwidth]{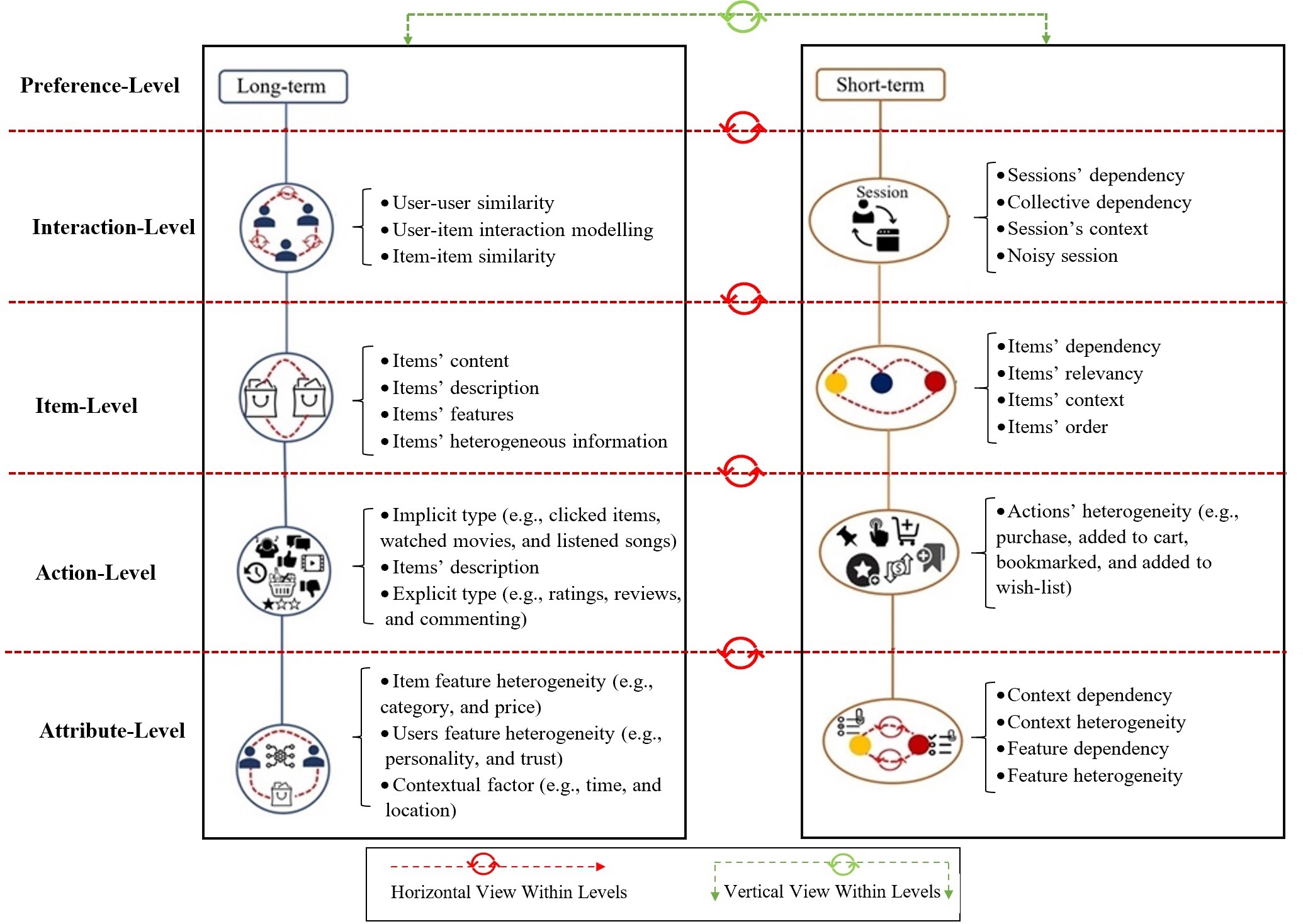}
\centering
  \caption{Hierarchical structure of key issues and challenges of recommender systems. This structure has two main pillars, one with the main goal of taking the long-term preference (e.g. general recommenders), and the other reserved for emphasising integration  short-term preference (e.g. sequential recommenders); there are two types of logical relations and views in this structure: $(1)$ a vertical view, dividing the problems of both the general and sequential recommenders into five levels separated by the dashed red arrows; $(2)$ a horizontal view which categorises all the sequential recommender challenges, on the right side and all the general recommender difficulties on the left side from the top to the bottom. }
  \label{ch1:figureRWcat}
  \end{figure*}

\subsection{Overview of Challenges and Issues}
Despite decades of research, RSs still have many challenges, and more investigations are required to  overcome the difficulties that they may face. In Figure~\ref{ch1:figureRWcat}, we summarise the main challenges in RSs. Given a hierarchical structure consisting of five levels from the preference level to the attribute level, we can see the main focus and goal of each level in the current RSs. Based on the explanation presented thus far, we can roughly classify RSs on the basis of the way they model user interest into two main classes of long-term preference-based models and short-term preference-based approaches. While long-term users' preferences are modelled by a general RS, short-term users' preferences are the main concern of a sequential RS. In general, there are two types of logical relations and views of different levels of this complex structure shown in Figure~\ref{ch1:figureRWcat}: (i) a horizontal view and (ii) a vertical view. First, if we take a horizontal look at this complex structure, we can realise that it contains two components which divide the existing challenges in RSs into two main challenges: approaches with the main concern of considering long-term users' preferences (i.e. general recommenders), and methods with the main focus of short-term users' preferences (i.e. sequential recommenders). Although each type of RS may have the same challenge as illustrated in Figure~\ref{ch1:figureRWcat} (e.g. dealing with the items-related issues at the third level or handling the attribute level at the fifth level), each type of the RSs need to solve them differently in their own way as they have different inputs and mechanisms. Second, if we vertically look at this design, we can find five levels of problems from the preference level to the attribute level in both the general and the sequential recommenders. At the item level, for instance, while general recommenders deal with how to take an item's content and its heterogeneous features into account, sequential recommenders deal with problems regarding the items' dependency and their order in a session. Both general and sequential RSs may be at the same vertical level, but their challenges, inputs, and adopted mechanisms will be different, as they have different natures and properties. Although the horizontal and vertical views mostly cover the main issues in RSs, coupling between all of these levels and views can be another logical relation in this design (e.g. at the preference level, there might be approaches that aim to consider both long and short-term users' preferences at the same time). In particular, a comprehensive explanation of the hierarchical structure is given in Chapter~\ref{ch:chapter2RW}. 

\subsection{Summary of Research Progress and Gaps in RSs}
In this section, we present a brief introduction to the current studies on both general and sequential recommenders and the existing gaps in each class of approaches.

\subsubsection{General Recommender Systems}

First, we start with a discussion of classic recommenders which  focus mainly on taking the long-term users' preferences divided in collaborative filtering, content-based, and hybrid approaches. In this thesis, we only focus on collaborative filtering types of general recommenders.
Collaborative filtering-based models can be classified into two main classes: memory-based and model-based~\cite{DBLP:conf/uai/BreeseHK98}, \cite{DBLP:journals/umuai/Burke02}. The main focus of memory-based approaches is to calculate the similarity between a pair of either users or items to discover a set of neighbourhoods for an active user and then make a prediction, accordingly~\cite{DBLP:journals/tois/AdomaviciusSST05}, \cite{GENG2015383}. The most widely adopted similarity metrics in this regards are Pearson correlation coefficient~\cite{DBLP:journals/kbs/GuoZY15} and cosine similarity~\cite{DBLP:books/mg/SaltonG83}. However, the main problems which these approaches may face are the data sparsity problem (i.e. when there is a lack of available information), the cold-start problem (i.e. when a new user
or item has just entered the system), and the scalability problem (i.e. when the numbers of existing users and items increase tremendously and computational resources go beyond the practical or acceptable levels)~\cite{DBLP:journals/advai/SuK09}.

While memory-based models are simple and easy to develop, their accuracy is not sufficiently good specifically for online systems~\cite{DBLP:journals/tweb/CachedaCFF11}, and thus, model-based algorithms motivated by statistical and machine learning methods have emerged to recognise complex patterns from the rating matrix~\cite{DBLP:journals/tkde/YuSTXK04}. The well-known model-based techniques are Bayesian-based models~\cite{DBLP:conf/uai/BreeseHK98}, \cite{DBLP:conf/pricai/MiyaharaP00}, \cite{DBLP:conf/ictai/SuK06}, clustering models~\cite{Ungar1998ClusteringMF}, \cite{DBLP:conf/dawak/CheeHW01}, and latent semantic models~\cite{DBLP:journals/tois/Hofmann04}. Among all the adopted techniques, MF is one of the most popular paradigms in the model-based type of collaborative filtering, which is also widely utilised by the existing works~\cite{DBLP:conf/icml/BillsusP98}. Although model-based approaches can show better performance than the memory-based class of algorithms, the data sparsity problem still remains the most important drawback of these approaches. Therefore, recommendation engines start to use side information to enrich their available data and to better overcome the data sparsity problem.

The increasing amount of generated data in each second motivates researchers to use these new data as the auxiliary information, which goes beyond the user-item matrix~\cite{DBLP:journals/csur/ShiLH14}. In general, there are different types of data which can be used as side information, such as an item's description, category, and price, which can better present its properties~\cite{DBLP:conf/kdd/AgarwalC09}, \cite{DBLP:conf/recsys/KoenigsteinDK11}. Moreover, users' additional information such as age, gender, and their attitudes and behaviours can be used by a recommender system to enrich the available information and make a better suggestion, accordingly~\cite{10.1007/978-3-642-40501-3_45}. Incorporating user demographic features into an RS cannot only result in performance improvement but also better solve
the cold-start problem~\cite{Safoury2013ExploitingUD}. Except for the above solutions for the data sparsity problem, at the attribute level, a user's personality type as the user's feature is incorporated into the RSs because of its considerable impact on the users’ attitudes, tastes, and behaviours~\cite{Costa}. It is widely recognised by psychologists that personality is a consistent
behaviour pattern that a person tends to show irrespective of her/his situation, and people with similar personality types tend to share similar interests~\cite{Ajzen}. Personality as a consistent behaviour pattern can be acquired either explicitly by filling out a questionnaire or
implicitly by observing the users’ behavioural patterns. Among all the personality models, the most influential one is the Five Factor Model (FFM) or the Big Five personality type model in both  psychology and computer science domains. FFM considers the five main personality traits as Openness to Experience, Conscientiousness, Extraversion, Agreeableness, and Neuroticism; it is typically abbreviated as OCEAN~\cite{McCrae}. Therefore, inspired by the excellent capability of personality in revealing human behaviour, some researchers have 
attempted to incorporate users’ big-five personality traits into the process of generating recommendations~\cite{DBLP:conf/recsys/HuP11}, \cite{Tkalcic}. Incorporating 
personality characteristics into RSs cannot only provide users with a diverse set of items~\cite{DBLP:conf/chi/McNeeRK06} but also provide a better group recommendation~\cite{DBLP:conf/um/KompanB14}, \cite{DBLP:conf/recsys/Recio-GarciaJSD09} and improve the accuracy of RSs for music, movies, e-learning, and web searches~\cite{DBLP:conf/recsys/HuP11}, \cite{DBLP:conf/hais/PaivaCS17}, \cite{DBLP:reference/sp/TkalcicC15}. For instance, Hu and Pu~\cite{DBLP:conf/recsys/HuP11} demonstrated the correlations between users’ personality types and their interests in music genres, and  then, they provided users with personality quizzes to explicitly detect their personality traits; they used Pearson correlation coefficient to understand how much these users are similar. Their proposed model achieved a significant improvement over the CF approaches that only take ratings into consideration for similarity detection. TWIN is another example of a recommender system that estimates users’ similarity in terms of their personality traits with the Euclidian distance~\cite{DBLP:conf/lrec/RoshchinaCR12}, \cite{DBLP:journals/jifs/RoshchinaCR15}. Furthermore, Elahi et al.~\cite{DBLP:conf/aiia/ElahiBRT13} developed a novel active learning strategy based on personality in order to deal with the cold-start problem. In this setting, users are asked to rate some items before they get recommendations. The authors illustrated that incorporating users' attributes such as age, gender, and, particularly, their personality traits can have an important impact on the users' rating behaviour. They provided the Ten-Item Personality Inventory (10-item TIPI) questionnaire in order to acquire the users' personality types explicitly, which may be relatively easy and accurate~\cite{DBLP:conf/um/DunnWHA09}. While effective, obtaining users' personality types is a challenging task, and more investigations on the same are required. 

\subsubsection{Sequential Recommender Systems}
In contrast to the general recommenders, which mainly consider the users’ long-term static preferences, SRSs have appeared as a new line of RSs which mainly focus on short-term transactional patterns~\cite{DBLP:journals/corr/abs-1902-04864}. The main goal of a sequential recommender is to model sequential dependencies, e.g. a customer who has recently purchased an iPhone is more likely to buy an iWatch next. SRSs can also refer to a session-based recommender system where the term of a session can be considered a set of consumed or interacted components at one time or within a certain period of time. For instance, a sequence of clicked items in one hour or in one Internet surfing session can be stored in one session. Recently, sequential recommenders have attracted more attention from both the academia and the industry. There are several reasons behind this fact. First, there is no need for the users to log in to the system, as they may not prefer to share their IDs or other source of information because of privacy concerns. Second, a sequential recommender can capture the current users' preferences and needs, which in turn results in personalised suggestions. Third, the dynamic users' taste is shown in a very recent session, which can help to provide accurate recommendations. Fourth, considering the short-term users' preferences can provide novel offers for users which they have not consumed already. 

Different from general recommenders, SRSs identify the current interactions between users and items to detect behavioural trends and interest shifts in the user community. From a business point of view, sequential recommender systems are more effective than the traditional recommender systems, as they are applicable in many real-world cases, and they have access to a wide range of data. However, a sequential recommender is relatively new and stands at its early stage. In general, given the properties of transactional data, the challenges of SRSs are two-fold from traditional models to the neural network-based methods. 

Sequential pattern mining (SPM) as one of the most popular techniques in the area of data mining, is one of the traditional solutions for SRSs, in which frequent patterns are discovered for further recommendations~\cite{DBLP:conf/dasfaa/YapLY12}. For instance, Morales et al.~\cite{Morales} used SPM to discover new, 
interesting, and useful knowledge from students' usage information. 
Their model can automatically provide links for new students while they are looking for a course. While SPM is an intuitive answer for the next item recommendation task, it may generate redundant patterns. Pattern/rule mining~\cite{DBLP:journals/dase/TabebordbarBBB20} as another traditional paradigm tries to discover item co-occurrence for generating recommendations~\cite{DBLP:journals/tsmc/WangC20}, while simple the items' context is ignored within a session, which may mislead recommendations~\cite{DBLP:conf/wise/GhafariYBO18}. In contrast to most of the traditional model-free models, the Markov Chain models are known as a straightforward model-based technique for modelling item dependency in SRSs; while the first-order MC models can capture  simple sequential dependencies~\cite{DBLP:conf/kdd/CadezHMSW00}, the higher-order ones can model complex relations~\cite{DBLP:conf/widm/EirinakiVK05}. MC can also be combined with other models to better learn the users' preferences. FPMC, as an example, is a combination of MC and MF, in which instead of using the same transition matrix for all the
users, an individual transition matrix is used for each user~\cite{DBLP:conf/www/RendleFS10}. 

Despite traditional methods, neural network models have gained considerable attention because of their success in modelling complex relations. RNNs, for instance, as a powerful paradigm in deep learning, have shown promising results in dealing with sequential data~\cite{DBLP:journals/corr/HidasiK17}. Although RNNs can exhibit impressive performance, they may not be applicable to unordered session data (e.g. when it does not matter which item is put in the cart first) because of their strong assumption that there is a rigid order between adjacent items~\cite{DBLP:conf/aaai/WangHCHL018}. Therefore, it is an important task to take a session's characteristics such as context, order, purpose, and length into consideration. As these characteristics may differ from one session to another, approaches at the interaction level try to handle these differences. Considering the session's context as an example, there may be some items in a session which are irrelevant to the session's context and may generate false dependencies. Therefore, in order to avoid the impact of noisy items in a session, SRSs need to pay more attention to contextual items. Some complex neural network techniques such as attention networks~\cite{DBLP:conf/aaai/WangHCHL018} have been proposed to emphasise more on the relevant items and reduce the impact of the irrelevant ones.  For instance, SHAN, a two-layer hierarchical network proposed by Ying et al.~\cite{DBLP:conf/ijcai/YingZZLXXX018}, uses attention networks to automatically assign different weights to different items in order to capture a session's context.   

The fourth level of the hierarchical diagram in Figure~\ref{ch1:figureRWcat} is associated with taking different types of actions into account. This is another important issue of SRSs that has been less explored. Usually, most of the current approaches treat users' actions in the same way with no difference among them, while there may be a difference between different action types such as purchasing and adding to cart. This may limit the recommendation performance, and it is important to take multi-behavioural sequences into consideration~\cite{DBLP:journals/tkde/LiuWW17}, \cite{DBLP:conf/momm/BeheshtiHY19}. Most of the current SRSs assume that there is only one action in each time step such as purchased items and they may ignore to take the other types of actions such as viewing and clicking items. The involvement of items' attributes, which is the focus of the attribute level, can help SRSs to better understand the item dependency and better deal with the cold-start problem, accordingly~\cite{DBLP:conf/um/KallooriR17}. Moreover, it may be possible that the context of the items affect the users' next behaviours. Here, the context refers to the specific explicit properties of an item when it is purchased, such as its popularity, discount, and community trends, or the implicit ones, such as the time and weather, season, and location, when the item was purchased~\cite{DBLP:conf/sac/JannachL17},~\cite{DBLP:conf/um/LercheJL16}.


\section{Overview of Challenges of Recommender Systems }
In this thesis, we focus on the following three challenges in recommender systems: dealing with the data sparsity problem, 
handling a noisy user-item interaction sequence, and modelling a collective dependency 
among items in a session. Next, we discuss these issues one by one in the following three subsections.

\subsection{Dealing with Sparsity of User-Item Interactions}
General recommenders have been widely studied in the last decade, and  the collaborative filtering technique is known as a typical solution in this regard.  
While the number of both users and items is very large in many large-scale applications, there are only a few available rated items among the total number of users. In particular, RSs may suffer from the data sparsity problem under up to a 99\%  sparsity rate~\cite{DBLP:conf/um/GuoZT12}. Lacking sufficient information regarding users and items may have a significant impact on the performance of a recommender system. Therefore, we explore a subclass of the data sparsity problem called the Data Sparsity With no Feedback on Common Items (DSW-n-FCI) problem, which refers to a situation when there is not much available information about user-item interactions. In this situation, the current CF models may fail as they mostly rely on rating patterns to find similar users and cannot find any connections among users under the data sparsity problem. To this end, we propose a personality-aware recommender system to discover similar users even under the DSW-n-FCI condition. We demonstrated that there were three main factors with a significant impact on the users' decision-making process in general recommenders, such as users' personality traits, users’ personal interests, and their level of knowledge. In Chapter~\ref{ch5:chapter5Personality}, we will discuss this framework and evaluate and compare it with state-of-the-art methods in recommender systems to demonstrate its effectiveness.

\subsection{Handling a Noisy User-Item Sequence }
Recently, SRSs have attracted considerable attention from researchers. While general recommenders are good at modelling a user's general preferences, SRSs try to model sequential dependencies among items. 
While considering a user’s sequential behaviour (i.e. short-term  preferences), which is reflected by the recently observed items, is good at discovering a user's current interest, a user's general taste which plays an important role in modelling her/his user's preferences is ignored by most of the SRSs. Therefore, better performance is achieved by approaches that combine both long and short-term users' preferences. Although the benefit from unifying these types of users' preferences is common and important and has shown potential in SRSs, it has not been sufficiently studied yet. For instance, Factorising Personalised Markov Chain (FPMC), as one of the pioneering model in this setting, integrates MC as a commonly used method in SRSs to model sequential behaviours and MF for modelling the users’ long-term preferences. FPMC  just  utilises a linear aggregation function to combine these two different types of users’ preferences~\cite{DBLP:conf/www/RendleFS10}. More importantly, most of the existing studies assume that there is a rigid order among items in a sequence, which may be a very simplistic assumption. There may be noisy items in a shopping basket, which may lead us to false dependency modelling in SRSs. In order to overcome the above-mentioned problems, in Chapter~\ref{ch6:chapterDAS}, we particularly focus on how to jointly model both long and short-term users' preferences and how to learn the different contributions of each item to both types of users' interests for the next item recommendation task.  

\subsection {Modelling Collective Sequential Dependencies}
Although unifying both general and sequential users' behaviour has shown a great result, the reported studies on SRSs have not paid considerable attention to this issue yet. Most of the current approaches either do no take both types of users' preferences or just linearly combine them together. Moreover, one of the drawbacks of the current unified approaches is that they assume that each user-item interaction in a session is independent and that there is no particular purpose in buying successive items. 
However, there may be a specific purpose behind purchasing a collective set of items. A common example is that of a user who buys butter, milk, and flour together for making a cake. In addition, different users may have different purposes and preferences,
and the same user may have various intentions. Thus, different users may
click on the same items with attention to a different purpose. Therefore, a user's behaviour pattern is not completely exploited by most of the current methods, and they ignore to take the distinction between the users' purposes and their preferences in both the users' long-term and short-term item sets.
Thus, in order to tackle the mentioned problems in Chapter~\ref{ch:chapter7CAN}, we will describe how to discover a user's purpose and her/his preferences over a set of items and the level of each item's contributions to both users' purposes and preferences.  

\section{Research Contributions}
The contributions of this thesis come from multiple folds and are summarised below:
\begin{itemize}
\item \textbf {Proposing an MF Model Incorporating Users’ Personality, Users’ Personal Interests, and Their Level of Knowledge}

The existing general recommender systems that use collaborative filtering highly depend on  exploiting the users’ feedback,
e.g. ratings and reviews on common items to detect similar users. Thus, they might fail when there are no common items of interest among users; this problem is called Data Sparsity With no Feedback on Common Items (DSW-n-FCI). The current general recommenders may fail to provide any suggestions when there are no common items of interest among users. In contrast, personality-based RSs have emerged as an interesting  paradigm in general recommenders without any need for looking for common items of interest to find similar users. Personality-based methods tend to discover similar users in terms of their personality traits. A personality trait is a consistent behaviour pattern that a person tends to show irrespective of her/his situation, and has a strong correlation with individuals’ interests. In addition, people with similar personality traits are more likely to consume similar products. The above observations motivate us to take the users' personality traits into account to make a better recommendation even under the DSW-n-FCI problem. In contrast to most of the personality-based approaches which detect the users' personality traits explicitly through a questionnaire, we ascertained the users' personality traits implicitly with no burden on the users (for more details, please refer to Chapter~\ref{ch5:chapter5Personality}).     

\item \textbf{Proposing a Deep Attention-Based Sequential Recommender System}

Most of the existing methods in SRSs assume that all the adjacent items in a sequence are highly dependent, which may not be practical in real-world scenarios because of the uncertainty of the customers’ shopping behaviour. Thus, they may not be able to truly model the items' dependencies. Moreover, while there can be a significant improvement in recommendation performance by capturing the long-term users' preferences~\cite{DBLP:conf/icwe/0002GZS19}, it is mostly ignored by the current SRSs. To overcome the mentioned problems, we propose a novel Deep Attention-based Sequential (DAS) model to differently attend to item contributions and discriminatively learn the dependencies among items in both users’ long-term and short-term item sets. We use an attention mechanism to emphasise the strongly context-relevant items and downplay the weakly correlated ones in a user-item interaction sequence. Furthermore, we apply a deep neural network to learn a combination of both long and short-term users' preferences (for more details, please see Chapter~\ref{ch6:chapterDAS}). 

\item \textbf{Proposing a Convolutional Attention Network for Unifying General and Sequential Recommenders}
Usually, people start shopping with a specific purpose and keep looking into different items until a preferred one is found. Moreover, different users may have different purposes for purchasing a set of items. For instance, suppose that Alice and Rose buy a new bag. While Alice is attracted to this bag because of its style and design which matches her shoes, Rose wants to buy it as a gift. Moreover, the same user may have different purposes and preferences by having the same actions on a set of items. Therefore, we propose to use convolutional neural network (CNN) in order to exploit the local context among a set of items in a long-term interacted item set. We apply the CNN network in the users' general preferences  in order to discover their main purposes. Next, we propose to use Purpose-Specific Attention Unit (PSAU) in order to discriminately learn the representations of the purpose and preference of the users. PSAU is used in the case of both long and short-term users' preferences to differently attend to different items and fully exploit the different informativeness of different items (for more details, please see Chapter~\ref{ch:chapter7CAN}).
\end{itemize}

\section{Thesis Organisation}
The remainder of this thesis is structured as follows: 

Chapter~\ref{ch:chapter2RW} provides a comprehensive review of the recommendation system approaches in two categories of classic recommenders and sequential recommenders. First, we will give an introduction to demonstrate the difference between these two categories of recommenders, followed by a further categorisation of each type of recommenders. Next, we will review all the categories one by one, from general recommenders to sequential recommenders.

Chapter~\ref{ch:chapter3PreliminariesandFoundations} first formalises the recommendation problem with a set of required definitions and notations. Then, we will introduce some preliminaries used in this thesis to clarify them for the readers. In the end, we will present
the commonly used datasets, evaluation metrics, and baseline methods used in this thesis.

Chapter~\ref{ch5:chapter5Personality} emphasises the significance of users' behaviour such as personality types for finding a neighbourhood in collaborative filtering-based methods. We will present the details of our novel matrix factorisation model, which considers three critical factors that generally affect a user's decision-making process. We will then discuss the importance of the users' personality types in finding similar users, followed by the users’ personal interests and their level of knowledge, as the key factors for increasing the  recommendations’acceptance rate. The empirical studies demonstrated the effectiveness of integrating these three main elements in general recommenders as compared to using the purely collaborative filtering-based models.

Chapter~\ref{ch6:chapterDAS} identifies the noisy user-item interaction sequences in sequential recommenders. The significance of taking the items' contributions into account is illustrated, and a framework to solve this issue is proposed. A corresponding algorithm based on the attention mechanism is designed to implement it. The experimental results over real-world transaction datasets demonstrated the advantages of this work as compared to the state-of-the-art models in this area of study.

Chapter~\ref{ch:chapter7CAN} focuses on modelling a collective dependency among items in a sequence in order to find a user's main purpose by clicking on a set of items. A framework together with a corresponding
algorithm is presented to show how to learn both the users' preferences and purposes simultaneously for the following session recommendation tasks. An empirical analysis of real-world datasets is provided to evaluate the algorithm's merits.

Chapter~\ref{ch:chapterConclusions and Future Work} concludes this thesis by summarising its contributions. We will also discuss some possible future directions related to this thesis. A hierarchy of the thesis is given below in Figure~\ref{ch1: figureThesisOutline} to better present the layout of this thesis.
\begin{figure*}
\includegraphics [width=0.9\textwidth]{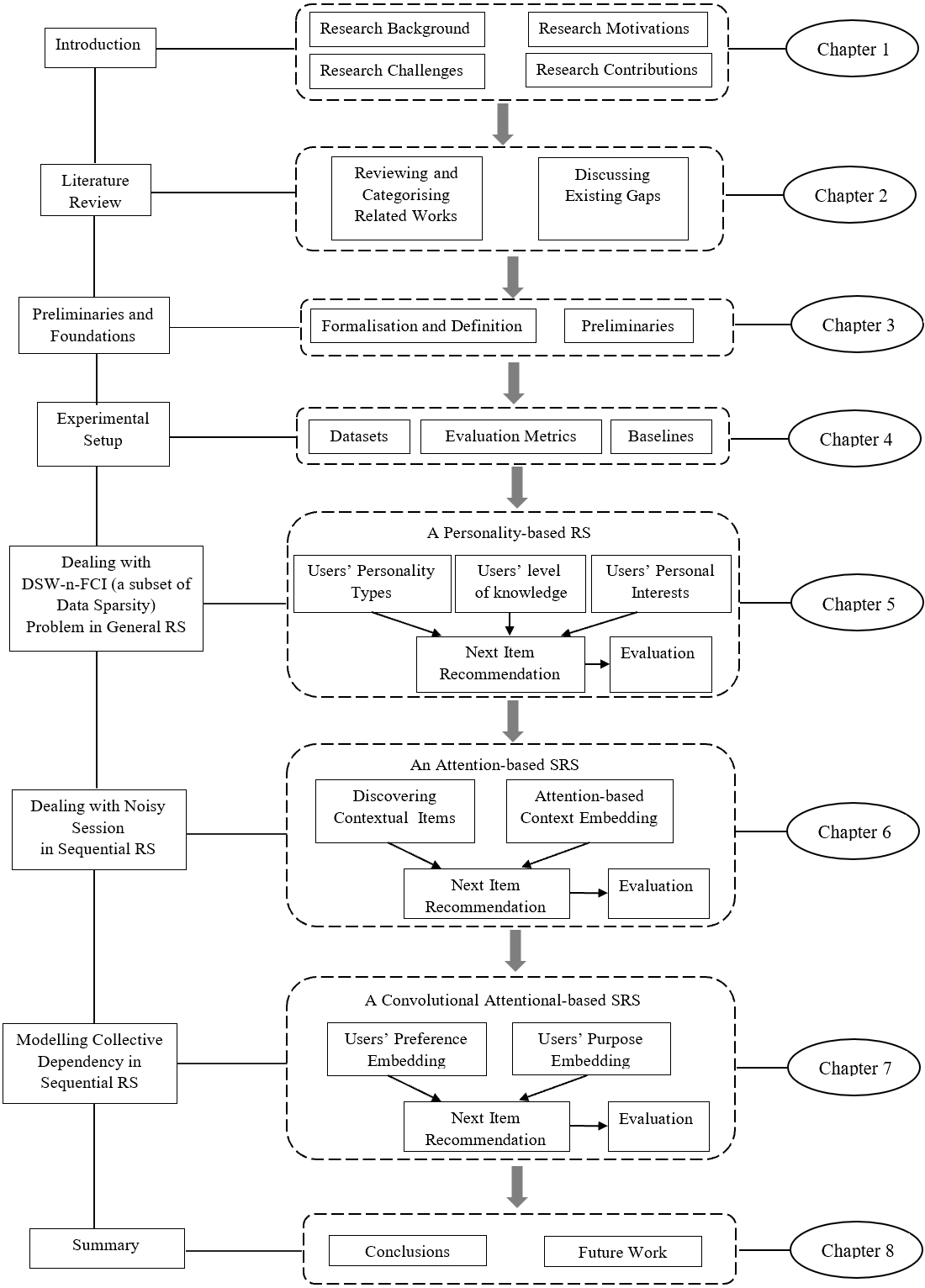}
\centering
  \caption{Structure of this thesis.}
  \label{ch1: figureThesisOutline}
  \end{figure*}

\chapter{Related Work}
\label{ch:chapter2RW}
In this chapter, we discuss the current general and sequential RSs and provide an extensive overview on state-of-the-art approaches.

\begin{figure*}
\centering
\includegraphics [width=0.8\textwidth, scale=1]{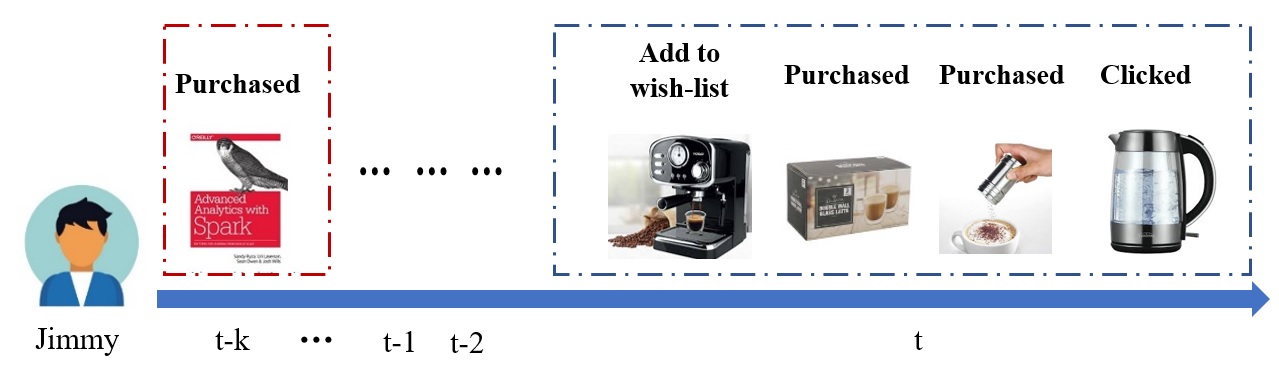}
\caption{User shopping basket in real world.}
\label{ch1: figureShoppingBasket}
\end{figure*}

Recommender Systems (RSs) are decision-making tools and techniques that provide suggestions for items in which a particular user may be interested~\cite{DBLP:reference/sp/RicciRS15}. Moreover, in the age of Big Data, many companies have started to trace the online activities of customers and capture every click of the mouse in the searching, browsing, comparing, and purchasing process. Given this huge amount of data and meta-data generated every second, RSs have an opportunity to make effective use of the captured data and provide a useful alternative to search algorithms. As shown in Figure~\ref{ch1: figureShoppingBasket}, Jimmy starts shopping at $t-k$; he first purchases a book from the computer science domain. Then, after a while, he makes a successive action at time $t$, by first putting a coffee machine onto his wishlist, then buying a set of double-walled glasses and cinnamon powder, and lastly clicking on an electrical kettle. Here, a recommender system should trace his activities and behaviours to determine the best set of items that meet his interests and needs. Therefore, recommender systems have appeared as a kind of a platform which automatically recommends a small set of items in order to help users find their desired items in online services.   

In this chapter, a comprehensive overview of the existing RS approaches is given. In particular, we discuss the attention of the community in two different lines of research, namely general recommenders and sequential recommenders. 

\section{Background}
\subsection{What is a Recommender System?}
The development of the Internet has resulted in the emerging growth of the generated information including, but not limited to, texts, images, and audios/videos. With this popularity, the number of users across the world is increasing at a rapid speed because of the various social communication networks such as Facebook~\footnote{https://www.facebook.com/} and Twitter~\footnote{https://twitter.com/}, e-commerce products websites such as Amazon~\footnote{https://www.amazon.com/}, and multimedia sharing platforms such as YouTube~\footnote{https://www.youtube.com/}. As a large number of online users have easy and instant access to the Internet, they can share and upload more information, which leads to the information overload problem. These data can be acquired explicitly by collecting a user's ratings or implicitly by capturing a user's behaviours, such as clicked items, downloaded applications, and watched movies. As stated by IBM, `Every day, Internet users generate 2.5 quintillion bytes of information so much that 90\% of the information on the planet today has been generated in the most recent 2 years alone'~\footnote{https://www.ibm.com/products/software}. 

Therefore, there is an urgent need for a system to be able to filter the generated information and to help people find their items of interest.
To deal with this issue, RS as an information retrieval system, has emerged to ease users with their decision-making process, particularly in the era of big data in which customers have access to more than countless products, services, movies, websites, and other options. In this context, personalisation  becomes an essential factor which needs to be accounted for in RSs in order to assist users for narrowing down their options, designing a better shopping experience, and boosting business benefits. In general, a recommendation engine generates a list of unobserved  items of interest for a user on the basis of the item's characteristics, user's descriptions, user's past behaviours and interactions with items, and some other sources of data such as spatial, temporal, social, and individual information.

\subsection{Why Recommender Systems?}
Recommendation Systems (RSs) are used to provide suggestions in environments such
as content-production industries and e-commerce platforms, e.g. Netflix, YouTube, and
LinkedIn, comparing member interests and creating recommendation items corresponding to the categorised topics or the other members. For example, 80\% of the movies watched
on Netflix~\cite{DBLP:journals/tmis/Gomez-UribeH16}, 60\% of the video clicks on YouTube~\cite{DBLP:conf/recsys/DavidsonLLNVGGHLLS10}, and more than 50\% of the
hiring in LinkedIn come from recommendations~\cite{DBLP:conf/kdd/Posse12}. A key component of every e-commerce service is its recommendation engine, which can help its customers with their options. Understanding users' preferences and their behaviours cannot only have a significant effect on making a right prediction for them and make recommendations more personalised but also improve the users’ satisfaction rate and boost business profits and users' loyalty. 

\subsection{How Does a Recommender System Work?}

In general, on the basis of the modelled users' preferences, there are two types of recommender systems: general recommenders and sequential recommenders. While the main goal of general recommenders  is to discover the users’ long-term preferences by exploiting their past items' interactions, sequential recommenders try to  understand the sequential user behaviours and model the short-term users' preferences by modelling the dependency between items. We will further discuss these two lines of recommenders and review the existing studies in detail.

\section{General Recommenders}
\label{ch2:GeneralRecommenders}
There are different categories of recommender systems in the literature according to multiple criteria such as the used techniques or based on a specific domain. However, the most common division is based on the type of data they use. Therefore, RSs can be broadly classified into two major categories: content-based and collaborative filtering. We comprehensively review each family of recommendation systems. At the end of this section, a more detailed comparison between the major classes of approaches in general recommender systems is provided in Table~\ref{ch2:tableGeneralRSComparison}.
\begin{figure*}[h!]
\centering
\includegraphics [width=\textwidth, scale=1]{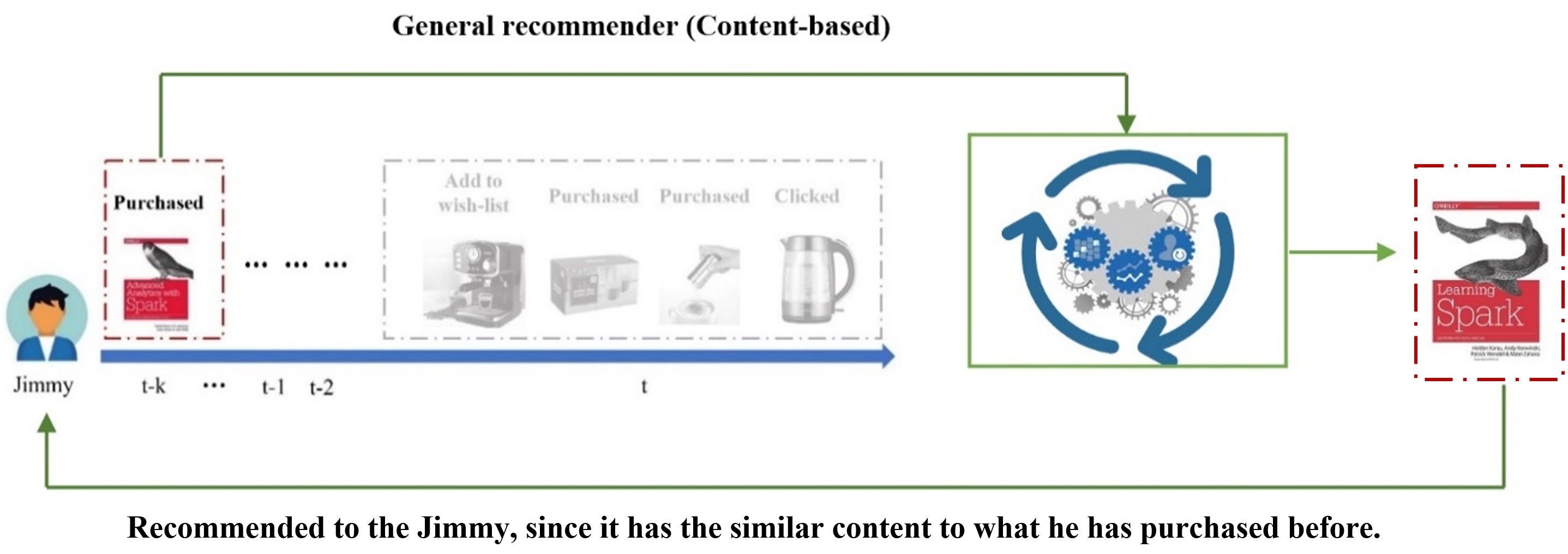}
 \caption{General framework of a content-based recommender system, which offers another book similar to the one that Jimmy has purchased before.}
\label{ch1:figureRWContentBased}
\end{figure*}
\subsection{Content-Based Recommendations}
A general framework of a content-based recommender system is presented in Figure~\ref{ch1:figureRWContentBased}. From this figure, we can observe that Jimmy is looking for a computer science-related book, and finally, he bought one at time $t-k$. Next, a content-based RS tries to find another book which is similar to the one that he purchased before in terms of its contents, and then, another similar book in the domain of computer science is recommended to him. The content-based category of RSs has its roots in information retrieval~\cite{DBLP:books/aw/Baeza-YatesR99}. In this category, an RS learns to recommend items, those with similar contents to what a user has liked in the past. As the name implies, content-based RSs mostly rely on analysing the items' contents and users' profiles in order to recommend items that are matched with the users’ interests~\cite{DBLP:reference/rsh/LopsGS11}. Contents can be a diverse range of information such as texts, images, and videos. 

The process of content-based approaches is to first analyse the contents of an item that a user has been interested in previously. Then, a user's profile is built on the basis of the extracted features from the previously interested items. This is an important step in a family of content-based approaches, and the better the profile construction is, the better is the users' preferences reflection and the more accurate is the recommendation. As the final step, a recommendation engine analyses the similarity between a user's profile and a set of recommended items to the user. In order to build a user's profile, the user's feedback needs to be collected and analysed. Users' feedback can be collected either explicitly, such as likes/dislikes, ratings, and writing comments, or implicitly by saving, discarding, printing, and bookmarking the interested items~\cite{DBLP:journals/cogsci/Rich79}. Most of the current content-based recommender systems rely on exploiting textual content for constructing a user profile. Term Frequency/Inverse Document Frequency (TF-IDF) weighting, as a well-known keyword matching technique in information retrieval, is one of the most common methods in this class which has been widely adopted by the approaches in content-based RSs~\cite{DBLP:books/aw/Salton89}.

There are several attempts in the literature in this direction by adopting one of the most popular techniques called, Word2Vec (W2V) to learn a low-level representation of the item descriptions~\cite{DBLP:journals/corr/abs-1301-3781}. ITR~\cite{DBLP:journals/umuai/DegemmisLS07}, \cite{DBLP:reference/dlhandbook/SemeraroBGL09} is an example of content-based recommendation system in which recommendations can be made in a wide range of domains such as movies, books and music. Re:Agent, proposed by Boone et al.~\cite{Reagent} is an intelligent email agent that uses examples and keywords provided by the user in order to extract features from documents. Then, Re:Agent can take suitable actions such as filtering, downloading to palmtops, and forwarding the email to voicemail, by using automatic feature extraction. There are other attempts in a different range of domains such as music~\cite{10.1007/11926078_67}, movies~\cite{DBLP:conf/webi/MakKP03}, and scientific search~\cite{CiteSeer}. FOAF~\cite{DBLP:conf/ismir/CelmaRH05}, \cite{DBLP:journals/ws/CelmaS08} for instance, is a music recommender system which builds a user profile based on the user’s listening habits. Then, it can collect the information regarding the listened songs such as an artist's name, the
song title, and a timestamp. In addition, FOAF can understand the user's psychological factors such as personality, demographic preferences, and social relationships. Letizia as a web assistance engine, is another example of a content-based recommender system which discovers a user's preference by collecting the user's implicit feedback such as bookmarking a page. Next, Letizia~\cite{DBLP:conf/ijcai/Lieberman95} tries to find a keyword related to the users' preferences.

Recently, deep learning has attracted significant attention because of its capability to deal with the problems of traditional models and achieve a high recommendation quality. For instance, in one study~\cite{DBLP:conf/nips/OordDS13}, a Convolutional Neural Network (CNN) was used to better learn the characteristics of a song that may affect the users' preferences. The authors have tried to extract some high-level properties such as genre, mood, instrumentation, and lyrical themes from audio signals in order to provide an accurate music recommendation. Ask
Me Any Rating (AMAR) is another work that uses a Long Short-Term Memory (LSTM) network to model textual content~\cite{DBLP:conf/um/Suglia0MGLS17}. AMAR embeds the users' preferences and item characteristics in low-dimensional representations, and then, a list of top-N items is recommended to a user according to the level of relevance, which is calculated by a logistic regression layer.
\begin{figure*}[h!]
\includegraphics [width=0.9\textwidth, scale=1]{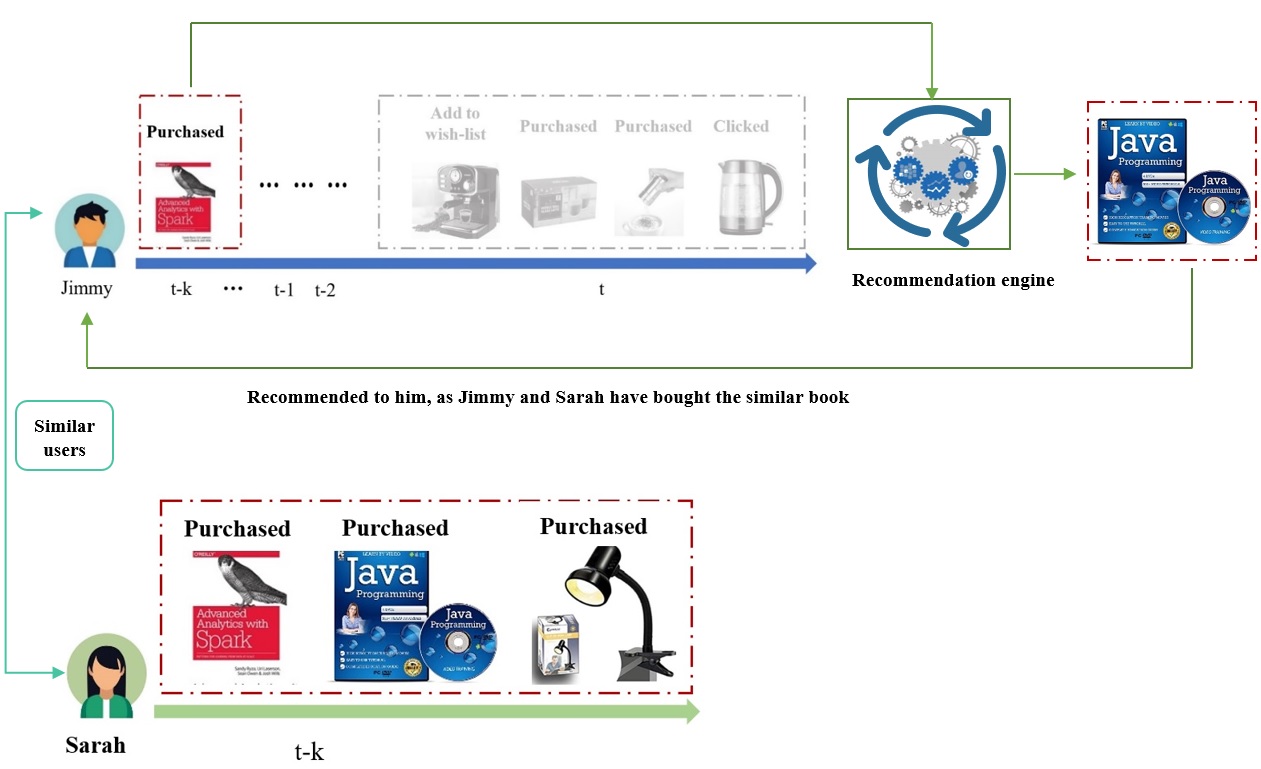}
\centering
  \caption{General framework of a collaborative-based recommender system  which recommends a set of CDs to Jimmy because of his similarity to Sarah.}
  \label{ch2:figureCollaborativeFiltering}
\end{figure*}

\subsection{Collaborative Filtering-Based Recommendations}
The pure CF recommender systems have better performance than the content-based approaches, as CF-based systems can make a recommendation irrespective of the content of an item, while content-based approaches highly depend on the descriptions of users and items for making a prediction~\cite{DBLP:conf/icml/SiJ03}. The core principle behind CF-based techniques is that among all the users in an online social network, those who had similar tastes in the past are more likely to share similar interests in the future. Figure~\ref{ch2:figureCollaborativeFiltering} presents a framework of a CF model, in which Jimmy and Sarah made a similar action at time $t-k$, by purchasing the same book, and thus, they may share similar interests in the future. Based on this observation, a collaborative filtering recommender system finds Jimmy and Sarah as similar users, and thus, it may suggest items that are liked by Sarah to Jimmy, accordingly.

\begin{figure*}
\includegraphics [width=\textwidth, scale=1]{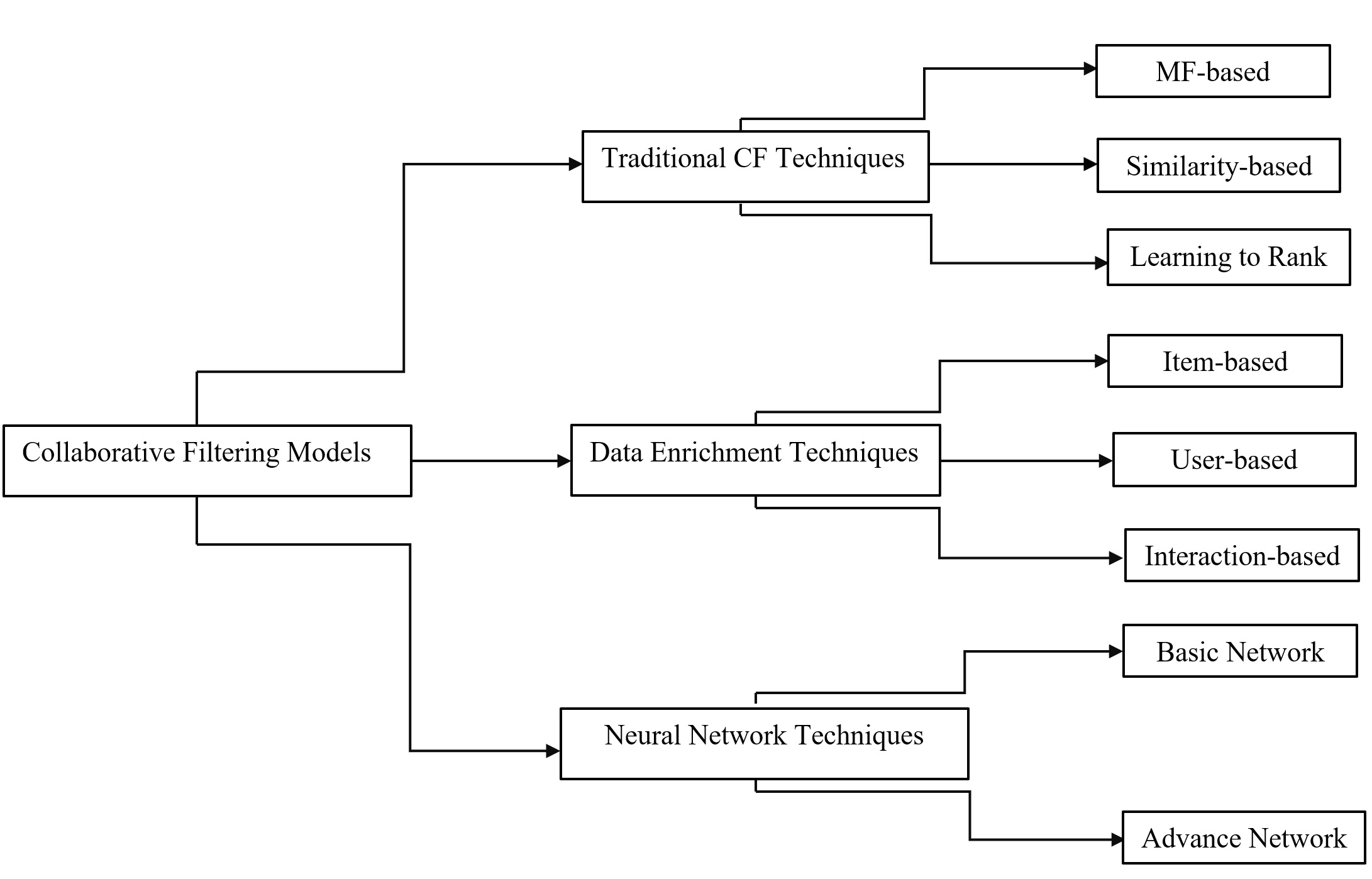}
\centering
  \caption{Categorisation of collaborative filtering approaches from the technical perspective.}
  \label{ch2:figureCF-basedCategorizes}
\end{figure*}

While there may be various types of CF-based categorisations in the literature, we present a new division of approaches in this class of models in Figure~\ref{ch2:figureCF-basedCategorizes}. In general, there are three main classes of techniques in CF-based RSs, namely $(a)$ traditional CF, $(b)$ data enrichment, and $(c)$ neural network techniques. `Tapestry', as the email filtering system, was the first CF-based recommender system, which was proposed by Goldberg et al.~\cite{DBLP:journals/cacm/GoldbergNOT92}, to coin the term `collaborative filtering (CF)' in 1992. Tapestry is an electronic message system that asks users for their feedback and then makes suggestions for a target user by collecting feedback from the other users~\cite{DBLP:conf/chi/ShardanandM95}, \cite{DBLP:conf/cscw/ResnickISBR94}. From the presented structure in Figure~\ref{ch2:figureCF-basedCategorizes}, we can see that Matrix Factorisation (MF) is a traditional solution for collaborative filtering models. MF is one of the most popular dimensionality reduction techniques in model-based CFs which factorises a user-item interaction matrix in latent features of the rating~\cite{DBLP:books/daglib/0033056}, \cite{DBLP:journals/computer/KorenBV09}. User-item (U-I) interactions are stored in the U-I matrix, which is used as an input to an appropriate prediction model in order to generate recommendations. MF-based methods use two different types of data: \textit{explicit feedback} and \textit{implicit feedback}. Models that focus on the explicit feedback (e.g. rating and review) deal with the rating prediction problem where users explicitly
express their preferences by providing a rating to particular items~\cite{DBLP:journals/computer/KorenBV09}. While approaches based on the implicit feedback, (e.g. clicking and purchasing history) aim to formulate recommendations as a ranking problem, this highly depends on the selection of the objective loss function to optimise~\cite{DBLP:conf/recsys/KaratzoglouBS13}.

Next, we introduce similarity-based collaborative filtering methods, which try to measure similarity among a set of items or users. If a similarity is calculated between users, it can be known as a neighbourhood-based collaborative filtering method, which looks for the best neighbourhood that may share similar interests~\cite{DBLP:reference/sp/2015rsh}. The term neighbours is used here to represent the users in OSNs who have a pattern of rating items which is similar to that of an active user. These approaches first capture the users' preferences in order to construct her/his profile by collecting their given ratings; a set of neighbours who have in common interested items is discovered next. In the final step, when neighbours are identified, a recommender engine provides a list of top-n interested items to the target user by analysing the items in which neighbours are interested the most~\cite{DBLP:journals/ijinfoman/KimKOR10}. User-based CF approaches predict users’ ratings on an unseen item by using similarity metrics such as the Pearson correlation or the cosine similarity~\cite{DBLP:journals/debu/Singhal01}. While, items-based collaborative filtering approaches calculate the similarity between items that a user has liked before for a recommendation task. The list of recommended items, can be easily updated as they only need to retrieve items that are similar to the previous ones with no need to re-train the model again. Thus, personalisation may be easily achieved~\cite{DBLP:journals/corr/BayerHKR16}. Although there is an extensive research study in the literature where statistical measures such as Pearson correlation and cosine similarity are used to calculate the similarity in item-based CFs, these models may find it difficult to adapt to a new dataset~\cite{DBLP:journals/tois/XueHWXLH19}. Recently, the sparse linear method (SLIM)~\cite{DBLP:conf/icdm/NingK11} and factored item similarity model (FISM)~\cite{DBLP:conf/kdd/KabburNK13} have taken one step forward in this direction by using data-driven methods to learn items similarity. The former, Sparse Linear Method (SLIM), measures the prediction score for a new item on the basis of the aggregation of other items in order to have a better top-N recommendation task. While the latter, Factored Item Similarity Model (FISM), calculates the product of two low-dimensional latent factor matrices in order to learn the similarity between items. A few years later, a Global and Local SLIM (GLSLIM)~\cite{DBLP:conf/recsys/Christakopoulou16} was proposed as an extension of SLIM to overcome the difficulties of SLIM in capturing users' preferences. GLSLIM improves the top-N recommendations task by providing a separate local item-item model for every user subset.

According to Ning and Karypis~\cite{wwwj2020}, learning to rank (L2R) is a `popular research area, as it directly models partial ordering relations between items, which happens to be consistent with the top-N recommendation
tasks. One key element of L2R methods is the objective measures, defined as either ranking
error functions or optimisation metric'. For a better understanding: there are two different types of actions in RSs, namely explicit actions and implicit actions, whose results can create different recommendations' inputs, accordingly. 
The explicit actions refer to a situation in which users explicitly present their preference on items by giving a rating, writing a comment, and liking. Each of the mentioned explicit actions provides different degrees of expressivity of the user’s preferences, but most of the existing RSs use explicit types of inputs as they are more convenient to collect and model~\cite{10.1145/1869446.1869453}. In contrast to the explicit type of inputs, an implicit type of inputs including purchase history, browsing history, search patterns, or even mouse clicks should be modelled differently, as their nature is different from that of the explicit ones. Modelling the implicit type of inputs may be more complicated than modelling explicit ones, as there is no negative feedback here. A user's clicks on the items may show her/his interests in them, but it may be difficult to understand a set of items that she/he may not be interested in. For instance, if a user has purchased a particular item, it can mean that she/he really likes this item, but it may be bought as a souvenir.
Hence, the implicit type of inputs is very noisy. Furthermore, users can give a rating to the items of their interest, which is a numerical value from 1 to 5.
In this setting, RSs can infer these values as the level of users' interest, where a higher value implies more pleasant items. For example, items which have received 5 star ratings can be considered to be the more preferred ones in which possibly the user is strongly interested, while items with 1 star ratings may represent the user's unsatisfaction. Therefore, numerical values in explicit feedback can be a good indicator of the users' preferences, while a numerical value in implicit feedback such as the time spent on a particular page indicates confidence. Lastly, the accuracy of approaches based on the implicit inputs differs from that of the methods based on the explicit ones~\cite{DBLP:conf/icdm/HuKV08}.

In contrast to the above-mentioned models, which are already known as the memory-based algorithms by the recommender systems' community, the model-based methods try to learn users’ preferences by using a certain recommendation model according to the concerned issues~\cite{DBLP:journals/kybernetes/AlyariN18}. At a high level, there are two main major steps that need to be taken for this class of approaches. The first step is to discover the relations between the items in order to build a model, and second a list of top-N suggestions should be generated for a target user. Different machine learning techniques are adopted in model-based CF, including Bayesian networks~\cite{DBLP:conf/ictai/SuK06}, clustering models~\cite{DBLP:conf/dawak/CheeHW01}, latent semantic models~\cite{DBLP:journals/tois/Hofmann04}, Markov decision processes (MDPs)~\cite{DBLP:journals/jmlr/ShaniHB05}, association rule-based, and dimensionality reduction techniques~\cite{DBLP:journals/tmis/Gomez-UribeH16}. The work proposed by Miyahara and Pazzani~\cite{DBLP:conf/pricai/MiyaharaP00} is one of the Bayesian-based models, which consists of two variants of collaborative filtering and shows a better performance improvement over a simple similarity-based CF. In this study, on the basis of the assumption that independent features are given to the class, the class with the highest probability was considered the predicted class. Clustering is another solution for CF-based techniques. For instance, K-means as a clustering method divides the data into the different clusters in which the entities in each cluster are similar to each other, while they are different from those of the other clusters~\cite{DBLP:books/mk/HanK2000}. In addition, MDP is another popular model adopted by the CF-based model. MDP-based CFs can be seen as a sequential optimisation problem by applying Markov decision processes~\cite{DBLP:journals/jmlr/ShaniHB05}; in this problem, an agent needs to take the best action according to the current state. Next, the association rule technique is a data mining technique which searches for the co-purchased items to be analysed in order to find items which are more probable to be purchased together in a customer transaction history~\cite{DBLP:journals/isci/KardanE13}. The association rule-based CF tries to discover the correlations between items and then recommend the strong associative items to the users~\cite{DBLP:conf/sigecom/SarwarKKR00}. CDAE is another recent work of the item-based CF model in which a denoising auto-encoder is used for obtaining a better top-N recommendations~\cite{DBLP:conf/wsdm/WuDZE16}. A user-item interaction in CDAE is considered a corrupted version which needs to be recovered during the training process. Because of the significant  advance of attention mechanisms in neural representation learning, a Neural Attentive Item Similarity Model (NAIS)~\cite{DBLP:journals/tkde/HeHSLJC18} was proposed very recently with an online personalisation item-based CF model. The key idea behind NIAS is to find the more important items according to the users' preferences.

The accuracy of recommendations highly depends on the U-I matrix, which may suffer from the data sparsity problem because of a lack of available information in real-world scenarios. In order to tackle the data sparsity problems, most of the CF-based recommender systems resort to the use of additional information to enrich their data, and this may generate a new line of research on RSs, called data enrichment in this thesis. Approaches in this category seek to use more available information in order to augment their data. Basically, there are different types of data which can be used as side information and can be placed into three categories according to the source of information: \textit{(i) item-based features}, \textit{(ii) user-based information}, and \textit{(iii) interaction-based}~\cite{DBLP:conf/kdd/AgarwalC09}, \cite{DBLP:conf/recsys/KoenigsteinDK11}, \cite{DBLP:conf/sigir/MoshfeghiPJ11}. This auxiliary information can be integrated into the user-item matrix in order to enrich this matrix for obtaining a better prediction. Meanwhile, item-related attributes such as content, category, and price can be added to the model to better describe the items' features. Moreover, user-based features can be divided into three main classes of information: as \textit{(a) user demographic features}, \textit{(b) user contributed information}, and \textit{(c) social networks information}. The first class is user demographic features including age, gender, personality and attitudes, behaviours, and location, which can help a recommender system to build an accurate user profile and make an accurate recommendation, accordingly~\cite{10.1007/978-3-642-40501-3_45}. Incorporating user demographic features into an RS cannot only result in performance improvement but also better solve the
cold-start problem~\cite{Safoury2013ExploitingUD}. Among all of the users' demographic features, personality as a consistent behaviour pattern with the strong relations with users' preferences has attracted more attention where there is an exclusive group of methods called personality-based recommender systems, which is  discussed later in this section. 

The second class of user-based information is user contributed data, including free-text reviews/comments, multimedia content, tags, and geo-tags~\cite{DBLP:conf/recsys/LeviMDT12}, \cite{articlejakob}, \cite{DBLP:conf/sigir/MoshfeghiPJ11}. Tags are short textual labels  assigned to the items by the users and can describe an item's properties~\cite{DBLP:journals/tweb/RobuHS09}. Tagommenders~\cite{DBLP:conf/www/SenVR09} is a good example of using the extracted information from tags. Based on the results of Tagommenders, tags can help recommender systems find the most important dimension of items. Moreover, social media platforms such as Flickr~\footnote{http://www.flickr.com/.} and YouTube~\footnote{https://www.youtube.com/} provide too many options for their users, such as sharing, posting, and liking images and videos. This opportunity can facilitate recommender systems to be aware of the categories of interest of photographs and videos and thus, may generate an accurate suggestion~\cite{DBLP:conf/imc/ZhouKG10}.

As exploiting tags as the extra source of information is beneficial for RSs, there is a separate line of research known as tag-based RSs, which was out of scope of this thesis~\cite{DBLP:conf/sac/Tso-SutterMS08}. Geo-tags provide explicit latitude and longitude coordinates, which can provide the location information for RSs. The geo-tags of a shared photograph can be used to trace the location of the users. The use of these kinds of information can facilitate a location recommendation, which was again out of the scope of this PhD thesis~\cite{DBLP:conf/cikm/ChengCL10}. Since the introduction of the Web 2.0 technology, people have been free to share their contents, such as videos, text, and images. Exploiting these shared contents on OSNs can create an opportunity to determine what type of items a user is interested in. These contents can also be semantically and sentimentally mined to better construct users' profile and learn their preferences~\cite{DBLP:conf/sigir/MoshfeghiPJ11}, \cite{article4}, \cite{DBLP:conf/recsys/LeviMDT12}. Recent techniques, such as image processing, natural language processing, and object detection, can contribute to the exploitation  of more information and make personalised recommendations to the users~\cite{DBLP:conf/recsys/DavidsonLLNVGGHLLS10}. 

Last, but not the least, the third subclass of user-based information is social network information, which can be provided by the relationships between two users in OSNs, such as trust relationships and `following' relationship between two users in OSNs~\cite{DBLP:conf/www/KwakLPM10}, \cite{DBLP:conf/sigir/KonstasSJ09}, \cite{ghafaridynamic}.  RSs can benefit from social information from different aspects, such as the social properties of users, social ties, communities, friendship, and social contents. For instance, the strong social ties which can be calculated by using some factors such as intimacy, emotional intensity, and amount of time, can be incorporated into the RSs~\cite{DBLP:journals/tmc/DalyH09}. Trust can be defined as `Trust
provides information about with whom we should share information, from whom we should
accept information and what considerations to give to information from people when aggregating or filtering data'~\cite{DBLP:conf/wsdm/TangGHL13}. TidalTrust~\cite{DBLP:phd/basesearch/Golbeck05} and MoleTrust~\cite{DBLP:conf/recsys/MassaA07} are the two the most well-known models in this domain that show that incorporating trust information in discovering the best set of neighbourhoods may improve the recommendation accuracy. TrustWalker~\cite{DBLP:conf/kdd/JamaliE09} is another example of integrating trust
information with the classic item-based CF, which shows a better result than the item-based CF. 

Context is another information which can help RSs to better understand users' preferences. There are different definitions of context in various domains, such as computer science, information retrieval,
cognitive science, linguistics, philosophy, social science, and psychology~\cite{DBLP:reference/rsh/AdomaviciusT11,CognitiveAugmentation,CognitivePrivacy2}, \cite{DBLP:conf/momm/GhafariJBPYO19}. According to Oxford Advanced
Learner’s Dictionary, context is `a situation in which something happens and that helps you to understand it'~\cite{DBLP:conf/dexa/Wilson93}, and according to Webster’s Dictionary `context is a situation in which something
happens: the group of conditions that exist where and when
something happens'~\cite{Webster}. From the computer science point of view, context is defined as `everything' that `affects the computation
except the explicit input and output'~\cite{DBLP:journals/ibmsj/LiebermanS00}. The concept of context was introduced in the computer science field in the late
1980s~\cite{DBLP:journals/eswa/HongSK09}, and Schilit et al.~\cite{context} advocated the introduction of context-awareness in computing in 1994. Among the different aspects that can be considered a context in recommender systems, time, location, and social relations have been shown to have great success in the literature. LARS is a location-aware RS, which exploits users' location to find the nearest items, such as restaurants, bars and cinema halls~\cite{DBLP:conf/icde/LevandoskiSEM12}, \cite{DBLP:conf/recsys/Rodriguez-Hernandez15}.
Time is defined as `a non-spatial continuum that is measured in terms of events that succeed one another from past, through present to future' or `the measured or measurable period during which an action, process, or condition exists or continues'~\cite{Webster}. Time is another dimension of context which can help the system to track the evolution of users' preferences over a period of time. Zimdars et al.~\cite{DBLP:conf/uai/ZimdarsCM01} considered taking time information as a context into their model. Baltrunas et al.~\cite{article2} considered morning and evening as a time of day factor and workday and weekend as a time of year for having an understanding of users' preferences for a particular period of time. In another study, seasons (e.g. fall or winter), day of week (e.g. Saturday or Sunday), and time of day (e.g. morning or midnight) were treated as a time factor in RSs~\cite{DBLP:series/sci/LeePKLL10}. Finally, besides users and items information, interaction-based data are the last type of side information, which refers to a set of associated information that comes along with the items that a user directly interacted with, such as the time of purchasing items~\cite{DBLP:journals/aim/AdomaviciusMRT11} and the location where a mobile application was downloaded~\cite{inproceedings2}.

\textbf{Personality-Aware Recommender Systems.} As we mentioned earlier, personality as one of the main characteristics of users has attracted more attention, and thus, we decided to discuss this class of approaches separately. Personality is described as a `consistent behaviour pattern and interpersonal processes originating within
the individual'~\cite{Burger}, which can explain the wide variety of human behaviour. Personality is known as the most important factor that plays a significant role in people's decision-making process~\cite{DBLP:conf/recsys/NunesH12}. People having a similar personality type tend to share similar tastes. According to Rentfrow et al.~\cite{Rentfrow12} people having the `reflective' personality type with a high score of openness really like to listen to the jazz, blues, and classical music, while `energetic' people with a high degree of extraversion and agreeableness usually prefer rap, hip-hop, funk, and electronic music. In addition, according to a study by Chausson~\cite{Chausson}, comedy and fantasy movies are two interesting genres of movies for people who have the openness personality type, while conscientious persons are more interested in action movies, and romantic movies are the favourite types of movies for the group of neurotic people. 

This observation motivates RSs to integrate this important factor into consideration. As illustrated by Braunhofer et al.~\cite{DBLP:conf/enter/BraunhoferE015}, among all of the users' characteristics, personality is the most effective one. Personality can be acquired by RSs either explicitly by filling out a questionnaire or implicitly by observing users’ behavioural patterns. From the psychological point of view, personality is the main individual characteristic which can explain `patterns of thought, emotion, and behaviour'~\cite{Personality}. Moreover, one of the key properties of personality is that it is a stable behavioural pattern which humans tend to show irrespective  of their situations. There are several personality
traits models that can explain human behaviours, and among them, the Five Factor Model (FFM) or the Big Five Model has drawn more
attention in both psychology and computer science research. FFM is `the dominant paradigm in personality research, and one of the most influential models in all of psychology'~\cite{DBLP:journals/taffco/VinciarelliM14}. On the basis of the FFM, people’s personality types can be categorised into five main traits, briefly called OCEAN~\cite{McCrae}.
\begin{itemize}
    \item Openness to Experience: creative, open-minded, curious, reflective, and not conventional;
    \item Conscientiousness: preserving, organized, and responsible;
    
    \item Extroversion: assertive, amicable, outgoing, sociable, active, not reserved or shy;
    \item Agreeableness: cooperative, trusting, generous, helpful, nurturing, not aggressive or cold;
    \item Neuroticism (Emotional Stability): relaxed, self-confident, not moody, easily upset, or easily
stressed.
\end{itemize}
One of the main difficulties of personality-based RSs is how to identify the users' personality type. Basically, there are two different ways of assessing people's personality types which can be can be grouped as follows:
\begin{itemize}
    \item  Explicit techniques (filling out a questionnaire according to the chosen model);
    \item  Implicit techniques (observing users’ behavioural patterns and then applying a regression/classification model).
\end{itemize}

There are several questionnaire types based on the FFM model, for instance NEO-Personality-Inventory Revised (NEOPI-R, 240 items), in which the participants’ personality types are revealed after they answer several questions~\cite{Costa}. For instance, Hellriegel and Slocum~\cite{Cengage} proposed 25 questions for a questionnaire in order to calculate each of the five dimensions of the FFM model. The score for each of these five factors was computed as the average score on the five associated questions to that particular dimension. If we take the `Openness to Experience' dimension of FFM as an example, the questions were based on assessing the `imagination', `artistic interests', `liberalism', `adventurousness', and `intellect' properties of a person. For instance, users answered questions which started with `I see Myself as Someone Who...'. The answers were based on a five-point Likert scale, ranging from `strongly disagree' to `strongly agree'. Besides the explicit ways of identifying users' personality types, user's personality types can be acquired implicitly by analysing their activities, such as written review texts, and posts from online social networks~\cite{DBLP:conf/aaaiss/SchwartzEDKBKSSU13}, \cite{DBLP:conf/ijcai/BeraRM17}. For instance,  Quercia et al.~\cite{DBLP:conf/socialcom/QuerciaKSC11} analysed the user generated contents on  micro-blogs and showed the capability of these contents to represent the user's personality types. The authors used the M5 rules regression as a machine learning tool to predict the users' personality types. In another study, the users' records of Facebook activities were collected by Kosinski et al.~\cite{DBLP:conf/mm/SegalinCPKSSCL17}, \cite{Kosinski2013PrivateTA}, and then, logistic regression was applied to predict the peoples' personality type. Although personality detection with questionnaires may reveal a better understanding of a user’s personality, it is a tedious and time-consuming task, and thus, users may be unwilling to undertake it. In contrast, implicit personality detection models  users' digital
footprints, and their behaviours and actions are exploited to be analysed with no extra burden on them~\cite{DBLP:conf/recsys/AzariaH16}. 

This observation has inspired some RSs to incorporate users' personality types as the side information into their model in order to boost their performance. Furthermore, aggregating the users' personality type cannot only help RSs to better deal with the data sparsity problem but also provide users with suggestions even in a cold-start scenario~\cite{DBLP:conf/enter/BraunhoferE015}. For instance, Hu and Pu~\cite{DBLP:conf/recsys/HuP11} detected the users' personality types by providing a questionnaire to the users and asking them to answer the questions; their results showed the better performance than the purely collaborative filtering methods. TWIN is another example of integrating personality in a recommender system,  which calculates the user personality profiles on the basis of the NEO-Personality Inventory-Revised classification scheme (also known as the Big Five model)~\cite{DBLP:journals/jifs/RoshchinaCR15}.

Personality as a domain-independent concept
along with individuals can also be used in a wide range of domains, such as music~\cite{DBLP:journals/jair/MairesseWMM07}, movies, and books~\cite{Rentfrow}. Therefore, some RSs try to adopt this factor into their model to not only help users with a diverse set of items~\cite{DBLP:conf/chi/McNeeRK06} but also provide a better group recommendation~\cite{DBLP:conf/um/KompanB14}, \cite{DBLP:conf/recsys/Recio-GarciaJSD09} and improve the accuracy of RSs for music, movies, e-learning, and web searches~\cite{DBLP:conf/um/HuP10}, \cite{DBLP:conf/hais/PaivaCS17}, \cite{DBLP:reference/sp/TkalcicC15}. In addition, the users' personality types can be seen as an important context that can be used in a Point 
Of Interest (POI) recommendation as well~\cite{DBLP:conf/ecweb/BraunhoferER14}. South Tyrol Suggests (STS) is an example of a context-aware POI recommender system which detects the users' personality type explicitly~\cite{DBLP:conf/ecweb/BraunhoferER14}, \cite{DBLP:conf/hci/BraunhoferEGR14}. As personality is a domain-independent factor, it can be used in cross-domain recommenders as well.  Cantador et al.~\cite{DBLP:conf/dmrs/CantadorF14} investigated the impact of users' personality types on different domains. According to another study ~\cite{DBLP:reference/sp/TkalcicC15}, `salsa-music lovers are dissimilar to science-fiction-books lovers or news-tv-show lovers are similar to mystery-books lovers'.

Thus far, we have discussed the benefits of incorporating personality factor into a recommender system. Beside these advantages, personality can offer a diverse set of items for users to discover unexpected items~\cite{DBLP:journals/tkde/AdomaviciusK12},  \cite{DBLP:journals/toit/HurleyZ11}. Brynjolfsson et al.~\cite{DBLP:journals/mansci/BrynjolfssonHS11} investigated the effect of different personality types on selecting diverse items. To do so, a survey with 181 subjects was conducted for users to choose their favourite movies. The individuals also needed to answer the questions to measure their personality types. Then, the researchers investigated the correlation between diversity and personality and realised that `reactive, excited and nervous persons (high in Neuroticism)' like to select movies from diverse directors, while `suspicious/antagonistic users (low in Agreeableness)`
prefer movies from different countries.

\textbf{Basic Deep Neural Network Models for CF.} In the past few years, numerous deep recommender systems have been proposed to model complex user-item relations~\cite{DBLP:journals/csur/ZhangYST19}. In particular, dealing with different types of data such as textual data (e.g. comments and tweets)~\cite{DBLP:conf/ijcai/GongZ16}, \cite{DBLP:journals/corr/ZhengNY17}, \cite{DBLP:conf/ijcnn/KhatamiNB0NZ20}, \cite{DBLP:conf/intellisys/SchiliroBM20}, image data (e.g. shared photographs and item images) has become a challenging task without the use of neural network paradigms. Therefore, most of the existing RSs which are based on the linear methods have become less attractive; consequently, RSs cannot take advantage of more general and abstract representations. As presented by Zhang et al.~\cite{DBLP:journals/csur/ZhangYST19}, the model proposed by Salakhutdinov et al.~\cite{DBLP:conf/icml/SalakhutdinovMH07} is the first neural network-based RS which uses Restricted Boltzmann Machine (RBM) as the core. Neural Network Matrix Factorisation (NNMF)~\cite{DBLP:journals/corr/DziugaiteR15} and Neural Collaborative Filtering (NCF)~\cite{DBLP:conf/wsdm/JingS17} are two representative works in CF which use Multiyear Perceptron (MLP). DeepFM is another neural network-based approach in which a deep component and a factorised matrix are jointly trained~\cite{DBLP:conf/ijcai/GuoTYLH17}. While the use of MLP is a very straightforward
solution, the other neural network techniques, such as CNNs and Auto Encoders, may be more expressive and highly efficient at identifying the main features of users and items. AutoRec~\cite{DBLP:conf/www/SedhainMSX15} and its extension, CFN~\cite{DenoisingAutoEncoders}, are examples of AutoEncoder-based models which try to reconstruct their input data in the output layer. The main difference between CFN and AutoRec is that CFN resorts to the additional information about users and items in order to mitigate the data sparsity problem, boost the prediction accuracy, and improve the model robustness. CNN is another viable deep neural network which is good at capturing local context. ONCF~\cite{DBLP:conf/ijcai/0001DWTTC18}, an outer product-based neural collaborative filtering (ONCF) framework, is an improved version of NCF. ONCF introduces the interaction map term, which uses an outer product operation in order to 
capture the pairwise correlations between the embedding dimensions. Next, the result of the interaction map is passed to the CNN in order to learn the high-order correlations among the embedding dimensions. Besides the mentioned models, Graph Convolutional
Networks (GCNs) is one of the most prominent deep learning models with the core principle of iteratively learning to aggregate feature information from local graph neighbourhoods~\cite{DBLP:journals/corr/abs-1806-01973,DREAM,GraphQLL,GraphClus,ProcessAtlas,GOLAP}. 

\textbf{Advanced Deep Neural Network Models for CF.} In recent years, the advancement of attention mechanisms for focusing more on the main features of the inputs has become more attractive in a wide range of domains, such as computer vision~\cite{DBLP:journals/corr/BaMK14}, natural language processing~\cite{DBLP:journals/corr/LuongPM15}, \cite{DBLP:conf/nips/VaswaniSPUJGKP17} and speech recognition~\cite{DBLP:journals/corr/ChorowskiBCB14}, \cite{DBLP:journals/corr/ChorowskiBSCB15}. In this context, among all types of deep learning techniques, CNNs and RNNs are the two commonly adopted architectures which can be used  with attention mechanisms~\cite{attaentionRah}. For example, Seo et al.~\cite{DBLP:conf/recsys/SeoHYL17} used a CNN owing to its significant capability to capture complex features of users and items properties by aggregating review texts. Moreover, the authors used dual local and global attention in their model to provide a better-learned representation of users
and items. ACF~\cite{DBLP:conf/sigir/ChenZ0NLC17} is another attention-based CF model which consists of a two-level attention mechanism: the component-level and item-level attention modules which can automatically calculate the scores to the two levels of feedback in a distant supervised manner. Attentive Contextual Denoising Autoencoder (ACDA)~\cite{DBLP:conf/ictir/JhambEF18} is an attention-based CF model in which the context of user-item interactions is incorporated by using attention mechanisms. ACDA is  an augmented personalised recommendation model based
on the denoising autoencoder. CML is another attention incorporated model for CF proposed by Tay et al.~\cite{DBLP:conf/www/TayTH18}. The main principle of CML is its operation in the
metric space, as it tries to minimise the distance between each user-item interaction in the Euclidean space. Here, an attention mechanism is applied to generate the latent relation vectors over an augmented memory module.

\subsection{Hybrid-Based Recommendations}
This category of RSs is a combination of the content-based and collaborative filtering-based RSs. Hybrid-based RSs take advantage of the benefits of both methods, simultaneously, in order to increase the accuracy of RSs. According to~\cite{DBLP:journals/tkde/AdomaviciusT05}, there are three different ways in which these methods can be combined. The first is to use the models that use collaborative filtering and content-based methods separately and then combine their predictions. The second type of hybrid models includes methods that incorporate either content-based or collaborative filtering characteristics into another model in order to make a hybrid recommendation. The third class of hybrid approaches combines both collaborative filtering and content-based recommendations into one single model in order to benefit from the advantages of both the models as a unified recommendation system. Next, we will review these three classes of hybrid recommendations models in detail.

\subsubsection{Combining Separate Recommenders} 

The very first step of approaches in this class of hybrid recommender systems is to separately construct collaborative and content-based recommender systems. Then, the final result of an RS can be generated either by combining the ratings from both individual recommenders~\cite{hybrid} or by selecting one of the recommenders which generates more accurate results~\cite{DBLP:journals/air/Pazzani99}. For example, a hybrid recommender system proposed by Tran et al.~\cite{Tran2000HybridRS} switches between two different types of recommendation engines by using the Interactive Interface Agent (IIA). IIA is responsible for choosing the more suitable subsystem for service according to the user information and behaviours. 

\subsubsection{Adding One Individual Recommender to Another One}
In the previous section, we saw that content-based RSs are successful in discovering items similar to those that a user was interested in before, in terms of their descriptions. Although, content-based RSs have shown a significant development in the mentioned task, CF-based studies may perform better because of several reasons. First, there is no need for CFs to analyse the items' descriptions, which may need a complex Natural Processing Technique (NLP) in order to provide an accurate prediction. Second, CF-based systems can generate a set of suggestions that are relevant for a user, but without the information from the user’s profile~\cite{DBLP:conf/sigir/HerlockerKBR99},~\cite{DBLP:journals/cacm/GoldbergNOT92}. On the basis of these observations, integrating users and items characteristics in a CF-based model is a more common and successful way to make a hybrid recommendation, which takes the advantages of both models and reduces their limitations. 

Fab is one of the first examples in this category of hybrid methods, which detects similar users who have similar website preferences by exploiting the content of users' profiles~\cite{DBLP:journals/cacm/BalabanovicS97}. Furthermore, a PTV system uses content-based techniques to analyse information about TV programs, and then, finds users who have the same preferences on TV shows. Next, it uses both content-based and collaborative filtering-based algorithms for making recommendations~\cite{DBLP:conf/aaai/CotterS00}. LIBRA, a book recommendation engine, is another example of a hybrid-based recommender system. LIBRA collects information from Amazon.com\footnote{https://www.amazon.com/} in order to make suggestions. Good et al.~\cite{DBLP:conf/aaai/GoodSKBSHR99} proposed a hybrid recommender system integrating multiple filter bots, which are learning agents, in order to help the collaborative filtering community make better recommendations. In addition, Content-Boosted Collaborative Filtering (CBCF)~\cite{DBLP:conf/aaai/MelvilleMN02} uses the contents of the items of interest to overcome the data sparsity problem, which by itself is a challenging problem for CF models.

Incorporating items' characteristics also can help this model to convert a sparse rating matrix into a full rating matrix. This solid framework can generate more accurate recommendations than the purely content-based or CF-based systems. Most of the mentioned hybrid models take advantage of using content-based approaches in CF systems. While there are other attempts in the literature to use the dimensionality reduction technique, latent semantic
indexing (LSI) is used to determine a set of neighbourhoods by analysing user profiles~\cite{hybrid}.   

\begin{table}[b!]
\centering 
\begin{center}
\begin{tabular}{||m{2cm} | m{7em} |m{7em} | m{7em} |m{7em} || } 
\hline
\small{ RS} & \small{Input} &  \small{Working mechanism} & \small{Pros} & \small{Cons} \\  
\hline 
\small{ Content-based} & \small{User characteristics , Item descriptions  } & \small{Recommend items similar to those that a user has liked before} & \small {Simple and straightforward, can deal with cold-start problem} &\small{Over-specialisation, limited content analysis} \\
\hline 
\small{ Collaborative Filtering} &\small {User-item rating matrix} & \small {Recommend items similar to those that a user's neighbours have liked} & \small {Can recommend serendipity items, effective and simple } & \small {Suffer from data sparsity and cold-start problems}\\
\hline 
 Hybrid  & \small{Item and user characteristics and their interactions }& \small{ Combines two or more types of recommendation
strategies} & \small{Benefits from the advantages of both content and collaborative -based approaches} & \small{Relatively complex to develop}\\
\hline 
\end{tabular}
\caption{Comparison of different models of general recommenders.}
\label{ch2:tableGeneralRSComparison}
\end {center}
\end{table}

\subsubsection{Developing a Single Unifying Recommendation Model}
Recently, among all the types of hybrid RSs, approaches that fall into this category have attracted more attention from researchers. According to~\cite{DBLP:journals/kbs/BobadillaOHG13}, hybrid-based recommenders are usually based on `probabilistic methods such as genetic algorithms~\cite{4680341}, \cite{DBLP:conf/icete/HoFY07}, fuzzy genetic~\cite{DBLP:journals/eswa/Al-ShamriB08}, neural networks~\cite{DBLP:conf/isda/ChristakouS05}, \cite{DBLP:conf/ah/LeeCW02}, \cite{DBLP:conf/fgcn/RenHGXW08}, Bayesian networks~\cite{DBLP:journals/ijar/CamposFHR10}, clustering~\cite{10.1016/j.eswa.2011.08.020} and
latent features~\cite{5136724}'. For instance, a Bayesian preference model was proposed by Ansari et al.~\cite{article22}, which made use of four types of information, such as a person's expressed preferences, preferences of other consumers and experts, items' characteristics, and individuals' characteristics. Their results represented an improvement in prediction ability compared to the RSs modelled only on collaborative filtering or content-based approaches. 
In another work, the authors used all the available information, such as users' ratings, users' features, and items' features in a single probabilistic unified framework to construct a hybrid recommendation system, which could address the cold-start, scalability, and sparsity problems~\cite{article21}.

\section{Sequential Recommenders}
\label{ch2:SequentialRecommendersq}
With the rapid growth of online platforms, many companies have started building their e-commerce websites and smartphone applications to encourage their customers to keep interacting with products and services. These platforms can be extremely helpful for users to narrow down their options, while a huge amount of interaction information can be generated. For instance, around 62 million trips with Uber\footnote{https://www.uber.com/au/en/} were recorded in July 2016~\cite{DBLP:conf/ijcai/YingZZLXXX018}. By analysing the huge amount of information about users' historical sequential behaviours, Sequential Recommender Systems (SRSs) can  predict the next interacted items. This can help users by easing their decision-making processes as well as increasing business profits for companies.

\begin{figure*}[h!]
\centering
\includegraphics [width=\textwidth, scale=1]{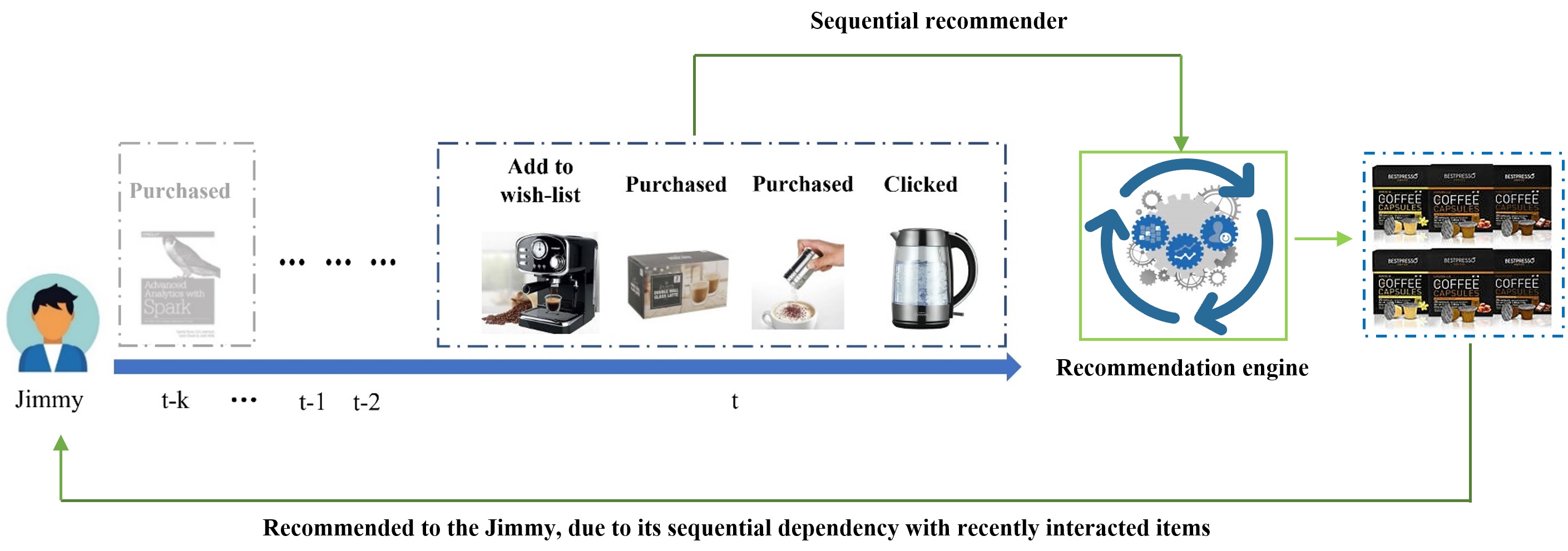}
\caption{General framework of a sequential recommender system.}
\label{ch2: figureSequentialRecommenders}
\end{figure*}

All the aforementioned traditional RSs mostly treat the recommendations task as the matrix completion problem, in which a sparse user-item rating matrix is given as an input. As already discussed in the previous section, this type of RSs (general recommenders) has been generally well-studied and different techniques have been applied in this setting. The main job of general recommenders is to predict the missing ratings value, which  can represent the general users' preferences by itself (i.e. long-term users' preferences). While traditional approaches are good at modelling the users' general taste, they usually ignore the short-term users' preferences, which
results in missing interest drifts of individual users over time~\cite{DBLP:conf/ismir/MooreCTJ13}. We present a general framework of a sequential recommender system in Figure~\ref{ch2: figureSequentialRecommenders}. 

As is clear from the Figure~\ref{ch2: figureSequentialRecommenders}, in a sequential recommender, the most recent user's activities are taken into account. Take Jimmy's shopping basket for example: a set of his current actions at time $t$ is passed to a sequential recommender. According to his transactional history at time $t$, Jimmy first looked for a coffee machine and added one into his wishlist for further shopping. Then, he purchased a set of double-walled glasses and cinnamon powder. Next, he made his final action at time $t$, by clicking on an electrical kettle. By analysing Jimmy's current behaviours, a sequential recommender system can realise that Jimmy is interested in coffee-related items. Hence, according to a sequential recommender, a box of coffee capsule may be a good suggestion for him. 

In contrast to the traditional RSs, Sequential Recommender Systems (SRSs) have emerged as a new line of RSs which differ from traditional RSs in terms of their inputs, outputs, computational tasks, and abstract problem characterisation~\cite{DBLP:journals/corr/abs-1802-08452}. Here, 
a set of purchased items in one basket can be considered a session, which can be used as an input. In contrast, SRSs take a sequence of user-item interactions to predict a set of next interacted user-item interactions, which is the SRSs' outputs. In general, the main purpose of SRSs is to model the sequential dependencies among user-item interaction sequences and discover the hidden information embedded from one sequence into another sequence. SRSs have attracted the attention of researchers on different applications such as POI recommendation~\cite{DBLP:conf/ijcai/ChengYLK13}, \cite{DBLP:conf/ijcai/FengLZCCY15}, music recommendation ~\cite{Hariri}, \cite{DBLP:conf/recsys/WuLCHLCH13}, and browsing
recommendation~\cite{DBLP:conf/bigdataconf/ZhangZLCZM15}.

Despite the significant improvement of SRSs from their emergence, there have been just a few attempts to categorise the existing studies in SRSs~\cite{DBLP:conf/www/QuadranaJC19}. Therefore, in this thesis, we have divided the current studies in SRSs from different points of view. As the basic unit of a sequential recommender system is a session, we will first start categorising SRSs regarding to the sessions' complex characteristics ~\cite{DBLP:journals/corr/abs-1902-04864}. Then, a summary of SRSs with respect to their adopted models will be given in Section~\ref{ch2:SRScategorization}. Next, we will provide a division based on the SRSs' input types in Section~\ref{ch2:SectionInputType}.   

\subsection{SRS Categorisation in terms of Session Characteristics and Challenges}

In this section, we aim to introduce the different characteristics of a session, which may bring various challenges in a sequential recommender system. We will discuss each of the related issues with the possible solutions, separately. As stated by Wang et al.~\cite{DBLP:journals/corr/abs-1902-04864}, users' shopping behaviours may not follow a certain pattern and may have complex characteristics which need to be solved. Therefore, we can divide the current studies on SRSs with respect to the main characteristics of each session of user-item interactions, such as dealing with the noisy items in a session, capturing a session's purpose, handling the items' order in a session, and dealing with the long user and item interaction session. In the following sections, we will comprehensively review all of them. 

\subsubsection { Dealing with Noisy Sessions} 
Usually, people's decision-making process is not very certain. Take a customer shopping basket for example: she/he may have a shopping list and collect all of the items on her/his basket, but at the last second of the shopping, she/he may make an instant decision and add several other items to her/his basket, which are totally different from what she/he has purchased already. The last-minute purchase may generate interference, which makes dependency modelling more complicated. While it is an important problem that needs to be solved, it has not been sufficiently studied as yet. Inspiring by the great capability of the attention mechanism in context learning, ATEM was proposed by Wang et al.~\cite{DBLP:conf/aaai/WangHCHL018}. ATEM applies the attention mechanism to discover the transactional context and pays more attention to the contextual items which are highly related to the next interested items and downgrades those which are not too relevant. An attention layer in ATEM is responsible for automatically calculating a different score for the contribution of different items. Thus, ATEM can avoid interference from irrelevant items in a long context. Another solution for dealing with a noisy session is memory networks. Because of the strong capability of external
memory networks (EMN)~\cite{DBLP:journals/corr/GravesWD14}, \cite{DBLP:journals/nature/GravesWRHDGCGRA16}, \cite{DBLP:journals/corr/WestonCB14} for storing historical
hidden states, they are used in a wide range of tasks such as question answering (QA)~\cite{DBLP:conf/icml/KumarIOIBGZPS16}, natural language transduction
(NLT)~\cite{DBLP:journals/corr/GrefenstetteHSB15}, and knowledge tracking (KT)~\cite{DBLP:conf/www/ZhangSKY17}. Memory-Augmented
Neural Network (MANN) was proposed by Chen et al.~\cite{DBLP:conf/wsdm/ChenXZT0QZ18} to use EMN in their system in order to make personalised recommendation scenarios. By doing so, MANN could use a memory matrix for storing the previous states instead of merging them. In particular, they proposed two versions of their
framework, namely item-level and feature-level Recommender system with external User Memory networks (RUMs), to better model
users' behaviours and improve the recommendation performance, accordingly.

\subsubsection{Capturing a Session's Purpose}
People usually start buying items with a specific intention in their mind. Take online shopping for example: a user may want to make a pizza. Considering this purpose, she/he may take successive actions, such as purchasing a pizza base, clicking on a pizza sauce, and  browsing for a light cheese. During these actions, the user may interact with some unrelated items because of curiosity, and thus, if sequential recommenders neglect to take the user’s main intention into account,  false dependencies will be modelled, leading to inappropriate suggestions. Therefore, it is an important job for a sequential recommender to capture the user’s main purpose. Neural Attentive Recommendation Machine (NARM) is a sequential recommender in which the user’s main purpose in the current session is emphasised~\cite{DBLP:conf/cikm/LiRCRLM17}. NARM consists of two encoders, namely a global encoder and a local encoder. The global encoder is responsible for summarising the whole sequential behaviour, while  the user’s main purpose in the current session is captured by the local encoder. Both the encoders use RNN with GRU as the basic component, and the recommendation scores for each potential item are calculated by using
a bilinear matching scheme. STAMP, proposed by Liu et al.~\cite{DBLP:conf/kdd/LiuZMZ18}, is another effective solution for showing the importance of taking users' purposes into account. These researches showed that the user’s intention may be more in response to the current action. STAMP proposes a novel short-term attention/memory priority model, which considers the users’ recent preferences from the last-clicks. Moreover, STAMP introduces an action priority mechanism to pay attention to the both long- and short-term user interests. As stated by the authors, STAMP is simpler and faster than NARM, which makes it more applicable in real world scenarios. Mixture-channel purpose routing networks (MCPRNs) is another work in this setting, proposed by Wang et al.~\cite{DBLP:conf/ijcai/Wang0WSOC19}, which assumes that a session may consist of several purposes. Recently, ASLI~ \cite{DBLP:conf/www/TanjimSBHHM20} was proposed as a sequential recommender system in which self-attention is used to identify similar items in a user-item interaction session in order to capture the users’ hidden intents. MCPRNs introduce a purpose routing network to discover the reason behind the appearance of each item and then, assign each item to its corresponding channels. Next, all the channels are integrated into a multi-purpose session to rank the candidate items. While capturing the users' main intention has shown its potential in SRSs, it is still in its early stages.

\subsubsection{Handling Items' Order in a Session} 
In the real world, it is more probable to purchase a set of products with no ordered assumption among them in the shopping basket. As stated by~\cite{DBLP:journals/corr/abs-2001-04830}, most of the grocery shoppers may just add products into their cart by random selection, which does not follow any strict order between them. For instance, it does not really matter whether users put bread or butter first into their shopping a bottle of jam. Moreover, they may suddenly be attracted by some items and then pick them up and add them into their basket. Therefore, it may be possible for a sequence of purchased items to relax the ordering assumption in order to have flexible user-item interaction sequences. While it is a simple and efficient observation, little attention has been paid to this effective fact by the current studies. The existing techniques on SRSs such as RNN, MC, SPM, and factorisation machines may not be good at modelling unordered items in a sequence, as point-wise dependency relationships are the key principle behind these models. However, although pattern/rule-based approaches are simple and effective techniques, they are suitable options for modelling a flexible session. The main intuition of these models is to discover frequent patterns and then use them for making further recommendations for other users. For instance, a general-purpose Semantic Web Service-based framework was proposed by Abel et al.~\cite{DBLP:journals/tlt/AbelBCHKV10} for e-learning systems. Their model based on a recommender architecture can allow for a flexible combination of components. 

\subsubsection{Dealing with Long User and Item Interaction Session} 
A long user-item interaction sequence is another important challenge of SRSs which needs to be tackled. This problem refers to a situation where there are too many items in a session, which may cause some difficulties in modelling. A higher-order Markov chain can be one intuitive choice to address this issue~\cite{DBLP:conf/icdm/HeM16}, which can model sequential dependencies among items in a long session. In addition, various types of RNNs have shown to work well with sequential data~\cite{DBLP:journals/corr/HidasiKBT15}. In contrast to RNN-based models, which may have difficulties to capture long-term dependencies, improved architectures such as Gated Recurrent Unit (GRU)~\cite{DBLP:journals/corr/ChoMGBSB14} and Long-Short Term Memory (LSTM)~\cite{DBLP:conf/nips/SutskeverVL14} have been introduced to better model long-range sequential data.

Another critical issue with long-range sequential data is how to model the dependencies between items which are placed far from each other in a session. In this context, a class of RNN approaches might not be a good answer, as they mainly assume that the nearby items in a session are highly dependent, which might not comply with the nature of this problem. Therefore, more robust techniques, such as mixture model~\cite{DBLP:conf/www/TangBJCBXC19}, memory network~\cite{DBLP:conf/wise/HuHSN19}, and attention mechanism~\cite{DBLP:conf/ijcai/YingZZLXXX018}, need to be used in order to model a long-range dependency. Mixture models try to combine different models that excel at solving their associated problems. M3~\cite{DBLP:conf/www/TangBJCBXC19} as a typical example of a mixture model applies different kinds of encoders to better deal with long sessions and provides effective recommendations. M3 can focus on different temporal dynamics and ranges depending on the applications' context. A memory network has shown to be effective in memorising long-term data characteristics~\cite{DBLP:conf/wsdm/ChenXZT0QZ18}. Hence, based on the reported recent progress on the memory mechanism of neural networks~\cite{DBLP:journals/corr/PerezL16}, \cite{DBLP:conf/emnlp/MillerFDKBW16}, \cite{DBLP:journals/corr/WestonCB14}, an augmented RNN-based
sequential recommender with external memories was proposed by Huang et al.~\cite{DBLP:conf/sigir/HuangZDWC18}. The authors combined RNN with Key-Value Memory Network (KV-MN) in order to build a hybrid model which can capture sequential users' preferences and attribute-level users' preferences, respectively. 

Besides the mentioned robust techniques, an attention mechanism with considerable capability to focus on important items is another good solution for long-range dependency modelling. For instance, SHAN~\cite{DBLP:conf/ijcai/YingZZLXXX018} as a two-layer hierarchical
attention-based structure tries to automatically assign different  weights to the items in a long sequence to capture the dynamic property of the users' preferences. MEANS~\cite{DBLP:conf/wise/HuHSN19}, which consists of a three-component memory module, an attention network, and a predictor layer is another example of the works in this category. While the memory module is responsible for storing the recent sessions, an attention mechanism is applied to learn the long-range dependency, and finally, a ranked list of items is provided by  MEANS at the predictor layer for the next-item recommendation task.

\subsection{SRS Categorisation in terms of Model Structure}
\label{ch2:SRScategorization}
In the previous section, we discussed issues and solutions for sequential recommender systems with respect to a session's characteristics. However, because of the popularity of different complex models and algorithms integrated in sequential recommendations, in this section, we have decided to divide the current SRSs in terms of their model structures. The reason behind this division is that algorithm adoption is a major step in a recommender system, where correct selection can  play an important role in modelling the users' behaviours and generate more accurate recommendations, accordingly.

\subsubsection{Sequential Pattern Mining} 
Sequential pattern mining (SPM)~\cite{DBLP:conf/icde/AgrawalS95}, as a powerful data mining technique, tries to discover repeated temporal patterns among users and is an intuitive solution for the next-items recommendation task. The main idea behind SPM is that if most of the users frequently use item \textit{x} after item \textit{y}, it is better to recommend item \textit{y} for a user who recently used item \textit{x}. SPM has shown significant improvement in various set of domains such as e-commerce~\cite{DBLP:journals/ecra/HuangH09},
web browsing~\cite{DBLP:journals/csur/ZobelM06}, and TV programs scheduling~\cite{DBLP:journals/mms/PyoKK13}. This observation has encouraged researchers to benefit from adopting SPM in their systems. Most of the current SPM-based approaches assume that there is no difference between sessions. In~\cite{10.1007/978-3-642-15766-0_25}, for instance, the proposed method mined sequential web
access patterns by collecting the users' log data to generate a pattern tree and then make recommendations, accordingly. Weighted sequential pattern mining is more efficient than simple SPM, as it can differentiate the level of importance among sequences~\cite{DBLP:journals/kbs/Yun08}. Furthermore, Capelle et al.~\cite{DBLP:conf/ideal/CapelleMB02} proposed the use of two constraints, namely a frequency constraint and a similarity constraint, for comparing the similarity of two patterns. Then, a threshold was defined in this study to consider two patterns as similar patterns only if the calculated similarity score between them was above a pre-defined threshold. Several years later, Yep et al.~\cite{DBLP:conf/dasfaa/YapLY12} introduced a personalised sequential pattern mining-based recommendation framework to emphasise the important sequences. In particular, user-specific sequential pattern mining was applied in their method to provide a personalised recommendation. To do so, the authors differentiated the sequences according to the available knowledge about different users.

Unlike the aforementioned models which allow the sequence data to be flexible, there are other studies which follow a strict order for items in a sequence of user-item interactions. The sequential pattern mining technique is a popular model that is commonly used by approaches in this domain. In education, for example, mining students’ usage information to find out the interesting information that can help a teacher to personalise the provided courses for her/his students. The most used path by the students is discovered by the use of sequential pattern mining and then is automatically recommended to the new students~\cite{Romero}. Web server log data are another source of information that can be mined to extract useful information. Niranjan et al.~\cite{DBLP:conf/ict/NiranjanSK10} proposed the use of this information to construct closed sequential patterns, and then, based on these patterns, pattern trees were built to make a personalised web recommendation.

\subsubsection{Pattern/Rule-Based Approaches} 
According to Wang et al.~\cite{DBLP:journals/corr/abs-1902-04864} `Pattern/rule-based RS mainly contain three stages: frequent pattern mining, session matching, and item recommendations.' The main idea behind pattern/rule-based approaches is to identify the frequent patterns by analysing the users and items interaction data to be used for further recommendations. For example, a chips and a coke are more likely to be purchased together, and this can be treated as a usual pattern in a shopping basket. In these approaches, there is no difference in which item should be added to a basket first; the pattern is the same. Hence, pattern/rule based approaches can be a good choice for dealing with unordered data. Apriori~\cite{DBLP:books/sp/fpm14/Aggarwal14} and FP-Tree~\cite{DBLP:conf/fskd/JiD07} can be two good examples of approaches in this domain. A rule-based framework proposed by Mobasher et al.~\cite{DBLP:conf/widm/MobasherDLN01}, clickstream data is captured from Web server logs to make a scalable recommender framework. The authors discuss that their proposed model can overcome some drawbacks of current rule-based RSs, low coverage for instance, which can result in a performance degradation. Moreover, LIN et al.~\cite{DBLP:journals/datamine/LinAR02} show the performance improvement of customised CF over pure CF and association-rule based models when the key task is to use  association rule mining as an underlying technology.

\subsubsection{Markov Chains Models}
Markov chains model have been known as a straightforward technique to model the transitions over user-item interactions in a sequence. As discussed in~\cite{DBLP:journals/tkde/CaoOY12}, MC technique is good at behaviour analysis due to its strong assumption that the current state is highly dependent only on the previous states. Based on the specific technique used, Markov chain-based RSs can be divided into basic Markov chain-based approaches and latent Markov embedding-based approaches. The former refers to the situation where the transition probability of every state is computed only based on the previous state. Following that, auxiliary information, e.g. context features, can be used to enrich the Markov chain model and improve the recommendation accuracy~\cite{DBLP:conf/pkdd/LeFL16}. Moreover, in one work~\cite{DBLP:conf/webi/ZhangN07}, the authors combine a mixed order Markov chains with the web content information to have a more accurate web-recommendation. They form a hybrid web-recommendation system in which Markov-based methods are used in order to model users’ Web browsing behaviour. Then, they combine it with content-based filtering models to extract contents from the links that users have visited online.
While the basic Markov chain-based approaches can only model the simple and basic dependencies between the states, the latent Markov embedding-based approaches are able to capture the complex relationships between the states as the transition probabilities are calculated based on the Euclidean distance of embedded Markov chains~\cite{DBLP:conf/kdd/ChenMTJ12}.

\subsubsection{Convolutional Neural Network-based Models} 
Recently, Convolutional Neural Networks (CNN) have shown a promising result in image-related tasks~\cite{DBLP:conf/cvpr/KarpathyTSLSF14,NabiFeature}, \cite{DBLP:conf/nips/KrizhevskySH12} and natural language processing domain~\cite{DBLP:conf/emnlp/Kim14}. Inspired by the success of
CNN in capturing local features, Caser~\cite{DBLP:journals/corr/abs-1809-07426,AdaptiveRule4} is proposed as the first CNN-based sequential recommender system. Caser stores all previous user-item interactions in a matrix and treats this matrix as an `image', which is fed into a CNN network. various convolutional filters are applied in Caser, such as horizontal filters and vertical filters to well extract local features at a different level from the matrix of the embedded previous items. Another CNN-based study is proposed by Tuan and Phuong~\cite{DBLP:conf/recsys/TuanP17}, which exploit the content of items into session clicks to generate personalised recommendations. They believe that current SRSs may not be able to make any recommendations when there are no available information regarding users' profiles or their past activities. Furthermore, they introduce a challenging problem called, add-to-cart prediction task, which may be more complicated than the next click prediction problem. To overcome the mentioned issues, Tuan and Phuong propose a 3D CNN model for SRS, which leverages the incorporating of the session associated content features such as item descriptions and item categories for accurate predictions. A 3D CNN input is able to capture spatio-temporal pattern which makes recommendation systems more robust. While Caser as a CNN-based recommender comparably performs well over the popular RNN-based model in the top-N sequential recommendation problems, it may lose some useful information due to using max pooling operations. Therefore, an efficient and highly effective
convolutional generative model is presented by Yuan et al.~\cite{DBLP:conf/wsdm/YuanKAJ019} for top-N item recommendation task in SRSs. In contrast to the basic filters in CNN, the dilated convolution filters are applied over an embedded matrix in order to include all useful information. The embedded matrix is padded with zero in order to make it larger than its original length to avoid losing any information, specifically in a long-range sequence data.

\subsubsection{Recurrent Neural Network-based Algorithms}
Recurrent Neural Networks (RNN), have been known as one of the widely adopted deep learning techniques in SRSs. One of the benefits of RNN is its great success in sequence modelling, which has been widely demonstrated in the field of natural language processing (NLP)~\cite{fang2019deep}. Following this, a novel RNN-based model is proposed to capture users' sequential behaviours for click prediction~\cite{DBLP:conf/aaai/ZhangDXFWBWL14}. While RNN-based sequential recommender systems have been reached to a significant results, they may not be able to model dependencies in
a longer sequence. Therefore, different from traditional RNN, several works have been introduced to modify classic RNN in order to better capture the whole historical user-item interaction sequences~\cite{HidasiKBT15}, \cite{DBLP:conf/um/DevooghtB17}. For instance, Wu et al.~\cite{DBLP:conf/wsdm/WuABSJ17} applies Long Short-Term Memory (LSTM)~\cite{lstm}, which is an updated version of RNN, to capture temporal dynamics of user profiles and movie attributes. Tan et al.~\cite{DBLP:conf/recsys/TanXL16} propose to use GRU, founded by~\cite{DBLP:journals/corr/HidasiKBT15}, and use data augmentation techniques to take temporal shifts in user behaviour.

\subsubsection{Graph Neural Network-based Models} 
With the rapid growth of GNN, an increasing interest is paid to GNN by the researchers in many applications such as natural language processing, computer vision, and recommender systems in particular~\cite{DBLP:conf/ijcai/LiDL18}, \cite{DBLP:conf/iccv/LiTLJUF17}. We et al.~\cite{DBLP:conf/aaai/WuT0WXT19} is one of the pioneering in adopting GNN in their proposed recommender system, SR-GNN. First, a directed graph is built with the help of transactional sequence data, and next the users and items embeddings are learned on the graph. This may help SR-GNN to benefit from more complex relations over the whole graph. While GNN as a strong deep learning technique has shown to be effective in recommender systems and generate promising results, a less attention is paid by the recommendations community and more investigation are required.

\subsubsection{Attention Mechanism-based Models}
Lately, researchers have employed attention mechanism due to its powerful capability in focusing on selective parts~\cite{DBLP:journals/corr/BahdanauCB14}. Attention mechanism is initially designed by Bahdanau et al.~\cite{DBLP:journals/corr/BahdanauCB14}, which is applied in machine translation task in order to emphasise on some selective parts of inputs with high impacts on the outputs sentence. The basic version of attention, named vanilla attention is used as a decoder in an RNN network. While vanilla attention which has been widely used in sequential recommender systems~\cite{DBLP:conf/cikm/LiRCRLM17}, has shown great success. An improved version of attention named, self-attention mechanism attract more attention as it can perform well without using either a recurrent or convolutional network~\cite{DBLP:conf/ijcai/XuZLSXZFZ19}. Self-attention network is originally introduced in transformer for neural machine translation task~\cite{DBLP:conf/nips/VaswaniSPUJGKP17}, but it is shown promising results in an RSs' community as well. An attention mechanism is successful to differentiate the contributions of different items in a session. In this regard, a two-layer hierarchical design called, SHAN, is proposed by Ying et al.~\cite{DBLP:conf/ijcai/YingZZLXXX018} as an attention-based SRS to focus on more relevant items in a session. Moreover, incorporating an attention network presents the superior performance in context learning in Wang et.al~\cite{DBLP:conf/aaai/WangHCHL018}. Attention mechanism also can be a good solution for dealing with a noisy session, as it is able to discover the most important items in  a session, and discard the irrelevant ones.

\begin{table}[t!]
  \includegraphics[width=\textwidth]{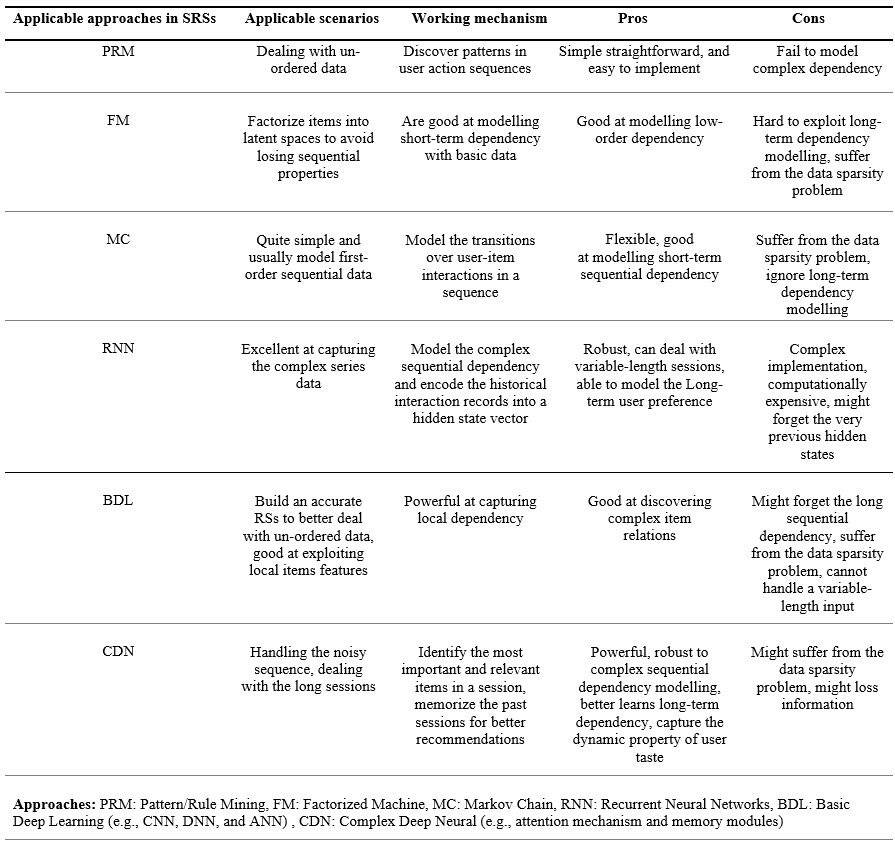}
  \caption{Comparison of approaches used by sequential recommenders.}
  \label{ComparisonSRSs}
\end{table}
\subsubsection{Factorised Machine-Based Models}  
Matrix factorisation or tensor factorisation are the most widely used techniques in recommendation systems, which map users and items into latent spaces~\cite{DBLP:journals/datamine/HidasiT16}. In sequential recommender systems, matrix or tensor is constituted of interactions rather than the ratings~\cite{DBLP:journals/corr/abs-2001-04830}. For instance, FPMC is built based on the combination of matrix factorisation and Markov chains to predict the next action~\cite{DBLP:conf/www/RendleFS10}. Another factorised model is FPMC-LR, which is an extended version of FPMC and proposed by Cheng et al.~\cite{DBLP:conf/ijcai/ChengYLK13}. The personalised Markov chain is applied in FPMC-LR for personalised point-of-interest (POI) recommendation task. FPMC-LR pays more attention to the users' movements around local region to provide users with more accurate suggestions. More importantly, FPMC-LR not only can offer locations which a user has not seen before, but also it can reduce the computation cost and ignore the
noisy information to increase recommendation performance.


\subsection{SRSs Categorisation In terms of Inputs}
\label{ch2:SectionInputType}
Sequential recommender systems may have different types of inputs including only user-item interaction, single or multiple user behaviour data, repeat purchasing behaviour, and incorporating side information. Therefore, in this section, we aim to categorise the existing sequential recommender systems in terms of their input. 

\subsubsection{Only User-item Interaction Session}
This is a very basic type of input data in sequential recommender systems. A session which records multiple actions is passed into a sequential recommender for further processing. To so so, a sequential recommender may not need to have the users' identifications as usually people may browse online without logging into the system. GRU4Rec which is proposed by Hidasi et al.~\cite{DBLP:journals/corr/HidasiKBT15} exploits a session with 1-of-N encoding ($N$ is the number of items) where an active item in a session is considered as 1, while inactive one is equal to 0. GRU4Rec generates a set of outputs which is a probability of the next interacted items. 
DRNN is another sequential recommender in which multiple hidden layers are used with temporal feedback loops in each layer to provide more personal and real-time recommendation~\cite{DBLP:conf/icde/WuRYCZZ16}. There are several more studies which only take transactional data and achieve a superior performance in SRSs~\cite{DBLP:conf/recsys/TanXL16}.

\subsubsection{Single Behaviour User Data}
In sequential recommendations, there are multiple sequence types such as `browse/buy/mark items, receive/use coupons, click ads, search keywords, write down reviews, or even watch videos by a particular brand'~\cite{DBLP:journals/corr/abs-1711-06632}. Obviously, each of the mentioned actions can represent different level of the users' preferences, and thus, need to be modelled differently. Usually users' behaviour are heterogeneous and polysemous, and thus, it is crucial task for a recommendation system to model each type of users' behaviour individually in order to generate better suggestions. Considering only one type of users' behaviour may be a simple type of input which can be taken by the most of the current SRSs~\cite{DBLP:conf/ijcai/LeLF18}. ATRank~\cite{DBLP:journals/corr/abs-1711-06632}, is an excellent example in which heterogeneous user behaviors are considered with using only the attention model. 

\subsubsection{Multiple Behaviour User Data}
As we have explained above, SRSs are dealing with a wide range of behaviours such as clicking on items, adding items to the wishlist, purchasing items, and putting items into a cart. Taking into account different types of actions is another issue of SRSs, which is less explored. Usually, most of the current approaches treat users' actions in the same way with no difference among them. This may limit the recommendation performance as there may be a difference between different action types, such as add-to-cart and clicked and it is important to take multi-behavioural sequences into consideration~\cite{DBLP:journals/tkde/LiuWW17}. Most of the current SRSs assume that there is only one action in each time step like purchased items and might ignore to take the other types like viewed items. In an online e-commerce website, a user may click on various items under consideration, add some of them to a shopping cart, and puts others
on a wishlist, and eventually just purchase a few of them. Here, we are dealing with multiple sequence types and behaviours which need to be addressed as clicked items also may indicate a user's preference. For instance, Le et al.~\cite{DBLP:conf/ijcai/LeLF18} pay different attention to a various behaviour type. They consider that purchasing action as the target behaviour type which is more important for modelling users' behaviour. Then the other type of behaviors like clicking can be seen as a supporting behaviour according to the target behaviour which can facilitate the next-item prediction task. Furthermore, discriminative behaviour learning also can help to learn the difference between actions' heterogeneity ~\cite{DBLP:conf/kdd/LiZLHMC18}.

The last level in Figure~\ref{ch1:figureRWcat}, Chapter~\ref{chap:introduction} in short-term users' preferences is related to the items' attributes, context and its associated further details and factors~\cite{fang2019deep}. Each item includes multiple heterogeneous attributes like category, price and producer which might be explicitly or implicitly interdependent. For instance, the same product may have different price based on the producers which can explain the explicit interdependence, or different products by the same category may be purchased together due to the implicit interdependence. How to model the interdependence and heterogeneity of items' feature is not an easy task as the data scales/types are totally different~\cite{DBLP:conf/aaai/WangCZW15}. The involvement of item attributes can help SRSs to better understand the item dependency and better deal with the cold-start problems, accordingly~\cite{DBLP:conf/um/KallooriR17}. Moreover, it may be possible that the context of the items affect the users' next behaviours. Here, the context refers to the specific explicit properties of an item when it is purchased, such as its popularity, discount, and community trends, or the implicit ones, such as the time and weather, season, and location, when the item was purchased~\cite{DBLP:conf/sac/JannachL17}, \cite{DBLP:conf/um/LercheJL16}. For instance, the time gaps between events and the time of day for each interaction are considered as a contextual factor in Contextual Recurrent Neural Networks for Recommendation (CRNNs), which is proposed by Smirnova and Vasile~\cite{DBLP:journals/corr/SmirnovaV17}.

Contextual information is an influential factor in both general and sequential recommenders, where there is a separate line of research in general recommenders known as context-aware recommenders~\cite{DBLP:journals/tkde/AdomaviciusT05}. CBS is an example of approaches in this context which is based on the analysing of contemporaneous basket sequences. Considering a session of a particular type of behaviours, then the target session is related to the more informative session~\cite{DBLP:conf/ijcai/LeLF18}. Behaviour-Intensive Neural Network (BINN) is proposed by Li et al.~\cite{DBLP:journals/corr/abs-1808-01075} to address the problem of personalised
next-item recommendations. Inspired by the item2vec~\cite{DBLP:conf/colt/DuchiHS10}, w-item2vec is introduced in BINN to discriminatively exploit user behaviours. As stated by NMRT, buying action may take place after clicking on an item~\cite{8930270}. Therefore, the cascading relationship among different types of behaviors are taken into account by NMRT. It means that the authors in NMRT believe that the `prediction of a high-level behaviour (i.e. purchase) comes from the prediction of the low-level behaviour (i.e. view)'. In this way, the behaviour semantics is incorporated by NMRT to better model multi-behaviour recommendations.  However, this important factor is not explored enough by the current SRSs.

\subsubsection{Repeat Purchasing Behaviour}
This type of user behaviour may be more common in the real-world cases, where the same item is repeatedly observed in a user transactional behaviour. While simple and effective, there is quite a few work which try to consider this type of user behaviour. As discussed by Ren et al.~\cite{DBLP:journals/corr/abs-1812-02646}, repeat consumption is very usual among people's behavioural habits. People are more likely to visit the same restaurant that they have been before, they may like to listen to the same songs and more probably there are some items that users may repeatedly purchase. Therefore, RepeatNet is proposed to adopt repeat-explore mechanism with an encoder-decoder structure. RepeatNet is able to discover repeat recommendation mechanism in order to select the best items for a user at the right time. Anderson et al.~\cite{DBLP:conf/www/AndersonKTV14} investigated the impact of the repeated consumption on several datasets and achieved an interesting outcome: that recency is the strongest predictor of repeated consumption. A recent attempt by Hu et al.~\cite{DBLP:conf/sigir/Hu0GZ20} demonstrated that although RNN is an effective solution for a sequential recommender system, it may not be able to model personalised item frequency (PIF). Hence, the authors propose a simple k-nearest neighbour model to benefit from the PIF in their next-basket recommendation task.

\subsubsection{Exploiting Side Information}
This property of sequential data refers to the incorporating items' characteristics as the  complementary information which makes the recommendations more accurate. As reported by Wang et al~\cite{DBLP:conf/pkdd/WangHC17}, the context of a transaction event needs to be taken in order to fully understand users' behaviours. They propose a neural network-based comprehensive transaction embedding model (NTEM) to learn the comprehensive embeddings of
both items and their features to  produce high quality recommendations. Generally, customers tend to try the newly-released items rather than always recommended by the popular or similar products, which in turn can increase business benefit as well. To address the mentioned problems, NTEM learns the embedding of items and their features of a transaction at the same time through a  wide-in-wide-out structure. While both categorical and numerical features are important, NTEM uses categorical features of items and learns the relations between item relevance~\cite{DBLP:journals/tnn/WangDZCC15}. NTEM  guarantees its efficiency to find the best next choices in terms of novelty and recommendation accuracy evaluation metrics. Items features also can be a good signal for revealing what types of items a user is interested in. A recent study demonstrates that instead of using ratings, item comparisons can be effectively used to compute recommendations~\cite{DBLP:conf/um/KallooriR17}. In this work, all items of interest for a user are collected and then will be compared  in terms of their features in order to construct the users' profiles. The extracted features are ranked according to their contribution to  system performance. Next, a similarity score between the two users is calculated. Additionally, the proposed model is able to handle the cold-stat problem and improve the performance of a recommender system. Item features involvement is also exploited by a factorisation-based model~\cite{DBLP:conf/recsys/ChouYJL16,TEXUS}, CNN~\cite{DBLP:conf/recsys/TuanP17}, and RNN~\cite{DBLP:conf/recsys/HidasiQKT16}. Parallel RNN (P-RNN) is a different version of RNN which is used by Hidasi et al.~\cite{DBLP:conf/recsys/HidasiQKT16} to show that `for each aspect/feature of the clicked item (e.g. the item-ID,
text description, image features) there is a separate RNN processing the input'. In the proposed model, the term `Parallel' means that they use multiple RNNs rather than distributed processing of the data or the algorithm computations. The enriched features are image and text, which are extracted from video thumbnails and product descriptions, respectively. Their proposed architecture achieves the higher recommendation accuracy over a simple SRS, showing the benefit of  using the item features (image data and text) in RNN-based sequential recommender.

\begin{table*}
\centering
        \centering 
        \begin{adjustbox}{width=\textwidth}
        \begin{tabular}[width=0.2\textwidth]{||c |c | c | c ||} 
 \hline
\small \textbf{Module} & \small \textbf{Factor} &  \small \textbf{Method} & \small \textbf{Papers}\\
\hline
\hline
    \multirow{5}{*}{Input} & Only user-item interaction session &  Utilising sequential data & \cite{DBLP:journals/corr/abs-1711-06632}  \\
                          & Single behaviour user data  & Simple behaviour embedding & ~\cite{DBLP:conf/ijcai/LeLF18}, \cite{DBLP:journals/corr/abs-1711-06632}\\
                          & Multiple behaviour user data & Temporal behaviour modelling  &\cite{DBLP:conf/ijcai/LeLF18}, \cite{DBLP:conf/kdd/LiZLHMC18} \\
                          & Repeat purchasing behaviour& Discovering users’ habits& ~\cite{DBLP:conf/sigir/Hu0GZ20}, ~\cite{DBLP:conf/aaai/RenCLR0R19}\\
                          &   Side information used & Data augmentation technique&~\cite{DBLP:conf/recsys/BoginaK17},\cite{ DBLP:journals/corr/DallmannGPZH17} \\
                        
 \hline

 \multirow{8}{*} {Adopted Model}& \multirow {4}{*}{ Model-based} & Recurrent  neural network & ~\cite{DBLP:journals/datamine/HidasiT16}, ~\cite{DBLP:journals/corr/HidasiKBT15}, \cite{DBLP:conf/recsys/QuadranaKHC17}, \cite{DBLP:journals/ki/BharadhwajJ18}, \cite{DBLP:conf/recsys/DonkersL017}, \cite{DBLP:conf/sigir/RenQF0ZBZXYZG19}, \cite{DBLP:conf/wsdm/WuABSJ17}\\
                                &                                & Convolution neural network & ~\cite{DBLP:conf/aaai/WuT0WXT19}, ~\cite{DBLP:conf/wsdm/TangW18}, ~\cite{DBLP:conf/wsdm/YuanKAJ019}, \cite{DBLP:conf/recsys/TuanP17}\\
                                &                                 & Graph neural network 
                                &    ~\cite{DBLP:journals/corr/abs-1812-08434}, ~\cite{DBLP:conf/aaai/WuT0WXT19}\\
                                &                                & Attention mechanism &~\cite{DBLP:conf/icdm/KangM18}, \cite{DBLP:conf/cikm/LiRCRLM17}, \cite{DBLP:conf/kdd/LiuZMZ18}, \cite{DBLP:conf/aaai/WangHCHL018}
                \\
                                                               \cline{2-4}

                                & \multirow {4}{*}{Model-free}  &  Sequential pattern mining & 
                                \cite{DBLP:conf/dasfaa/YapLY12}, \cite{DBLP:conf/pakdd/ZhangC13}, \cite{DBLP:journals/kes/ZhouHF06}

                                \\
                                 &                                   & Factorised Machine-based& 
                                 
                                 \cite{DBLP:conf/recsys/LiangACB16}, \cite{DBLP:conf/www/RendleFS10}, \cite{DBLP:journals/advai/SuK09}\\
                                 &                                 & Markov Chain & ~\cite{DBLP:conf/widm/EirinakiVK05}, \cite{DBLP:journals/jmlr/ShaniHB05}\\
                                  &                                &  Pattern/Rule-based & ~\cite{DBLP:journals/tsmc/WangC20},   \cite{DBLP:conf/ausai/YanL06}, \cite{4795935},  \cite{DBLP:journals/datamine/LinAR02}\\
                                                                    \cline{2-4}

   \hline
 
 \multirow{4}{*}{Item Characteristics} & Session’s purpose & Convolutional Neural Network & \cite{DBLP:conf/wsdm/TangW18},\cite{DBLP:conf/www/TanjimSBHHM20}\\
 
                                 & Noisy session &  Attention mechanism, Memory network &~\cite{DBLP:conf/aaai/WangHCHL018}, \cite{DBLP:conf/wsdm/ChenXZT0QZ18} \\
                                  & Session's order & Pattern/Rule-based & \cite{DBLP:conf/ah/AbelBHKV08},  \cite{DBLP:conf/widm/MobasherDLN01}\\
                                   & Long session& LSTM,GRU, RNN, Mixture models &\cite{DBLP:conf/www/TangBJCBXC19},  \cite{DBLP:conf/wsdm/WuABSJ17}, \cite{DBLP:conf/cikm/XiaJSZWS18}, \cite{DBLP:journals/corr/HidasiKBT15}, \cite{DBLP:conf/recsys/TanXL16}, \cite{DBLP:conf/www/TangBJCBXC19}\\
                                   \hline
 
\end{tabular}
  \end{adjustbox}
  \caption{\centering Summary of the influential factors of sequential recommender systems.}
\label{ch2:tableSRSInfluentialFactors}
 \end{table*}
\section{Unified Recommenders}
Nowadays, RSs are the core components of many of the modern online services including movies, products, news, songs and more. Each of these services has its own characteristics, issues and recommendation scenarios which need its associated techniques to be adopted, accordingly. While the key idea behind each recommender system is to provide accurate suggestions, satisfying users’ preferences and needs in a quick response, increase the business profit and ease user with their decision-making process. Therefore, learning accurate users' preferences is a critical task for the recommendation problem. While the users' preferences may be very diverse, some type of user interests may last for a long
time and are consistent for the same user~\cite{DBLP:journals/eswa/LiZYL14}. For example, a user may be a fan of the popular singer $Justin Bieber$ and thus she/he is more likely to follow her/his songs for several years, even if she/he listens to the other singer as well. This kind of users' preferences can be regarded as a long-term preference (i.e. general taste), which  is mainly modelled by general recommenders. In addition, user interests may keep evolving over time and may be affected by specific contexts such as time, location, and friend. For example, a user may become interested in sad music since she/he may be upset because of her/his current loss. This kind of user interest can be known as a short-term preference. Although modelling long-term users' preferences is important, short-term users' preferences also play a significant role in building an accurate user profile. Hence, unlike general recommenders, sequential recommenders are introduced as a new paradigm of recommenders with the capability of taking users' current needs and interests.

Although modelling each of which users' preferences in a separate manner may be easier, recent progress is captured by approaches that combine both types of users' preferences in a unique framework. FPMC for instance, as a unified framework exploits MF for modelling user general preference and uses the MC technique in order to model sequential behaviour and then linearly combine them to make a recommendation~\cite{DBLP:conf/www/RendleFS10}. According to the~\cite{DBLP:conf/icdm/HeM16}, while FPMC records superior results compared to both the state-of-the-art MF and MC models, it may not be able to deal with the sparsity issue and the long-tailed distribution of many
datasets, and thus it may need more improvement. Therefore, inspired by the strength of FISM in modelling long-term users' preferences through using item-item similarity matrix specifically on sparse data, Fossil~\cite{DBLP:conf/icdm/HeM16}  proposed to learn a personalised weighting scheme over the sequence of items. Fossil was able to make a personalised item recommendation even on a large and sparse dataset.  CoFactor~\cite{DBLP:conf/recsys/LiangACB16}, which jointly uses MF and the item-item co-occurrence matrix with shared item latent factors, is another example of a unified recommender system. CoFactort outperforms pure MF-based models, where the rare
items can be discovered for recommendation through exploiting the co-occurrence patterns, unlike standard MF.

Although the mentioned approaches model both type of users' preferences, the performance of their model may be limited as both types of users' preferences are linearly combined. Soon after, HRM  partially solves the problem of modelling high-level user-item interactions through adopting a nonlinear aggregation function~\cite{DBLP:conf/sigir/WangGLXWC15}. Specifically, HRM builds a hybrid user and item representations by employing a two-layer structure. Moreover, the benefits of different aggregation functions are incorporated among multiple factors by HRM which results in large flexibility and promising potential.

Learning users' behaviour is an essential task for a recommender system as it can help the system to better understand users' needs and interests. Therefore, BINN~\cite{DBLP:conf/kdd/LiZLHMC18} is proposed which contains two main components: Neural Item Embedding and Discriminative Behaviours Learning. Particularly, a novel item embedding technique called w-item2vec method is introduced in BINN to embed items and the interactive behaviours over item sequences. W-item2vec is a combination of a Skip-gram model with a Negative Sampling method~\cite{DBLP:conf/nips/MikolovSCCD13}. Reported results can demonstrate the performance improvement of BINN in  recommendations of the next items for the target users. 
LSDM~\cite{DBLP:journals/corr/abs-1903-00066} is another hierarchical architecture, in which  users’ purchase preferences over time are modelled with an LSTM network. The main assumption behind LSDM is to cluster the successive products together and deal with  users’ repeated purchase demands of items at different multi-time scales.

Differently, inspired by the great success of the GAN mechanism presented by Goodfellow et al.~\cite{DBLP:journals/corr/GoodfellowPMXWOCB14}, in a wide range of task such as image generation~\cite{DBLP:conf/icml/ReedAYLSL16}, image captioning~\cite{DBLP:journals/corr/ChenLCHFS17}, and sequence
generation~\cite{DBLP:conf/aaai/LiuLYQZL18}, PLASTIC is introduced by Zhao et al.~\cite{DBLP:conf/ijcai/ZhaoWYGYC18} to use a generative adversarial network (GAN) in order to discriminatively exploit the performance contributions of combining the MF approach and the RNN approach. Before PLASTIC, IRGAN~\cite{DBLP:conf/sigir/WangYZGXWZZ17} shows the capability of exploiting  GAN in their proposed  model in order to iteratively optimise a generative retrieval component and a discriminative retrieval component.

Inspired by Convolutional Neural Networks (CNNs), Caser~\cite{DBLP:conf/wsdm/TangW18} has been introduced as a sequential recommender that treats user-item interactions as an image and then learns sequential patterns as local features of the image by using convolutional filters. Furthermore, RNNs  also have attracted more attention in modelling sequential dependencies in SRSs~\cite{HidasiKBT15}, \cite{DBLP:conf/recsys/Twardowski16}. For instance, SLi-Rec improves the classic RNN structure such as Long Short-Term Memory (LSTM) by proposing
time-aware and content-aware controllers to fully exploit user modelling. Then the attention-based framework is applied to combine general and sequential recommender~\cite{DBLP:conf/ijcai/YuLML019}. While taking the advantages of both general and sequential recommenders is a strong solution for user representations learning, most of the mentioned approaches try to adopt one recommendation paradigm to help the other one and thus may not be able to deal with both tasks in a unified way. Hence, Recurrent Collaborative Filtering (RCF) as a multi-task learning approach is presented by Dong et al.~\cite{DBLP:conf/ijcai/DongZZW18} which
jointly models both types of users' preferences. RCF models user general taste under  MF-based collaborative filtering and captures user sequential behaviour by utilising RNN. RCF can act as a general recommender and a sequential recommender at the same time and deliver superior performance. Due to the great success of both CNN and RNN in capturing local sequential patterns and complex long-term dependencies, respectively, Xu et al.~\cite{DBLP:conf/www/XuZLXSCZX19} propose a novel Recurrent Convolutional Neural Network model (RCNN) to better generate recommendations. Apart from basic RNNs, improved architectures such as Gated Recurrent Unit (GRU)~\cite{DBLP:journals/corr/ChoMGBSB14} and Long-Short Term Memory (LSTM)~\cite{DBLP:conf/nips/SutskeverVL14} have been developed to model sequential dependencies.

Lately, researchers have employed an attention mechanism due to its powerful capability in focusing on selective parts~\cite{DBLP:journals/corr/BahdanauCB14}. Although incorporating an attention network presents the superior performance in context learning in the work by Wang et.al~\cite{DBLP:conf/aaai/WangHCHL018}, this model ignores users' general taste. Instead, a two-layer hierarchical design called, SHAN, has been proposed by Ying et al.~\cite{DBLP:conf/ijcai/YingZZLXXX018} as an attention-based SRS to incorporate both users' general tastes and short-term preferences in a unified manner. SHAN calculates the attention score which is guided by the user embedding, and thus it may not completely discover the contributions of each item, and it may not be able to find noisy items.  
According to Alyari et al.~\cite{DBLP:journals/kybernetes/AlyariN18}, a significant improvement on a probabilistic classifier is achieved by Maron et al.~\cite{DBLP:journals/jacm/Maron61} which has been recently known as a Naïve Bayesian classifier~\cite{DBLP:conf/adaptive/PazzaniB07}.
\begin{table}
  \includegraphics[width=\textwidth]{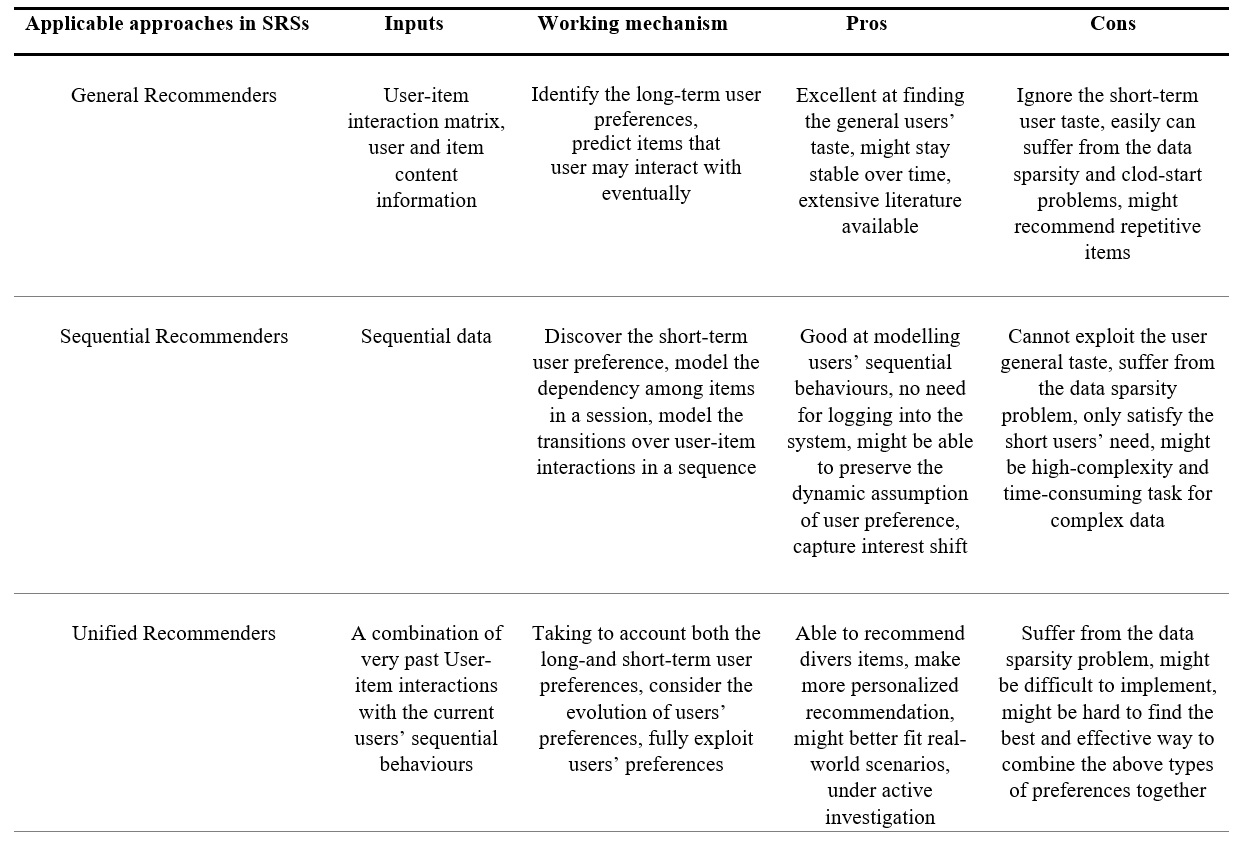}
  \caption{Comparison of approaches in recommender systems.}
  \label{ComparisonSRSs}
\end{table}

\section{Summary}
In this chapter, we presented an overview of the work related to recommender systems, and categorising them based on their structure. First, we provided a brief background of recommender systems, and then discussed the approaches that
only focused on the general recommender systems, followed by those with the main focus of sequential recommender systems. Finally, we reviewed unified recommender systems, which incorporate general and sequential recommender systems. In the following chapters, we propose a series of novel recommender systems, including a general recommender system which analyses users' personality traits for providing personalised recommendations, and two sequential recommender systems, in which the first one pays different attention to different items clicked by the users, and the second one aims to discover the users' main purposes and preferences of buying successive items.





\chapter{Fundamental Concepts and Preliminaries}
\label{ch:chapter3PreliminariesandFoundations}

In this chapter, we first introduce important notions and definitions in both general and sequential recommender systems which are the key concepts of this thesis. Then, we will formulate both the mentioned tasks and their relevant information for giving a better view of the other chapters of this thesis. 

\section{Problem Definition for General Recommenders}
In this section, we will first introduce the most important concepts and definitions in related to general recommenders, which will help a reader to better understand the recommendation problem.

\textbf{User-Item interaction matrix (Utility matrix $R$)}. In the context of recommender systems, users and items are the two main concepts. 
There is a set of $|u|$ users and $|v|$ items, where $U= \{u_1,u_2,...,u_{|u|}\}$ denotes the user set and $V=\{v_1, v_2,...,v_{|v|}\}$ indicates the item set.
The users' preferences for a set of items can be denoted by a utility rating matrix $R^{|u| \times |v|}$, where $R_{ij}$ denotes the preference of user $u_i$ over item $v_j$. In this matrix, each row is reserved for a user and each column can show each item.

\textbf{Type of interactions.} There are two types of user-item interactions, namely explicit feedback and implicit feedback, which can be used as the recommendation inputs. According to the different types of inputs the recommendation problem and the adopted solution may change. Sometimes, users can explicitly express their level of preferences for interested items (e.g. music, movie, and products) with a given rating on the Likert scale (from 1 to 5). The higher rating can indicate the higher level of a user's preference for an item. This is an explicit type of users' feedback. Under this setting, a recommendation system should use the rating matrix $R$, which is usually sparse, to complete the U-I matrix by predicting the ratings for unseen items. Besides for explicit feedback, users can indirectly show their interests in items by clicking, viewing or purchasing the items. This type of feedback is known as implicit feedback in which $R=\{0,1\}$. There is a binary value in this type of feedback, where if $R=0$, it means that the user has not interacted with this item, while the value of $R$ is equal to 1 if the user has watched, clicked, or purchased the item. Dealing with implicit feedback is much more complicated than dealing with explicit feedback, as it is not clear whether users dislike the unobserved items or just do not realise them yet. Therefore, unlike the previous RSs which use explicit feedback, the main goal of approaches fed with the implicit feedback is to provide a ranking list of items rather than the real preference scores.

\textbf{Rating recommendation task}. Based on the adopted technique, the recommendation task may differ from content-based, collaborative filtering, and hybrid-based. As only model-based collaborative filtering was applied for developing a pure general recommender in this thesis, we will only introduce the mechanism of model-based CF approaches. The core idea behind the item recommendation task is to compute the recommendation score and then compare it with the actual rating in order to minimise the point-wise comparison loss. In this paradigm, mostly, a U-I matrix which records the users' explicit feedback, is given to the recommendation engine to calculate the ratings for an unobserved item.

\textbf{Ranking recommendation task.} Different from point-wise methods, a pair-wise ranking structure is proposed to personalise recommendations. The top-n recommendation task produces a ranking ordered list of preferred items where items at the higher order at the top of the list may better satisfy the users' interests. The core assumption behind the pair-wise ranking function is that a user may prefer the clicked items to the unclicked ones. A ranking order set of triplets $(i_1,u_1,i_2)$ is defined, which indicates that user $u_1$ might prefer item $i_1$ to $i_2$. Finally, the pair-wise strategy is mostly adopted in order to better deal with implicit feedback.

\section{Problem Definition for Sequential Recommenders}
Here, we define the concept of session and  provide a set of useful definitions for sequential recommenders. Next, we will formulate the task of session-aware recommender systems and explain how the SRSs work.  

\textbf{Session.} A session can be defined as a set of items which a user has interacted with (e.g. purchasing, clicking, viewing or adding to cart) in one transaction or during a certain period of time (e.g. one hour or one day). A session consists of a group of interactions with items during a given time frame, which can happen on the same day or
over several days. 

In other words, a session can be seen as a set of purchased items such as bread, cheese, and butter together in one transaction, or successive clicks on a set of items in one hour, or a list of watched movies by a user in one day. In this thesis, when we talk about the session, it refers to a set of bought items in one shopping basket. A session includes transactional  data which record the IDs of the purchased items together.

\textbf{Session-aware recommender system.}  Recently, this line of RSs has emerged and attracted more attention from both the industry and academia, as it can better fit the real-world scenarios. The terms session-based recommender system, session-aware recommender system, sequential recommender, and sequence-based recommender system can be used, interchangeably. However, we have mostly used sequential recommender system throughout this thesis for consistency reason. As is clear from the name, the basic unit of SRSs is the transactional data stored in a session. The main goal of a session-aware RS is to model the sequential dependencies among items in a session and then predict the most probable item/items that a user may interact with in the near future. In this regard, there are two main branches of session-aware recommender systems based on whether the recommended items are a part of the current session (i.e. next-item/s recommendation task) or the predicted items are completely placed in the next session (i.e. next-basket(session) recommendation task). We will formally define these two types of SRSs, but the main focus of this thesis is on the first structure (next item/s recommendation problem).

\textbf{Session's context.} Context is a multifaceted concept that has been studied across different research disciplines. However, we define it as an important concept which specifies the situation in which a user makes a decision. The contextual factor such as time, weather, location, season, time of day, and venue may have a significant effect on the user's final choice. This critical parameter may be obvious and can be extracted explicitly from a session, while it may be hidden in a user's transaction history. For instance when a user consistently clicks on a playlist of multiple romantic songs, clearly she/he may be interested in this genre of  music, and this contextual information can be extracted from her/his behaviours. While in a shopping basket of a user if she/he puts butter, milk, eggs, and baking powder, she/he may want to make a cake, and this contextual factor is hidden in her/his past actions. Therefore, identifying a contextual factor is an important task, which plays an important role in better understanding the users' needs and interests and to thus make a right prediction, accordingly. As it is an important parameter particularly in terms of recommender systems, there is a separate line of research on general recommenders, known as Context-Aware Recommender Systems (CARS); however, it is in its early stage.

\textbf{Next-item recommendation task.} The main goal of the next-item recommendation task is to take the context of the current session and then predict the best next item which this user may interact with in the future. Most of the current sequential recommenders fall into this category and only consider the last session for making a prediction~\cite{DBLP:conf/recsys/QuadranaKHC17}, \cite{DBLP:conf/recsys/ChouYJL16}, \cite{DBLP:journals/corr/HidasiKBT15}. There are only a few works which take the context factor into account from multiple sessions, which is a complicated task by itself and requires more investigation.

\textbf{Next-basket (session) recommendation task.} In contrast to the previous structure of sequential recommenders, this line of research mainly focuses on finding the most probable items that a user may prefer in the next session. Hence, under this setting, all the $n$ number of previous sessions are taken into account and the shared context among them is modelled for providing recommendations in the next session.

As mentioned earlier, in this thesis, we have only paid attention to the next-item recommendation problem and have left the next-basket recommendation task for future work. Below, we provide some notations which may be useful for a frequent reader to better understand this domain.
\begin{table}[h!]
\centering
 \begin{tabular}{c||c} 
 \hline
 Notation & Meaning \\ [0.5ex] 
 \hline\hline
$u$ &  user\\
$U$ &  user set\\
$v$ &  item\\
$V$ &  item set\\
$T$ &  transaction history of a shopping basket of an user  \\
$s$ & session including a set of multiple items in a certain period of time\\
$S$ &  session set  \\
$c$ &  session context\\
$G$ & long-term interacted items set\\
$Q$ &  collection of all previous sessions\\
$R$ &  recommendation utility matrix\\
$r$ &  recommendation score\\
$k$ &  latent dimension\\
$h$ &  embedding vector \\
$\alpha$ &  attention score\\
$\beta$ &  attention score\\
$b$ &  bias of neural network \\
$e$ &  latent vector \\
$W$ &  neural network weighting matrix \\
$\lambda$ & neural network regularisation\\
$c$ & latent representation vector \\
$K$ & kernel of CNN\\
$p$ & user latent vector\\
$q$ &  item latent vector\\
 \hline
 \end{tabular}
 \caption{List of commonly used notations in this thesis.}
\end{table}
\section{Preliminaries}
In this thesis, various embedding, data mining, deep neural networks, and machine learning models were applied to both general and sequential recommender systems in order to build the approaches for overcoming the potential challenges. In particular, we introduce matrix factorisation model, convolutional neural network, and attention mechanism. Matrix factorisation is one of the widely used classic techniques in recommender systems, which embeds user-item interactions into low-dimensional spaces. In addition to matrix factorisation, we explain convolutional neural network as one of the most powerful designs in deep neural networks. Capturing the local context among a set of items is one of the main capabilities of convolutional neural networks, which attracts the attention of the researchers in this community. Furthermore, attention mechanism is recently introduced as one of the successful technologies in learning information from the input data. Therefore, we discuss the core concepts and procedures of these algorithms in order to make it more easier for the readers to follow the other chapters.


\begin{figure*}[t!]
\includegraphics [width=0.7\textwidth, scale=1]{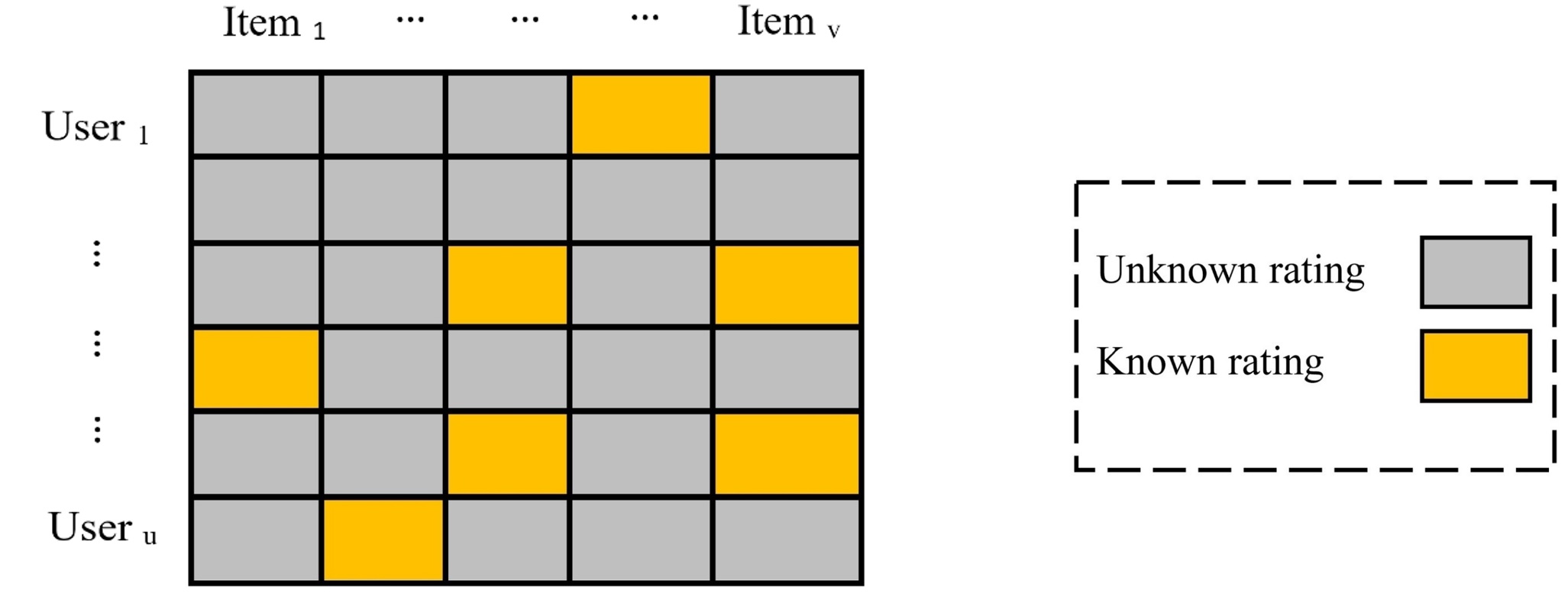}
\centering
\caption{Overview of a rating matrix.} 
\label{MatrixFactorization}
\end{figure*}

\subsection{Matrix Factorisation}
\label{ch:chapter4MF}
Matrix factorisation, which has been widely adopted by the current recommender systems, is one of the most popular methods for performing model-based collaborative filtering. As shown in Figure~\ref{MatrixFactorization}, this matrix records the users' interactions with the items. Matrix factorisation is a latent factor model using the U-I rating matrix to embed the hidden factors of both the items and users~\cite{DBLP:journals/computer/KorenBV09}. Matrix factorisation is a mathematical tool that works with the  matrices to simply discover the hidden relationships between users and items, and thus, it can be used in a wide range of domains. In real-world applications, we have too many users and items, while just a small portion of items are rated by even very active users. Moreover, even very popular items are rated by only a few users. This problem refers to the data sparsity problem, with which most of the current RSs are confronted. Under this setting, there is a lack of available information which makes a recommendation more challenging. Therefore, matrix factorisation as a dimensionality reduction method can be one of the best solutions for the data sparsity problem. 

Mathematically speaking, we have a set of users ($U$) and a set of items ($V$). The rating matrix $R$ contains all the available ratings  given to all the items. The goal of matrix factorisation is to convert this matrix into two latent vectors $u$ and $v$ to first deal with the data sparsity problem and then learn the hidden features embedded in the rating matrix. To do so, it is required for matrix factorisation to identify $k$ latent features. Next, matrix factorisation predicts the ratings for unseen items to complete the U-I  matrix by computing the inner products of a user's and item's latent vectors together. 

\begin{figure*}[t!]
\includegraphics [width=1.1\textwidth, scale=1]{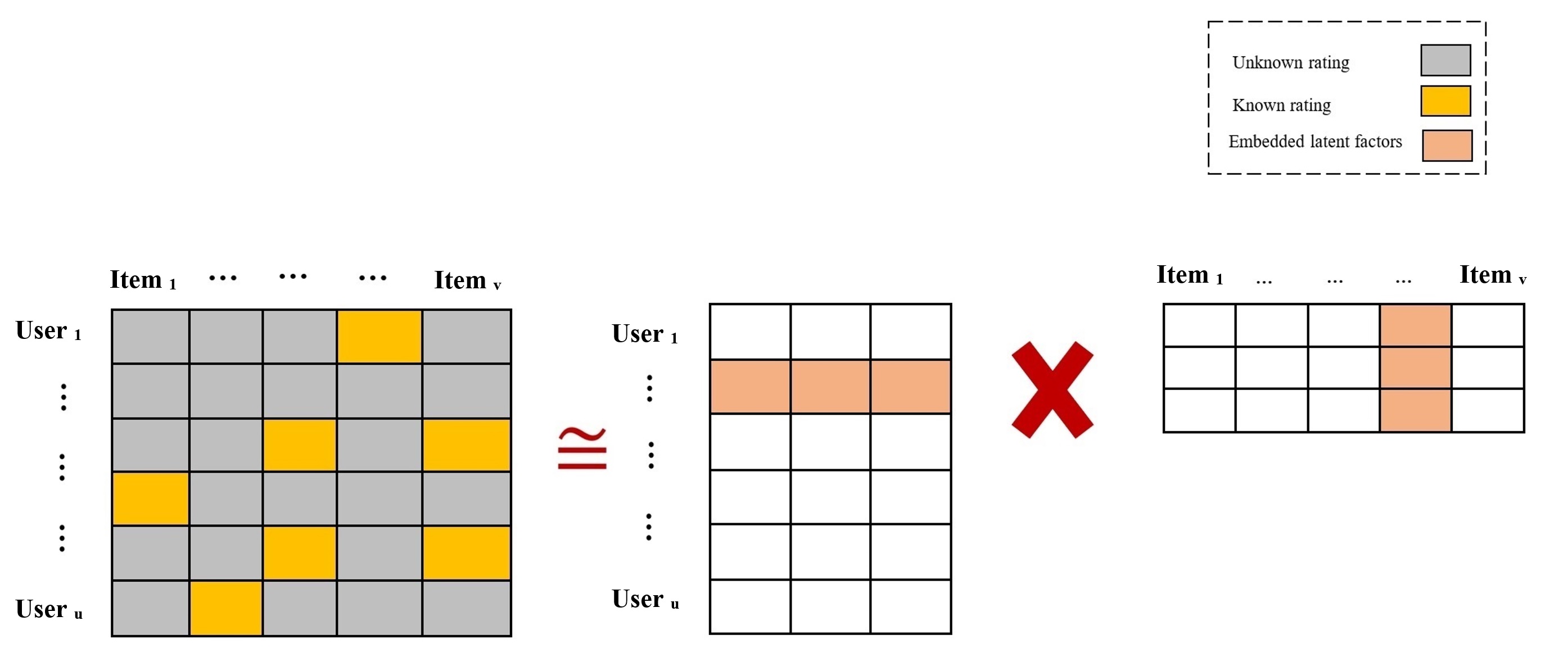}
\centering
\caption{Matrix factorization structure.} 
\label{MatrixFactorization2}
\end{figure*}

The basic model of matrix factorisation is given in Formula~\ref{MF}. As is clear from this formula, a user's interest for a particular item is calculated by the inner product of a user $u_i$ and item latent $v_j$ vectors as follows:

\begin{equation}
\label{MF}
   r_{ij}= {q^T_j} {p_i}
\end{equation}

Then, in order to learn the latent vectors, the RSs need to minimise the regularised squared error on the set of known ratings as follows:

\begin{equation}
\label{loss}
  {min \sum ( { r_{ij}^'}- {r_{ij}}) + \lambda {|| q_j||}^2 +\lambda {||p_i||}^2}
\end{equation}
where $r_{ij}^'$ is the real rating value and $r_{ij}$ is the predicted value by Formula~\ref{MF}.

\subsection{Convolutional Neural Networks}

Convolutional Neural Networks (CNNs) were introduced by David Hubel and Torsten Wiesel~\cite{CNN} in the 1990s. They found that the neurons in the primary visual cortex are sensitive to the specific regions of the visual field features in the visual environment, particularly the oriented edges. In the late 1950s and the early 1960s, David Hubel and Torsten Wiesel found that the neuronal cells in the brain fired only in the presence of the edges of a certain orientation. This region of the visual space is known as the  receptive field. Hence, inspired by the concept of visual cortex from biology, CNNs have emerged as a powerful technique in computer vision, where the first project related to the handwritten digits recognition showed a promising result~\cite{726791}. Years after, `ImageNet Large Scale Visual Recognition Challenge' was introduced by Russakovsky et al.~\cite{DBLP:journals/corr/RussakovskyDSKSMHKKBBF14} to demonstrate the considerable potential of these networks. `AlexNet' as a CNN model proposed by Alex Krizhevsky and his collaborators achieved the first place and thus has become a state-of-the-art model in the field of computer vision. Although CNN was initially designed to work with images and has achieved excellent results for image classification~\cite{DBLP:conf/nips/KrizhevskySH12}, it has been shown to be effective for sentence
classification~\cite{DBLP:conf/emnlp/Kim14}, sentence modelling~\cite{DBLP:journals/corr/KalchbrennerGB14}, and speech recognition~\cite{DBLP:journals/taslp/Abdel-HamidMJDPY14}.

\begin{figure*}[b!]
\includegraphics [width=\textwidth, scale=1]{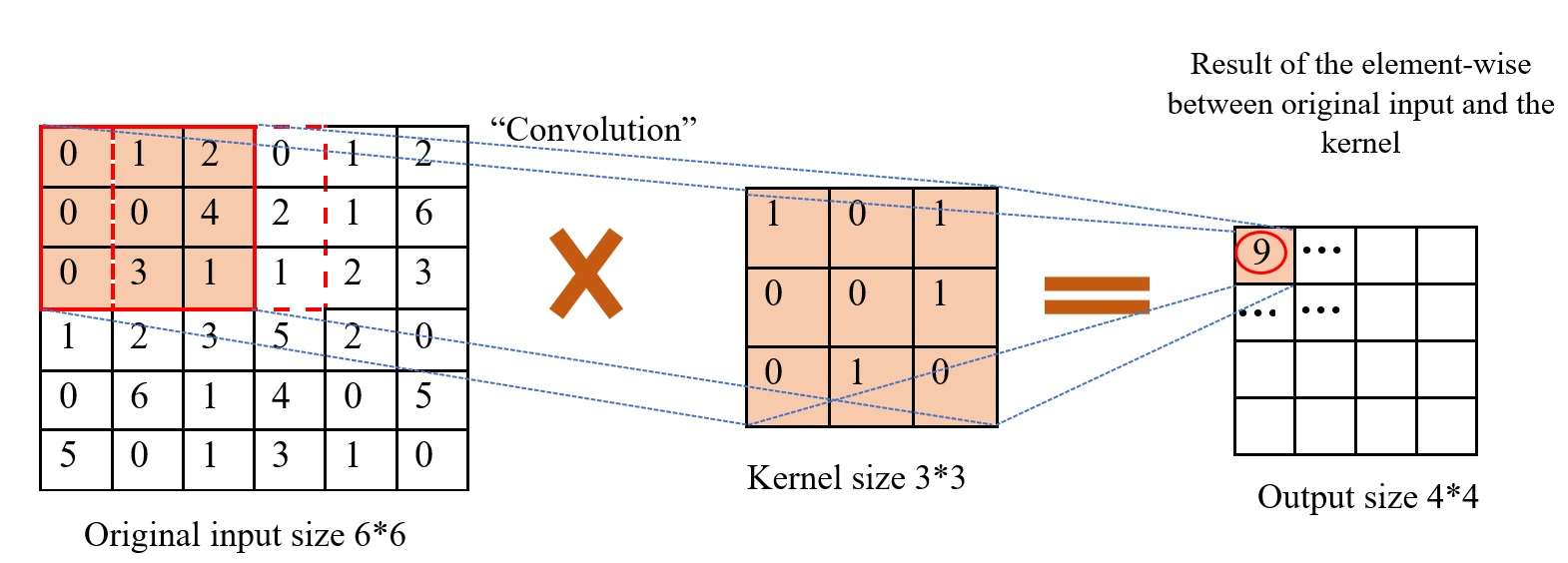}
\centering
\caption{Example for convolution operation on a 6*6 image with a 3*3 kernel.} 
\label{Conv}
\end{figure*}

A CNN is a feed-forward neural network which consists of several layers with different aims and jobs: such as a convolutional layer, a nonlinearity layer, a pooling layer, and a fully connected layer. The first layer (convolutional layer) takes a matrix of inputs and then performs a convolution operation between a set of kernels and the inputs in order to identify features. There are various sizes of kernels (filters) responsible for detecting  different parts of the inputs.  As shown above in Figure~\ref{Conv}, an element-wise product between the input and the kernel is applied, and then, the value is summed to build an output from this layer. Then, the kernel is stridden to the right and the element-wise product is calculated again in this new part. This procedure is continuously repeated from left to right and from top to bottom until it covers all the inputs.  

The next step is to use a nonlinear activation function, when the output of the convolutional layer is ready. There are different types of activation functions, such as tanh or sigmoid~\cite{DBLP:conf/iwann/HanM95}, but the ReLU
function is the most widely adopted one as it is faster to compute than the other functions~\cite{LOPEZPINAYA2020173}. 

\begin{equation}
    c_i= ReLU (K \times e_{{\lfloor i-\frac{k-1}{2}\rfloor}:{\lfloor i+ \frac{k-1}{2}\rfloor}} + b)
\end{equation}

\begin{figure*}[b!]
\includegraphics [width=0.8\textwidth, scale=1]{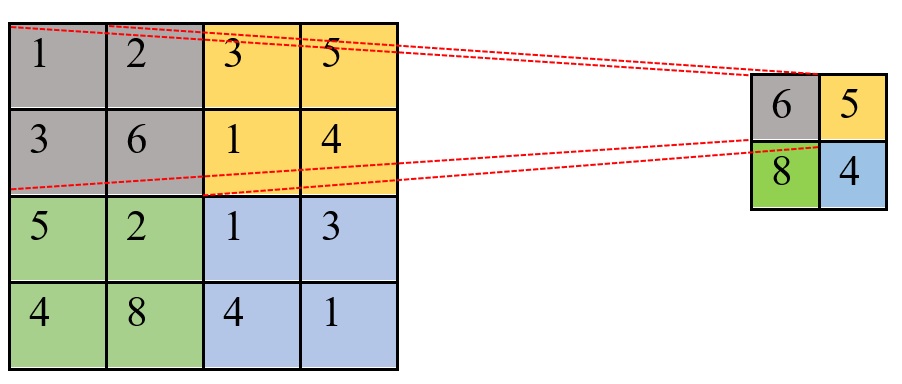}
\centering
\caption{Example of image portion for a max-pooling operation.} 
\label{Maxpooling}
\end{figure*}

Then, the pooling operation is applied to the output of the previous layer to reduce the spatial size of the representation captured by the convolutional layer and thus speeds up the computation time. Max-pooling is one of the commonly used functions in the pooling layer. In this layer, a window strides over the convolved feature matrix and takes the maximum value information. 

The last layer in CNN is a fully connected layer. The outputs from the previous layer are first converted into a 1D format and then are inputted into this layer. The main job of this layer is to calculate the probability score for each class of objects in order to perform a classification task. 
\begin{figure*}[h!]
\includegraphics [width=\textwidth, scale=1]{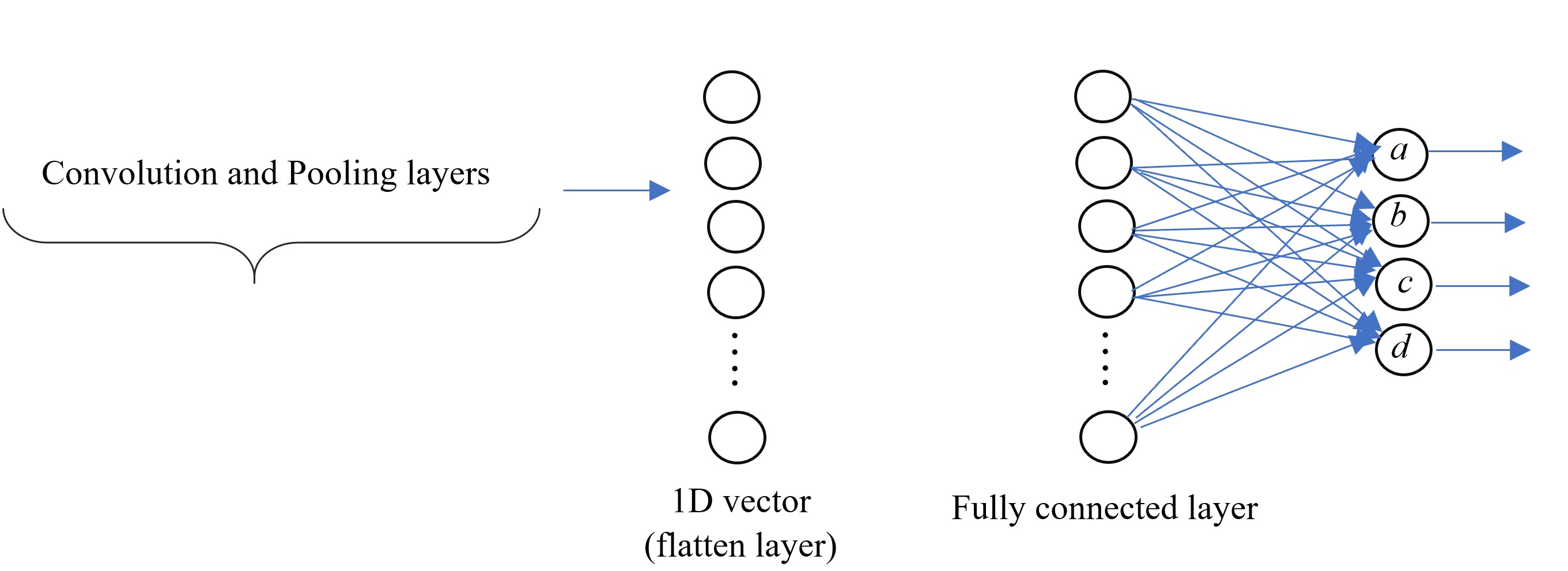}
\centering
\caption{Fully connected layer.} 
\label{Fullyconnectedlayer}
\end{figure*}

\subsection{Attention Mechanisms}
Attention is, to some extent, motivated by the basic function of a human brain, which  pays visual attention to different parts of an image. Usually, humans focus more on the selective part of an image with `high resolution' and ignore the `low resolution' parts. In other words, when we look at a picture, the whole picture may be shallowly scanned at first. Then, we may like to see more details of the picture and thus pay more attention to a small portion of an image. This observation shows that our brain can differentiate all the regions in a picture and give more weight to the important parts of the picture. Motivated by this observation, an attention mechanism is introduced to discriminately learn the information from the input data~\cite{DBLP:journals/corr/BahdanauCB14}. While an attention mechanism was initially applied in natural language processing (NLP) tasks~\cite{DBLP:journals/corr/LuongPM15}, \cite{DBLP:journals/corr/VaswaniSPUJGKP17}, it has also been shown to perform well in a wide range of domains such as machine translation~\cite{DBLP:conf/aaai/ChenWUSZ18}, image captioning~\cite{DBLP:journals/corr/ChenZXNSC16}, and recommendations~\cite{DBLP:journals/tkde/HeHSLJC18}. In addition, the considerable success of  attention mechanisms motivates other tasks such as dependency modelling and sequential recommenders~\cite{10.1145/3285029}. 

Various types of attention mechanisms have been reported in the literature, but here, we applied the most popular one, which was proposed by Bahdanau et al.~\cite{DBLP:journals/corr/BahdanauCB14} for an NLP task. In this model, the attention mechanism is used for neural machine translation tasks. In such system, a sentence is inputted into the system and the goal of the system is to translate this sentence to the correct version of translation, which is the output of the neural machine translation. To do so, encoder-decoder modules are required for translating between different languages. Here, the main issue is that these modules convert a sentence to a fixed-length vector, which may result in information loss and performance reduction, accordingly. Therefore, an extension of the encoder–decoder model was proposed by Bahdanau et al.~\cite{DBLP:journals/corr/BahdanauCB14} to concentrate more on the most relevant information in order to learn the context of a source sentence. Thus, both the length of a source sentence and the squashing of the information in a source sentence in a fixed-length vector are not required,  which may avoid information loss. 

If we consider $X=\{x_1,..., x_{T_x}\}$ as an input sentence in the source language, $h_t$ as the hidden state at time $t$, $c_i$ as the context vector, $Y=\{y_1,...,y_{T_x}\}$ as the translated output sentence in the target language, then the task is to maximise the conditional probability defined in Formula~\ref{conprobability} as follows:
\begin{equation}
    \label{conprobability}
    P(y_i| y_1,...,y_{i-1}, X)=g(y_{i-1},h_i, c_i)
\end{equation}
where $h_i$ is an hidden state at time $i$ and is calculated as follows:
 
\begin{equation}
    h_i=f(h_{i-1}, y_{i-1}, c_i)
\end{equation}
where $c_i$ is a context vector built on a sequence of annotations $(e_1,...,e_{T_x})$; at each time point, $e_i$ contains the information of the whole sequence of inputs with the main emphasis on the selective parts around the $i-th$ words from the input sequence. $c_i$ as the weighted sum is computed as follows:
\begin{equation}
c_i=  \sum_{j=1} ^ {T_{x}} \alpha _{ij} e_j
\end{equation}
where the weight $\alpha_{ij}$ of each $e_j$ is calculated as follows:
\begin{equation}
    \alpha_{ij}=  \frac {exp(s _{ij})}{\sum_{k=1}^{T_x} exp(s_{ik})}
\end{equation}
\begin{equation}
    s_{ij}= a(h_{i-1}, e_j)
\end{equation}
where $a$ is the parameterised weight to jointly train all the other components in a feedforward neural network. In the end, the attention mechanism can assign different weights to the different parts of the inputs according to their level of importance to the current word.

\chapter{Experimental Setup}
\label{ch4:chapterExperimentalSetup}

In this chapter, we will discuss the experimental setup, including the datasets used in our experiments, the evaluation metrics, and the baseline methods, for a comparison with the sets of approaches proposed in this PhD
thesis. 
\section{Datasets}
Here, we will first introduce the datasets that were used in general recommender systems, followed by the datasets for sequential recommender systems. 

\subsection{Amazon Instant Video Dataset}
\begin{figure*}[h!]
\includegraphics [width=0.8\textwidth, scale=1]{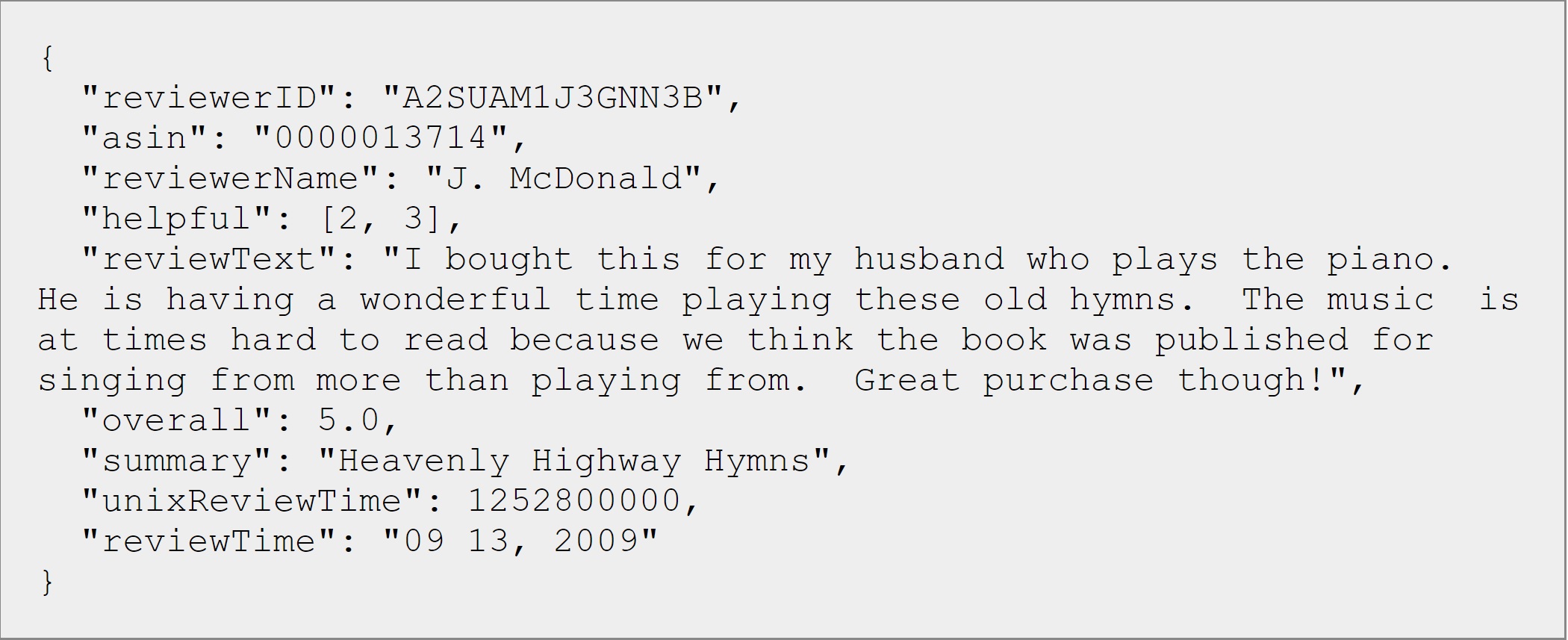}
\centering
\caption{Sample review of Amazon dataset.} 
\label{Amazon Datset}
\end{figure*}

As the first proposed model is for a general recommender system, we have decided to select an appropriate dataset from this domain. Therefore, in this first approach, we have needed to detect the users' personality types and selected a dataset that contained textual information that would help with the personality detection task. We have selected the Amazon dataset, which has been widely used in RSs~\cite{DBLP:conf/www/HeM16}, \cite{DBLP:journals/corr/McAuleyTSH15}. The Amazon~\footnote{http://jmcauley.ucsd.edu/data/amazon/} dataset is a rich source of reviews  and  provides a wide range of useful information (e.g. ratings and reviews), which consists of 2000 users, 1500 items, 86690
reviews, 7219 number ratings, 3.6113 average number of rates per user, 0.2166 average number of ratings per item, and the user rating density of 0.0024. Figure~\ref{Amazon Datset} shows a sample of this dataset, where reviewerID is the id of the reviewer, e.g. A2SUAM1J3GNN3B;
asin is the id of a product, e.g. 0000013714; reviewerName is the name of the reviewer; 
helpful denotes the helpfulness rating of the review, e.g. 2/3; reviewText is the text of the review; overall is the given rating to the product; summary is a summary of the review; unixReviewTime is the time of the review (unix time); and  reviewTime is the time of the raw review. We have selected a subset of the Amazon dataset called Instantvideos, as it includes an important source of knowledge that can be extracted for a different purpose. In addition, because of the strong relations between a person's personality type and her/his choices of videos, we have selected the Instantvideos dataset to test the proposed model on it, and have left the other domains for future works. We have selected users who left more than three reviews in this dataset in order to obtain a better analysis of their generated content. 

\subsection{Tmall, Tafeng, and Gowalla Datasets}
\label{Tmall , Tafeng and Gowalla Datasets}

For sequential recommender systems, we have selected three real-world datasets, namely Gowalla~\cite{DBLP:conf/kdd/ChoML11}, Tafeng~\cite{brijs99using}, and Tmall~\cite{DBLP:conf/ijcai/HuCWXCG17}, to compare the performance of the proposed model with that of the baseline approaches. Gowalla~\footnote{https://snap.stanford.edu/data/loc-gowalla.html} aggregates the users' check-in information from the location-based social networking website, Gowalla. Tmall~\footnote{https://ijcai-15.org/index.php/repeat-buyers-prediction-competition} records the users' transactions on the largest online shopping website in China, where each session (transaction) consists of multiple items. Tafeng~\footnote{http://www.bigdatalab.ac.cn/benchmark/bm/dd?data=Ta-Feng} is a grocery shopping dataset, which covers products ranging from food to office supplies to furniture.   
\begin{figure*}[h!]
\includegraphics [width=0.8\textwidth, scale=1]{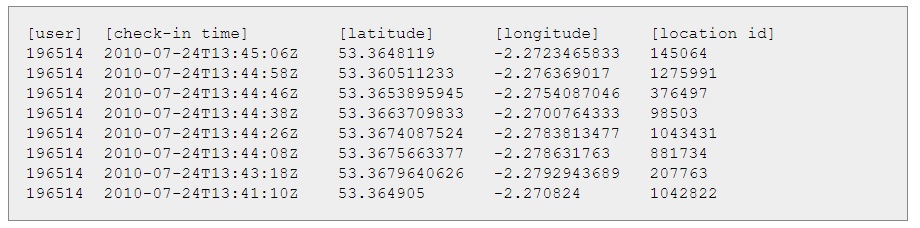}
\centering
\label{Gowalla Datset}
\caption{Sample review of Gowalla dataset.} 
\end{figure*}

 Note that, similar to another work~\cite{DBLP:conf/ijcai/YingZZLXXX018}, in both the Tmall and the Gowalla datasets, we only considered the data for the previous  seven months and removed the sessions with only one item and with items with less than 20 observations. Then, to better represent the users' sequential behaviours (i.e. short-term preference), transactions in one day were considered one session. While similar to the work done by Hu et al.~\cite{DBLP:conf/wise/HuHSN19}, in the Tafeng dataset, all the users, who have consumed less than 10 items were removed from the datasets. Thus, we only kept sessions with at least 10 items, which might cause a longer average session. The reason behind the limitation of 10 items, is that there are a fewer number of users and items than in both the Tmall and the Gowalla datasets. The characteristics of these datasets obtained after the pre-processing step are shown in Table~\ref{datasets}.

\begin{table}[ht]
\centering 
\begin{tabular}{c c c c } 
\hline\hline %
Dataset & Tmall & Gowalla  & Tafeng\\ [0.5ex] 
\hline 
 Number of users & 20,716 & 15,254 & 12095\\
 Number of items & 25,143 & 13,052 & 11024 \\
Avg.session length & 2.81 & 2.99 & 3.01\\
Train session & 71,998 &128,115 &131,102 \\
Test session & 3565 & 3611 & 3984\\[1ex] 
\hline 
\end{tabular}
\caption{Statistics of Tmall, Tafeng, and Gowalla datasets.}
\label{datasets}
\end{table}

\section{Evaluation Metrics}
In this section, we will define the evaluation metrics used throughout this thesis. As we have proposed novel models for both general and sequential recommender systems, we had to select the appropriate evaluation metrics to compare each of these models with the current studies. In Chapter~\ref{ch5:chapter5Personality}, wherein the main focus is on improving the accuracy of a general recommender, we have decided to similarly use the accuracy metrics adopted by the existing approaches for the general recommenders. In this chapter, we propose an algorithm to calculate a rating score for the unobserved items for a particular user, and thus, how accurately our model can predict this score should be assessed. Therefore, we have used the mean absolute error (MAE) and root mean squared error (RMSE) evaluation metrics, which will be introduced in Section~\ref{Accuracy}. As the main goal of  Chapters~\ref{ch6:chapterDAS} and~\ref{ch:chapter7CAN}
will be the performance improvement of sequential recommender, we will evaluate the performance of the proposed models in terms of accuracy and ranking metrics, just like the other sequential recommender systems.

\subsection{Accuracy Metrics}
\label{Accuracy}

In general, in the domain of recommender systems, typically, there are two broad ways for measuring the performance of a recommendation system: rating prediction and ranking prediction. As we have already discussed in Chapter~\ref{ch:chapter3PreliminariesandFoundations}, the main focus of each of the mentioned task was different, and hence, the corresponding performance evaluation metrics had to be selected accordingly. While the rating prediction-based approaches are only concerned about the observed ratings, the unobserved items are taken into consideration by ranking the prediction-based models~\cite{DBLP:conf/recsys/Steck13}.

\subsubsection{Rating-Based Evaluation Metrics}

The two widely adopted accuracy metrics for recommender systems are the mean absolute error (MAE) and the root mean squared error (RMSE). These metrics are mostly applied in  ratings-based recommender systems, wherein the predicted score is compared with the actual value in order to calculate the error between them. The smaller values of MAE and RMSE denote a better recommendation accuracy, as they imply that the predicted rating score for an item is close to its real value.
\begin{itemize}
    \item 
\textbf{MAE.} Mean absolute error is a metric that computes the average of all the absolute differences between the real and the predicted rating values~\cite{Patrous2016EvaluatingPA}. The lower the MAE is,
the better is the accuracy. MAE can be measured as follows:

\begin{equation}
    MAE= \frac{1}{|n|} \sum_{j=1} ^{n} |r_{ij}- r_{ij}^{'}|
    \end{equation}
where $n$ is the total number of users and $r_{ij}$ and $ r_{ij}^{'}$ are the real and the predicted values of the ratings, respectively.

\item \textbf{RMSE.} Root mean square error, which was proposed by Herlocker et al.~\cite{DBLP:journals/tois/HerlockerKTR04}, calculates the square root of the result of the 
mean value of all the differences squared between the true and the predicted ratings values as follows:

\begin{equation}
RMSE=\sqrt {\frac{1}{|n|} \sum_{j=1}^{n}(r_{ij}- r_{ij}^')^2}
\end{equation}

where similar to those in the case of the MAE, $n$ is the total number of users, and $r_{ij}$ and $ r_{ij}^{'}$ are the real and the predicted values of the ratings, respectively. 

\end{itemize}

\subsubsection{Ranking-Based Evaluation Metrics}
\begin{figure*}[h!]
\includegraphics [width=0.8\textwidth, scale=1]{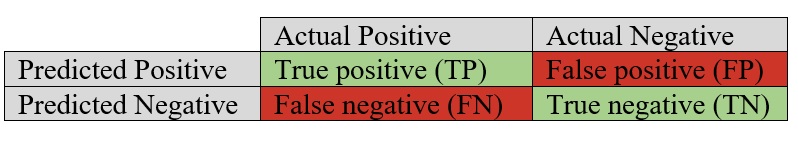}
\centering
\caption{Overview of a confusion matrix.}
\label{ConfusionMatrix}
\end{figure*}

Recommender systems can be usually considered a special case of information retrieval systems, which try to retrieve the related information from the user's historical data~\cite{PerformanceEvaluation}. In this case, the recommendation performance should be evaluated with respect to the order of the recommended items. The top-n recommendation problem, which recommends a list of top $n$ items with the highest popularity for a particular user, can be a good example of the ranking-based approach. In this prediction schema, where the output is a ranked list of items, we have needed evaluation metrics with the capability of testing the relevancy of the recommended items in this list over the users' interests. According to the observations, we have selected precision and recall, which are two popular evaluation metrics for ranking-based RSs. However, before we dive into more details, we need to explain the confusion matrix, which is the basic concept for understanding precision and recall. In general, the main idea behind a confusion matrix is to count the number of correct and incorrect predictions. A confusion matrix is a  $2 \times 2$ table which summarises the prediction results of a model. As is shown in Figure~\ref{ConfusionMatrix}, each column represents an actual class, while each row  represents a predicted class.

The outcomes of the RS model belong to one of four classes: True Positive (TP), False Positive (FP), True Negative (TN), and False Negative (FN) which are defined below:
\begin{itemize}
    \item True positive (TP): when a model correctly predicts the positive class.
     \item False positive (FP): when a model incorrectly predicts the positive class.
     \item True negative (TN): when a model correctly predicts the negative class.
     \item False negative (FN): when a model incorrectly predicts the negative class.

\end{itemize}

\textbf{Precision.} This parameter evaluates how well the proposed model can capture  the relevant items among all the predicted items. It assesses the ability of the model to find all the relevant top-N items within a dataset as follows:
    
    \begin{equation}
    \label{Precision}
        Precision=\frac{TP}{TP+FP}
    \end{equation}
According to Equation~\ref{Precision}, precision is calculated as the ratio of the correct positive predictions to the total predicted positives.
    
\textbf{Recall.} This metric quantifies the ability of a model to correctly retrieve positive instances from all the observations in  the actual class.   
    
        \begin{equation}
    \label{Recall}
       Recall=\frac{TP}{TP+FN}
    \end{equation}
    
According to Equation~\ref{Recall}, recall describes the
proportion of the user's favourite items that are not missing. Note that larger precision and recall values indicate better performance.
    
\begin{figure*}[b!]
\includegraphics [width=0.5\textwidth, scale=1]{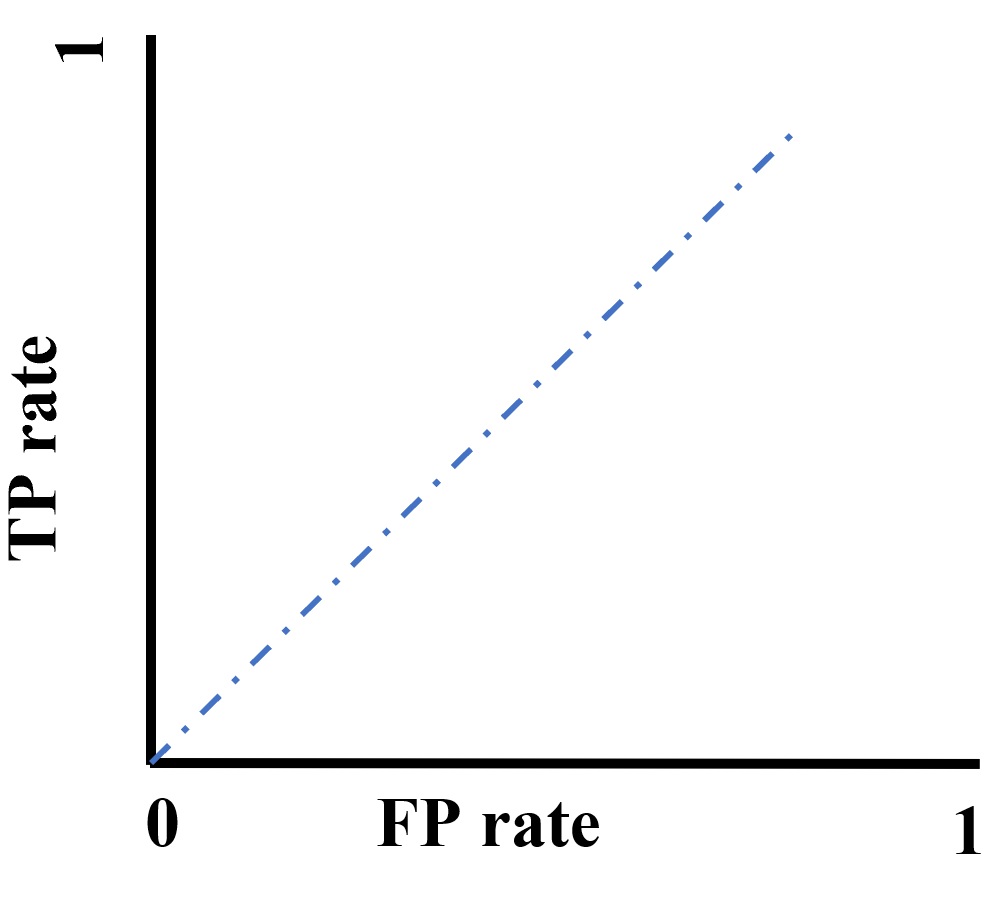}
\centering
\caption{ROC curve.} 
\label{AUC}
\end{figure*}
\textbf{Area Under Curve (AUC).} This is another evaluation metric that aims to calculate the overall performance of a model. For calculating AUC, first, we need to introduce ROC, which stands for `Receiver Operating Characteristics' and was originally used as an analysis tool for signal detection to separate the `signal' from the `noise'.  Recently, it has become more popular, particularly in the field of information retrieval and evaluation of machine learning algorithm~\cite{DBLP:journals/prl/Fawcett06}. As shown in Figure~\ref{AUC}, the `ROC curve is a two-dimensional coordinate graph, the X-axis is the false positive rate (FPR)
and the Y-axis is the true positive rate (TPR). The ROC curve shows the correspondence between FPR and TPR'~\cite{PerformanceEvaluation}.

Therefore, AUC as the area under the ROC curve can be a good parameter to evaluate the performance of different recommender systems. AUC as a summary of the ROC curve can measure the ability of a model to rank a randomly chosen positive instance higher than a randomly chosen negative instance. Hence, it can be a better measure than a classification error rate based upon a single prior probability. As can be seen in Figure~\ref{AUC}, the higher the TP rate and the lower the FP rate for each threshold, the better is the accuracy, and thus, models that have curves standing taller on the top-left-side are better. In other words, it means that a perfect system would have an AUC of 1.

\subsection{Novelty Metric}

Besides for the mentioned evaluation metrics that we have explained above, the novelty factor is another important parameter to test the recommendation's performance. Considering that recommending items similar to those that a user has already purchased may not be satisfying and users may be more willing to be recommended new items, similar to the work done by Wang et al.~\cite{DBLP:conf/aaai/WangHCHL018}, we have considered the novelty metric as another evaluation parameter to show the capability of the proposed model to recommend novel items. The novelty metric can measure the difference between contextual items in a shopping basket and a set of recommended items, where a larger difference can represent the novel items~\cite{DBLP:conf/pkdd/WangHC17}. Novel items provide an opportunity for users to try unexpected items, which may leverage the satisfaction rate as well.


\textbf{Context Aware Novelty (CAN).} This evaluation metric can mainly represent how different the recommended items are compared to the previously purchased ones. Therefore, the more distance between this range can demonstrate the higher novelty metric and result in suggesting more different items to the users. As it is shown in Equation~\ref{CAN}, items which are purchased in context $c$ can be used for recommendation $R$. The more the overlap between recommended list $R$ and purchased items in context $C$, the less the novelty. Hence, CAN can be defined as follows:

\begin{equation} 
CAN= 1 - \frac{|R\cap C|}{|R|}
\label{CAN} 
\end{equation}

\textbf{MCAN@K}. This is the mean of CAN (MCAN), which is somehow similar to CAN. MCAN measures the mean of unseen items corresponding to context $C$ over all $N$ top-K recommended items. Similar to the work done by Hu et al.~\cite{DBLP:conf/ijcai/HuCWXCG17}, we compared the proposed model by using the MCAN evaluation metric, which can be defined as follows:
\begin{equation} 
MCAN= \frac{1}{N}\sum CAN
\label{MCAN}
\end{equation}

Noted that we have considered $k\in \{5,100\}$ in this thesis, as in reality, people are usually interested in the first few pages of the recommendations. Therefore, we have evaluated the performance of the proposed models in this thesis under this condition as it may be more challenging to discover the exact actual items which the users may prefer among a million available options.

\section{Baselines}
We have compared the proposed approaches with the state-of-the-art recommender systems for both general and sequential recommenders. Below, we list the baselines used to compare the proposed models with in most of the following chapters. First, we will introduce the baselines with respect  to the general recommender systems followed by several representative models for sequential recommender systems. 

\subsection{Baselines for General Recommenders}
In this section, we will introduce the state-of-the-art traditional RSs used as the baselines for comparison.

\subsubsection{Personality-Based RSs} As in Chapter~\ref{ch5:chapter5Personality}, we propose a personality-based recommender system, present the existing representative models in this regard, and then discuss the other classic 
RSs.

\begin{itemize}
\begin{figure*}[t!]
\includegraphics [width=0.7\textwidth, scale=1]{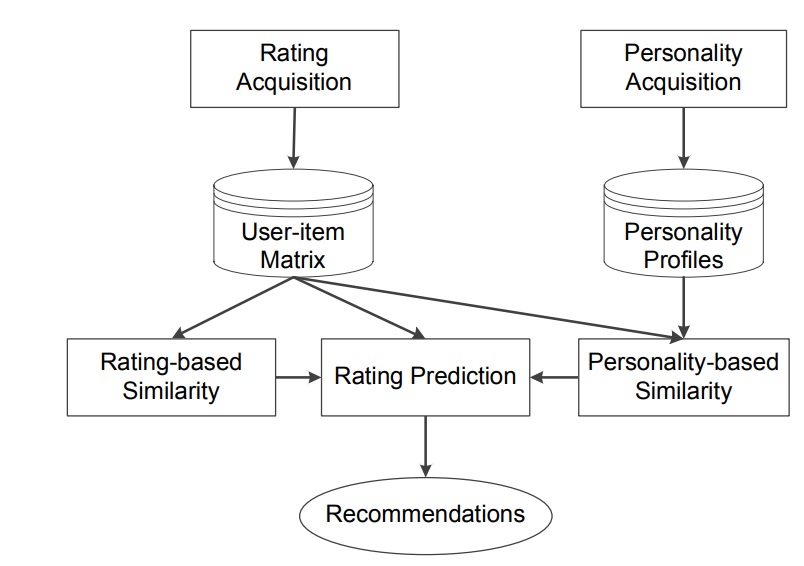}
\centering
\caption{Overall framework of Hu~\cite{DBLP:conf/recsys/HuP11}, which consists of two main parts, namely rating-based and personality-based components.} 
\label{Hu}
\end{figure*}
    \item \textbf{Hu.} This is a personality-based 
    RS in which the correlations between users’ personality types and their interests in music genres are explored~\cite{DBLP:conf/recsys/HuP11}. The framework of Hu is illustrated  in Figure~\ref{Hu} and includes two core components, namely the rating-based part on the left side of this figure and the personality-based part on the right side of the figure. Rating-based modules try to identify the most similar users in terms of their rating pattern, while the main focus of the personality-based module is on discovering a personality-based neighbourhood. Then, a final score is computed using Equation~\ref{HUEQ}. The authors provided users with personality quizzes to explicitly detect their personality traits and then used Pearson correlation coefficient to understand how similar these users were. In this model, a user's personalities with the ratings information is linearly combined under the CF framework. In particular, the similarity between user $u$ and user $v$
    can be measured by using the following formula:
    \begin{equation}
        sim (u,v)= \alpha simr(u,v) + (1-\alpha)simp(u,v)
        \label{HUEQ}
    \end{equation}
    
where $simr(u,v)$ and $simp(u,v)$ represent  the item-based and  personality-based similarity between user $u$ and user $v$, respectively. In this equation, $\alpha$ is the weighting parameter which controls the contribution of each of the two components in the final score and is set it during their experiment.  

\begin{figure*}[b!]
\includegraphics [width=0.7\textwidth, scale=1]{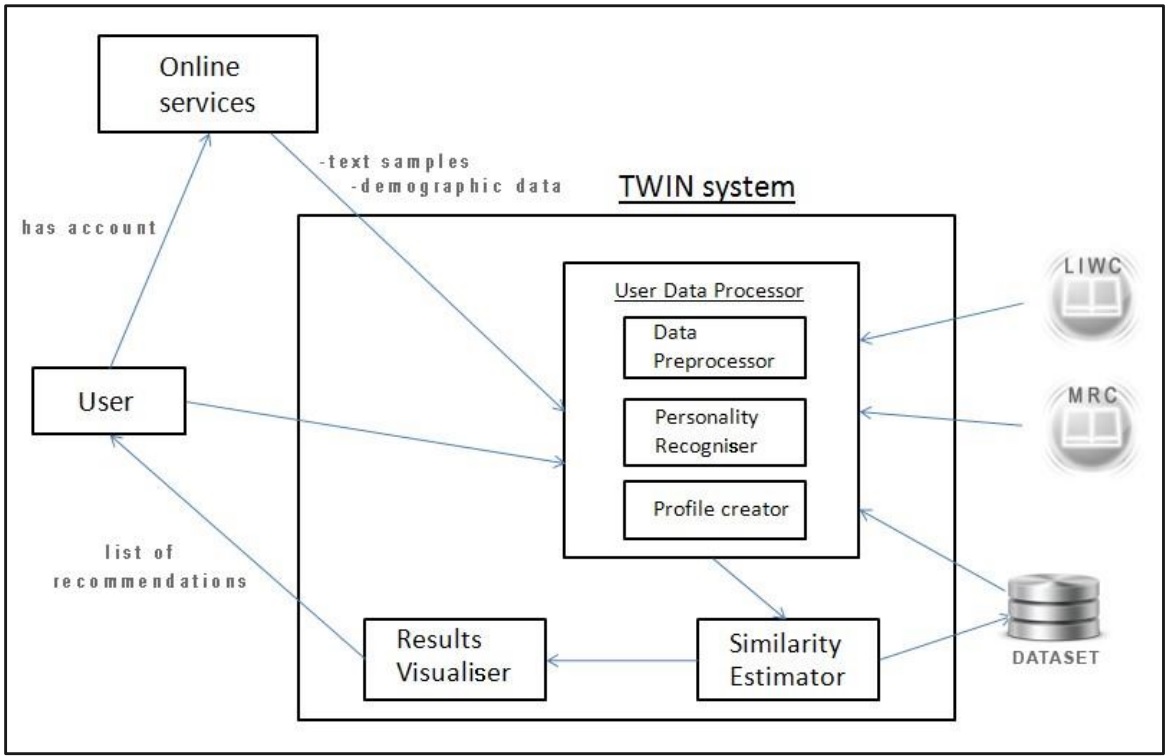}
\centering
\caption{Overall framework of TWIN~\cite{DBLP:journals/jifs/RoshchinaCR15} including data processor, personality recogniser, profile creator, similarity estimator, and results visualiser.} 
\label{TWIN}
\end{figure*}
     \item \textbf{TWIN.} `Tell me What
I Need' (TWIN) is an example of a recommender system that estimates the users’ similarity in
terms of their personality traits with the Euclidian distance~\cite{DBLP:journals/jifs/RoshchinaCR15}. Similar to the work done by Mairesse et al.~\cite{DBLP:journals/jair/MairesseWMM07}, in this study, the user's personality types were estimated implicitly on the basis of their written texts. The researchers have crawled all the resources that a person had already provided such as generating personal information, giving rate, and answering any questions. Then, they have evaluated this information regarding 24 important features related to the user's personality type, as discussed by Mairesse et al.~\cite{DBLP:journals/jair/MairesseWMM07}. Next, they have identified a set of neighbours for a target user with the k-nearest neighbour approach (KNN). Finally, Euclidian distance was applied to estimate the similarity between the users' profiles as follows:

\begin{equation}
\label{Euclidian distance}
  \sqrt{(a_1^{(1)} - a_1^{(2)})^2}  +  (a_2^{(1)} - a_2^{(2)})^2 + ...+ (a_n^{(1)} - a_n^{(2)})^2
\end{equation}
 
where $n$ is the number of attributes and $a_n$ are the attribute values. TWIN is applied in the online travelling domain, TripAdvisor~\footnote{https://www.tripadvisor.com/}, to test the impact of incorporating a user's personality type in the recommender system and has shown promising results. 
\end{itemize}

\subsubsection{Traditional Collaborative Filtering-Based RSs}
In Chapter~\ref{ch5:chapter5Personality}, we use personality information as additional information to alleviate the data sparsity problem and increase the accuracy of the recommendation systems. Therefore, in order to test the performance of the proposed model, we not only compared the proposed model with the personality-based recommender systems but also compared the model with classic recommender systems that incorporated side information. Below, we list those models as follows:

\begin{figure*}[b!]
\includegraphics [width=0.5\textwidth, scale=1]{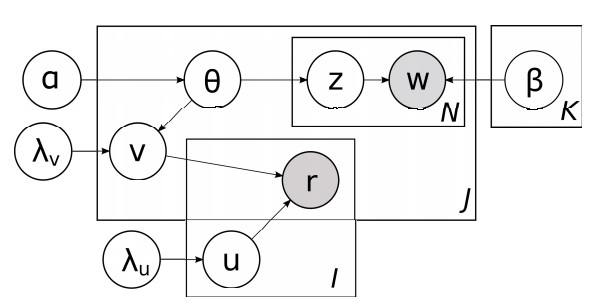}
\centering
\caption{Graphical model for the CTR framework~\cite{DBLP:conf/kdd/WangB11}.} 
\label{CTR}
\end{figure*}
\begin {itemize}
\item \textbf{CTR.} The collaborative topic regression
(CTR) model was proposed by Wang et al.~\cite{DBLP:conf/kdd/WangB11} to combine traditional collaborative filtering and probabilistic topic modelling for recommending scientific articles to users of an online community. The graphical structure of this model is presented in Figure~\ref{CTR}, where latent Dirichlet allocation (LDA)~\cite{DBLP:journals/jmlr/BleiNJ03} is used for the topic modelling purposes. In this figure, for each word $n$, there is a topic assignment $z_{jn}$ $\sim$ $Mult$ $(\theta_j)$, and for each word in article $w_j$, $w_{jn}$ $\approx$ $Mult$ $(\beta _{z_{jn}})$.

Wang et al.~\cite{DBLP:conf/kdd/WangB11} updated the matrix factorisation technique which is basically used for purely CF models and then used the EM style algorithm (i.e. expectation–maximisation (EM) algorithm as an iterative method to find (local) maximum likelihood; see~\cite{EMAlgorithm} for more detail) to learn the maximum a posteriori (MAP) estimates. Next, the following updating rules are generated for $u$ and $v$:
\begin{equation}
    u \leftarrow {(VCV^T + \lambda I^K)}^{-1}VCR  \quad\mathrm { and} \quad {    v \leftarrow {(UCU^T + \lambda I^K)}^{-1}(UCR + \lambda \theta)}
\end{equation}
where $U$ and $V$ are the user and the item matrices, respectively; $C$ is a diagonal matrix, $R$ is a rating matrix, $I$ is an identity matrix, $\lambda$ is the regularisation parameter,  and $\theta$ is the topic proportion.

\item \textbf{SVD++.} This is one of the representative approaches in recommender systems, which combines the benefits of both the neighbourhood approach and the latent factor models. SVD++~\cite{DBLP:conf/kdd/Koren08} also achieves a significant improvement by exploiting both the explicit and the implicit feedback by the users. The final proposed formula for this combination is presented below:

\begin {dmath}
    r^{\land}_{ui}= \mu +b_{i}+ q_{i}^{T} \left( p_{u} +|N(u)|^{\frac{-1}{2}} \sum _{j \in N_{(u)}}y_i \right) +|R^k (i; u)|^{\frac{-1}{2}} \sum _{j \in R ^k (i;u)} (r_{ui}-b_{uj}) w_{ij} + |N^k (i;u)|^{\frac{-1}{2}} \sum _{j \in N ^k (i;u)} c_{ij}
    \end {dmath}

where $p_u$ and $q_i$ are user $u$'s and item $i$'s factors vectors, respectively; $y_i$ is the item $i$' associated parameter; $r_{ui}$ is the rating score for user $u$ for item $i$ and $b_{ui}$ is its baseline; and $b_i$ is the item $i$ baseline. The weight from item $j$ to item $i$ is denoted by $w_{ij}$, and $c_{ij}$ is the offset added to the baseline estimation. $R(u)$ denotes all the explicit preferences for user $u$, while $N(u)$ is its implicit ones. In order to avoid confusion, we will not delve into the details of the more complicated mathematical concept here. However, for those who are willing to understand more details of this model, we will encourage them to deep-dive into this algorithm.

\item \textbf{UserMean.} This is a basic model in which the mean of all the ratings that a user has given to items is computed.
\begin{equation}
    r_{ij}=\frac{1}{N} \sum _{i=1} ^n r_{i}
\end{equation}
where $r_{ij}$ is a given rating to item $i$ by user $j$ and $n$ is the total number of items.

\item \textbf{ItemMean.} This model is similar to  ItemMean; however, this time, the mean of all the ratings that an item has received by the users is calculated as follows:
\begin{equation}
    r_{ij}=\frac{1}{K} \sum _{j=1} ^k r_{j}
\end{equation}

where $k$ is the total number of users.
\item \textbf{Random.} This model randomly assigns the rating values.

\item \textbf{Top.} This model recommends the top popular items determined during the training process.

\end{itemize}

\subsection{Baselines for Sequential Recommenders}
In this section, we will introduce the representative session-based recommender systems used in this thesis as the comparison methods. 

\begin{itemize}
\begin{figure*}[t!]
\includegraphics [width=0.6\textwidth, scale=1]{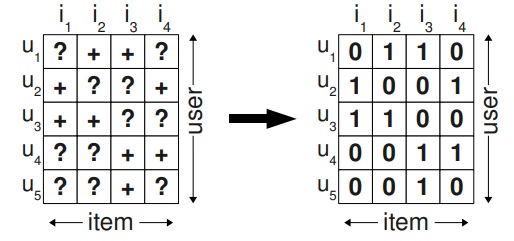}
\centering
\caption{Left matrix represents the observed items. The right matrix stores positive feedback as 1, and the unobserved items as negative feedback, which is equal to 0~\cite{DBLP:conf/uai/RendleFGS09}.} 
\label{BPR-part1}
\end{figure*}
\begin{figure*}[b!]
\includegraphics [width=0.6\textwidth, scale=1]{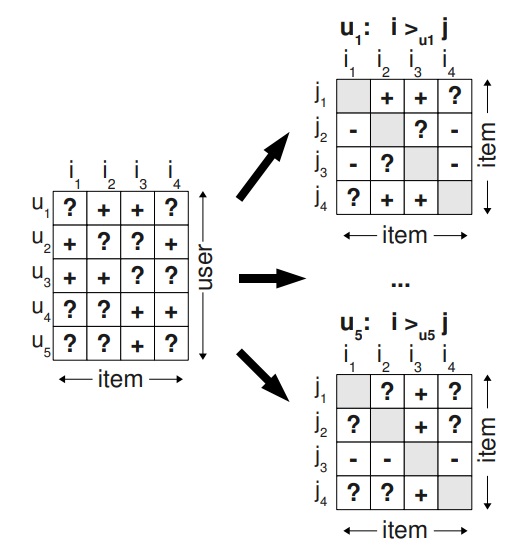}
\centering
\caption{Left matrix records observed items by a user, while the right matrices are inferred by BPR which applies a pair-wise ranking~\cite{DBLP:conf/uai/RendleFGS09}.} 
\label{BPR-part2}
\end{figure*}
\item \textbf{BPR.} This is a state-of-the-art method for item recommendation  with a pair-wise ranking loss through implicit feedback~\cite{DBLP:conf/uai/RendleFGS09}. Since the implicit feedback is different from the explicit one, typical machine learning models may find it difficult to deal with it. In implicit feedback systems, there is no difference between the real negative feedback (items that the user may not prefer to purchase) and the missing values (items that the user may not be aware of). Thus, it is difficult for a recommender system to directly learn the users' preferences from this matrix. As is clear from Figure~\ref{BPR-part1}, the positive feedback is treated as $1$, and the negative feedback is considered $0$. While the right matrices in Figure~\ref{BPR-part2} are inferred by BPR, in which a ranking order such as $i>_u j$ between items is calculated to find the users' preferences correctly. On the right side of this figure, $(+)$ means that a user prefers item $i$ over item $j$, while $(-)$ indicates that a user prefers item $j$ over item $i$. In this study, a generic optimisation criterion BPR-Opt for personalised ranking is presented to recommend a personalised ranked list of items which the user may prefer to consume. Given $\theta$ as the model parameter, the Bayesian formulation is to personalise the ranking for all items $i \in I $ by maximising the following posterior probability:
    
    \begin{equation}
        p(\theta| >_u)\propto P(>_UU | \theta) P(\theta)
    \end{equation}
 
where $>u$ is the desired but latent preference structure for user $u$.

\item \textbf{FPMC.} Factorising Personalised Markov Chain (FPMC), which was proposed by Rendel et al.~\cite{DBLP:conf/www/RendleFS10}, integrates the Markov Chain (MC) as a commonly used method in SRSs to model sequential behaviours and  Matrix Factorisation (MF) for modelling users' long-term preferences. FPMC  just utilises a linear aggregation function to combine these two different types of users' preferences. 

\begin{equation}
    p(i \in B_t^u|B_{t-1}^u)= \frac{1}{|B_{t-1}^u|} \sum _{l \in B_{t-1}}^u p(i \in B_t^u |l \in B^u _{t-1})
\end{equation}

where $B_t^u$ is the purchase history of user $u$ at time $t$, which consists of the total $i$ number of items; $l$ is the last item in the basket. 
\item \textbf{Fossil.} Factorised Sequential Prediction with Item
SImilarity ModeLs (or Fossil in short) is a combination of similarity-based methods and Markov chains for the sequential prediction task~\cite{DBLP:conf/icdm/HeM16}. One of the benefits of Fossil is its ability to deal with the cold-start problem. The basic form of Fossil can be expressed  as follows:

\begin{equation}
       p_u(j|i)\propto \sum _{j^\prime \in I_u^+/ \{j\} } <P_{j^\prime }, Q_{j}> + (\eta +\eta_u). <M_i, N_j>
\end{equation}
where $I_u^+$ is the set of items that user $u$ prefers, $\eta$ is the weighting parameter shared by all the users, and each item $i$ is parameterised  with four vectors $M$ , $N$, $P$,and $Q$. Then, in order to reduce the parameters, two matrices are forced to $P = M $  and $Q = N$. Thus, the new formulation can be expressed as follows:
\begin{equation}
     p_u(j|i)\propto \beta _j + <\frac{1}{|I ^+_u / {j}|^ \alpha} \sum _{j^ \prime \in I^ +_u / \{j\}} P_{j^ \prime} + (\eta +\eta_u). P_i, Q_j >
\end{equation}
 where $\beta _j$ is the bias term for item $j$.

\begin{figure*}[t!]
\includegraphics [width=0.6\textwidth, scale=1]{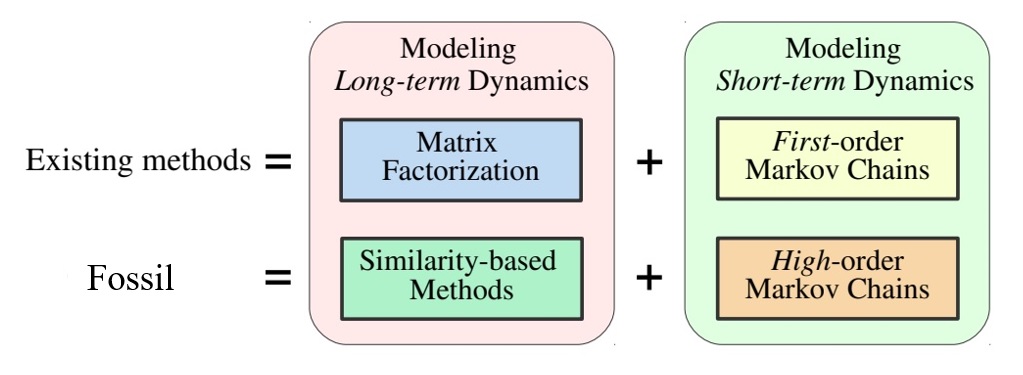}
\centering
\caption{Difference between Fossil and the current models, in which Fossil combines two different methods to provide a better suggestion~\cite{DBLP:conf/icdm/HeM16}.} 
\label{Fossil}
\end{figure*}
\begin{figure*}[b!]
\includegraphics [width=0.7\textwidth, scale=1]{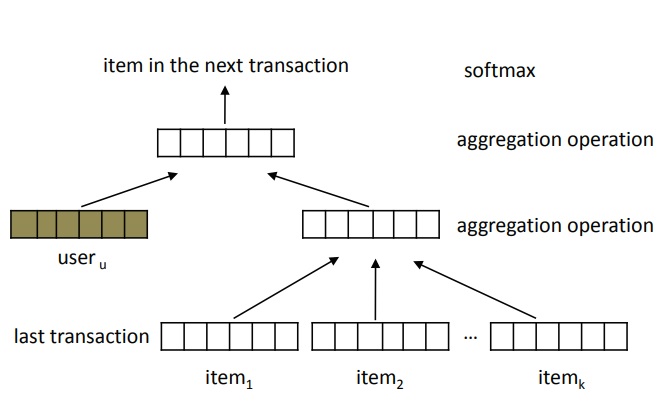}
\centering
\caption{Framework of HRM model: a two-layer hierarchical structure which builds a hybrid representation of the user and item interactions to predict the next purchased items~\cite{DBLP:conf/sigir/WangGLXWC15}.} 
\label{HRM}
\end{figure*}
\item \textbf{ HRM.} This model introduces a hierarchical representation to learn both the users' general taste and their sequential behaviour~\cite{DBLP:conf/sigir/WangGLXWC15}.  Given a user $u$ and two consecutive transactions $T_{t-1}^u$ and $T_t^u$, the probability of purchasing the next item $i$ is defined by HRM, as follows: 

\begin{equation}
    p(i \in T^u_t|u,T^u_{t-1}) = \frac{exp ({\vec{v_i}}^I. \vec{v}_{u,t-1}^{\small Hybrid}) }{\sum _{j=1}^{|I|} ({\vec{v_j}}^I. \vec{v}_{u,t-1}^{\small Hybrid})}
\end{equation}
where $\vec{v}_{u,t-1}^{\small Hybrid}$ is the hybrid representation which is defined as follows:

\begin{equation}
    \vec{v}_{u,t-1}^{\small Hybrid} := f_2(\vec{v}_u^U,f_1(\vec{v}_l^I \in T^u_{t-1}))
\end{equation}
where $f_1(.)$ and $f_2(.)$ are the aggregation operations at different layers of the HRM structure.

\begin{figure*}[b!]
\includegraphics [width=\textwidth, scale=1]{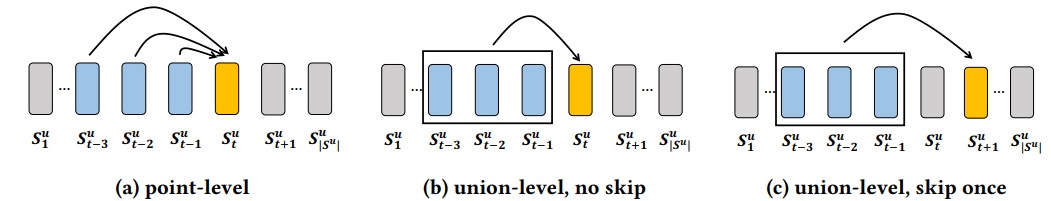}
\centering
\caption{Example of point-level, union-level, and skip-level behaviours in~\cite{DBLP:conf/wsdm/TangW18}.} 
\label{Caser}
\end{figure*}
\item \textbf{ Caser.} Convolutional Sequence
Embedding Recommendation Model is a state-of-the-art model, which uses CNN for sequence embedding~\cite{DBLP:conf/wsdm/TangW18}. As stated by Tang et al.~\cite{DBLP:conf/wsdm/TangW18}, most of the current Markov chain-based approaches only model point-level sequential patterns in which each previous action (blue part in Figure~\ref{Caser}) is considered individually, whereas a collection of actions (yellow part in Figure~\ref{Caser}) can have a significant effect on the target action. Those approaches which only consider modelling point-level behaviour may not be able to model the union-level and skip-level patterns shown in Figure~\ref{Caser}. While in a the union-level pattern, a set of joint several previous actions can have an effect on the target action, while skip behaviours mean that previous actions still have their impact on the target action even if a few previous steps are skipped. Caser proposes to use CNN filters to capture sequential patterns at the point-level, at the union-level, and of skip-level behaviour. It pays equal attention to both long- and short-term users' preferences;  considering $C^u=\{L+1, L+2, ..., |S^u|\}$ to be the collection of the previous actions of user $u$ at different time steps, the likelihood of all the sequences can be defined as follows:

\begin{equation} 
    p(S|\Theta)= \prod _u \prod _{t \in C^u} \sigma (y_{S^u_t}^{(u,t)}) \prod _{j\ne S_t^u} (1-\sigma (y_{j}^{(u,t)}) )
\end{equation}

where $\sigma =\frac{1}{1+e ^{1-x}}$ is the sigmoid function, $\Theta$ is the model parameters, and $y^{(u,t)}$ is the value of the output layer.

\begin{figure*}[b!]
\includegraphics [width=\textwidth, scale=1]{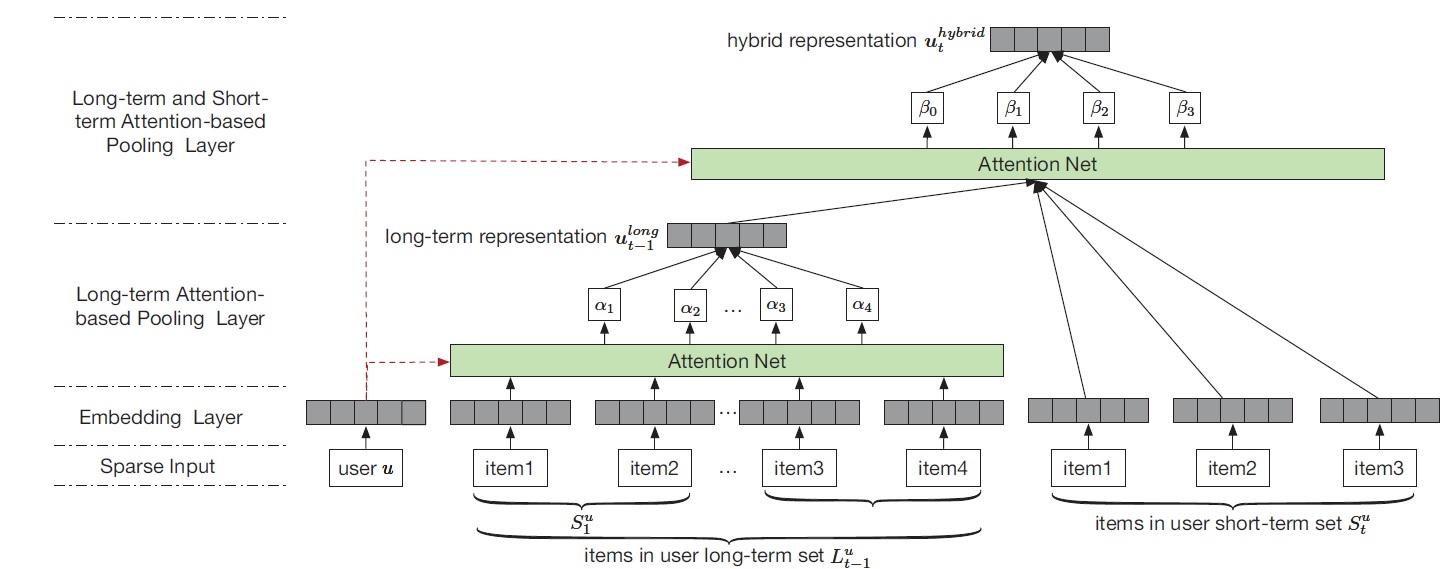}
\centering
\caption{Architecture of SHAN in which a combination of long- and short-term users' preferences are learned by a sequential hierarchical structure~\cite{DBLP:conf/ijcai/YingZZLXXX018}.} 
\label{SHAN}
\end{figure*}
\item \textbf{SHAN.} Sequential Hierarchical Attention Network is a state-of-the-art method in SRSs, which combines sequential behaviours with the users' general taste through the hierarchical attention-based structure~\cite{DBLP:conf/ijcai/YingZZLXXX018}. In this study, an attention mechanism was applied to both long- and short-term users' preferences to retain their dynamic properties. The attention mechanism can automatically assign different weights to the items to capture the dynamic property. The hybrid user representation is calculated by SHAN as follows:
\begin{equation}
    u_t^{hybrid}=\beta _0 u_{t-1}^{long} + \sum _{j \in S _t ^u} \beta _j v_j
\end{equation}
 where $\beta _0$ is the weight of long-term users' preferences and  $\beta _j $ is the attention score of different items in a short-term item set, which is computed as follows:
 
 \begin{equation}
 \label{SHANBeta}
    \beta _j = \frac{exp (u^Th_{2j})}{\sum _{p \in S _t ^u \cup \{0\}}   exp (u^T h_{2p})}
 \end{equation}

where $ h_{2p}$ is the high-level representation of item $j$ in the short-term item set for user $u$, and as is clear from Figure~\ref{SHAN}, $S _t$ and $L_{t-1}$ are the short-term and long-term items set, respectively.

\begin{figure*}[b!]
\includegraphics [width=0.6\textwidth, scale=1]{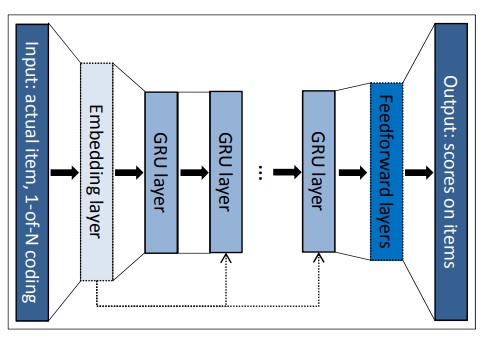}
\centering
\caption{Very general design of the GRU4Rec network~\cite{DBLP:journals/corr/HidasiKBT15}.} 
\label{GRU4Rec}
\end{figure*}
\item \textbf{GRU4Rec.} This is a state-of-the-art sequential recommender, which applies a modern recurrent neural network (GRU) to model the whole session~\cite{DBLP:journals/corr/HidasiKBT15}. Recurrent Neural Networks (RNNs) have demonstrated their success in working with variable-length sequence data. However, an improved version of RNNs such as a Gated Recurrent Unit (GRU)~\cite{DBLP:journals/corr/ChoMBB14} shows the strength of the vanishing gradient problem and memorising the very long-distance previous states by using the update and reset gates, which are defined as follows:

\begin{equation}
\label{update}
    z_t= \sigma (W_z x_t+ U_z h_{t-1})
\end{equation}

\begin{equation}
\label{reset}
    r_t= \sigma (W_r x_t+ U_r h_{t-1})
\end{equation}

where Equations~\ref{update} and~\ref{reset} are used for computing the update gate and the reset gate. $U$ projects the input into a hidden space; $W_z$, and $W_r$ are the weighting matrices for the update gate and the reset gate, respectively; and $x_t$ is the input at time $t$.

The general form of the GRU4Rec architecture is shown in Figure~\ref{GRU4Rec}, where the actual state of the sessions is projected into the network, and then, a set of interested items in a next session is predicted. There are varying numbers of GRU layers between the input layer and the output layer. Despite the normal form of RNNs which uses a fixed vector for input, here, a normalised weighted sum of items' representations is fed into the neural network which can reinforce the memory effect to capture very local ordering constraints. Finally, in this design, additional feed-forward layers are added between the last layer and the output layer.

\begin{figure*}[b!]
\includegraphics [width=0.8\textwidth, scale=1]{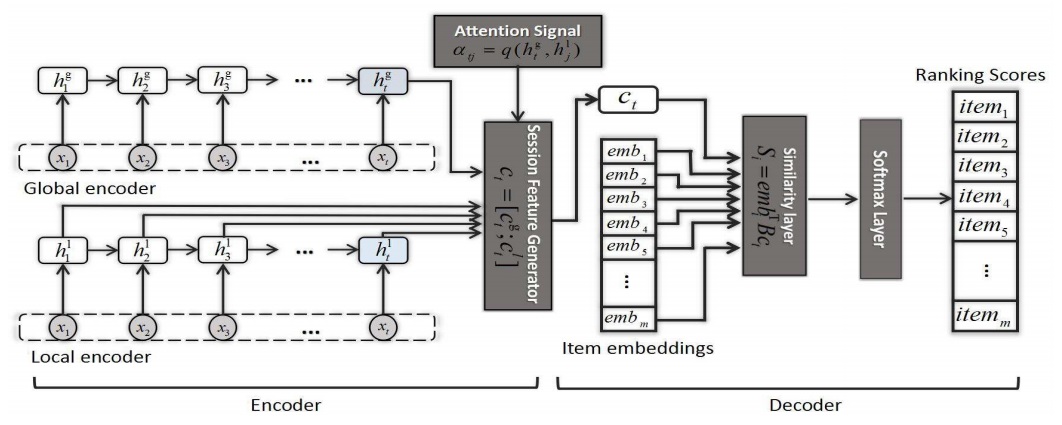}
\centering
\caption{Encoder-decoder architecture of NARM model, where the user’s main purpose in the current session is captured by the local encoder, and the global encoder is used to model the user’s sequential behaviour; next, the combination of the results of both of these components is used to compute the recommendation score for each item~\cite{DBLP:conf/cikm/LiRCRLM17}.} 
\label{NARM}
\end{figure*}
\item \textbf{NARM.} Neural Attentive Recommendation Machine is a neural encoder-decoder sequential recommender, which combines a recurrent neural network with an attention network~\cite{DBLP:conf/cikm/LiRCRLM17}. The authors stated that a user’s main purpose in the current session is ignored in most of the current RSs. The global encoder and the local encoder are used in the NARM structure to capture the user’s sequential behaviour and the main purpose of the current session, respectively. As is shown in Figure~\ref{NARM}, $c_t$ is a unified representation of both the user’s sequential behaviour
and the main purpose in the current session, which is calculated as follows:
\begin{equation}
    c_t=[h_t^g ; \sum _{j=1}^t \alpha _{tj}h_t^l]
\end{equation}

while $h_t^g$ is the last hidden state of the global encoder and $h_t^l$ is the  hidden state of the local encoder, aiming to compute attention weight $\alpha$, where $\alpha_{tj}$ calculates the alignment between the final hidden state $h_t$ and the representation of the previously interacted item $h_j$.

\begin{figure*}[b!]
\includegraphics [width=0.7\textwidth, scale=1]{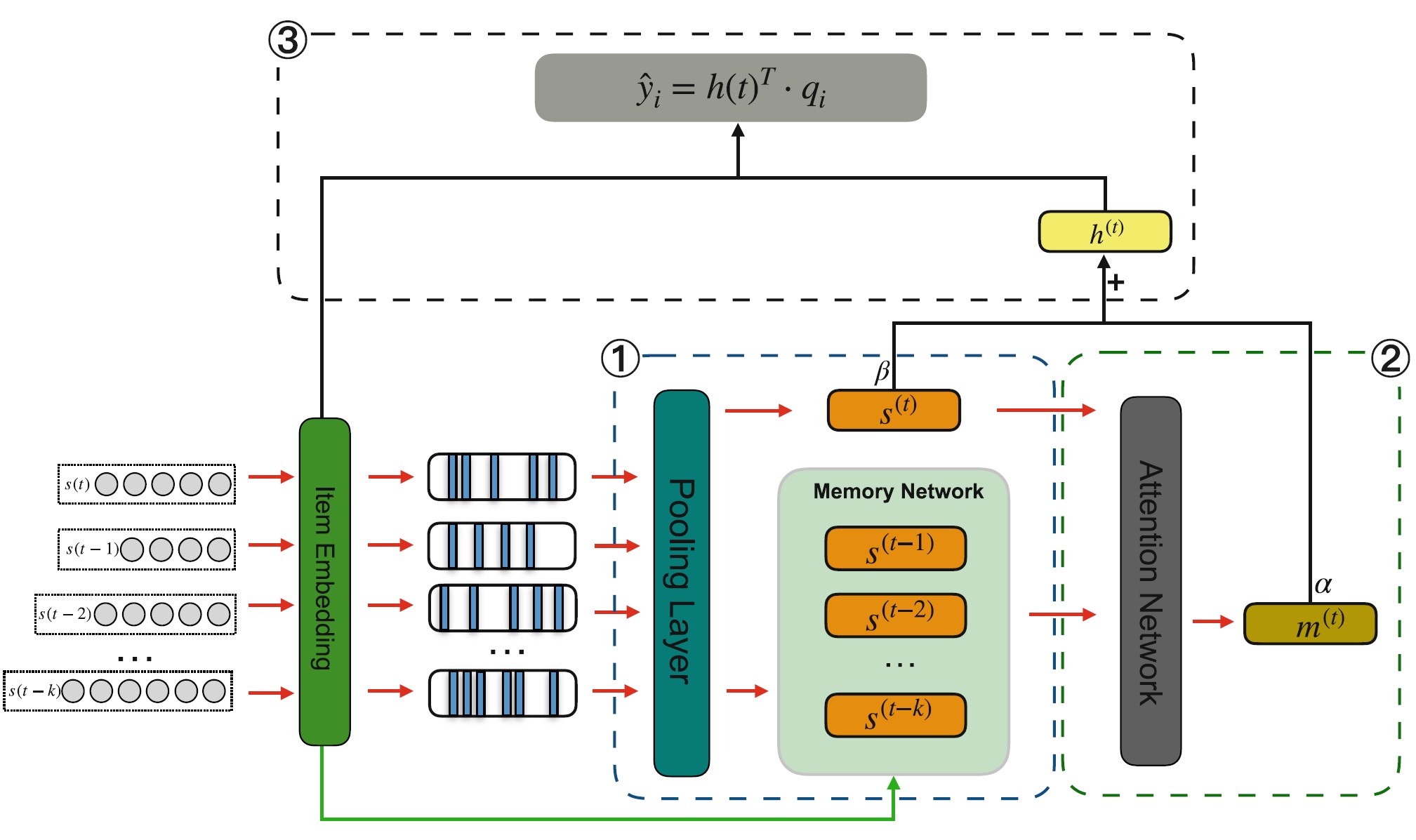}
\centering
\caption{Graphical structure of MEANS~\cite{DBLP:conf/wise/HuHSN19}.} 
\label{MEANS}
\end{figure*}
\item \textbf{MEANS.} Memory-Augmented Attention Network
for Sequential Recommendation is a recommender system in which a mixture of long-term and short-term preferences is learnt for making recommendations~\cite{DBLP:conf/wise/HuHSN19}. The authors believed that the local patterns within the current session are missed by most of the current RSs, as they may assume each item to be a separate entity in a session. Therefore, MEANS was proposed to capture the high-level information in the sessions. This structure consists of three main components: $(1)$ a memory module to store the most current sessions, $(2)$ an attention mechanism to learn long-term users' preferences, and $(3)$ a prediction layer in which a recommendation is made by learning a mixture of long- and short-term preferences. MEANS first  operates a max-pooling technique on the most recent sessions to extract the more salient features from item embedding, and then, the results are stored in an external memory.  

Given $S_u^{(t)}=\{v_1^{(t)}, v_2^{(t)},..., v_n^{(t)}\}$ as a session at time $t$ for user $u$ which stores the interaction of user $u$ with item$v$, $s_u^{(t)}$ as the current session representation can be computed  as follows:
 
\begin{equation}
    s_u^{(t)}=Max-Pooling (v_1^{(t)}, v_2^{(t)},..., v_n^{(t)})
\end{equation}

Next, the long-term user representation $m_u^{(t)}$ can be acquired as a weighted sum of the most recent $k$ session as follows:

\begin{equation}
    m_u^{(t)}=\sum _{i=1} ^k \alpha _i s_u ^{(t-i)}
\end{equation}

where $\alpha_i$ is the attention score which can reflect the level of contribution of item $i$ in the long-term users' preferences. Finally, the user’s representation $h_u^{(t)}$, which is a combination of both the long- and the short-term users' preferences can be obtained as follows:

\begin{equation}
    h_u^{(t)}= \alpha m_u ^{(t)} + \beta s_u ^{(t)}
\end{equation}

where $\alpha$ and $\beta$ are the weighting parameters to control the weights of the long-term and the short-term users' preferences, respectively.

\end{itemize}

\chapter{Enabling the Analysis of Personality Aspects
in Recommender Systems}
\label{ch5:chapter5Personality}

\section{Introduction}
In this chapter, we will focus on one of the main challenges at the interaction level in general recommender systems: the data sparsity of user-item interaction, when there is a lack of availability of common feedbacks among them, as shown in the hierarchical structure in Figure~\ref{ch1:figureRWcat} in Chapter~\ref{chap:introduction}. In this chapter, we present a personality-based recommender system to deal with the problem of sparse user-item interactions. We will introduce a problem called DSW-n-FCI, which is a subset of the data sparsity problem. We propose a novel algorithm to identify the users’ personality types implicitly with no burden on the users and incorporate this important feature along with the users’ personal interests and their level of knowledge. This chapter is also related to the first challenge that we have addressed, as shown in the thesis structure in Figure~\ref{ch1: figureThesisOutline} in the first chapter. The results of the proposed approach in this chapter are accepted and published in several papers by $13^{th}$ ACM International WSDM Conference 2020~\cite{DBLP:conf/wsdm/BeheshtiHYMG020} (Core Rank A*~\footnote{https://www.core.edu.au/conference-portal}), the Pacific Asia Conference on Information Systems (PACIS) 2019~\cite{Yakhchi} (Core Rank A), and Web Information Systems Engineering Society (WISE) Workshop 2018~\cite{DBLP:conf/wise/YakhchiGB18} (Core Rank A).

\subsection{The Target Problem and The Motivation}
In the last two decades, we have witnessed the emerging growth of the generated information by people’s daily activities (e.g. browsing, clicking, listening to music, and purchasing items). Because of this information explosion, people are surrounded by a large number of options and services. Recommender systems have appeared as filtering tools which  can help customers with their decision-making process, find items of their interest, and alleviate the information overload problem. As discussed in Chapter~\ref{ch:chapter3PreliminariesandFoundations}, Section~\ref{ch:chapter4MF} in particular, the matrix factorisation  technique on Collaborative Filtering (CF) is one of the widely adopted solutions for general recommenders. Collaborative filtering is a method for making automatic predictions about the interests of a user by collecting preferences
from many users. Although pure collaborative filtering-based recommender systems have been studied well in the literature, the data sparsity problem still remains as one of the major difficulties of these models. CF-based techniques try to predict the interests of a particular user in items by discovering the previously rated similar items. In this situation, they may
fail when there is no captured previous feedback on common items. There are several attempts to alleviate the data sparsity problem by incorporating additional side information such as the users’ generated contents~\cite{DBLP:conf/kdd/WangB11}, \cite{DBLP:conf/pacis/HuLCTD16}, \cite{DBLP:conf/aaai/BaoFZ14} or social network information~\cite{DBLP:conf/wsdm/MaZLLK11}. The semantic correlation between users and items can be captured to help users with their options even in the case of the data sparsity problem~\cite{DBLP:conf/aaai/ChenZWHC16}. Moreover, rule-based techniques can generate a rule by exploiting the users’ reviews from different aspects~\cite{DBLP:conf/pacis/ChenWL17}.
While the mentioned approaches have been proposed to overcome the data sparsity problem, they may fail in the case of the DSW-n-FCI problem, as they considerably rely on discovering a set of users who previously rated similar items. Moreover, they resort to external information to deal with the data sparsity problem. Therefore, most of the existing models such as CTR~\cite{DBLP:conf/kdd/WangB11} and SVD++~\cite{DBLP:conf/kdd/Koren08} may not be able to provide any suggestions for users in the DSW-n-FCI situation, as they cannot find any connections among users.

In contrast  to such approaches, personality-based RSs do not need to look up  common items of interest to find similar users, and thus, they may provide better results even with sparse data. Psychologically speaking, personality is a consistent behaviour pattern that a person tends to show irrespective of her/his situation. Personality  also has a strong correlation with the individuals’ interests, and people with similar personality types tend to share similar interests. Because of the strong correlation between personality traits and users’ preferences, recommender systems have inspired to incorporate
personality characteristics into their model to not only help users with a diverse set of items~\cite{DBLP:conf/chi/McNeeRK06} but also provide a better group recommendation~\cite{DBLP:conf/um/KompanB14} ,\cite{DBLP:conf/recsys/Recio-GarciaJSD09} and improve the accuracy of recommendations for music, movies, e-learning, and web searches~\cite{DBLP:conf/um/HuP10}, \cite{DBLP:conf/hais/PaivaCS17}.

The Five Factor Model (FFM) is one of the widely adopted personality models in both the psychology and the computer science domains. FFM consists of five main personality traits, namely as Openness to Experience, Conscientiousness, Extraversion, Agreeableness, and Neuroticism, and is briefly known as OCEAN. As stated by psychologists, people with similar personality types, more or less, tend to consume similar items (e.g. music, movies, services, and products)~\cite{DBLP:journals/taffco/VinciarelliM14}. For instance, people with the Openness personality type mostly prefer to watch the comedy and fantasy genres of movies, while romantic movies are more likely to be watched by people with the Neurotic personality type~\cite{DBLP:conf/um/CantadorFB13}. We also adopted one of the leading text analysis tools with a lexical root, known as Linguistic Inquiry and Word Count (LIWC). LIWC can categorise the given textual content in  more than 88 psychological categories related to the FFM personality traits~\cite{Pennebaker}, \cite{DBLP:conf/chi/GolbeckRT11}. However, the main drawback of most of the existing studies is that they rely considerably  on explicit personality detection approaches, such as filling out a questionnaire. In an explicit personality detection task, users will be asked to answer some questions and conduct a survey. This is a time-consuming task, and users may be unwilling to participate because of privacy concerns~\cite{DBLP:reference/sp/TkalcicC15}, \cite{DBLP:conf/um/HuP10}, \cite{DBLP:conf/recsys/NunesH12}, \cite{Sitaraman}. Furthermore, people usually do not disclose much about themselves and probably will answer questions incorrectly, which can have a negative effect on the accuracy of a recommender system. Despite the above-mentioned benefits of incorporating personality types in recommender systems, only a few personality-based recommender systems have been reported in the literature. 

Besides the users' personality types, their level of knowledge in a particular domain can be another important factor that has a considerable  impact on the recommendations' acceptance rate. For example, assume that Mark is personally interested in the cooking and music domains. Mark is a music expert with a low level of knowledge of cooking. On the basis of this observation, he is less likely to be influenced by the other users’ suggestions in the music domain, while he is more willing to seek other users’ recommendations from the cooking domain as he has no expertise in this domain. Therefore, most of the aforementioned personality-aware methods not only detect the users’ personality types explicitly by having them fill out a questionnaire, which is an unfeasible  task in many real-world applications, but also ignore the differences between the users’ level of knowledge in different domains. The users' personal interests is another factor, which is mostly ignored by the current personality-based recommender systems. Let us make it more clear by giving an example, imagine Sarah and Joe are two users who give 2 and 5 stars (on a Likert scale of 0-5) to a particular item, respectively. This indicates that both the users are interested in this item, but perhaps, Joe likes it considerably more than Sarah as he gives a higher rating to this item. Therefore, taking the users' personal interests into consideration can help to make personalised recommendations, improving the users’ satisfaction rate and boost the business benefit.

In order to tackle the above-mentioned problems, in contrast to personality-based recommender systems that simply use personality as a similarity measure, we directly considered  this important factor in our mathematical model and propose a novel matrix factorisation model. The
assumption  of our method is that the users’ decision-making process consists of three major factors: users’ personality types, their level of knowledge in a particular domain, and their personal interests. We built a user-item interaction matrix with the real values of ratings in order to preserve the users’ actual interests in items and show their level of personal interest.

\subsection{Proposed Design and Main Contributions}
The above observation shows the importance of incorporating personality in RSs, general ones in particular. It is a factor in the users' decision-making process, as it is relatively stable over time and is relatively consistent over situations. Here, we propose a novel model, APAR, to investigate the effect of a user's personality type in recommender systems. In this chapter, we will identify the users’ personality types implicitly with no burden on users
and incorporate it along with the users’ personal interests and their level of knowledge. A two-step framework of APAR is illustrated in  Figure~\ref{APAR}. APAR analyses the pattern of users' activities and behaviours in order to find their personality types. The main contributions of the proposed model can be summarised as follows:

\begin{figure*}[b!]
\includegraphics [width=\textwidth, scale=1]{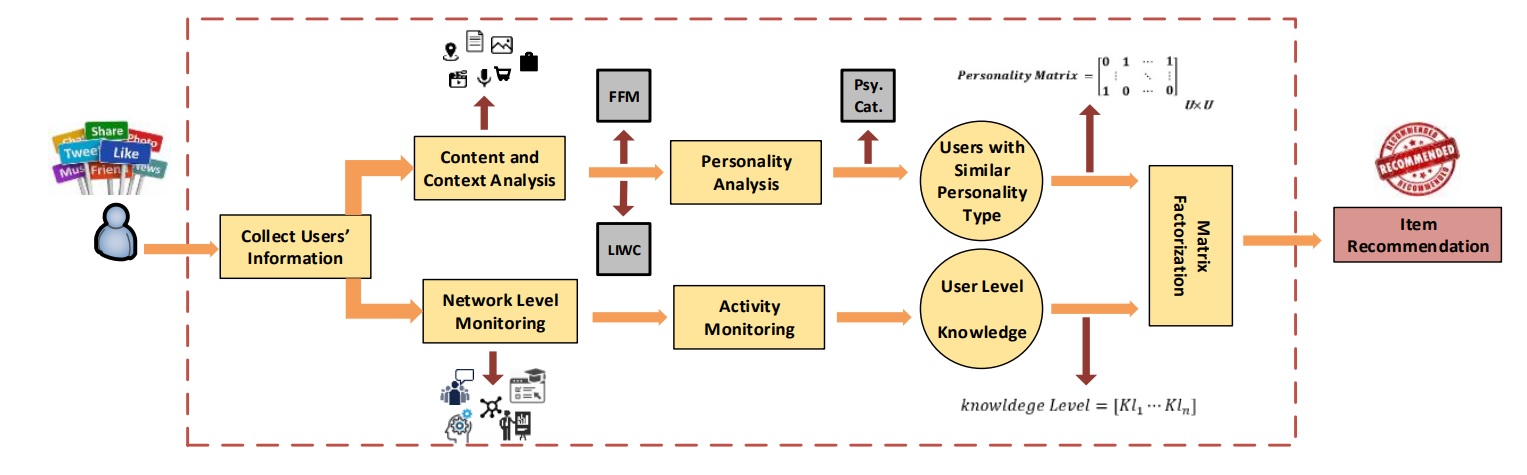}
\centering
\caption{Framework of APAR, which consists of two major components: users' personality detection and their level of knowledge extraction. In the personality detection module, `Psy. Cat' refers to psychology categories which we extract from the users' provided contents.} 
\label{APAR}
\end{figure*}



\begin{itemize}
    \item We develop an algorithm to construct a user-item interaction matrix with the actual ratings scores to show the level of users’ personal interest in the items.
    \item We propose a novel recommender system called APAR, integrating three main factors which may affect a user’s decision-making process such as the users’ personal interests, their personality types, and their level of knowledge.
    \item We detect the users’ personality types implicitly with no need for the users’ efforts by analysing their online-generated contents. 
\end{itemize}

The rest of this chapter is organised as follows: we present the overview and the framework in Section~\ref{ch:Chapter5OverviewandFramework}. The experiment and the evaluation are presented in Section~\ref{ch:Chapter5ExperimentandEvaluation}, before concluding the chapter in Section~\ref{ch:Chapter5summary}.

\section{Overview and Framework}
\label{ch:Chapter5OverviewandFramework}
In this section, we present the proposed  personality-based recommender system. We first constructed a user-item interaction matrix with the real values of ratings in order to preserve the degree of the users’ interests and their personal interests. Figure~\ref{APAR} depicts the proposed  novel framework called APAR, which consists of two major steps of the users' personality detection, and  their level of knowledge extraction.  The upper part of this structure is reserved for personality detection in which the users' textual contents are analysed, and `Psy. Cat.' in the middle indicates the extracted psychology categories from LIWC. At this stage, the generated users' raw data are collected and contextualised in order to discover the users’ main characteristics, such as their personality types. For the next phase, as shown in the bottom part of Figure~\ref{APAR}, all the activities of a user and her/his neighbours in a particular domain are monitored to ascertain her/his level of knowledge by finding how the other users are influenced by this user's opinions/comments. Then, we propose a novel matrix factorisation model that incorporates the users’ personality types, their level of knowledge, and their personal interests. 
In the following sections, we provide more detail
about each of the main phases of APAR.


\subsection{Problem Statement}
Suppose that there are $|v|$ total number of items $V=\{v_1,v_2,\dotsm,v_v\}$ and $|u|$ number of users $U=\{u_1,u_2,\dotsm,u_u\}$. Let $R \in R^{|u|\times |v|}$represents
the rating matrix, and $R_{ij}$ indicates the given rating to item \textit{i} by user \textit {j}, and if this item has not been seen by the user, there is no available rating and we show this unrated item with \textit{unr}. Let us consider $L \in R^{|v|\times |v|}$ as the personality matrix, where $L_{jj^ \prime}=\{0,1\}$; there is a direct connection between $u_j$ and $u_j^ \prime$ if they have the same personality type. In other words, $L_{j j^ \prime}=0$ denotes that $u_j$ and $u_ {j^ \prime}$ do not have similar personality types.\\
There are two different ways of constructing the user-item matrix from the users' given ratings. One solution, which is followed by most of the existing approaches, is to treat all the rated items as the same and equal to 1, while the rest of the unrated items are 0. Thus, there is no difference among all the rated items, while the rating value can indicate the level of the users' preferences for an item~\cite{DBLP:conf/sigir/HeZKC16},~\cite{DBLP:conf/www/HeLZNHC17}. However, in order to capture the users' personal interests completely and express the level of users' preferences, we constructed this matrix according to Equation~\ref{equation}~\cite{DBLP:conf/ijcai/XueDZHC17}.
\begin{equation}
\label{equation}
  W_{ij}=\begin{cases}
    0, & \text{if $R_{ij}=unr $}\\
    R_{ij}, & \text{otherwise}
  \end{cases}
\end{equation}
where $W \in R ^{|u|\times |v|}$,  $W_{ij} >0$ can represent the actual interest of user $i$ in item $j$, and if $W_{ij}=0$, there is no preference for this item by this user.

\subsection{Personality Acquisition}

In contrast to the traditional CF-based approaches, the users’ personality types, which can explain the wide variety of human behaviour, have inspired some recommender systems. Personality is a domain-independent concept along with individuals in a wide range of domains, such as music~\cite{DBLP:journals/jair/MairesseWMM07}, movies, and books~\cite{Rentfrow}. Moreover, personality can be acquired by analysing people's nonverbal behaviour such as their shared posts, their online activities, and their written review texts~\cite{DBLP:conf/aaaiss/SchwartzEDKBKSSU13}, \cite{DBLP:conf/ijcai/BeraRM17}. While there are too many theories that try to define the concept of human personality in a wide range of domains, including the
cognitive, biological, learning, and humanistic perspective~\cite{HandbookofPersonality}, the trait theory is one of the most widely adopted personality theories~\cite{traittheory}. According to this theory, human personality traits are habitual patterns of behaviours, thoughts, and emotions, and the Big-five personality trait model is one of the popular methods in trait theories. The Big-five model is also known as the five-factor model (FFM)
and is based on the common language description of personality. Therefore, FFM can be a good choice to work within various tasks including `computing technologies, natural language processing, machine learning, and semantic technologies'~\cite{DBLP:journals/tcss/NingDA19}.

Personality is a context-independent factor, which normally does not change over time and can be extracted either explicitly by using a questionnaire or implicitly by applying regression~\cite{DBLP:reference/sp/TkalcicC15}. There are several questionnaires based on the FFM model, NEO-Personality-Inventory Revised (NEOPI-R, 240 items) for instance, in which the participants’ personality types are revealed after they answer several questions~\cite{Costa}. While the explicit personality detection task may reveal a better understanding of a user’s personality type, it is a tedious and time-consuming task and the users may be unwilling to perform it. Instead, implicit personality detection models exploit the users' digital footprints and actions to analyse their behaviours with no extra burden on the users~\cite{DBLP:conf/recsys/AzariaH16}. Unlike most of the current methods which detect the users' personality types by asking them to fill out a questionnaire, in the present study, the users' personality types were measured implicitly with no need for their efforts. To do so, we collected all the written users’ reviews to categorise them with the LIWC tool. A sample of the categories of LIWC and the corresponding words is given in Table~\ref{table: liwc}. 

\begin{itemize}
    \item  \textbf{Openness to Experience:} from cautious/consistent to curious/inventive. People with the openness to experience personality tend to have intellectual curiosity and creativity and are open to having a variety of experiences. A higher score of openness means a higher degree of imagination, artistic interest, emotionality, adventurousness, intellect, and liberalism.
    
    \item \textbf{Conscientiousness:} from careless/easy-going to organised/efficient. Conscientiousness people are self-disciplined and focus on their personal achievements. These types of people are organised, and a high score of conscientiousness shows strong degrees of self-efficacy, orderliness, dutifulness, achievement-striving, and cautiousness.

    \item \textbf{Extraversion:} from solitary/reserved to outgoing/energetic. This type of person is more social and talkative and aims to find positive emotions, such as happiness, satisfaction, and excitation. A high score of extraversion denotes a higher tendency for friendliness, gregariousness, activity level, excitement-seeking, and cheerfulness.
    
    \item \textbf{Agreeableness:} from cold/unkind to friendly/compassionate. The main characteristics of people with this personality type are kindness, concern, truthfulness, and cooperativeness toward others. A high score of agreeableness reflects strong degrees of morality, altruism, sympathy, modesty, trust, cooperation, and conciliation.
    
    \item \textbf{Neuroticism:} from secure/calm to unconfident/nervous.  Anger, anxiety, depression, and vulnerability are the key features of this type of people, who lack emotional stability and impulse control. Individuals with a higher score of neuroticism demonstrate a higher level of hostility, social anxiety, depression, immoderation, vulnerability, and impulse.
\end{itemize}
\begin{table}[h]
\centering
\begin{tabular}{c c }
\hline\hline
LIWC category & Words\\ [1ex]
\hline
Anger words	& hate, kill, pissed \\
Metaphysical issues	& God, heaven, coffin\\
Friends & buddy, friend, neighbour\\
Family members & mom, brother, cousin\\
Past tense verbs & walked, were, had\\
References to friends & pal, buddy, coworker\\
Positive emotions & love, nice, sweet\\
Negative emotions & hurt, ugly, nasty\\
Sadness & crying, grief, sad\\
Prepositions & to, with, above\\
Family & daughter, husband\\
Humans & adult, baby, boy\\
Physical state/function & ache, breast, sleep \\
Cognitive process & cause, know, ought\\
Tentative & maybe, perhaps, guess\\
Insight & think, consider\\
Social processes & talk, us, friend \\ 
Achievement & hero, win, earn\\
Inclusive words & with, and, include\\
Motion & arrive, car, go\\
\hline
\end{tabular}
\caption{Examples of LIWC categories and corresponding words in each category.}
\label{table: liwc}
\end{table}
In order to determine the  users' personality types, we first collected all the users’ textual content such as reviews/tweets/posts, and then analysed them using the Linguistic Inquiry and Word Count (LIWC) tool. We used LIWC to understand how many words of the users’ textual content are related to each of the 88 categories of LIWC (such as positive emotions, cognitive process, and social processes). From Table~\ref{ch5:tbaleLIWCCategorise}, we can see how each of the five FFM personality traits is related to the categories of LIWC. For instance, according to this table, people of the openness to experience personality type are more likely to use the punctuation category (e.g. periods, commas, colons, and semicolons) and affect words (e.g. anger and sad). 

Inspired by Roschina et al.~\cite{DBLP:journals/jifs/RoshchinaCR15}, we used a linear regression model to measure a user’s personality traits as follows:
\begin{table*}[t!]
\centering
\begin{adjustbox}{width=\textwidth}
\begin{tabular}{lllll}
\toprule
\textbf{Open.} & \textbf{Consc.} & \textbf{Extra.} & \textbf{Agree.} & \textbf{Neuro.}\\  
\midrule
Punctuation &  Affect Words &Total pronouns     & Exclamation marks & Affect Words \\
 Affect Words& Death        & Exclamation marks   & Dictionary words  & Anger \\
 Apostrophes & Future       & Article   &Feel     & Anxiety \\
Achievement  & Home        & Friends             & Home    & Article \\
 Anger      & Prepositions  & Periods             & Singular Pronoun & Feel \\
 Home       & Anger         & Pronoun            & Anger            & Leisure \\
 Article    &  Body          & Question marks      & Negative emotion & Music \\
 Positive Feeling & Hear    & Positive Emotion    & Positive emotion & Number \\
 Assent     & Apostrophes    & Punctuation         & Body             & Apostrophes \\ Causation  & Certainty     &\ Apostrophes         & Family           &Exclamation marks \\
 Death      & Hear          & Parentheses         & Motion           & Family \\
 Family     & Job           & Body                & Negations        & Friends \\ Feel       & Music         & Certainty           & Parentheses      &Singular Pronoun \\ Friends    & Negations     & Family              & Pronoun          & Negations \\
 Singular Pronoun& Negative emotion & Fillers      & Future           & Negative emotion \\ Job        & Prepositions  & Other punctuation   & Periods          & Total pronouns \\
 Motion     & Question marks&  Singular Pronoun    & Achievement      & Prepositions \\
 Music      & Nonfluencies  & Music  & Anxiety          & Present focus  \\
\bottomrule
\end{tabular}
  \end{adjustbox}
  \caption{Relationship between the Big Five personality traits and different LIWC categories~\cite{DBLP:journals/jair/MairesseWMM07} (in no particular order),  based on the analysis from Pennebaker and King~\cite{LIWC}. The Linguistic Inquiry and Word Count (LIWC) categorises words into more than 88 categories. Here, the categories of LIWC related to the Big Five personality traits, namely `Open.', `Consc.', `Extra.', `Agree.', and `Neuro.', represent Openness to experience, Conscientiousness, Extraversion, Agreeableness, and Neuroticism, respectively.}
   \label{ch5:tbaleLIWCCategorise}
\end{table*}

\begin{equation}
\label{eq:personality}
E = w_1 X_1+ w_2 X_2+ w_3 X_3+ \dotsm + w_i X_i
\end{equation}
where $X_i$ and $w_i$ denote each of the LIWC's categories and its corresponding category weight, respectively, $i \in \{1, 2, ..., n\}$ and $n =88$, which is the total number of LIWC categories. The final score $E$ can show the level of each of the five personality traits, if we only place their correlated categories form LIWC and their corresponding weights, which can be extracted by using the method proposed by  Mairesse et al.~\cite{DBLP:journals/jair/MairesseWMM07}.




\subsection{ Acquiring Users’ Level of Knowledge }

The level of an individual’s knowledge is one of the main criteria  to determine the acceptance rate of her/his recommendations, which we termed as the users’ level of knowledge. In the real world, individuals may be an expert in one or two domains and  have a different level of knowledge in various domains. Therefore, it is important to take a user's level of knowledge into consideration, in order to provide a better recommendation. 

We denoted a user's  level of knowledge with $u_i$ in domain $d$, as $kl_i^d$, which could be computed as follows:

\begin{equation}
    kl_i^d= \frac{1}{n_i^d} \sum _{p=1} ^ {n_i^d} h_p ^ {i,d}
    \label{ch5:EquationKnowledge}
\end{equation}
where $n_i ^d$ is the total number of reviews left by user $u_i$ in domain $d$, and $h_p ^ {i,d}$ denotes the number of ratings that a user  $u_i$ has received by writing a review $p$ from the other users.

\subsection{Our Framework}

We computed user $u_i$'s interest in item $j$ in domain $d$ by using an inner product between the latent feature vector of item $j$, as $q_j^{(d)}$ and the latent feature vector of user $i$, as $p_i^{(d)}$. Matrix $ L$ contains the personality information, and $ \varphi _i ^{+ (d)}$ is the set of users who have the same personality type as that of user $i$, where $l_{ik}=1$ means that user $u_i$ and user $u_k$ have the same personality type. For the sake of simplicity, we denote $\gamma _i ^d$ $= \beta +  kl_i^d $, where $\beta$ is the controlling parameter that controls the weight of users' preferences. Finally, the rating score for the unobserved items can be calculated as follows:

\begin{equation}
    R_{ij}^{(d)}= \gamma_i^d p_i^{(d)^T}q_j^{(d)}+(1-{\gamma_i^d})\sum_{k \in \varphi_ i ^{+ (d)}} L_{ik}p_k^{(d)^T}q_j^{(d)}
\end{equation}
where $ R_{ij}^{(d)}$ predicts the rating value for the unobserved items. It is a combination of three critical factors of the users' decision-making process, namely users’ personality types, their level of knowledge, and their personal interests.

\begin{equation}
\begin{aligned}
\small
&\ min {\frac{1}{2}}\sum_{i=1}^{|v|}\sum_{j=1}^{|u|}I_{ij}^{(d)}\Bigg({R_{ij}^d}-\bigg(\gamma_i^d p_i^{(d)^T}q_j^{(d)} + \Big((1-\gamma_i^d) \sum_{k \in \varphi_i ^{+ (d)}} L_{ik}p_k^{(d)^T}q_j^{(d)}\Big)\bigg)\Bigg)^2 +\\
& \alpha_1\Vert P^{(d)}\Vert_F^2 + \alpha_2\Vert Q^{(d)}\Vert_F^2
\end{aligned}
\end{equation}
where $I_{ij}^{(d)}$ = 1, if user $i$ has rated item $j$; otherwise, $I_{ij}^{(d)}$ = 0. In order to prevent overfitting, we introduced $ \eta(j,j ^ \prime)$ as the personality coefficient between $u_j$ and $u_{j ^ \prime}$ with some features: $(1)$ $\eta (j, j ^ \prime)$ $\in \{0,1\}$, ~$(2)$ ~$\eta(j, j ^ \prime)= \eta(j, j ^ \prime)$, and $(3)$ if $\eta(j, j ^ \prime)=1$, indicating that \textit {$u_j$} and \textit {$u_{j ^ \prime}$} are more likely to have some common interests. Then, as in~\cite{DBLP:conf/wsdm/TangGHL13}, we performed the personality regularisation as follows:

\begin{equation}
min \sum_{i=1}^{|v|}\sum_{j=1}^{|u|} \eta(j, j ^ \prime)\Vert P(i,:)-Q(j,:)\Vert_2^2
\end{equation}
where $P$ and $Q$ are the users and items latent matrices, respectively. After some derivations for a specific user $u_i$, we obtain the following regularisation:

\begin{equation}
\begin{split}
&\frac{1}{2} \sum_{i=1}^{|v|}\sum_{j=1}^{|u|} \eta(j, j ^ \prime)\Vert P(i,:)-Q(j,:)\Vert_2^2= \frac{1}{2} \sum_{i=1}^{|v|}\sum_{j=1}^{|u|}\sum_{k=1}^{d} \eta(j, j ^ \prime) \Big(P(i,:)-Q(j,:)\Big)^2\\
&=\frac{1}{2} \sum_{i=1}^{|v|}\sum_{j=1}^{|u|}\sum_{k=1}^{d} \eta(j, j ^ \prime)P^2(i,k)+ \frac{1}{2} \sum_{i=1}^{|v|}\sum_{j=1}^{|u|}\sum_{k=1}^{d} \eta(j, j ^ \prime)Q^2(j,k)\\
&-\sum_{i=1}^{|v|}\sum_{j=1}^{|u|}\sum_{k=1}^{d} \eta(j, j ^ \prime)P(i,k)-Q(j,k)= \sum_{k=1}^{d} P^T(:,k)(D-Z)Q(:,k)= Tr(P^TYQ)
\end{split}
\end{equation}
where $Tr$ is short for showing a trace of a matrix in linear algebra. Next, we used the alternating least squares algorithm and updating rule presented by Krompaas et al.~\cite{Nonnegativetensor} to update $P$ and $Q$ as follows:

\begin{equation}
 \label{Krompaas}
 \Theta_i=\Theta_i \Bigg({\frac{\frac{\partial C(\Theta)^-}{\partial \Theta _i}}{\frac{\partial C(\Theta)^+}{\partial \Theta _i}}}\Bigg)^a
\end{equation}
where $\Theta$ is a non-negative variable and $C(\Theta)$ is the negative part of the derivation. We took the  partial derivative of Equation~\ref{Krompaas} with respect to $P$ and $Q$ and make it equal to zero. Next, on the basis of the Karush Kuhn Tucker complementary condition ~\cite{TRIFUNOVIC/2010}, \cite{Lenhart/2010} and the model presented by Tang et al.~\cite{TangGHL13}, we have the updating rules as follows: 

\begin{equation}
P(i,j)\leftarrow P(i,j)\sqrt{\frac{A(i,j)}{B(i,j)}},\quad \quad Q(i,j)\leftarrow Q(i,j)\sqrt{\frac{C(i,j)}{D(i,j)}}
\end{equation}
where $A, B, C,$ and $D$ can be expressed as follows:

\begin{equation}\label{a}
A=q^{{(d)}^T}R^{(d)}\gamma ^{(d)}+(1-\gamma_i^{(d)})L^TR^{{(d)}^T}q^{{(d)}^T}+\gamma ^{(d)}R^{(d)}q^{{(d)}^T}
\end{equation}

\begin{equation}\label{b}
\begin{aligned}
\begin{split}
&\ B=\gamma ^{(d)}q^{(d)}p^{(d)}q^{{(d)}^T}+q^{{(d)}^T}\gamma ^{(d)}p^{(d)}q^{(d)}+\gamma ^{(d)}p^{(d)}L^Tq^{{(d)}^T}+\\
&(1-\gamma_i^{(d)})L^Tq^{{(d)}^T}p^{{(d)}^T}+(1+\gamma_i^{(d)})L^TR^{(d)}q^{{(d)}^T}+(1-\gamma_i^{(d)})L^Tq^{(d)}p^{(d)}q^{{(d)}^T}+\\
&(1-\gamma_i^{(d)})q^{{(d)}^T}Lp^{(d)}q^{(d)}+(1-\gamma_i^{(d)})L^Tp^{{(d)}^T}q^{{(d)}^T}+(1-\gamma_i^{(d)})L^Tq^{{(d)}^T}p^{{(d)}^T}\\
&+YP+Y^Tp+\alpha_1 2p
\end{split}
\end{aligned}
\end{equation}

\begin{equation}\label{c}
C=R^{(d)}\gamma ^{(d)}p^{{(d)}^T}+(1-\gamma_i^{(d)})p^{{(d)}^T}L^TR^{{(d)}^T}+\gamma ^{(d)}p^{{(d)}^T}R^{(d)}
\end{equation}

\begin{equation}\label{d}
\begin{aligned}
\begin{split}
&\ D=\gamma ^{(d)}p^{{(d)}^T}q^{(d)}p^{(d)}+\gamma ^{(d)}p^{(d)}q^{(d)}p^{{(d)}^T}+(1-\gamma_i^{(d)})p^{{(d)}^T}q^{{(d)}^T}p^{{(d)}^T}L^T+\\
&1-\gamma_i^{(d)})p^{{(d)}^T}L^Tq^{{(d)}^T}+(1+\gamma_i^{(d)})p^{{(d)}^T}L^TR^{(d)}+(1-\gamma_i^{(d)})p^{{(d)}^T}L^Tq^{(d)}p^{(d)}+\\
&(1-\gamma_i^{(d)})Lp^{(d)}q^{(d)}p^{{(d)}^T}+(1-\gamma_i^{(d)})p^{{(d)}^T}L^Tq^{{(d)}^T}p^{{(d)}^T}L^T+\\
&(1-\gamma_i^{(d)})p^{{(d)}^T}L^Tq^{{(d)}^T}p^{{(d)}^T}L^T+\alpha_2 2q
\end{split}
\end{aligned}
\end{equation}

 The details of the proposed model are presented in Algorithm~\ref{APARaalgorithm}. 

\begin{algorithm}
\begin{algorithmic}[1]
 \State Input: L, d, $\gamma$, $\alpha_1$ and $\alpha_2$
  \State Randomly initialising $P,Q$

\State While Not convergent
 \State \quad Set A  as Equation~\ref{a} and Set B as Equation~\ref{b}
 \State \quad for j = 1 to |u| do
 \State  \quad    \quad  for i = 1 to |v| do
 \State  \quad   \quad \quad   $P(i,j)\leftarrow P(i,j)\sqrt{\frac{A(i,j)}{B(i,j)}}$
 \State \quad       \quad end for
 \State \quad   end for
  \State \quad Set C as Equation~\ref{c} and Set D as Equation~\ref{d}
 \State \quad for j = 1 to d do
 \State  \quad \quad fori= 1 to d do
 \State  \quad  \quad \quad   $  Q(i,j)\leftarrow Q(i,j)\sqrt{\frac{C(i,j)}{D(i,j)}}$
 \State  \quad  \quad end for
 \State \quad end for
 \State \quad end for
 \State end While
 \State return $P, Q$
 \caption{Framework of APAR with Personality Regularisation}
 \label{APARaalgorithm}
 \end{algorithmic}
\end{algorithm}

\section{Experimental Evaluation of APAR and Analysis}
\label{ch:Chapter5ExperimentandEvaluation}
In order to evaluate the proposed model and compare the performance of this model with that of the state-of-the-art methods, we used the dataset (Amazon Instant Video), evaluation metrics (MAE and RMSE), and baseline methods introduced in Chapter~\ref{ch4:chapterExperimentalSetup}.

\subsection{Experimental Setup}
We used the five-fold cross-validation method and considered the average performance of all the folds as the final performance value. Then, experimentally set the controlling parameters of the proposed model as follows: $\gamma = 0.5$, $\alpha_1 = 0.1$, $\alpha_2 = 0.1$, and $d=100$.

\subsection{Performance Comparison and Analysis}
To evaluate the performance of the proposed approach, we compared it with the following three categories: (i) purely rating-based recommender systems such as SVD$++$~\cite{DBLP:conf/kdd/Koren08}, UserMean and ItemMean~\cite{ma2011recommenderitemmean}, and Random~\cite{DBLP:conf/wise/GhafariYBO18}~\cite{DBLP:conf/ic3/GhafariFPR13}~\cite{7210303}~\cite{7152209}; (ii) recommender systems that incorporate side information, such as CTR~\cite{DBLP:conf/kdd/WangB11}; and (iii) personality-based recommender systems such as  TWIN~\cite{DBLP:journals/jifs/RoshchinaCR15} and Hu~\cite{DBLP:conf/recsys/HuP11}. We discussed the compared methods in detail in Chapter~\ref{ch4:chapterExperimentalSetup}.

As is clear from Table~\ref{ch5:tableAPARAnalysis}, we used different training dataset sizes (60\%, 70\%, 80\%, and 90\%). From Table~\ref{ch5:tableAPARAnalysis}, we can infer that when we increase the volume of the training data, the
accuracy of all the models will be improved, accordingly. All the approaches reached their best performance when they were trained with 90\% of the data. Therefore, to have a fair comparison, we considered the results of all the baselines related to the 90\% training dataset size. Among all the compared methods, the proposed model, APAR, achieved the best performance in terms of both RMSE and MAE. This might be because of paying more attention to the real values of ratings, which can better explain the level of a user's interests on an item. Additionally, APAR considers the users’ personality types and their level of knowledge in addition to their personal interests, which results in making more personalised suggestions for a particular user. Next, SVD++ performs better than UserMean and ItemMean, because they simply use the mean rating value of a user and an item, respectively. 

\begin{table}[h!]
\centering
\begin{adjustbox}{width=\textwidth}
 \begin{tabular}{|c c c c c c c c c c|}
 \hline
\small { Training Data} & \small { Metrics} & \small {Random } & \small {Hu} & \small {ItemMean} & \small {UserMean} & \small {SVD++ } & \small { CTR} & \small{TWIN} & \small {\textbf{APAR}}\\ [2 ex]
 \hline\hline
 60\% & \small {MAE} & 7.759 & 2.983 & 2.889 & 2.770 & 2.033 & 1.887 & 1.353 & \textbf {1.165}\\
 \hline
 60\% & \small {RMSE} & 8.124 &	3.32 &	3.084  &	2.896 & 2.314 & 2.129& 2.074	& \textbf {1.554}\\
 \hline
 70\% & \small {MAE} & 7.467	& 2.763 & 2.648 &	2.54 &	1.864 &	1.69 & 1.215	& \textbf {1.005}\\
 \hline
 70\% & \small {RMSE} & 7.985 &	3.152 &	2.815 &	2.785	& 2.29 & 2.101 &1.898	& \textbf {1.295}\\
 \hline
 80\% &\small {MAE} & 7.058	 & 2.581  &	2.354	 & 2.321	 & 1.719	 &1.52 & 1.132	& \textbf{0.936} \\
 \hline
 80\% & \small {RMSE} &7.654	 & 2.978 &	2.642	& 2.763 &	2.011	& 1.986 & 1.620	& \textbf {1.058}\\
 \hline
 90\% & \small {MAE} &	6.589 &	2.426 &	2.224 &	2.203 &	1.547 &	1.391 & .95	& \textbf {0.850}\\
 \hline
 90\% &\small {RMSE} &7.236 &	2.649	&	 2.545 &		2.539	&	 1.99 &	1.718	& 1.428	& \textbf {0.995}\\ [1ex]
 \hline
\end{tabular}
\end{adjustbox}
\caption{Comparison of the performance of APAR with other baseline approaches with respect to MAE and RMSE on the Amazon dataset.}
\label{ch5:tableAPARAnalysis}
\end{table}

However, the recommendation performance of UserMean was higher than that of ItemMean; this may be because of having a higher number of average ratings per user than the average number of ratings per item in the used dataset. Moreover, the recommendation performance of CTR was higher than that of SVD++ by
around 11\% and 14\% in terms of MAE and RMSE, respectively, as SVD++ only took the user’s rating into consideration while CTR sought  to use additional information. In the context of personality-based recommender systems, TWIN performed better than Hu, which integrated both the rating and the personality information. The reason behind this performance degradation might be the use of two sources of information, ratings in particular, which might increase the chance of suffering from the data sparsity problem.
Finally, APAR, which detects the users’ personality types implicitly, exhibited superior performance to the other methods because it integrated the users’ personality types, their personal interests, and their level of knowledge in a particular domain. In addition, unlike a majority of the existing approaches, APAR constructed a matrix with the actual rating score, which could help to better understand the users’ interests. It outperformed CTR by 39\% and 42\%, and TWIN by 11\% and 31\% in terms of MAE and RMSE, respectively.

\begin{figure}[h!]
  \centering    \includegraphics[width=0.6\textwidth]{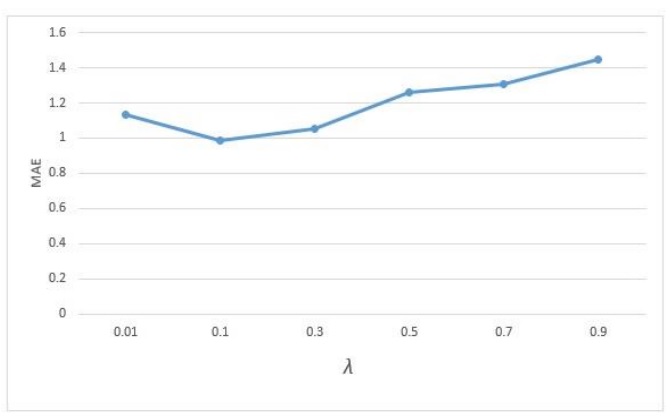}
   \caption{Impact of personality regularisation with respect to MAE evaluation metric on the Amazon dataset.}
   \label{ch5: figureRegularizationMAE}
  \end{figure}
  
  \begin{figure}[h!]
  \centering
    \includegraphics[width=0.6\textwidth]{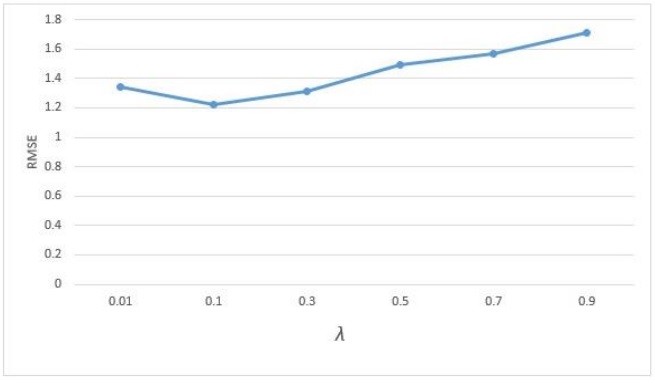}
  \caption{Impact of personality regularisation with respect to RMSE evaluation metric on the Amazon dataset.}
\label{ch5: figureRegularizationRMSE}
\end{figure}


\subsection{Impact of Personality Regularisation }
In this section, we will evaluate the impact of the proposed personality regularisation factor on the effectiveness of the proposed  model in terms of RMSE and MAE. Considering $\lambda$ to be a controlling parameter, we tested the proposed model under different values $\lambda= \{0.01, 0.1, 0.3, 0.5, 0.7, 0.9\}$; the corresponding  results are shown in Figure~\ref{ch5: figureRegularizationMAE} and Figure~\ref{ch5: figureRegularizationRMSE}. According to this figure, increasing the value of $\lambda$ gradually increased the performance of APAR; APAR reached its best performance (MAE = 1 and RMSE = 1.2), when $\lambda = 0.9$.  

\begin{figure}[b]
  \centering
    \includegraphics[width=0.60\textwidth]{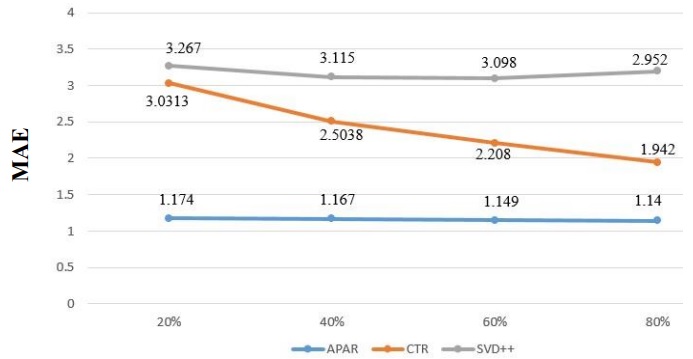}
    \caption{Recommendation performance in the presence of DSW-n-FCI with respect to MAE evaluation metric on the Amazon dataset.}
    \label{ch5: figureDSW-n-FCIMAE}
    \end{figure}
  \begin{figure}[b]
  \centering
    \includegraphics[width=0.60\textwidth]{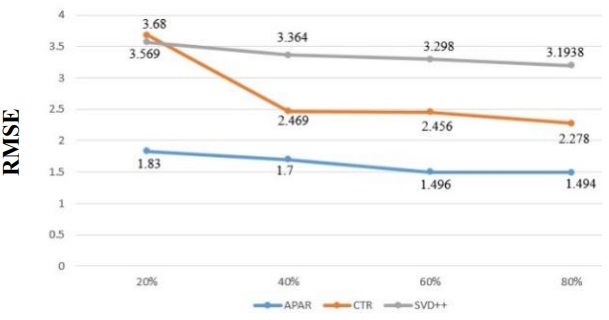}
   \caption{Recommendation performance in the presence of DSW-n-FCI with respect to RMSE evaluation metric on the Amazon dataset.}
  \label{ch5: figureDSW-n-FCIRMSE}
\end{figure}


\subsection{DSW-n-FCI Experiments}

In this section, we will investigate the performance of the proposed model and other baseline methods in the case of the Data Sparsity With no Feedback on Common Items (DSW-n-FCI) problem. To do so, we randomly divided the dataset into four different sub-datasets with the DSW-n-FCI degree of 20\% (SD1), 40\% (SD2), 60\% (SD3), and 80\% (SD4), where DSW-n-FCI= 20\% means that only 20\% of the users did not have any feedback on the common items out of all the users. We computed the degree of DSW-n-FCI as follows:

\begin{equation}
d=\frac{number-of-users-without-any-similar-feedback}{total-users}
\end{equation}

We compared the proposed model with other methods on each sub-dataset already explained above. From Figure~\ref{ch5: figureDSW-n-FCIMAE} and Figure~\ref{ch5: figureDSW-n-FCIRMSE} , we can see the superiority of APAR in all the sub-datasets and under different degrees of DSW-n-FCI. This indicates the ability of APAR to deal with the DSW-n-FCI problem, and having recommendations even when there is no recorded feedback on the common items among users. We only compared APAR with CTR, and SVD++ under the DSW-n-FCI situation, as they demonstrated great results in the experiments. On the SD1, SD2, SD3, and SD4 sub-datasets, on average, APAR showed a performance improvement of 51\%, 63\% , 39\%, and 52\% as compared to CTR and SVD++ in terms of MAE and RMSE, respectively. This implies that discovering the users' personality types can help APAR better handle the DSW-n-FCI  problem.

\section{Summary}
\label{ch:Chapter5summary}
Collaborative filtering based on matrix factorisation (MF), which embeds  user and item latent features  to learn the users' preferences over a set of items, is one of the most successful and straightforward solutions to general recommender systems. Collaborative filtering-based models are still valuable in some real-world organisations as they can capture the users' general tastes. However, most of the current approaches considerably  depend on analysing the users’ feedback, e.g. ratings and reviews of common items, to discover similar behaviour patterns among users. Therefore, they may fail when there are no common items of interest among users, a problem which we have called the Data Sparsity With no Feedback on Common Items (DSW-n-FCI) problem in this thesis. Personality as a stable behavioural pattern is one of the key factors with the ability to explain why people are different in their choices and decisions. Inspired by the significance of using personality for discovering the users' ultimate interests, some general recommenders have started adopting personality in their models. However, while personality-aware methods detect the users’ personality types explicitly by having the users fill out a questionnaire, which is an unfeasible task in many real-world applications, they nevertheless ignore the differences among the users’ level of knowledge in different domains. Most of the existing methods exploit the users’ personality types to simply use them as a similarity measure. In contrast to those approaches, we
directly considered the personality factor in our mathematical model and proposed a novel matrix factorisation model. The assumption of the proposed method is that the users’ decision-making can be affected by three major factors: users’ personality types, their personal interests, and their level of knowledge in a particular domain. Moreover, to the best of our knowledge, no study has been reported  in the extant literature on the assessment of the capability of personality–based RSs to deal with the Data Sparsity With no Feedback on Common Items (DSW-n-FCI) problem. Experimental results on a real-world dataset showed that the proposed model could achieve very promising results even in the case of the  DSW-n-FCI problem.

\chapter{Toward A Deep Attention-Based Sequential Recommender System}
\label{ch6:chapterDAS}
\section{Introduction}
In this chapter, the main focus is on improving the performance of sequential recommender systems. As is clear from the hierarchical structure in Figure~\ref{ch1:figureRWcat} in Chapter~\ref{chap:introduction}, recommender systems can be categorised  into two major classes of approaches: those which emphasise long-term users' preferences (i.e. general recommenders), and those that emphasise short-term users' preferences  (i.e. sequential recommenders). While in Chapter~\ref{ch5:chapter5Personality}, our main goal was to improve the performance of general recommenders, in this chapter, we pay more attention to addressing the challenges of sequential recommenders. In this chapter, we aim to handle one of the main issues at the interaction level of the graphical structure presented in the second level (i.e. interaction level) of Figure~\ref{ch1:figureRWcat}, which is dealing with a noisy session. Noisy session refers to a case when in a transaction consisting of multiple items, it is more likely that either some of these items are irrelevant to the following items or they may be relevant but in different scales. 
However, most of the existing methods assume that all the adjacent items in a sequence are highly dependent, which may not be practical in real-world scenarios because of the uncertainty of the customers' shopping behaviours. A user-item interaction sequence may contain some irrelevant items, which in turn may lead to false dependencies. Moreover, the long-term users' preferences are ignored by most of the current studies. To fill this gap, in this chapter, we propose a novel Deep Attention-based Sequential (DAS)~\cite{9123874} model to overcome the above-mentioned problems. The results of the approach proposed in this chapter have been reported in a paper published by IEEE Access Journal (IF: 4.1), 2020.

\subsection{The Target Problem and The Motivation}
As discussed in Chapter~\ref{ch:chapter2RW},  Section~\ref{ch2:GeneralRecommenders}, compared with those in  the general recommender systems, new challenges have emerged in SRSs. First, the aggregated form of user-item interactions in SRSs is a type of implicit feedback (e.g. check-in behaviours and purchased items). It is also quite a challenging task for approaches based on the implicit data to predict whether a user is not interested in the unpurchased items or just not aware of them. Therefore, a traditional RS which treats recommendation as a prediction task and only optimises one-class prediction score (i.e. `1' or `0') may not be appropriate~\cite{DBLP:conf/ijcai/YingZZLXXX018}, \cite{DBLP:conf/www/BayerHKR17}. Second, while a user's sequential behaviour (i.e. short-term preference) reflects the recently observed items, the user's general taste (i.e. long-term preference) plays an important role in forming the user's preferences~\cite{DBLP:conf/ijcai/WangHWCSO19}, which is another challenge of SRSs. Most of the existing sequential recommenders either do not take both types of users' preferences into account or mostly combine them linearly. Although there have been  a few attempts to consider both types of users' preferences, they may have some difficulties in both sequential and general parts of their model. In the general part, they mainly learn the users' long-term preferences through a static low-dimensional representation, which implies  that a user's general taste always remains the same. However, this may not be practical in many real-world cases. Besides the  general part, in the sequential part, they may not be able to deal with a noisy session in which there are some irrelevant items. Most of the current SRSs have the same overly strong assumption that each item is contextually dependent on the next one and has the same contribution in predicting the next target item. Therefore, they underestimate the impact of irrelevant items in dependency modelling and neglect the context of each item in a shopping basket, which can be defined by selecting a set of items in a transaction.

Let us make the above problem more clear with an example. Assume that $S_1=\{\textit{burger}, \textit{coke},$ $\textit{chips}\}$ is a customer shopping list, to which at the last minute, a \textit{shampoo} is added to $S_1=\{\textit{burger}, \textit{coke}, \textit{chips},\textit{shampoo}\}$. While the first three purchased items are strongly correlated, the last item is pretty much irrelevant to them and may generate interference for the next predicted items. In this example, the customer may be more willing to buy \textit{ketchup} as the next item,  which is contextually dependent on \textit{burger}, \textit{coke}, and \textit{chips}. In this transaction history, if we consider the first three purchased items as a context, a \textit{shampoo} has no correlation with them and does not share a similar context.
The existing methods that may not distinguish irrelevant items within a contextual sequence may recommend a \textit{conditioner} as a next item because of its high correlation to the last purchased item. This example demonstrates the importance of identifying irrelevant items in a transaction for the next-item recommendation task. Hence, this can explain that user-item interaction sequences may not follow a strict order and shopping behaviours may contain noisy items that are not contextually dependent on the previous items.

Thus, we can infer that it is quite hard to learn both long- and short-term users' preferences. One typical solution for approaches in this setting is to use Markov chain models for modelling the time-series data in the sequential part of their recommender systems. In these models, all items in a session are taken into consideration to capture the comprehensive sequential dependency. As an example, Factorising Personalised Markov Chain (FPMC), which was proposed by Rendel et al.~\cite{DBLP:conf/www/RendleFS10}, integrates a Markov Chain (MC) as a commonly used method in SRSs to model sequential behaviours and  Matrix Factorisation (MF) for modelling long-term users' preferences. Despite the considerable  success of MF-based methods in capturing long-term users' preferences, their performance may be limited as they ignore to consider  short-term users' preferences. FPMC  just utilises a linear aggregation function to combine these two different types of users' preferences. Hierarchical Representation Model (HRM) partially solves the problem of modelling high-level user-item interactions by adopting a non-linear aggregation function. However, it may lose a considerable amount of  information because of the use of the max-pooling function as an aggregation function~\cite{DBLP:conf/sigir/WangGLXWC15}. Inspired by the word embedding technique~\cite{DBLP:conf/nips/MikolovSCCD13}, Prod2Vec was proposed to utilise information from a sequence of interacted items to improve the performance of MF~\cite{DBLP:conf/recsys/LiangACB16}.  Motivated by the success of Convolutional Neural Networks (CNNs) in capturing the local context, Caser~\cite{DBLP:conf/wsdm/TangW18} was introduced as a sequential recommender system in which user-item interactions are treated as an image. Caser learns sequential patterns as local features of the image by using convolutional filters. Recently, Recurrent Neural Networks (RNNs), as one of the most popular deep learning-based methods, have drawn more attention in modelling sequential dependencies in SRSs~\cite{HidasiKBT15}, \cite{DBLP:conf/recsys/Twardowski16}. 

Different from traditional RNNs, several works have been introduced to modify a classic RNN in order to better capture the whole historical user-item interaction sequences~\cite{HidasiKBT15}, \cite{DBLP:conf/um/DevooghtB17}. For instance, SLi-Rec improves the classic RNN structure such as Long Short-Term Memory (LSTM) by proposing
time-aware and content-aware controllers to fully exploit the user modelling. Then, an attention-based framework is applied to combine the general and the sequential recommenders~\cite{DBLP:conf/ijcai/YuLML019}. Because of the considerable  success of CNNs in capturing local sequential patterns, and RNNs in dealing with complex long-term dependencies, Xu et al.~\cite{DBLP:conf/www/XuZLXSCZX19} proposed a novel Recurrent
Convolutional Neural Network model (RCNN) to better generate recommendations. Lately, researchers have used an attention mechanism because of its powerful ability to focus on selective parts~\cite{DBLP:journals/corr/BahdanauCB14}. Although Wang  et al.~\cite{DBLP:conf/aaai/WangHCHL018} used an attention network in their model and successfully achieved superior performance in context learning tasks, the users' general tastes were not taken into consideration by them. Instead, a two-layer hierarchical design called, SHAN, was proposed by Ying et al.~\cite{DBLP:conf/ijcai/YingZZLXXX018} as an attention-based SRS to incorporate both the users' general tastes and their short-term preferences in a unified manner. The main difference between the proposed  work and SHAN can be discussed from three different aspects: firstly, in modelling each of the user's long and short-term presences, SHAN calculates the attention score which is guided by the user embedding. Therefore, it may not completely discover the contributions of each item, and it may not be able to find noisy items. Secondly, SHAN ignores the context of the users' shopping basket, which in turn can play an important role in predicting the next-item recommendation. Thirdly, unlike in SHAN, we added the user's embedding vector in the final layer, which, as stated in~\cite{DBLP:conf/www/HeLZNHC17}, could improve the model performance by using the pre-training model’s parameters.

To sum up, most of the existing studies do not consider both types of users' preferences in their recommender systems. Little attention has been paid to combine these two users' preferences, in which  they mostly assume that there is a rigid order between two adjacent items in a sequence. This may not be true in many real-world cases, and there may be some noisy items in a sequence that generate fake dependencies. While there are strong techniques, which have already been discussed in Chapter~\ref{ch:chapter2RW}, such as PRM, MC, SPM, FM, RNN, and CNN, to work with sequential data, they may not be able to fit with a flexible session.

\subsection{Proposed Design and Main Contributions}
This chapter addresses the issue of handling the problem of user-item interaction sequences with noise in the long- and short-term item sets by proposing a novel Deep Attention-based Sequential recommender system (DAS). DAS consists of three different blocks: $(i)$ \textit{An embedding block:} in this block, we aimed to take raw user-item and item-item interactions as inputs and embed them into low-dimensional spaces. $(ii)$ \textit{An attention block:} here, we used two attention networks to differentiate the importance of each item in both long- and short-term users' preferences without a rigid ordering assumption. The main purpose of using the attention network is to assign different weights to the items. This could emphasise the strongly context-relevant items and downplay the weakly correlated ones in a user-item interaction sequence. $(iii)$ \textit{A fully connected block:} in this block, we used two deep network structures, to which we first fed a combination of the outputs of two attention networks to the first deep network to learn a mixture of users' preferences. Next, we combined a learned users' mixture preferences with the users' embedding vectors and projected them to the second deep network to produce a personalised recommendation.  As a result, DAS was more effective and robust in predicting the best next interacted items in a transaction. The main contributions of the present work can be summarised as follows:
 
\begin{itemize}
     \item Proposing a novel recommender system in which both types of users' preferences are taken into consideration.
   
     \item Introducing an attention network to discriminate the importance of items in the long- and short-term users' preferences. 
    \item Designing a deep network structure consisting of three different blocks to generate a mixture of long- and short-term users' preferences.  
     \item Conducting empirical evaluations on two real-world datasets. The results demonstrated the superiority of the proposed  method to the state-of-the-art methods in terms of Area Under Curve (AUC), Precision,  Recall, and novelty evaluation metrics.
\end{itemize}

\begin{figure*}[t]
\includegraphics [width=1 \textwidth, scale=1]{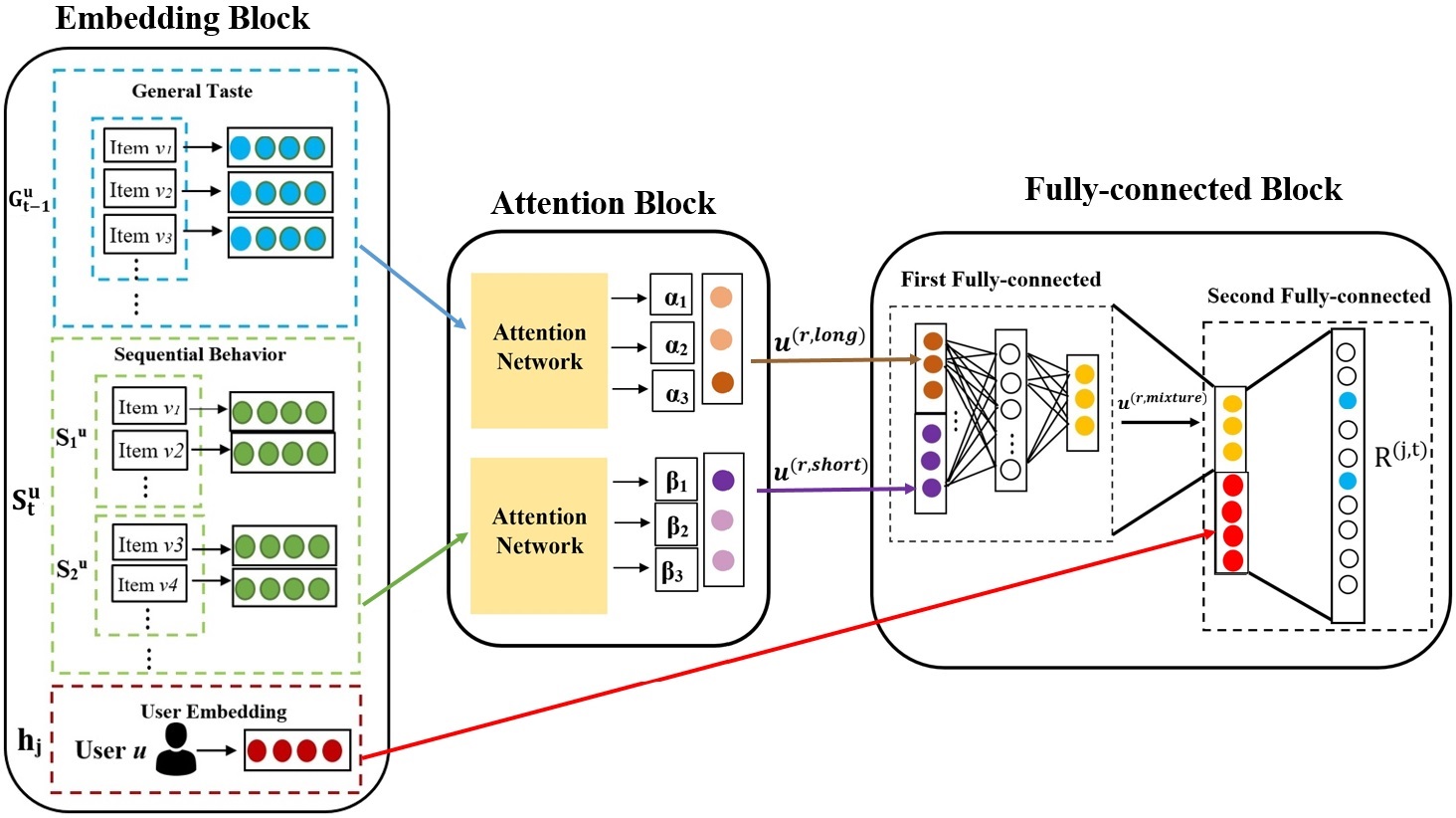}
\centering
\caption{Graphical structure of DAS model, which consists of three main blocks: (i) Embedding Block to embed user and item into low-dimensional representations; (ii) Attention Block to emphasise on the importance of each item, $u^{(r, long)}$ and $u^{(r, short)}$ are the long-term and short-term users preferences, respectively, which are fed to the next block; and (iii) Fully Connected Block to first learn a mixture of users' preferences and then predict the best next interacted items.} 
\label{ch6:figureDAS}
\end{figure*}

The rest of this chapter is organised as follows: we will discuss the overview and framework in Section~\ref{ch:Chapter6OverviewandFramework}. The experiment and evaluation are presented in Section~\ref{ch:Chapter6ExperimentandEvaluation}, and the chapter is concluded in Section~\ref{ch:Chapter6Summary}.

\section{Overview and Framework}
\label{ch:Chapter6OverviewandFramework}
Figure~\ref{ch6:figureDAS} depicts a framework of DAS, which consists of three different blocks: $(i)$ \textit{an embedding block} which embeds users and items into low-dimensional spaces; $(ii)$ \textit{an attention block} which aims to discriminatively learn dependencies among the items in both the users' long-term and short-term item sets; and $(iii)$ \textit{a fully connected block} which first learns a mixture of users’ preferences representation in a non-linear manner and then combines it with the users’ embeddings to provide a personalised recommendation.

\subsection{Problem Statement}
Let $U= \{u_1,u_2,...,u_{|u|}\}$ denote the user set and $V=\{v_1, v_2,...,v_{|v|}\}$ indicate the item set, where $|u|$ and $|v|$ are the total number of users and items, respectively. For a given user $u$, $Q^u=  G^u_{t-1}\cup S^u_t$ represents the total users' sessions. At a certain timestamp $t$, $S^u_t=\{v_1, v_2,..., v_i\}$ is $u$'s sequential behaviour, where $i\in V$, and reflects the user's short-term preference. Following this, $G^u_{t-1}=\{S^u_1, S^u_2, ..., S^u_{t-1}\}$ is a set of interacted items at timestamp $t-1$ (e.g. purchasing history, clicked items, and check-in behaviours), which represents a user's general taste (i.e. long-term preference). For the sake of simplification, $G^u_{t-1}$ and $S^u_t$ represent the $u$'s long-term and short-term interacted item sets, respectively.

\subsection{Embedding Block}

In sequential recommender systems, predicting the next interesting items in a session is similar to predicting a relevant word in Natural Language Processing (NLP), in which the sizes of both items and vocabularies are very large. Inspired by NLP techniques, if we consider words as items or users, which are indexed by meaningless IDs, we need to transform them from IDs space to a more informative representation space rather than IDs. Therefore, a fully connected layer is used to embed the user
and item IDs (i.e. one-hot representations) into two continuous low-dimensional spaces as $U\in {R}^{k\times|u|}$ and $V\in {R}^{k\times|v|}$ , where $k$ is the latent dimension of the embedding spaces.

\subsubsection{Item Embedding}
Given the original one-hot encoded item set as an input (where in this vector, item $i$ $(i \in V)$ at position $i$ is equal to 1, and the rest of the positions are set to 0) may limit the derived information, in DAS, we used the embedding block to map these sparse vectors into an informative representation. We now have a vector with length $|v|$, which represents the context of each item. We used $h_i\in R^k$ to represent the context of item $i$, where $k$ is the latent dimension, namely:
\begin{equation} 
\label{ch6:equationItemembedding}
 h_i=\sigma (W^1_{:,i})
\end{equation}
where $W^1_{:,i} \in {R}^{k\times|v|}$ is a weight matrix and the $i^{th}$ column of matrix $W^1_{:,i}$ endows one-hot vector item $i$ to its embedding $h_i$. Here, we used the logistic function $\sigma$(.) as the activation function to better model the non-linearities.

\subsubsection{User Embedding}
Similar to item embedding, we took user $j$'s one-hot vector $(j \in U)$ in this block to embed it into $h_j \in{R}^{k}$. Formally:
\begin{equation} 
\label{ch6:equationUserembedding}
 h_j=\sigma (W^2_{:,j})
\end{equation}
where $W^2_{:,j} \in {R}^{k\times|u|}$ is a weight matrix and the $j^{th}$ column endows the one-hot vector of user $j$ to its embedding $h_j$. For the sake of simplification, we set the same size for the user and the item embedding.

\subsection{Attention Block}
\label{ch6:sectionAttentionBlock}
The main purpose of this block is to focus more on the relevant items in a session and deal with a noisy session. To avoid a noisy session, we need a solid technique with the ability to assign different weights to different items according to their contributions to the existence of the next item. Therefore, inspired by the considerable  success of the attention network in many tasks, such as machine translation~\cite{DBLP:conf/aaai/ChenWUSZ18}, image captioning~\cite{DBLP:journals/corr/ChenZXNSC16}, and recommendation~\cite{DBLP:journals/tkde/HeHSLJC18}, we were motivated to use an attention network here in order to discriminate the contributions of different items in predicting the next items. Despite most of the existing studies, which assume that all items are contextually dependent and there is no noise in a session, in this block, we used two attention networks to identify the highly context-dependent items in both long- and short-term users' preferences. The benefits of applying the attention mechanism can be denoted as two fold. First, an attention mechanism in the general part (i.e. long-term users' preferences) can capture the dynamic properties of the users' preferences, and second, the use of an attention network in the sequential part (i.e. short-term users' preferences) enables a session  to relax the ordering assumption. The attention network automatically assigns different weights to different items to downplay the impact of irrelevant items which may overwhelm the impact of the relevant ones.

\subsubsection{Long-Term User's Preference Representation}
A set of interacted items in the long-term item set $G^u_{t-1}$, can reflect a user's general taste. However, considering a fixed weight for modelling a user's general taste may not reflect the user's real preferences because of a simple assumption that each item plays an equal role in the long-term users' preferences. To fulfil the above requirements, for a given user $u$, we first measured the alignment of item $v_i$'s embedding vector, $h_i$, with respect to the context matrix $W^\alpha$. Then, we compute the attention score $\alpha_{ip}$ through Equation~\ref{ch6:equationAttention}, which indicated the level of contribution of contextual item $v_i$ in the occurrence of item $v_p$ in a long-term interacted item set. Then, the softmax function was applied to normalise the attention score $\alpha_{ip}$; a higher score represented a larger item contribution. Finally, a weighted sum over the attentive context embeddings in a long-term item set $G^u_{t-1}$ was computed to build a user's long-term representation. Formally,

\begin{equation} 
\alpha_{ip}= \frac{exp(e(h_i))}{{\sum_{p\in G_{t-1}^u}}exp(e(h_p))}
\label{ch6:equationAttention}
\end{equation}

\begin{equation}
\label{ch6:equationAttentionscore}
    e(h_i)= W^\alpha h_i^T
\end{equation}

\begin{equation} 
\label{ch6:equationUserLong}
u^{(r, long)}= {\sum_{i\in G_{t-1}^u}\alpha_{ip}} h_i  \quad\mathrm{w.r.t}\quad {\sum_{i\in G_{t-1}^u}\alpha_{ip}}=1
\end{equation} 
where $W^ \alpha $ is a shared weight over the first attention network, which is randomly initialised and will be learned during the training process. Similar to Wang et al.~\cite{DBLP:conf/aaai/WangHCHL018}, we considered $W^ \alpha$ to be an item-level context matrix shared by all the items to be able more informative items among a set of items such as the one used in a memory network~\cite{DBLP:conf/icml/KumarIOIBGZPS16}. Finally, the long-term users' preference representations, $u^{(r, long)}$, could be calculated with the help of the attentive context,  which is represented in Equation~\ref{ch6:equationAttentionscore}.

\subsubsection{Short-Term User's Preference Representation}
Rather than the long-term users' preferences, the most recent interacted items in a session, which form  the short-term users' preferences, are also important for predicting the next items. Therefore, similar to the users' long-term item sets, an attention network was applied here to focus more on the key items in the users' short-term item sets.

\begin{equation} 
\label{ch6:equationShort1}
\beta_{sn}= \frac{exp(e(h_s))}{{\sum_{n\in S^u_t}}exp(e(h_n))}
\end{equation}
\begin{equation}
e(h_s)=W^\beta h_s^T
\end{equation}
where similar to the long-term users' preferences, we first measured the similarity weights of each item $v_s$ embedding as $h_s$, with the context matrix $W^ \beta$, to find the most relevant items in the users' short-term item set. Then, we normalised this attentive context through the softmax function to find the level of contribution of item $v_s$'s context with respect to the target item $v_n$~\cite{DBLP:conf/naacl/YangYDHSH16}. Finally, to form a user’s short-term representation, 
$u^{(r, short)}$, a weighted sum over contextual item embedding in the short-term item set $S^u_t$ was computed, where the weights could be inferred from the attention network.

\begin{equation} 
\label{ch6:equationShort3}
u^{(r, short)}= {\sum_{s\in S^u_t}}\beta_{sn} h_s   \quad\mathrm{w.r.t}\quad {\sum_{s\in S^u_t}}\beta_{sn}=1
\end{equation} 

\subsection{Fully Connected Block}
In this block, we adopted two deep network structures: one for learning a mixture of users' preferences and the other for making a personalised item recommendation. Unlike most of the existing studies that linearly combine long- and short-term users' preferences, which may limit model performance~\cite{DBLP:conf/www/RendleFS10}, \cite{DBLP:conf/icdm/HeM16}, we concatenated the outputs from the attention block and fed it to the first deep network in order to learn a mixture of users' preferences. 
\begin{align}
\label{ch6:equationMixturePrefrence}
    u ^ {\textit{mixture}}&=\phi_a (W\begin{bmatrix}
           u^{(r, long)} \\
           u^{(r, short)} \\
           \end{bmatrix}+b) \end{align}
where $\phi_a$ is an activation function; $W\in {R}^{{d}\times{2d}}$ and $b\in {R}^{d}$ are the weight matrix and bias, respectively; and $d$ is the dimension of the hidden layers. We called $u^ {\textit{mixture}} \in {R}^d$ the mixture of long- and short-term users' preferences, which encodes two types of users' preferences in the user-item interaction item set $Q^u$. Next, when $u^{\textit{mixture}}$ was ready, we combine dit with a user's embedding vector $h_j$ and projected it onto the final output network with $|I|$ number of nodes, to capture the high-level representation of user $u_j$, namely: 

\begin{align}
\label{ch6:equationRating}
    R^{(i, t)}&=W'\begin{bmatrix}
     u^{\textit{mixture}} \\
     h_j\\
     \end{bmatrix}+b' \end{align}
where $W'\in {R}^{|I|\times 2d}$ and $b'\in {R}^{|I|}$ are the weight matrix and bias in the final deep network, respectively. In this layer, $ R^{(i, t)}$ denotes the probability that a user will interact with item $i$ at time $t$. Similar to Wang et al.~\cite{DBLP:conf/wsdm/TangW18}, we added the user's embedding vector $h_j$ to the final layer for two main reasons: (1) to make a more personalised recommendation and (2) to improve the model performance by using the pre-training model's parameters, as stated in~\cite{DBLP:conf/www/HeLZNHC17}.

\subsection{Network Training}
As discussed in Chapter~\ref{ch:chapter2RW}, working with implicit data may be more challenging. Therefore, in this work our goal was to focus on providing a ranked list of items~\cite{DBLP:conf/www/BayerHKR17}, as we had the implicit feedback (e.g. check-ins and purchase transactions). After computing  the users' mixture preferences, we proposed to use a pair-wise ranking objective function to rank the observed entries higher than the unobserved ones~\cite{DBLP:conf/wsdm/TangW18}. To do so, we transformed the outputs of the final deep network, $R^{(i,t)}$, into the probabilities score.

\begin{equation}
\label{ch6:equationProbability}
p(S^u_t|S^u_1, S^u_2,..., S^u_{t-1})= \sigma  (R^{(i, t)})
\end{equation}
where $\sigma $ is the sigmoid function, {$\sigma=1/(1+e^{(-x)})$}. For each positive pair $(u,i)$, in {$I^+_{(u,t)}=\{i | R^{(i, t)} =1\}$}, we randomly kept one item in each session as an unobserved item at time $t$ which had to be predicted by the proposed model \small{$I^-_{(u,t)}$}. Then, we updated the binary cross-entropy loss function as follows:
\begin{equation}
\textit{l}= -\sum _{(u,t)}\Big( \sum_{i\in I^+_{(u,t)}} log \sigma  (R^{(ui, t)}) 
\sum_{i'\in {I^-_{(u,t)}}}\\
log(1-\sigma(R^{(ui', t)})\Big)+ \lambda _{ui} ||\Theta_{ui}||^2+  \lambda _{at}||\Theta_{at}||^2
\end{equation}
where $\Theta$ = \{$\Theta_{ui}$, $\Theta_{at}$, $\Theta_{d}$, $b$, $b'$\} is the set of model parameters; \small{$\Theta_{ui}$ = \{$W^1, W^2$ \}} is the set of weights for item and user embeddings, respectively; and $\Theta_{at}$ = $\{W^\alpha,W^\beta\}$ is the set of weights in the attention networks. Further, {$\Theta_{d}=\{W,W'\}$} is the set of weights for two deep networks, and $\lambda$  = $\{\lambda _{ui}, \lambda _{at}\}$ is a set of the considered regularisation parameters. We also adopted Stochastic Gradient Descent (SGD) for updating the considered parameters. Algorithm~\ref{ch:Chapter5DASAlgo} presents the details of the proposed model.

 \textbf{Recommendation}. We fed our network with the user's embedding vector $h_j$ and user-item interaction set \small{$Q^u= G^u_{t-1} \cup S^u_t$}. Next, we predicted $R^{(i, t)}$, which denoted the probabilities of the user's interaction with item $v_i$ at timestamp $t$. Then, $N$ items with the highest values were recommended to this user. Therefore, the complexity of the proposed  model for making a recommendation to all the users was $O(|U||I|d)$.

\begin{algorithm}[H]
  {\textbf{Input:} long- and short-term item sets\{$G^u_{t-1}$, $S^u_{t}$\}, learning rate $\eta$, $\lambda$, $K$}
  {\textbf{Output:} a set of parameter  
  $\Theta$}
  
Initialise $\Theta_{ui}$, $\Theta_{d}$ with Normal Distribution $N(0, 0.01)$ 

Initialise $\Theta_{at}$ with Uniform Distribution[$-\sqrt{\frac{3}{K}},\sqrt{\frac{3}{K}}$]

 While \textit{converge}{
 
  \quad  For $u$ $\in$ $U$ do:
    
    \quad \quad Randomly pick an item $i$ from $Q^u$

 \quad \quad $h_i$, $h_j$ $\leftarrow$ arrange embeddings Equations~\ref{ch6:equationItemembedding} and~\ref{ch6:equationUserembedding}
            
 \quad \quad compute a user's long-term representation $u^{(r,long)}$ based on Equations~\ref{ch6:equationAttention}-\ref{ch6:equationUserLong}
            
 \quad \quad compute a user's short-term representation $u^{(r,short)}$ based on  Equations~\ref{ch6:equationShort1}-\ref{ch6:equationShort3}
           
 \quad \quad compute a user's mixture of preferences  representation  $u^{mixture}$ based on            Equation~\ref{ch6:equationMixturePrefrence}

\quad \quad $R^{(i, t)}$ $\leftarrow \sigma$ $(u^{mixture}$. $Concat$ $h_j$) $based on Equations$~\ref{ch6:equationRating}~  $and$~\ref{ch6:equationProbability}

        \quad \quad update $\Theta$ with gradient descent
      
      \quad  \textbf{end for}
      
    \textbf{end while}}
    
 \textbf{return} $\Theta$
  
\caption{DAS Algorithm.}
\label{ch:Chapter5DASAlgo}
\end{algorithm}

\section{Experiment and Evaluation}
\label{ch:Chapter6ExperimentandEvaluation}

The empirical study of the proposed DAS is described in this section. In particular, we will first describe the preparation of the datasets, evaluation metrics, and baseline methods required for the experiments  introduced in Chapter~\ref{ch4:chapterExperimentalSetup}. Next, we will discuss the impact of different hyper-parameters on the proposed model, DAS, and then verify the performance of DAS in terms of the recommendation accuracy and the novelty. 

\begin{table*}
  \centering
\begin{tabular}[hbt!]{|c|c|c|c|c|c|c|c|c|c|}
\hline
  Dataset & Metric & DAS & SHAN & HRM & FPMC & Caser & Fossil & BPR & Top  \\
  \hline
  Gowalla 
 &\begin{tabular}{@{}c@{}}Prec@1 \\  Prec@5 \\  Prec@10\end{tabular}
 &\begin{tabular}{@{}c@{}}\textbf{0.2887} \\ \textbf{0.2984 } \\  \textbf{0.3012}\end{tabular}
 &\begin{tabular}{@{}c@{}}0.1941 \\ 0.1962\\  0.1977\end{tabular} 
 &\begin{tabular}{@{}c@{}}0.1401 \\  0.1410\\  0.1483\end{tabular}
 &\begin{tabular}{@{}c@{}} 0.1341\\  0.1374\\  0.1398\end{tabular}
 &\begin{tabular}{@{}c@{}}0.1131\\  0.1148 \\  0.1194\end{tabular}
 &\begin{tabular}{@{}c@{}} 0.0917\\  0.0951\\  0.0974\\ \end{tabular}
 &\begin{tabular}{@{}c@{}}0.0884 \\  0.0890 \\  0.0899\end{tabular}
 &\begin{tabular}{@{}c@{}}0.0655 \\  0.0671 \\  0.0698\end{tabular}\\

 \hline
 Tmall 
  &\begin{tabular}{@{}c@{}}Prec@1 \\  Prec@5 \\  Prec@10 \end{tabular}
 &\begin{tabular}{@{}c@{}}\textbf{0.2541} \\  \textbf{0.2712} \\ \textbf{0.2917} \end{tabular}
 &\begin{tabular}{@{}c@{}}0.1812 \\ 0.1847 \\ 0.1853 \end{tabular} 
 &\begin{tabular}{@{}c@{}}0.1324 \\  0.1371 \\ 0.1394 \end{tabular}
 &\begin{tabular}{@{}c@{}}0.1211 \\ 0.1284 \\ 0.1296 \end{tabular}
 &\begin{tabular}{@{}c@{}}0.0998 \\0.1001\\  0.1131 \end{tabular}
 &\begin{tabular}{@{}c@{}}0.0811\\  0.0841\\  0.0891 \end{tabular}
 &\begin{tabular}{@{}c@{}}0.0711 \\  0.0725 \\  0.0741 \end{tabular}
 &\begin{tabular}{@{}c@{}}0.0601 \\ 0.0611 \\  0.0642 \end{tabular}\\
 \hline
\end{tabular}
\caption{Performance comparison on Gowalla and Tmall datasets in terms of Precision.}
  \label{ch6:tableComparison}
\end{table*}

\subsection{Experimental Setup}
 We evaluated the proposed  model on two real-world datasets, Gowalla~\cite{DBLP:conf/kdd/ChoML11} and Tmall~\cite{DBLP:conf/ijcai/HuCWXCG17}, to compare the performance of DAS with that of the baseline approaches. Gowalla aggregates the users' check-in information from the location-based social networking website, Gowalla, while Tmall records the users' transactions on the largest online shopping website in China, where each session (transaction) consists of multiple items. Note that, similar to~\cite{DBLP:conf/ijcai/YingZZLXXX018}, in both the datasets, we only considered the data of the previous  seven months and removed the sessions with only one item and with items with less than 20 observations. Then, to better represent the users' sequential behaviours (i.e. short-term preference), transactions in one day were considered to be one session. Following~\cite{DBLP:conf/ijcai/HuCWXCG17} and~\cite{DBLP:conf/ijcai/YingZZLXXX018}, we randomly divided the datasets into $20$\% and $80$\% for test and training, respectively. To better evaluate DAS, we randomly kept one item in each session to be predicted by the proposed model.

We set the size of the item and the user embedding to 100, which was initialised randomly with normal distribution $N (0, 0.01)$; the weight parameters in the attention network were initialised from the uniform distribution U${\small{(-\sqrt{\frac{3}{K}},\sqrt{\frac{3}{K}})}}$.

We used Stochastic Gradient Descent (SGD) as the optimisation technique to update our parameters, the learning rate $\eta$ was set to 0.001, and the training epoch size was set to 10. We also empirically set the batch size to 50 and considered $\lambda_{uv}=\{0.01, 0.001, 0.0001\}$ as the user and item embedding regularisation; further $\lambda_a=\{0, 1, 10, 50\}$ was set as the attention network regularisation.

We tested the performance of DAS in terms of the recommendation accuracy and novelty. Therefore, we used three widely adopted metrics for measuring the recommendation accuracy in SRSs, namely Recall@N, $Precision@N$, and AUC, where a larger value indicated better performance. Recall and Precision evaluated the ability of the proposed model to find all the relevant top-N items within a dataset, where $N \in \{5,100\}$, while AUC showed how well the proposed model could rank the ground truth items.  We compared the proposed model with three different types of baseline methods, including approaches in the classic next item recommendation, models that combine both long- and short-term users' preferences, and attention-based sequential recommender systems. For the evaluations, we used the same datasets as SHAN, FPMC, Caser, Fossil, and HRM. Therefore, for implementing these approaches, we considered the corresponding experimental setup as explained in their papers. Furthermore, for implementing the BPR and Top approaches, we tune them in a manner to achieve their best performance in order to have a fair comparison.

\begin{figure*}[t!]
\includegraphics[width=0.8\textwidth,scale=1]{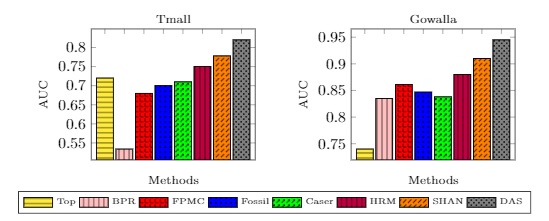}
\centering
\caption{Performance comparison of the Gowalla and Tmall  datasets in terms of AUC.}
  \label{ch6:figureAUC}
\end{figure*}

\subsection{Performance Evaluation}
\begin{figure*}[t!]
\centering
\includegraphics[width=0.8\textwidth,scale=1]{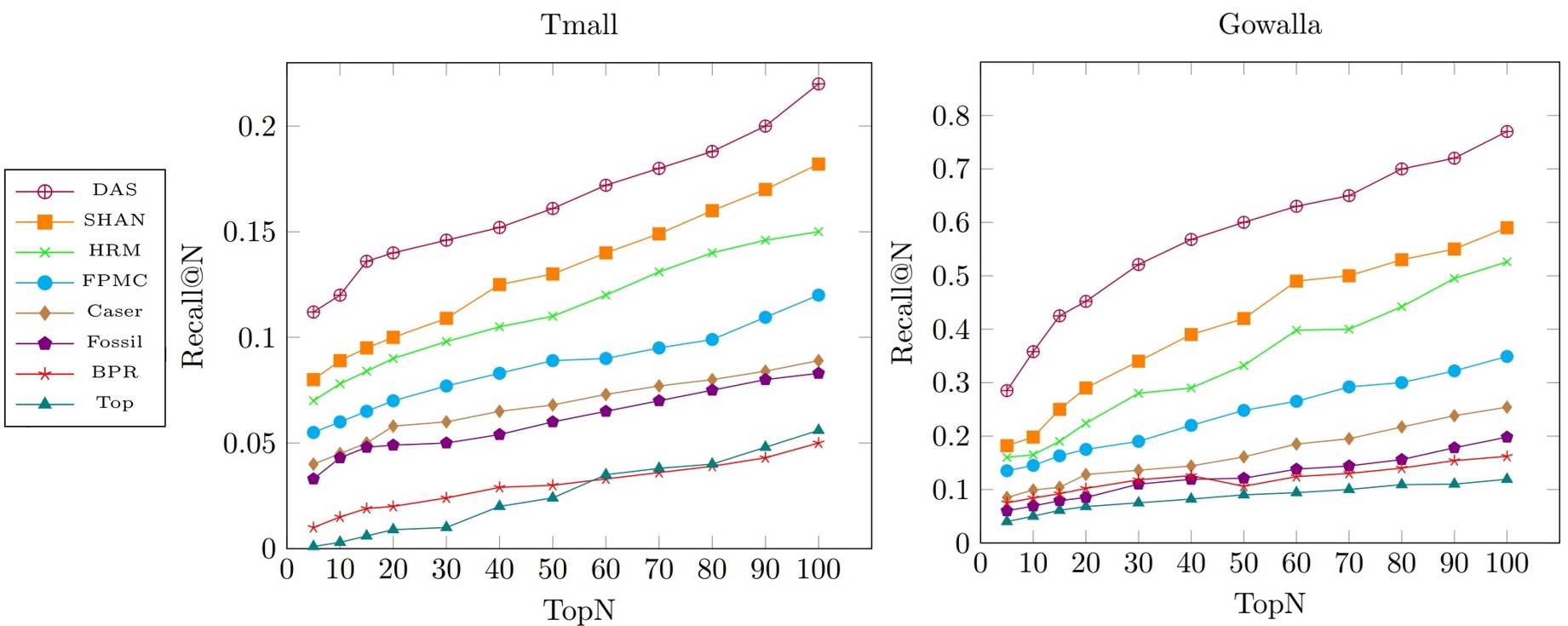}
\caption{Performance comparison on the Gowalla and Tmall  datasets in terms of Recall@N.}
 \label{ch6:figureRecall}
\centering
\end{figure*}

From Figures~\ref{ch6:figureAUC} and~\ref{ch6:figureRecall} and Table~\ref{ch6:tableComparison}, we can infer the following: (\textit{i}) DAS significantly and consistently outperformed all the compared approaches with respect to Precision, Recall, and AUC evaluation metrics on both the Gowalla and the Tmall datasets. From the results, we can infer the superiority of DAS to SHAN, which is a state-of-the-art method in SRSs. There might be two reasons behind this result. First, as stated in Section~\ref{ch6:sectionAttentionBlock}, we used the attention mechanism in DAS to measure the importance of each item as the similarity of its embedding with the item-level context vector shared by all the contextual items. While SHAN computed the attention score as the similarity between the embedding of the item and the user. Therefore, unlike SHAN, the proposed model which treated all the contextual items as a whole might better model the complex dependency relations. Second, we added the embedding of the user at the final fully connected layer to  make  a  more  personalised  recommendation. Therefore, compared with SHAN which is the second-best model, DAS showed 21.5\%  and 35.5\%  improvements on the Tmall dataset with respect to Recall@20, and Precision@10, respectively. We also observed that on the Gowalla dataset, DAS showed promising results, where it achieved 37.5\% and 0.35.4\% improvements in terms of Recall@20 and Precesion@10, respectively. 

According to Table~\ref{ch6:tableComparison}, where the best result in each row is highlighted in boldface, DAS showed the best performance. This demonstrated the effectiveness of DAS in capturing  the most important and relevant items in  the long- and short-term users' preferences by using the attention mechanism; (\textit{ii}) in comparison with the RSs that modelled both the users' preferences (e.g. HRM, FPMC, Caser, and Fossil), after DAS, SHAN outperformed the other models. On Recall@50, SHAN achieved 16\% and 27\% performance improvement as compared to that of the HRM on Tmall and Gowalla datasets, respectively.

These observations demonstrated that selecting the attention mechanism as an aggregation function performed well despite the use of simple max-pooling. Compared with that of  the HRM, the performance of FPMC was limited because of the use of the linear aggregation function. The performance of Caser and that of Fossil were close, but Caser showed a slight improvement in the Precision and Recall evaluation metrics, which might be attributed to the use of various convolutional filters. DAS achieved better performance than all the above-mentioned approaches, indicating the superiority of DAS in truly capturing users' preferences; and (\textit{iii}) among all compared methods, the approaches which combined the  users' sequential behaviours with their general taste (e.g. DAS, SHAN, HRM, FPMC, and Caser and Fossil) generally outperformed the traditional methods where the users' short-term preferences (i.e. BPR and Top) were ignored. This indicated that it was difficult for general recommender systems to recommend items that complied with recent user-item interactions.
Surprisingly, in the Tmall dataset, the Top method achieved better performance than  BPR with respect to Recall, particularly  when $N$ started increasing from 60. According to Figure~\ref{ch6:figureAUC}, a minor improvement was recorded for BRP as compared to the FPMC in terms of AUC. This indicated the trend of purchasing popular items in online shopping. In contrast to the Tmall dataset, Top achieved the lowest Recall for different $N$s on the Gowalla dataset, possibly because of the property that the Gowalla dataset had more personalised information. Finally, according to Table~\ref{ch6:tableComparison}, the proposed approach suppressed all the compared baselines in terms of Recall, Precision, and AUC in both the Gowalla and Tmall datasets.
\begin{figure*}[t!]
\centering
\includegraphics[scale=1, width=0.8\textwidth ]{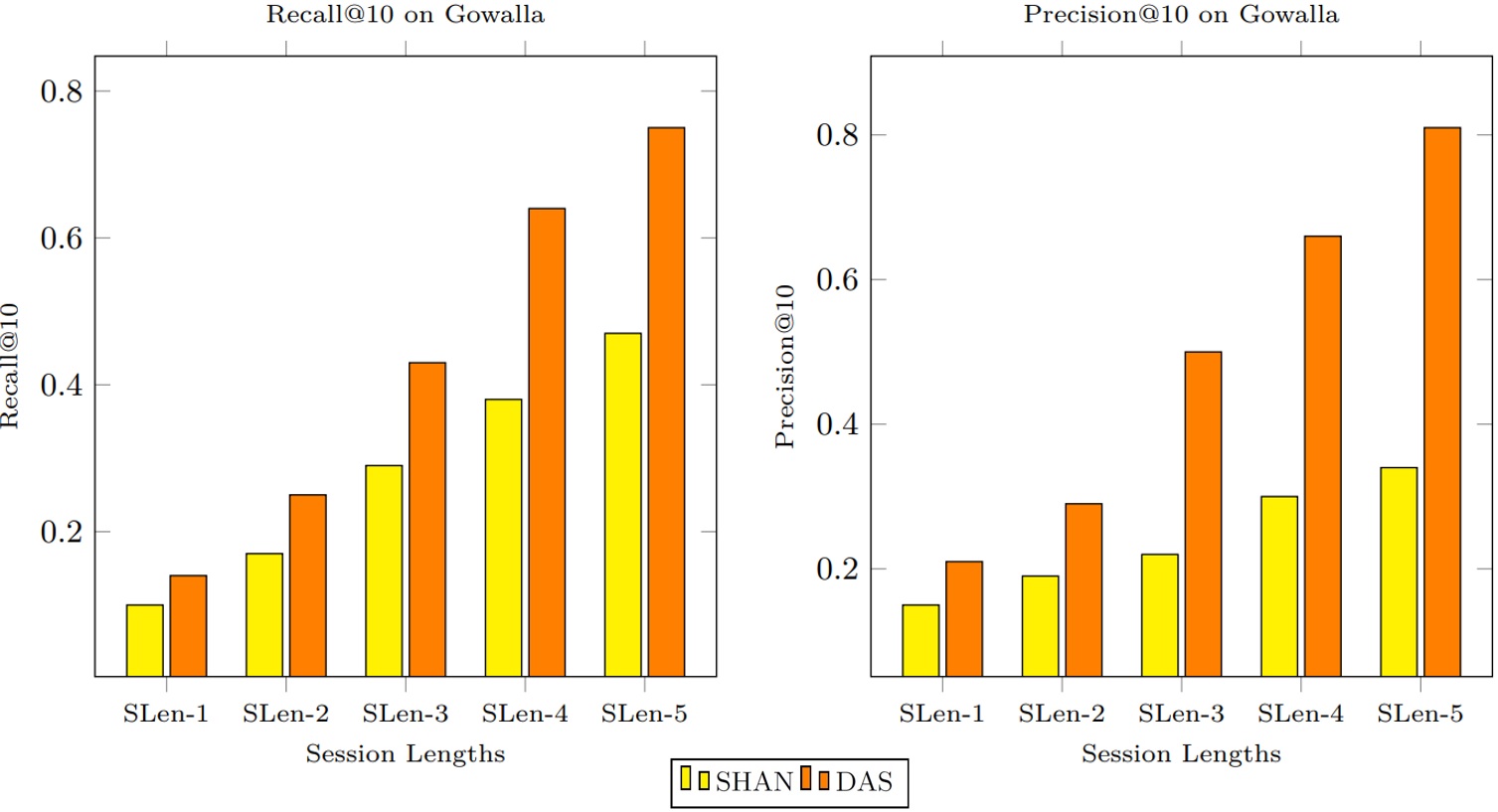}
\caption{Effect of different session lengths.}
\label{ch6:figureSLEN}
\end{figure*}
\subsection{Evaluating User-Item Interaction Sequences with Noise}
In this section, we will describe the testing of the ability of the proposed model to deal with the long and noisy sessions. To do so, first, the performance evaluation with different session lengths is discussed, followed by a discussion of  the experiments on the effect of the disorder items.
\subsubsection{Effect of Session Lengths}
Here, we will examine the performance of DAS as compared to that of the other baselines to show the advantages of the proposed model to handle the user-item interaction sequences with noise for different sequence lengths. To do so, we tested the proposed model for different session lengths, which we call SLen for simplicity. SLen denotes the number of existing items in one session. For example, SLen-5 means that there were five contextual items in a session. The longer session lengths might contain more irrelevant items which in turn could reduce the recommendation's accuracy, if the items were not recognised. From Figure~\ref{ch6:figureSLEN}, we can infer that DAS consistently outperformed the others for different session lengths. Moreover, increasing the session lengths resulted in better recommendation accuracy, where SLen-5 achieved the best results with respect to the evaluation metrics Recall@10 and Precision@10. We only tested the effect of different session lengths on DAS and SHAN, which were sensitive to the session lengths and left other models out as they mostly modelled the first-order dependency. Moreover, we only examined the effect of session lengths on the Gowalla dataset, as on average, it had longer session lengths.

\subsubsection{Effect of Disorder Items}
In this section, we aim to evaluate the behaviour of DAS under discorded data. DAS can capture the attentive context and thus focus more on the most important items in a session and ignore the irrelevant ones irrespective  of their order. To prove this, we make the Tmall dataset disordered and tested the performance of DAS under this condition. 
\begin{figure*}[t!]
  \centering
\includegraphics[scale=1,width=0.7\textwidth]{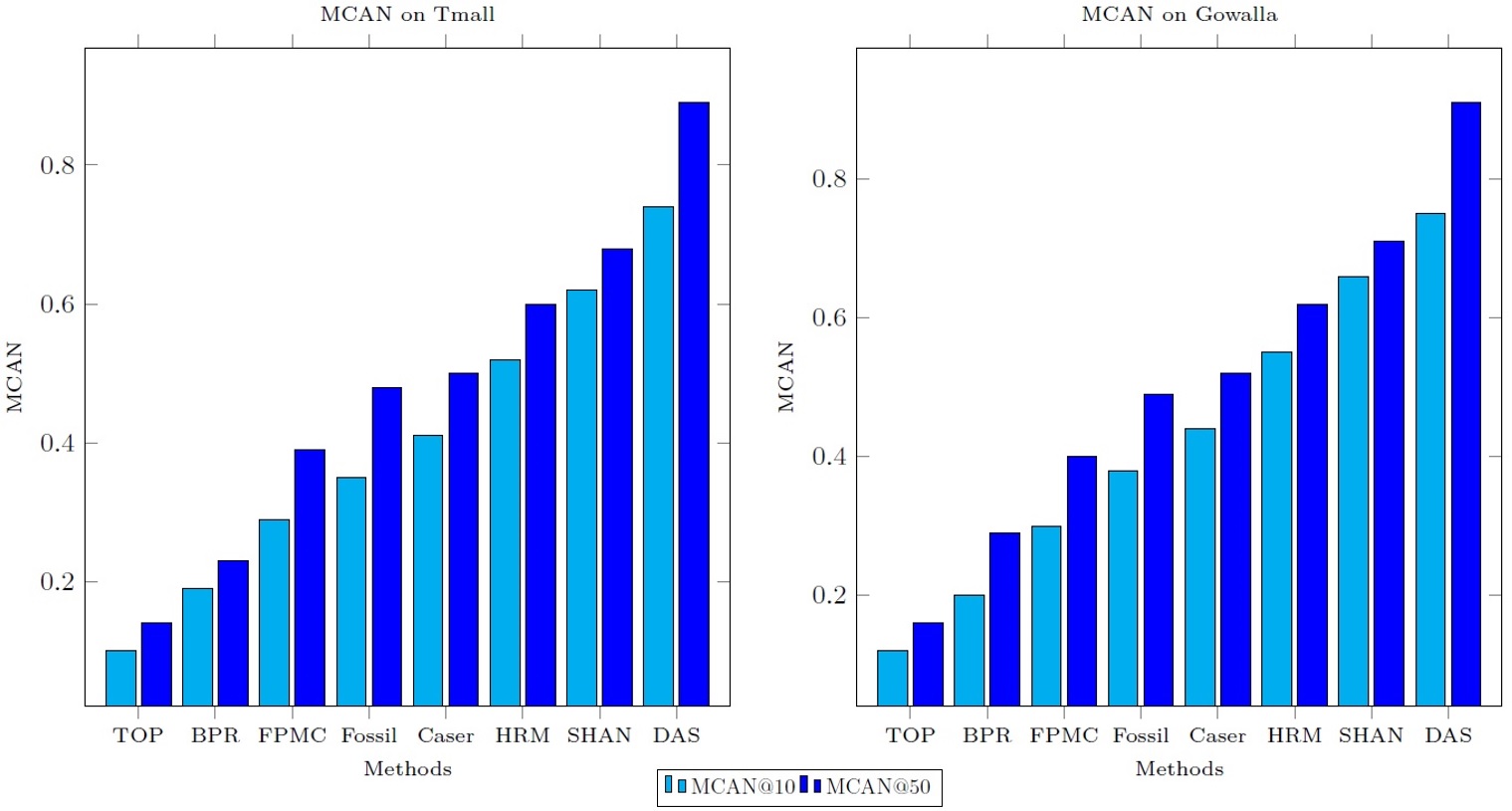}
\caption{Evaluation under novelty metric.}
  \label{ch6:figureNovelty}
\end{figure*}

\begin{table}
\centering 
\begin{tabular}{c c c c } 
\hline\hline
Model & AUC & Recall@10 & Precision@10 \\ [0.5ex]
\hline 
\textbf {DAS} & \textbf{0.8102}& \textbf{0.1043} & \textbf{0.2802}\\
SHAN & 0.6851 &0.0701 &0.1684 \\
HRM & 0.6114 & 0.0541 & 0.1211\\
FPMC & 0.5885 & 0.0521 &0.1124\\
Caser& 0.5311 & 0.0321 & 0.1021\\
Fossil& 0.5047 & 0.0294 &0.0747 \\
BPR & 0.4987 & 0.01020 & 0.0665 \\
TOP &0.3851 &0.0019 & 0.06024\\[1ex] 
\hline 
\end{tabular}
\caption{Performance evaluation on disordered Tmall dataset.}
\label{ch6:tableDisordereddata}
\label{disorder}
\end{table}

To do so, as in~\cite{DBLP:conf/aaai/WangHCHL018}, we randomised the default item's order in the Tmall datasest to create a new disordered dataset. Table~\ref{ch6:tableDisordereddata} demonstrates the superiority of DAS on the new dataset. According to  Table~\ref{ch6:tableDisordereddata}, unlike DAS, the performance of the rest of the compared methods deteriorated in a disordered dataset, indicating the strong ability of DAS to emphasise on the important items irrespective of where they were in a transaction.

\subsubsection{Novelty Evaluation} 
Besides the above-mentioned evaluation metrics, we compared the performance of DAS with that of the existing approaches in terms of the novelty metric. Considering that recommending items similar to those that a user has already purchased, may not be satisfactory and users may be more willing to be recommended by new items. As in~\cite{DBLP:conf/aaai/WangHCHL018}, we considered a novelty metric as another evaluation parameter to show the capability of the proposed model to recommend novel items. The novelty metric could measure the difference between contextual items in a shopping basket and a set of recommended items, where the larger difference can represent the novel items~\cite{DBLP:conf/pkdd/WangHC17}. The proposed model by discovering an attentive context and measuring the contributions of each item in both long- and short-term item sets, can provide users with new and unseen items and avoid recommending duplicated items.

\textbf{MCAN@K}. In the proposed model, items purchased in the context \textit{c} can be used for recommendation $R$. A larger overlap between the recommended and the purchased items denotes the less novelty. Subsequently, the novelty measures the mean of the unseen items corresponding to the context $c$ over all $N$ top-K recommended items.
\begin{equation}
MCAN= \frac{1}{N}\sum_{i=1}^{N}(1-\frac{|R_i\cap c_i|}{|R_i|})
\end{equation}

Figure~\ref{ch6:figureNovelty} illustrates the performance of DAS as compared to that of the other methods in terms of the novelty metric on both Tmall and Gowalla datasets. From Figure~\ref{ch6:figureNovelty}, we can infer that both the Tmall and the Gowalla datasets showed a similar trend. Among the approaches that only considered the users' long-term preferences (e.g. BPR and Top), Top recommended the top  popular items to a user, and hence, it was more likely that those items had been already observed  by the user. Thus, TOP achieved the lowest novelty score. Although FPMC took both types of users' preferences into account, it could not well learn the parameters on such sparse datasets. Therefore, the recommended items might be relatively random ones, and accordingly, users could not experience a novel item in the case of FPMC method, resulting in a low novelty score. Besides SHAN, compared with the models that combined both long and short-term users' preferences, HRM reported a high level of novelty may be because of the use of the different aggregation functions as well as optimisation criteria. After DAS, SHAN had a higher novelty score than all of the aforementioned methods because of the use of the attention mechanism that considered the dynamic properties in the users' long and short-term preferences. To sum up, the reported results indicated that DAS might better capture the most influential and contextual items through the attentive context in both the long and the short-term users' preferences, and thus represent a promising result in SRSs.

\begin{table*}[h!]
    \begin{minipage}{.5\linewidth}
    \centering
    \caption{Tmall}
    
        \begin{tabular}{||c | c | c ||} 
\hline Methods & AUC & Recall@20\\
\hline
\hline
DAS-long & 0.720 & 0.032 \\
\hline
DAS-short & 0.781 &  \textbf{0.175}   \\
\hline
DAS  &  \textbf{0.821} & 0.149 \\
\hline
\end{tabular}
    \end{minipage}
    \begin{minipage}{.5\linewidth}
       \caption{Gowalla}
       \centering
     
           \begin{tabular}{||c | c |  c||} 
\hline Methods & AUC & Recall@20\\
\hline
\hline
DAS-long & 0.910 & 0.245  \\
\hline
DAS-short & 0.934 &  0.425 \\
\hline
DAS  & \textbf{0.945} & \textbf{0.443}\\
\hline
\end{tabular}
    \end{minipage} 
    \caption{Impact of different types of users' preferences on Recall@20.}
 \label{ch6:tableDifferentpreferences}
    \end{table*}
    \begin{table*}[h!]
    \begin{minipage}{.5\linewidth}
    \centering
      \caption{ Tmall}
      
            \begin{tabular}{||c | c | c | c||} 
 \hline
\backslashbox{$\lambda_ \alpha$ }{$\lambda _{uv}$ }  & 0.01 & 0.001 & 0.0001\\
\hline
\hline
0 & 0.091 & 0.084 & 0.077  \\
\hline
1  & 0.132 & 0.122 & 0.109 \\
\hline
10  & 0.142 & 0.131 & 0.119 \\
\hline
50 & 0.149 & 0.139 & 0.128\\
\hline
\end{tabular}
    \end{minipage}%
    \begin{minipage}{.5\linewidth}
    \centering
        \caption{Gowalla}
          
        \begin{tabular}{||c | c | c | c||} 
 \hline
\backslashbox{$\lambda_ \alpha$ }{$\lambda _{uv}$ }  & 0.01 & 0.001 & 0.0001\\
\hline
\hline
0 & 0.242 & 0.365 & 0.384  \\
\hline
1  & 0.347 & 0.422 & 0.458 \\
\hline
10  & 0.354 & 0.414 & 0.449 \\
\hline
50 & 0.361 & 0.443 & 0.452\\
\hline
\end{tabular}
    \end{minipage}\caption{Impact of different regularisation at Recall@20.}
\label{ch6:tableDifferentRegularization} 
\end{table*}

\subsection{Impact of Different Types of Users' Preferences}

In this section, we aim to evaluate the impact of each of the long and short-term users' preferences on the next item recommendation problem, separately. Therefore, we will consider a version of DAS, called DAS-long, for situations wherein only a user's long-term preference is modelled and another version, called DAS-short, which only considers the user's short-term preferences. A comparison of the performance of DAS with that of DAS-long and DAS-short is presented in Table~\ref{ch6:tableDifferentpreferences}. Compared with the approaches which only model a user's general tastes, DAS-long performed better than BPR by 17\% and 13\% with respect to Recall@20 on the Tmall and Gowalla datasets, respectively. This could explain why the use of the attention mechanism could truly identify a set of context-relevant items in the user's general taste, leading to a better presentation of the user's mixture of preferences. On both the datasets, DAS-short achieved better performance than DAS-long, indicating the importance of sequential behaviours in the next item recommendation problem. Remarkably, DAS-short represented better performance than SHAN on both the datasets, as SHAN achieved AUC of 0.778 and 0.901 on the Tmall and the Gowalla datasets, respectively. This might explain although the main focus of DAS-short was on identifying the most important items in short-term interacted item set, a part of the user's general taste was learned during the training process.

In a nutshell, Table~\ref{ch6:tableDifferentpreferences} indicates the superior performance of DAS to that of DAS-long and DAS-short. This demonstrated that considering both types of users' preferences could result in better users' preferences modelling and increase the recommendation's accuracy, accordingly.

\subsection{Impact of Hyper-Parameters}
In this section, we will investigate the impact of the hyper-parameters on the performance of DAS. As in~\cite{DBLP:conf/ijcai/YingZZLXXX018}, we demonstrated the results just under Recall@20. The size of the item and user embeddings was the same and was set to 100. We also empirically set the batch size to 50. We considered $\lambda_{uv}=\{0.01, 0.001, 0.0001\}$ as the user and item embedding regularisation, and $\lambda_a=\{0, 1, 10, 50\}$ as the attention network regularisation. As shown in Table~\ref{ch6:tableDifferentRegularization}, the performance of DAS gradually improved when $\lambda_a$ started moving from 0 to 50 on both the Tmall and the Gowalla datasets, which implied that using the attention mechanism could help us to better discover the users' preferences.

\section{Summary}
\label{ch:Chapter6Summary}
To avoid a noisy session and suggest items that satisfy both types of users' preferences, which are related to the session's context, in this chapter, we proposed the attention-based recommender model DAS. While MC and RNN were originally designed for time-series data, they assumed a rigid natural order in a session which limited their applications in SRSs. The current SRSs have the same overly strong assumption that there is a strict order between items in a shopping basket. This observation may not be true in some real-world applications, and there may be irrelevant items in a session which in turn may generate a false dependency. Furthermore, the existing  attempts either do not take the long-term users' preferences into account or consider a static low-dimensional representation when modelling a user's representation. To fill this gap and provide users with novel items, in this chapter, we have proposed a Deep Attention-based Sequential model, DAS, for the next item recommendation problem. We have first embedded the users and the items into the latent dimensional spaces and then passed them into two attention networks to discriminate the contributions of each item to both long- and short-term users’ preferences. Next, we have combined these two types of users' preferences to learn a mixture of the users' preferences through a deep network. Finally, we have fed a concatenation of the learned  users' mixture preferences with the users' embedding to the final deep neural network to make a personalised item recommendation. In particular, the proposed method could model complex and abstract user-item interactions by using a non-linear aggregation function. Extensive experiments on two real-world datasets have demonstrated the superiority of the proposed model to the state-of-the-art methods in terms of the adopted evaluation metrics.

\chapter{A Convolutional Attention Network for Unifying General and Sequential Recommenders}
\label{ch:chapter7CAN}
\section{Introduction}
In this chapter, we will focus on another critical challenge in SRSs: modelling collective dependency. As shown by the complex and comprehensive design in Figure~\ref{ch1:figureRWcat} in Chapter~\ref{chap:introduction}, this issue is placed on the interaction level, which is the second level of this structure. Collective dependency refers to the case of a transaction consisting of multiple items where, there may be a specific purpose behind purchasing a collective set of items. However, the current SRSs regard each item to be a separate entity with no particular purpose, and each user-item interaction in a sequence is independent. This may be a too simplistic an assumption, as there may be a particular purpose behind buying the successive items in a sequence. In fact, a user makes a decision through two sequential processes, i.e. start shopping with a particular intention and then selecting a specific item which satisfies her/his preferences under this intention. Moreover, different users usually have different purposes and preferences, and the same user may have various intentions. Thus, different users may click on the same items with attention to different purposes. Therefore, a user's behaviour patterns and preferences may not be completely exploited in most of the current methods and these methods neglect the distinction between the users' purposes and their preferences. To bridge this gap, we propose a novel method named CAN, which takes both the users' purposes and preferences into account for the next-item recommendation task. 

\section{The Target Problem and The Motivation}
Because of the information explosion, people are surrounded by a large number of options and services. Therefore, there is a need for a tool to help customers with their decision-making process, find items of their interest, and alleviate the information overload problem. Recommendation systems have emerged as a platform which automatically recommends a small set of items in order to help users find their desired items in online services. On the basis of how the users' preferences are modelled, there are two major types of recommenders: general recommenders which focus on modelling the general users' taste and sequential recommenders with the aim of modelling the users' sequential behaviour. Although there is an extensive repository of methods in RSs that try to overcome the potential obstacles of RSs and make a more accurate recommendation, some issues are still challenging and more efforts are required. One of the main  difficulties is how to exploit the users' decision-making pattern in order to make a personalised recommendation. This in turn may lead to boost the recommendation's performance and increase the business profit and the customer's satisfaction and loyalty. To do so, we need to pay more attention to both long and short-term users' preferences, as they have a significant effect on the users' final decision. Moreover, while a few studies have tried to mix both types of users' preferences, they have mainly taken each user-item interaction independently and considered each item in a sequence as a separate entity. Because of this assumption, most of the current studies fail to capture the local contexts in a session and ignore a user’s purpose, which is reflected by a set of clicked successive items in a session. Additionally, the same user may have various purposes, and different users may have different purposes by clicking on the same items. For instance, while Alice buys a new bag for herself, Rose purchases it as a gift for her sister. Furthermore, different items within a session may have different informativeness for revealing the users' purposes and preferences. To make it clear, imagine $S=\{butter, milk, eggs, cake powder\}$ as a shopping basket of Mike for making a cake. Here, \textit{cake powder} can better represent the purpose of Mike's shopping, and \textit{butter} may be less representative of his purpose. Therefore, most of the previous works have neglected the hierarchical distinction between the users' purposes and the users' preferences, which in turn makes it a challenging task to fully exploit the users’ preferences and their decision-making patterns.

Let us make the above problem more clear with an example. Usually, a user's decision-making process is a combination of two sequential steps: a user's main purpose and her/his preference. Take the shopping event of a user for example: she/he starts shopping with a specific purpose and then keeps looking into different items until she/he finds items that satisfy her/his preference. Suppose that Alice is a PhD student and her previous actions are mostly related to her field of study, such as looking for a workshop and finding an article. Alice has a plan to travel overseas for presenting her work at an international conference. She starts booking her flight and hotel, and her next action may be visiting some universities or institutions. The current RSs may recommend tourist attractions or car rental companies to her because many users may look for them after booking a hotel and a flight, ignoring her educational purpose of this travel, which is hidden inside her long-term interacted item set. This example can show the importance of taking the users' purposes and their preferences into account to better capture the users' behaviours and generate personalised suggestions.

According to the mentioned observation, we can observe that the users' decision-making patterns have not been exploited thoroughly by the existing models. To model the users' preferences, some studies have tried to only consider long-term users' preferences by modelling the users' past behaviours~\cite{DBLP:journals/computer/KorenBV09}, \cite{DBLP:conf/nips/SalakhutdinovM07}. For instance, matrix factorisation is one of the most widely used methods in this setting, which learns the user-item interactions in a latent vector space to model the general users' preferences~\cite{DBLP:reference/rsh/KorenB11}, \cite{DBLP:conf/www/SarwarKKR01}. 
In contrast, sequential recommenders focus on capturing sequential patterns from the previously visited items~\cite{DBLP:conf/ijcai/WangHWCSO19}.  A conventional sequential recommender is based on Markov Chains (MCs), which assumes that a user’s next item interactions considerably depend on only one or several of the most recent interactions~\cite{DBLP:conf/kdd/GrbovicRDBSBS15}. For instance, SPMC exploits both sequential and social information to make a more personalised recommendation model~\cite{DBLP:journals/corr/abs-1708-04497}. Owing to the deep learning methods which have shown great capability in modelling the complex interactions between users and items, among the deep neural network techniques, Recurrent Neural Network (RNN) has become a dominant paradigm in sequential recommenders because of its capability for sequence modelling. Apart from using basic RNN~\cite{DBLP:journals/corr/HidasiKBT15}, \cite{DBLP:journals/corr/ZhangDXFWBWL14}, improved architectures such as Long Short-Term-Memory (LSTM)~\cite{DBLP:conf/wsdm/WuABSJ17} and Gated Recurrent Unit (GRU)~\cite{DBLP:conf/recsys/HidasiQKT16} have
also been introduced to better model dependencies in a longer sequence. Different from RNN, convolutional neural networks (CNNs) are introduced to model complex relations over the user-item interaction sequences~\cite{DBLP:journals/corr/HidasiKBT15}, \cite{DBLP:conf/wsdm/TangW18}, \cite{DBLP:conf/wsdm/YuanKAJ019}. CNN stores the embedding of the user-item interaction sequences in a matrix and then treats this matrix as an image~\cite{DBLP:conf/wsdm/TangW18}, \cite{DBLP:conf/wsdm/YuanKAJ019}. 

Although the basic deep neural networks (i.e. RNN and CNN) have shown considerable success in modelling sequential dependencies, they may have some shortcomings in modelling complex relations between users and items. Thus, three advanced models have been introduced to overcome this problem: (i)\textit{attention mechanism}: by more focusing on relevant and important interactions in a sequence~\cite{DBLP:journals/corr/abs-1808-10031}, \cite{DBLP:conf/ijcai/YingZZLXXX018}; (ii) \textit{memory networks}: by incorporating an external memory matrix~\cite{DBLP:conf/wsdm/ChenXZT0QZ18}, \cite{DBLP:conf/wise/HuHSN19}; and (iii) \textit{mixture models}: by combining the strength of the current deep neural models~\cite{DBLP:conf/www/TangBJCBXC19}. While sequential recommender models are good at capturing the sequential dependency, they mostly recommend items similar to those that a user currently visited and the general users' preferences are ignored.
Both the aforementioned classes of approaches have their strengths and shortcomings~\cite{DBLP:conf/sigir/WangGLXWC15}. Although general recommenders have been widely adopted to capture the long-term users' preferences, their performance is limited as they ignore the short-term users' preferences. A major advantage of the sequential recommenders is their capability to model sequential dependencies, e.g. a customer who has recently purchased an iPhone is more likely to buy an iWatch next. However, sequential recommenders discard  prior user-item interactions within user behaviours and thus, fail to capture the general users' preferences~\cite{DBLP:conf/ijcai/DongZZW18}.

Therefore, it is better to build a recommender system which benefits from the advantages of both general and sequential recommenders. For instance, FPMC is one of the pioneering works in the literature, which fuses MF and MC into one model in order to learn both the users' long and short-term preferences~\cite{DBLP:conf/www/RendleFS10}. Soon after, Hierarchical Representation Model (HRM) was proposed by Wang et al.~\cite{DBLP:conf/sigir/WangGLXWC15} which non-linearly models both sequential behaviours and the users’ general taste to make a better recommendation. While FPMC and HRM have exploited  the users' long-term preferences to improve the performance of sequential recommenders, CoFactor benefits from integrating a co-occurrence item-to-item matrix into an MF model
~\cite{DBLP:conf/recsys/LiangACB16}. Moreover, BINN, which was proposed by Li et al.~\cite{DBLP:conf/kdd/LiZLHMC18}, is another attempt in unifying both types of users' preferences. The authors stated that different types of users' actions (e.g. browse, click, collect, cart, and purchase) need to be treated differently. Their proposed model consists of two main components: Neural Item Embedding and Discriminative Behaviours Learning. The first component of BINN tries to find the item similarities
by analysing the users' sequential behaviours, while in the second component, two alignments, namely Session Behaviours Learning (SBL) and Preference Behaviours Learning (PBL), are introduced to learn discriminative behaviours~\cite{DBLP:conf/kdd/LiZLHMC18}. Although BINN can record a significant improvement over several state-of-the-art models, it uses LSTM for the discriminative behaviours learning part, which may limit the performance of their recommender system, as it may not be able to capture the dynamic property of the users' preferences. Moreover, BINN only considers the purchase behaviour for modelling the users’ historical preferences. This  may not only result in the lose of some useful information by exploiting other types of user behaviours (e.g. click and add to cart) but also fail to learn the latent users' purpose, which is hidden in a collection of successive user-item interactions. Loyola et al.~\cite{DBLP:conf/recsys/LoyolaLH17} proposed an encoder-decoder attention-based architecture for modelling the users' sessions. While the attention mechanism has shown better result than the traditional deep learning models (e.g. RNN and LSTM), it uses a fixed query vector~\cite{DBLP:conf/ijcai/YingZZLXXX018} and cannot be personalised to different users' preferences. Furthermore, although the authors have stated that they could model the users' intents by analysing a set of user activities, they only considered the browsing and purchasing behaviours as the users' intents. In other words, their model could predict what the intent of a user's next action on items would be (e.g. browsing items or purchasing items). Thus, it may not be sufficiently suitable to simply assume  these two types of users' future behaviours as their intentions, as the users' purposes may be hidden in a previous set of user-item interactions.  


The above illustrations reveal the difficulty of capturing the collective dependency in a session. In other words, the next choice of item may not be affected by a part of the current session, but all the items will need to be taken into consideration as a collective of the interacted items may have a particular purpose. Moreover, most of these works have taken the user-item relationships into consideration from the static views, and the dynamic property of the users' preferences are ignored. More importantly, not only are the users' main purposes forgotten ,but there is also
no difference between the contributions of the same items in modelling the  preferences of different users. Therefore, how to fully exploit the users' decision-making process and simultaneously consider both the users' motivations and their current interests is still largely unexplored.

Therefore, in this chapter, we aim to fully exploit the users' decision-making process to provide a better recommendation. To do so, we will discover a user’s main purpose and her/his preferences over a set of interacted items. We will apply CNN to the long-term item set to identify the main purpose of purchasing a set of items, and  then in order to discriminate the contribution of each item to both the long and the short-term item sets, a PSAU cell is proposed. Finally, in the prediction layer, a recommendation is made by learning a mixture of long and short-term preferences.

\section{Proposed Design and Main Contributions}
To address the above-mentioned issues, we propose a novel model called CAN, which unifies the benefits of both general and sequential recommenders. CAN consists of two main modules: purpose encoder and preference encoder. In the purpose encoder, we first embed users and items into low-dimensional vectors and then use  a CNN to identify the user purposes by capturing the local and high-level information of the long-term interacted item set. Then, we propose to use a Purpose-Specific Attention Unit (PASU) to differently attend to different items and fully exploit the different informativeness of different items. Next, at the preference encoder, we also utilise PASU  to learn the items' informativeness in the short-term interacted item set to better understand the users' preferences. Lastly, a final user representation is learned by coupling the users' long-term and short-term
preferences. The model’s parameters are learned  by using the Bayesian personalised ranking optimisation criterion to generate a pair-wise loss function~\cite{DBLP:conf/uai/RendleFGS09}. From the experiments, inferred the superiority of the proposed model to the state-of-the-art algorithms on two datasets. The main contributions of this work are summarised as follows:

\begin{itemize}
    \item We propose a PASU network to take user embedding as the query vector of the purpose-level and personal preference-level attention networks to differentially attend to important items according to the users' purposes and preferences.
    \item We introduce a unified framework, named CAN, integrating a CNN and a PSAU component to model a user's purpose and personal preferences.
    \item We use PSAU in both the long-term and short-term interacted item sets to generate a high-level hybrid user representation.
    \item We conducte extensive experiments on two real-world datasets. The experimental results demonstrated the superiority of the proposed model to the state-of-the-art methods.
\end{itemize}
\begin{figure*}
\centering
\includegraphics[width=0.9\textwidth,scale=1]{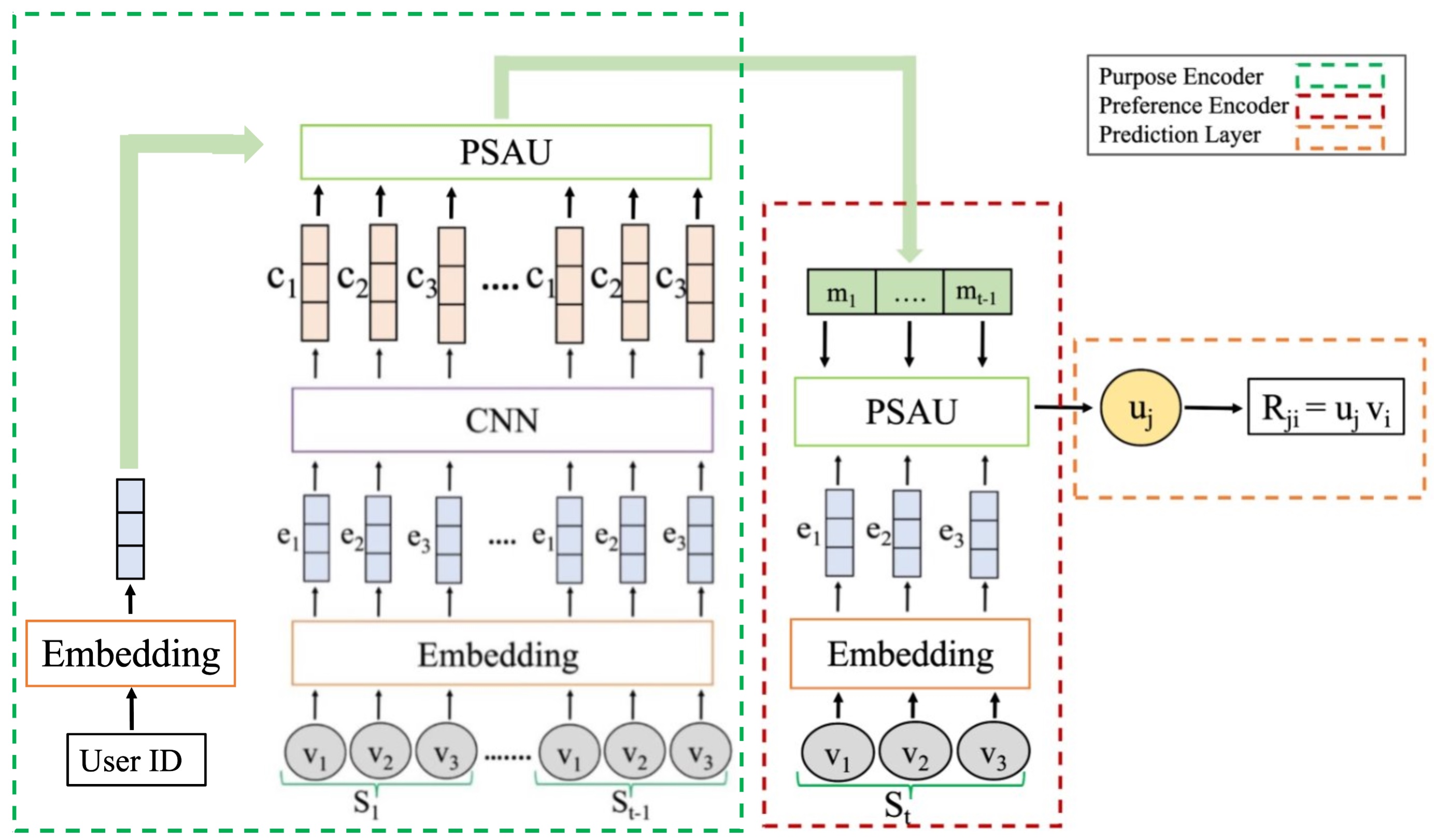}
\caption{Architecture of CAN, which consists of two main modules, namely purpose encoder and preference encoder.}
\label{Ch7:figureCAN}
\end{figure*}

The rest of this chapter is organised as follows: we discuss the overview and the framework in Section~\ref{ch:Chapter7OverviewandFramework}. The experiment and evaluation are presented in Section~\ref{ch:Chapter7ExperimentandEvaluation}, before concluding the chapter in Section~\ref{ch:Chapter7Summary}.

\section{Overview and Framework}
\label{ch:Chapter7OverviewandFramework}

The framework of CAN is illustrated in Figure~\ref{Ch7:figureCAN}. As shown in Figure~\ref{Ch7:figureCAN}, the proposed model consists of two main modules: (1)~\textit{purpose encoder} and (2)~\textit{preference encoder}. The first module aimed to learn the main purpose of the users' long-term interacted item set. Considering that all the users' historical motivations may be implied in a long-term user's behaviours, we used the purpose encoder to take a set of user-item interactions in the long-term item set and embed it into low-dimensional vector representations. Then, these vectors were passed to a CNN to effectively capture the local contextual information hidden in a sequence of user-item interactions in order to identify the user's main purpose. We propose to use a Purpose-Specific Attention Unit (PSAU) to differentially attend to the users' main purposes and emphasise the informativeness of purchasing the same item for different users. The reason behind using PSAU in the purpose encoder is that different users may have different purposes for buying the same items. 
The next module is the (2)~\textit{preference encoder}, which aims to learn the users' current preferences. The same user may have different preferences, and each item may be more or less informative for that specific preference. Hence, PSAU is  applied here to discriminate each item's informativeness level.

\subsection{Problem Statement}
In this section, we investigate the next-item recommendation problem with implicit feedback data. Let us consider $U= \{u_1,u_2,...,u_{|u|}\}$ as the user set and $V= \{v_1,v_2,...,v_{|v|}\}$ as the item set, where $|u|$  and $|v|$ are the total number of users and items, respectively. For each user $u$, we defined $G^u=\{S^u_1,S^u_2,..., S^u_T\}$ as her/his transaction history, where $T$ is the total number of sessions and each session $S^u_t \subseteq V (t \in [1, T])$, where $S^u_t$ represents a set of interacted items for users $u$ at time step $t$. We denoted $S^u_t$ as the short-term preference item set of user $u$  (i.e. her/his sequential behaviours) at a specific time step $t$. In addition to the short-term preference, the long-term preference item set of user $u$ was important for identifying items that users would interact with in the near future. Therefore, we considered $G^u_{t-1}= \bigcup_{t=1}^{t-1}S_t^u$ to reflect the long-term preference of user $u$ (i.e. general preference), where $G^u_{t-1}$ is a set of interacted items before time step $t$. For the rest of this chapter, we will call $G^u_{t-1}$ and $S^u_t$ as the long-term and short-term interacted item sets regarding time step $t$, respectively. Given $G^u$ as $u$'s transaction history, we aim to predict the next items which this user will most likely purchase by learning her/his historical motivations and present consumption.

\subsection{Purpose Encoder}
Usually, the users' decision-making process consists of two sequential and critical steps, namely, a user's historical purposes and her/his recent users' preferences. Normally, people start shopping with a motivation and then browse for different items until they find items of interest  that satisfy their preferences. The above-mentioned  purpose encoder module has three core components: (i) embedding look-up, (ii) convolutional neural network, and (iii) Purpose-Specific Attention Unit (PSAU).

In this block, we aim to first convert a session of items into a sequence of low-dimensional
dense vectors. Then, we will use a convolutional neural network for capturing the local information, as the local contexts within a set of interacted items may imply a user's purpose. For instance, Julia wants to have a Halloween party. She goes shopping and puts a set of \{\textit{hanging ghost}, \textit{pumpkin},  \textit{lollipop}, \textit{plastic blood bag}\} together. In this collection of items, the local combination of the `hanging ghost', and `plastic blood bag' may be more important to show the user's main intention of this shopping expedition. Therefore, we use a CNN here to learn the contextual
representations of a set of items. Finally, at this block, the PSAU is applied to distinguish the level of informativeness of different items for revealing the users' motivations for purchasing a set of items together. The reason behind using a PSAU in the purpose encoder is that different items may have different levels of contributions in presenting a user's main purpose, and the same words may have different informativeness for the recommendation of different users. Therefore, we need to identify important items for demonstrating a shopping session's purpose for different users, and thus, the personalised attention-based network is proposed for application in this block.


\subsubsection{Embedding Look-Up}
First, we used embedding look-up to the embed user and item IDs (i.e. one-hot representations) into two continuous low-dimensional spaces, where $e_i$ represents the item embedding vector of item $i$, and $u_j$ denotes the user embedding vector of user $j$. The embedding matrix is denoted by $E=[e_1, e_2, ..., e_i]$, $E \in {R}^{|V| \times D}$, where $D$ and $|V| $ represent  the embedding dimension and the total number of items, respectively. The matrix $U \in {R}^{D\times|U|}$ is the user embedding matrix, where $u_j$ denotes the user embedding vector of user $j$.

\subsubsection{Convolutional Neural Network (CNN)}
Second, we used CNN to learn the contextual
information of user-item interactions~\cite{DBLP:conf/emnlp/Kim14}. CNN is one of the deep learning techniques with an excellent capability to capture local information~\cite{DBLP:conf/wsdm/WuWLHHX19}. Therefore, we used CNN to capture the user’s main purpose in the long-term item set. Next, we used a convolution operator on the matrix $E$ as the concatenation of the items' embedding vectors. Let $K_w \in {R}^{N_f\times(2K+1) D}$, and $b_w \in {R}^{N_f}$ denote the parameters of the CNN, in which $K_w$ is the kernel and $b_w$ represents the bias parameters. $N_f$ is the number of CNN filters, and $2K+1$ is the window size of the CNN. Then, $c_i$ illustrates the contextual representation of item $i$:

\begin{equation}
    c_i= ReLU (K_w \times e_{{\lfloor i-k\rfloor}:{\lfloor i+ k\rfloor}} + b_w)
\end{equation}
where $e_{{\lfloor i-k\rfloor}:{\lfloor i+ k\rfloor}} \in G^u_{t-1}$ is the combination of the embedding vectors of items from position ${\lfloor i-k\rfloor}$ to position ${\lfloor i+ k\rfloor}$. We used ReLU as  the non-linear activation function.

\subsubsection{Purpose-Specific Attention Unit (PSAU)}
The last component in the \textit{purpose encoder} is the Purpose-Specific Attention Unit (PSAU), to differentially attend to important items according to the user purposes. In a sequence of user-item interactions, each item may be more or less informative for learning the users' purpose representation. For instance, imagine \{\textit {pizza bread}, \textit{pepperoni}, \textit{cheese}\} as a set of items purchased together for making a pizza. In this shopping basket, \textit {pizza bread} is more informative for representing the users' purposes than \textit{cheese}. Furthermore, different users may purchase the same items for different purposes. Therefore, identifying the contributions of different items for different users plays an important role in personalised recommendation. However, most of the current approaches use a classic attention network which computes the attention score as a weighted sum over the embeddings of items and a fixed attention query vector, ignoring the users' main purposes. To learn the informativeness of each item for different users, we propose the use of the PSAU cell to identify the most informative items related to the users' main purpose within a user-item interaction sequence. PSAU first takes the embedded user-ID vector $u_j^ {'}\in R^{D_u}$, where $D_u$ is the user embedding dimension. Then, we used a dense non-linear layer to transform the embedding vector $u_j^ {'}$ to the purpose-level user preference vector $p_j$, which could be formulated as follows: 

\begin{equation}
p_j= ReLU(W_1 \times u_j^ {'} + b_1),
\end{equation}
where $W_1 \in {R} ^{D_u \times D_p}$ and $b_1 \in {R} ^{D_p \times 1} $ are the model parameters and $D_p$ is the preference vector dimension. Next, we denoted $\alpha _j$ as the attention score of item $j$, which could extract the level of informativeness of each item according to the users' main purpose. The attention score  $\alpha _j$, was calculated on the basis of the interaction between the user preference vector and the contextual item representations, as follows :
\begin{equation}
a_i= c_i^T tanh (W_2 \times p_j + b_2),
\end{equation}
\begin{equation}
\alpha_i=  \frac {exp(a_i)}{\sum_{i \in G^u_{t-1}} exp(a_i)}
\end {equation}
where $W_2 \in {R} ^{D_p \times N_f}$ and $b_2 \in {R} ^{N_f\times 1} $ are the model parameters. Next, the user's main purpose representation $m_i$ was modelled as a weighted sum of the contextual representation of item $i$ with its attention score. Formally, this representation can be formulated as follows:
\begin{equation}
 m_i= {\sum_{i \in G^u_{t-1}} {\alpha_i}{c_i}} 
\end{equation}

\subsubsection{Preference Encoder}
As is clear from Figure~\ref{Ch7:figureCAN}, PSAU is also used in the preference encoder module to learn an informative user short-term preference representation. Different users may have different preferences by clicking on the same items, and different items are more or less informative for modelling the users' preferences. Hence, we used PSAU  here as well to model the different informativeness of the same items for different users. We first took the item embedding $e_i \in S^u _t$ in a short-term interacted item set to model a user preference vector $p_d$, as follows: 

\begin{equation}
p_d= ReLU(W_3 \times e_i + b_3),
\end{equation}
where $W_3 \in {R}^{D_u \times D_q}$ and  $b_3 \in {R}^{D_q \times 1}$ and $D_q$ is the preference query size. Next, the attention weight $\alpha ^{'}_i$ represented the level of informativeness of item $i$ in the short-term users' preferences, which could be computed by the interactions between the user's purpose representation and the user preference vector. Then, the softmax function was used to normalise the attention weight, which was calculated as follows:

\begin{equation}
a^{'}_i= m_i^T tanh (W_4 \times p_d+ b_4),
\end{equation}
\begin{equation}
\alpha ^{'}_i=  \frac {exp(a_i)}{\sum_{i \in S^u_{t}} exp(a_i)} 
\end{equation}
where $W_4 \in  {R}^{D_q \times N_f}$ and $b_4 \in {R}^{N_f \times 1}$ are the model parameters. 
Finally, the contextual user representation $u_j$ was computed as follows:

\begin{equation}
u_j= {\sum_{i \in S^u_{t}} {a ^{'}_i}{m_i}} 
\end{equation}

\subsubsection{Prediction Layer}
After the final user representation $u_j$ was learned, we calculated its inner product and the item representation $v_i$ in order to compute the user preference score $R_{ij}$ as follows:
\begin{equation}
R_{ij}= u_jv_i
\end{equation}

Next, as in~\cite{DBLP:conf/uai/RendleFGS09}, we used a pair-wise loss function  to train the proposed  model. We aimed to provide a ranked list of the next items to be recommended, where the observed items had to have a higher
score than the unobserved ones. Let $D=\{(u, v_i, v_j): u \in U, v_i\in G^{u},v_j \in V/ G^{u} \}$ denote the set of pair-wise training instances. Then, we trained the proposed model by maximising a posterior (MAP) as follows:

\begin{equation}
\arg \min_{\Theta} \sum _{(u, v_i, v_j) \in D}-\ln\sigma (R_i^u- R_j^u)+\lambda _{uv}||\theta_{uv}||^2 + \lambda _{a}||\theta_{a}||^2
\end{equation}
where $\theta _{uv}$ =\{U, V\} is the set of user and item embedding parameters, $ \theta_{a}=\{W_1, W_2, W_3, W_4\}$ is the set of weights of attention networks, $\lambda_{uv}$ and $\lambda_{a}$ are the regularisation parameters, and $\sigma$ is a logistic function. 

\section{Experiment and Evaluation}
\label{ch:Chapter7ExperimentandEvaluation}
We conducted our experiments on two widely used datasets, Tmall and Gowalla and compared the proposed model in terms of the evaluation metrics with the baseline models discussed in Chapter~\ref{ch4:chapterExperimentalSetup}, Section~\ref{Tmall , Tafeng and Gowalla Datasets}, in particular. We followed the same preprocessing procedure as that in SHAN~\cite{DBLP:conf/ijcai/YingZZLXXX018} and treated the user records of one day as a session. We randomly selected the sessions in the previous  week as the test set, and the rest were used for training. In addition, we randomly kept one item in each session as the next item to be predicted by the proposed model.

\subsection{Experimental Setup}
We set the item embedding and user embedding dimensions, $D$, to 100, which was a trade-off between the performance of the recommendation and the computation cost for both the datasets. As in~\cite{DBLP:conf/kdd/WuWAHHX19}, we set the number of CNN kernels $N_f$ and the window size to 400 and 3, respectively. We applied the  dropout strategy~\cite{DBLP:journals/jmlr/SrivastavaHKSS14} to
each layer of the CNN in order to avoid overfitting. The dropout rate was set to 0.2, the batch size was empirically set to 50, and the sizes of both the user purpose query $D_p$ and the preference query $D_q$ were set to 200. The learning rate $\eta$ was $0.01$. The items and users dimensions were randomly initialised with normal distribution $N(0, 0.01)$ and then learned during the training process. The attention parameters were initialised with the $U (-\sqrt{\frac{3}{k}}, \sqrt{\frac{3}{k}})$.

\begin{figure*}[t!]
\centering
\includegraphics[width=0.7\textwidth,scale=1]{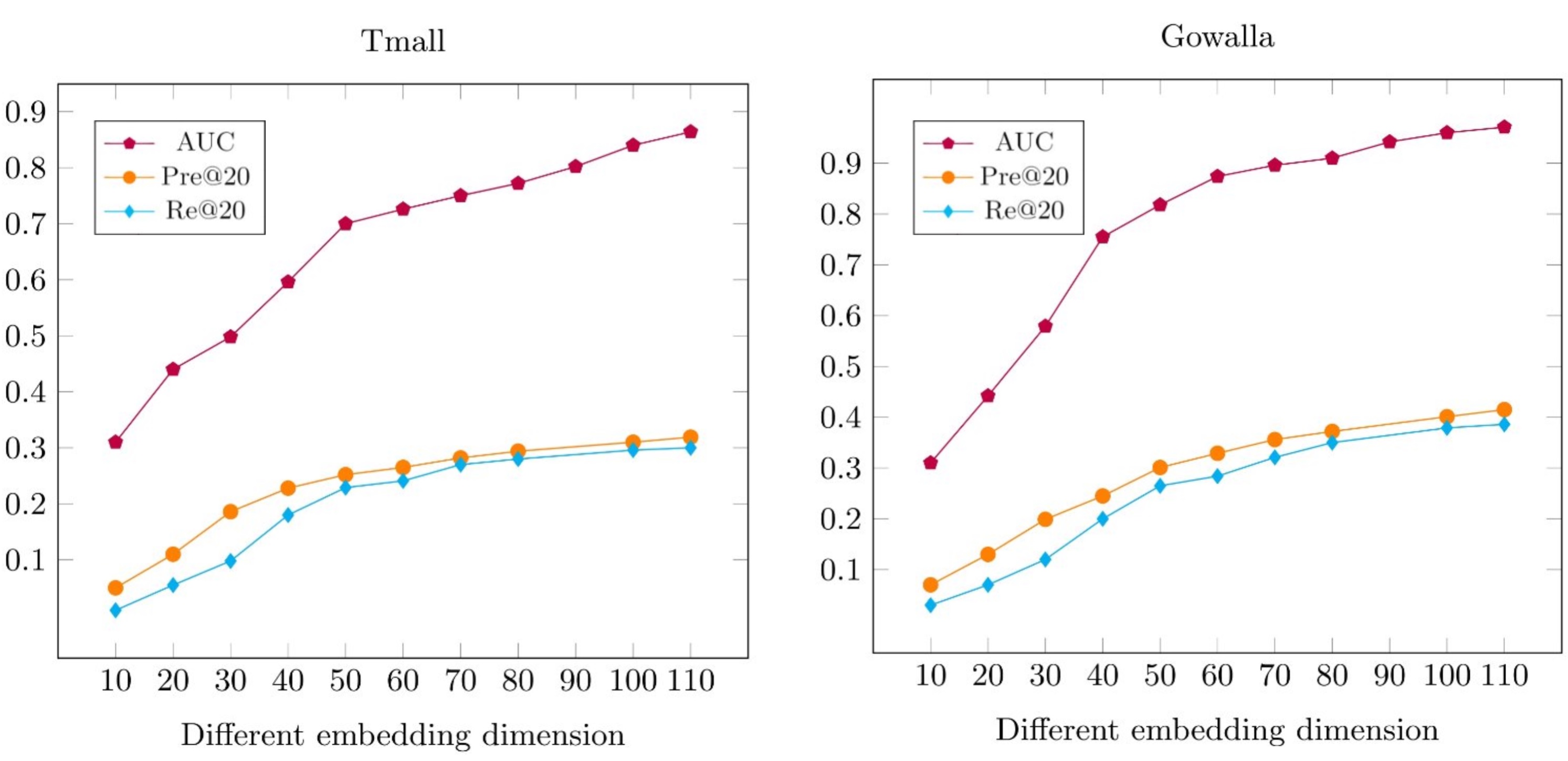}
\caption{Impact of different embedding dimensions on  Gowalla and Tmall  datasets.}
\label{ch7:figureHyperparameter}
\end{figure*}

\begin{table*}[b!]
\centering
    \begin{minipage}{0.55\textwidth}
        \centering
  
            \begin{tabular}{||c |c | c | c | c | c||} 
 \hline
Dataset & \backslashbox{$\lambda_ {uv}$ }{$\lambda _\alpha$ }  & 0 & 1 & 10 & 50\\
\hline
\hline
    \multirow{3}{*}{Tmall} & 0.01 &  0.085 & 0.126& 0.143 & 0.146  \\
                          & 0.001  & 0.079 & 0.124& 0.138 & 0.139 \\
                          & 0.0001 & 0.073  & 0.111 & 0.129 &0.133 \\
                        
 \hline
 
 \multirow{3}{*}{Gowalla} & 0.01 & 0.250&0.344 &0.355 & 0.372\\
                          & 0.001 & 0.321 & 0.397 &0.423 &0.432 \\
                          & 0.0001 & 0.342 &  0.421 & 0.452&0.461\\                       
  \hline           
\end{tabular}
\caption{\centering Impact of different regularisation at Recall@20.}
\label{ch7:tableRegularization}
    \end{minipage}
    \end{table*}

\subsection{Impact of Hyper-parameters}
In this section, we investigate the impact of hyper-parameters on the performance of CAN, and as in~\cite{DBLP:conf/ijcai/YingZZLXXX018}, we will only show the results for Recall@20. We considered $\lambda_{uv}=\{0.01, 0.001, 0.0001\}$ as the user and item embedding regularisation, and $\lambda_a=\{0, 1, 10, 50\}$ as the attention network regularisation. According to Table~\ref{ch7:tableRegularization}, the performance of CAN gradually increased when $\lambda_a>0$ in both Tmall and Gowalla datasets, which indicated the effectiveness of applying the attention mechanism in to the proposed model. We also tested the impact of different embedding dimensions, $D$, related to the user, item, and hidden layer parameters in the attention network. As is clear from Figure~\ref{ch7:figureHyperparameter}, a higher embedding dimension could result in better AUC, Recall@20, and Precision@20, as it could learn  more latent features from the user and the item as well as their interactions through the attention mechanism. From this figure, we can infer that  slight improvement was recorded when the embedding dimension was increased from 100, and thus, we set the embedding size to 100.

\begin{table*}[t!]
\begin{minipage}{.5\linewidth}
\centering
      \caption{Tmall}
      
       \begin{tabular}{||c | c | c|  c ||} 
\hline Methods & AUC & Re@20 & Pre@20 \\
\hline
CAN-S & 0.745 & 0.196  & 0.213   \\
\hline
CAN-L  &  0.889  & 0.221 & 0.282  \\
\hline
\end{tabular}
    \end{minipage}
    \begin{minipage}{.5\linewidth}
           \caption{Gowalla}
             \centering
           \begin{tabular}{||c | c |  c| c||} 
\hline Methods & AUC & Re@20 & Pre@20 \\
\hline
CAN-S  & 0.814 & 0.219 & 0.263 \\
\hline
CAN-L  & 0.916 &  0.298 & 0.342 \\
\hline
\end{tabular}
\end{minipage} 
\caption{Impact of different session lengths.}
\label{ch7:tableSessionLengths}
\end{table*}
\begin{table*}[b!]
\begin{minipage}{.5\linewidth}
\caption{Tmall}
\centering
    \begin{tabular}{||c | c | c|  c ||} 
\hline Methods & AUC & Re@20 & Pre@20 \\
\hline
\hline
CAN-PurEn & 0.817 & 0.256 & 0.278  \\
\hline
CAN-PreEn  & 0.781 & 0.210 & 0.264  \\
\hline
CAN  &  \textbf{0.915 } & \textbf {0.317} & \textbf{0.322} \\
\hline
\end{tabular}
\end{minipage}%
\begin{minipage}{.5\linewidth}
\caption{Gowalla}
\centering
\begin{tabular}{||c | c |  c| c||} 
\hline Methods & AUC & Re@20 & Pre@20 \\
\hline
\hline
CAN-PurEn   & 0.924 & 0.284 & 0.312 \\
\hline
CAN-PreEn  & 0.899 &  0.256 & 0.299 \\
\hline
CAN & \textbf{0.989} &\textbf{0.392 } & \textbf{0.401} \\
\hline
\end{tabular}
\end{minipage} 
\caption{Impact of CAN modules}
 \label{ch7:tableImpactofCAN}
    \end{table*}
\subsection{Impact of  Different Session Lengths}
We examined the performance of CAN under different sequence lengths as the local features captured by the CNN might be different. Table~\ref{ch7:tableSessionLengths} presents the results of our investigation. We considered sessions with less than three items as  short sessions and treated sessions with more than three items as long sessions. The percentage of short and long sessions was $90\%$ and $10\%$, and $83\%$ and $17\%$ in both the Tmall and the Gowalla datasets, respectively. In Table~\ref{ch7:tableSessionLengths}, CAN-S refers to a situation where the short sessions were modelled, while only the long sessions were considered in CAN-L. From this table, we can draw several observations. First, the performance of both CAN-L and CAN-S was very close. Second, CAN-L performed slightly better than CAN-S with respect to AUC, Pre@20, and Re@20 on both the Tmall and the Gowalla datasets. This was probably because of the capturing of  more contextual features through the long sessions. Third, the performance of CAN-L was still too close to the overall performance of the proposed model.

\subsection{Impact of Different Modules of CAN}
In this experiment, we aimed to test the performance of two modules:  purpose encoder and preference encoder, as shown in Table~\ref{ch7:tableImpactofCAN}. CAN-PurEn means only the user purpose module was used, while CAN-PreEn only considered the use of the preference component. In Table~\ref{ch7:tableImpactofCAN}, we can notice that first, CAN-PurEn can effectively improve the performance of the proposed approach. This might be attributed to the capturing of the local patterns in a long-term interacted item set through CNN and highlighting the important items according to the users' preferences by a PSAU cell. Second, CAN-PreEn was another effective module in the proposed model, which indicated a significant improvement in the performance of CAN. This was probably because items in a short-term interacted item set usually had different informativeness and recognising the important ones could help to better model the users' representations. Third, generally, CAN performs better than  CAN-PurEn and CAN-PreEn, when only a single module is incorporated in CAN. This demonstrates the benefit of combining these two modules for learning the users' representations and predicting the next items, eventually.

\begin{figure*}[b!]
\includegraphics[width=0.85\textwidth,scale=1]{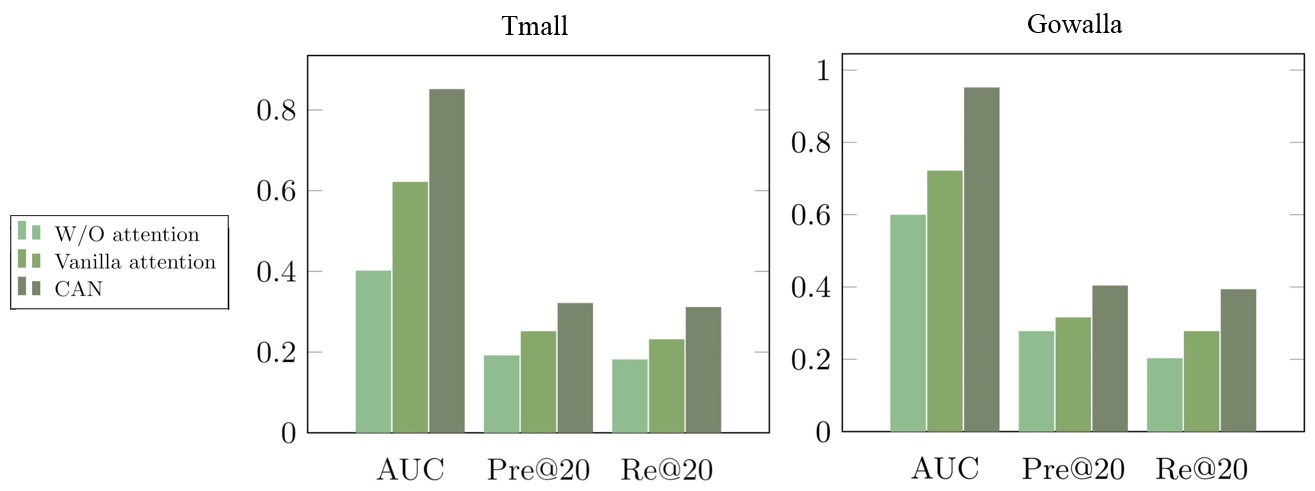}
\centering
\caption{Impact of PSAU on  Gowalla and Tmall datasets. W/O attention means no attention mechanism was used.}
\label{ch7:figurePSAU}
\end{figure*}
\subsection{Impact of PSAU Components}
In order to verify the effectiveness of the PSAU component in the proposed model, we compared the performance of the proposed model in the presence and the absence of the PSAU cell. As is clear from Figure~\ref{ch7:figurePSAU}, we noted different findings: (1) the application of the attention mechanism could show better performance than the model without attention. The reasons behind this observation might be the assignment of a different weight to different items and the capability of the attention mechanism to discover the important items in a user-item interaction. (2) The proposed model CAN dynamically outperformed the model without the attention mechanism and vanilla attention. This was more likely because of the assignment of  different scores to the same items for modelling the purposes and the preferences of different users. In contrast, vanilla attention networks use a fixed query vector, and thus, they may not be able to differentiate the importance of the same items in modelling different users' preferences. Unlike vanilla attention, the proposed model can highlight important items in the users' purposes according to their personal preferences, which in turn can help to obtain better users' representations learning. Thus, we could validate the effectiveness of the PSAU cell in the proposed approach.

\begin{table}[t!]
\begin{center}
\begin{tabular}{||l|lllllll||}
\hline
Dataset & \multicolumn{7}{c|}{Tmall} \\
\hline \hline
Metrics & Re5& Re10 & Re20 &Pre5 &Pre10 &Pre20& AUC  \\\hline
Top & 0.021 & 0.052 & 0.084 & 0.051 &0.062 &0.074 & 0.392 \\ 
BPR& 0.024 & 0.090 & 0.122 & 0.062 &0.069 & 0.074 &0.481 \\ 
Fossil& 0.110 & 0.120 &0.125 & 0.083& 0.088 & 0.092&  0.691 \\   
Caser & 0.041 & 0.049 & 0.052 &  0.100& 0.108& 0.115 &0.701 \\ 
FPMC &0.050 & 0.055& 0.061 & 0.118 & 0.125 & 0.130 &0.742 \\ 
HRM &0.060  & 0.065 & 0.070 &0.121 & 0.129 & 0.133& 0.751 \\  
GRU4Rec& 0.062 & 0.065& 0.069&0.138 & 0.145& 0.149& 0.762 \\
NARM & 0.063 & 0.068 &0.073 & 0.141& 0.149 & 0.159 &0.781 \\ 
SHAN & 0.071 & 0.076 & 0.079& 0.155& 0.160& 0.166& 0.789 \\
MEANS&0.074  & 0.079& 0.082 &0.163 & 0.172&0.177& 0.790\\ \hline
\textbf {CAN} & \textbf {0.201 } & \textbf{ 0.278} &\textbf {0.317}& \textbf {0.200}& \textbf {0.260} & \textbf {0.322} & \textbf {0.915}  \\ \hline
\end{tabular}
\caption{Performance of different methods on Tmall dataset.}
\label{ch7:tablePerformanceTmall}
\end{center}
\end{table}
\begin{table}[b!]
\begin{center}
\begin{tabular}{||l|lllllll||}
\hline
Dataset & \multicolumn{7}{c|}{Gowalla} \\
\hline \hline
Metrics & Re5& Re10 & Re20 &Pre5 &Pre10 &Pre20& AUC  \\\hline
Top & 0.038 & 0.048 & 0.059 &  0.061 &0.066 & 0.071 & 0.711 \\ 
BPR& 0.069 & 0.074 & 0.081 & 0.077 &0.082 &0.089 & 0.800   \\ 
Fossil& 0.215 & 0.298& 0.312 & 0.091& 0.095 &0.099 & 0.810  \\   
Caser & 0.075 & 0.083 & 0.089 & 0.114 & 0.119 &0.124 & 0.815  \\ 
FPMC & 0.115 &0.129 & 0.138& 0.127& 0.133& 0.142&0.820  \\ 
HRM &0.119 &0.125 &0.145 &0.150 & 0.157& 0.161& 0.824  \\  
GRU4Rec & 0.121& 0.135& 0.141&0.155 &0.160 & 0.165&0.828  \\
NARM & 0.130 & 0.136& 0.140& 0.156& 0.159& 0.163 & 0.830 \\ 
SHAN &0.135& 0.140 &0.144 &0.163 &0.169 & 0.175& 0.832  \\
MEANS& 0.142& 0.150& 0.158& 0.170& 0.175& 0.180 &0.840\\ \hline
\textbf {CAN} & \textbf {0.250} &\textbf{ 0.312 }& \textbf {0.392} &\textbf{ 0.360} & \textbf{0.399} & \textbf {0.401} & \textbf {0.989} \\ \hline
\end{tabular}
\caption{Performance of different methods on Gowalla dataset.}
\label{ch7:tablePerformanceGowalla}
\end{center}
\end{table}

\subsection{Overall Performance Comparison}
In this section, we will compare the results of the proposed model, which is summarised in Table~\ref{ch7:tablePerformanceGowalla},  with those of the state-of-the-art approaches for both the Tmall and the Gowalla datasets. From this table, the following inferences could be drawn: 
\begin{enumerate}
\item In the mentioned tables, where the
best result in each row is highlighted in boldface, the proposed model significantly and consistently outperformed all of the state-of-the-art models in terms of Precision@N, Recall@N, and AUC in different $Ns$ on both the Tmall and the Gowalla datasets. In particular,
compared with MEANS, which was the best baseline in terms of all the evaluation metrics, CAN
showed 14\% and 16\% improvements with respect to the AUC on the Tmall and the Gowalla datasets, respectively. This indicated the ability of CAN to recognise important items in the users' purposes according to their preferences through the CNN and the PSAU component.
\item Deep learning-based methods that used an attention network (CAN, MEANS, SHAN, and NARM) showed better performance than the methods without an attention mechanism. The reason could be  the capability of the attention mechanism to emphasise  the most important items in the user-item interaction set.
\item Overall, all the unified approaches (CAN, MEANS, SHAN, NARM, HRM, FPMC, and Fossil) outperformed the best general and sequential recommenders, such as BPR and GRU4Rec, respectively. This might encourage other researchers to use both types of users' preferences in their model to provide personalised suggestions to the users. 

\item Among all the unified approaches, after CAN, MEANS outperformed the other methods such as SHAN, NARM, HRM, FPMC, and Fossil. While the performance of MEANS and SHAN was very close, MEANS could not achieve around 5\% and 9\% improvement as compared to SHAN in terms of Recall@20 on the Tmall and the Gowalla datasets, respectively. This indicated the effect of using external memory to store a long-term user-item interaction set after a max-pooling operation. However, MEANS could not effectively model the local contexts in the long term users' preferences and could not find important items for revealing the purposes and preferences of different users. Moreover, although MEANS used an attention mechanism, it could not discover the level of informativeness for different items. Different from all of the mentioned approaches, the proposed model could dynamically find important items  according to the users' purposes and  preferences. 
\end{enumerate}

\section{Summary}
\label{ch:Chapter7Summary}
In this chapter, we proposed a novel model to unify both the general and the sequential recommenders. The proposed model called CAN could model the users’ decision-making process and exploit their purposes and preferences. The main idea behind CAN was to fully capture the users' interests and provide a list of preferred items to them. The proposed CAN framework have contained two main components: Purpose Encoder and Preference Encoder. At the first component, CAN aimed to discover a user's motivation behind interacting with a set of successive items. The second component focused on finding the most important items in a set of user-item interactions. A CNN was applied at the first module to capture the local and contextual information of a long-term interacted item set. As different users may have different purposes for clicking on the same items, and the same users may have different purposes and preferences,  we proposed a personalised attention mechanism, PSAU, to do the job for CAN. PSAU exploited the embeddings of the user IDs as the queries of the purpose- and preference-level attention networks. PSAU was applied in both the modules to differentiate the level of contributions of different items to both the users' purposes and  preferences. PSAU could recognise important items in the users' preferences according to their purposes. The extensive experimental results on real-world datasets validated the effectiveness of the proposed approach as compared to that of several representative state-of-the-art methods.


\chapter{Conclusions and Future Work}
\label{ch:chapterConclusions and Future Work}
This thesis has started with a  comprehensive and extensive analysis of both general and sequential recommenders in Chapter~\ref{ch:chapter2RW}. To be specific, we have addressed several different challenges of both general and sequential recommenders, such as dealing with the data sparsity problem, handling a noisy session, and handling  collective sequential dependencies. We have provided comprehensive solutions to each of these mentioned issues in separate chapters from Chapters~\ref{ch5:chapter5Personality} to~\ref{ch:chapter7CAN}, as summarised in the next part. In the following, we provide a summary of the contributions of the thesis followed by a detailed discussion of further research directions.  

\section{Conclusion}
Chapter~\ref{ch:chapter2RW} offered a systematic literature review on the two major types of RSs, the recent progress, and the existing gaps in this domain. We have introduced the detailed mechanism of general recommender systems, sequential recommender systems, and unified recommender systems (which combined both general and sequential recommender systems into one model). Moreover, we have pointed out the differences among the three mentioned RSs and explained why each them was important for a recommendations task.

In Chapter~\ref{ch5:chapter5Personality}, the main focus was on general recommenders. We have introduced one of the main challenges that the current RSs are confronted with: dealing with the data sparsity problem. The data sparsity problem which refers to the lack of available information regarding users or items is one of the main difficulties of the current RSs, especially the general ones. Most of the existing approaches rely on the users’ ratings to discover their preferences and identify similar users. Thus, they might
fail when there are no common items of interest among users, a problem which we call the Data Sparsity With no Feedback on Common Items (DSW-n-FCI). To fill this gap, we have proposed a personality-based recommender system in which the users' personality types were discovered implicitly with no extra burden on the users. We have introduced DSW-n-FCI as a subset of the data sparsity problem, because of which the current RSs, even the personality-based ones, might fail to have any suggestions for users. We have discussed that a user's decision-making  process might be affected by three main factors, namely  users’ personality types, their personal interests, and their level of knowledge in a particular
domain. Therefore, in this chapter, we have proposed a mathematical algorithm to integrate these parameters into a pure matrix factorisation model. Empirical evaluations on a real-world dataset have demonstrated the effectiveness of the proposed
model, particularly in DSW-n-FCI situations.

In Chapter~\ref{ch6:chapterDAS}, we have addressed the problem of a noisy session (a session including a set of multiple items) in sequential recommender systems. Each item had a different level of contribution to the occurrence of the next items, and not all the adjacent items in a session were highly dependent. Therefore, in a session, there might be some irrelevant items that might create a noisy session and a false dependency. Dependency modelling is a  critical step in session-based recommender systems and has a significant effect on the performance of an RS.
In addition, less attention is paid to the consideration of  long-term users' preferences by the existing SRSs, as they may mostly focus on short-term users' preferences. While considering  short-term users' preferences might satisfy the users' current needs and interests, long-term users' preferences might provide users with the items with which they would probably interact, eventually. To overcome the mentioned issues, in this chapter, we have proposed to use an attention mechanism in order to differentiate the importance of each item in both long and short-term users' preferences. An attention network, by assigning different weights to the items, could put more emphasis on items which were strongly context-relevant and downplay the weakly correlated ones in a user-item interaction sequence. The proposed model, DAS, consisted of three main blocks: (1) an embedding block to convert raw user-item and item-item interactions to low-dimensional spaces, (2) an attention block to focus on the contextually relevant items in a long and short-term item sets, and (3) a fully connected block to first learn a combination of both long and short-term users' preferences through the first feed-forward neural network, and then pass the users' mixture preferences along with the users' embedding vectors to the second fully connected network in order to produce a personalised recommendation. Furthermore, we have conducted empirical evaluations on two real-world datasets to demonstrate the superiority of the proposed method to that of  state-of-the-art methods in terms of the Area Under Curve (AUC), Precision,  Recall, and novelty evaluation metrics.

Chapter~\ref{ch:chapter7CAN} has discovered another drawback from which most of the current SRSs suffer: collective dependency. Collective dependency means that there may be a possible and particular dependency behind a collection of purchased items in a session. This observation has mostly been ignored by the current SRSs, as they assume each item to be a separate entity and each user-item interaction to be independent. This might not be true in real-world cases, as there might be a particular purpose behind interacting with successive items
in a sequence. People make their decision through two sequential processes, i.e. start shopping with a particular intention and then select a specific item which satisfies their preferences under this intention. Furthermore, different users usually have different purposes and preferences, and the same user may have various intentions. Thus, different users may click on the same items with an attention
to a different purpose. Therefore, we inferred that the distinction between the users’ purposes and their preferences needs to be taken into consideration in order to fully capture the users' behaviour patterns and their preferences. To bridge this gap, in this chapter, we have proposed a novel method named CAN, which takes both the users’ purposes and their
preferences into account for the next-item recommendation task. CAN consists of two main modules, namely a purpose encoder and a preference encoder, and unifies the benefits of both general and sequential recommenders. The purpose encoder is responsible for identifying the users' purposes by capturing the local and high-level information from the long-term interacted item set with the help of a CNN. Next, at the preference encoder, we have proposed to use a Purpose-Specific Attention Unit (PASU) to differently attend to different items and fully exploit the different informativeness of different items. Through the experiments, we have verified the superiority of the proposed model to the  state-of-the-art algorithms on two datasets in terms of AUC, Recall, and Precision.

In short, in this thesis, we have systematically reviewed the less explored and more important issues in both general and sequential recommender systems, such as DSW-n-FCI problem, noisy session, and collective dependency. In each chapter, the problem and the proposed solution were given, and we have evaluated the performance of the proposed model on real-world datasets to illustrate the efficiency of our work as compared to that of the existing studies. Each chapter (from Chapter~\ref{ch:chapter2RW} to Chapter~\ref{ch:chapter7CAN}) of this thesis is supported by a research paper~\footnote{The papers based on the results of Chapters~\ref{ch5:chapter5Personality} and ~\ref{ch6:chapterDAS} have been published, while the papers based on the results of Chapters~\ref{ch:chapter2RW} and ~\ref{ch:chapter7CAN} are
under review.} 
. Hence, the  approaches proposed in this thesis can have a good effect on the improvement of the performance of both general recommenders and sequential recommenders in the related research community  and applications in this area of RSs. Moreover, considering the fact that a recommender system is dealing with large datasets, it could be important to perform an analysis on the impact of the proposed approaches on the computational performance of the methods, which we leave it as a direction for our future work.

\section{Future Work}

Although recommender systems have been extensively studied as well as used in real-world applications, there are still several open issues which need to be tackled by the RS community. After a deep exploration of the current studies in recommender systems, we have identified a collection of challenges which may provide a good direction for the future works of researchers in this  community.

\subsection{General Recommender Systems' Open Issues}
In this section, we discuss the some of open issues of general recommender systems which still remain less explored. This might also represent future research directions. We will first explain the significance of the problems and then introduce the remaining challenges in this domain. 

\subsubsection{Taking More Heterogeneous Users' Features}
\textbf{Significance.} As we have discussed in Chapter~\ref{chap:introduction}, there are different levels of issues in general recommender systems. At the attribute level, there are different users' attributes (e.g. trust, personality, and attitudes) that can have a significant and positive effect on the performance of a recommender system. Incorporating more features can help a recommender system  better understand the users' preferences and provide more personalised options to them. General recommender systems aim to discover the users' general taste, and thus, considering  more user-relevant features can help to make recommendations more accurate. As one of the critical steps in memory-based RSs is how to discover an accurate set of neighbours, incorporating more users' features can do the job for RSs. For instance, trust, which can be defined as a context-dependent factor~\cite{DBLP:conf/wise/GhafariYBO18a},  \cite{DBLP:journals/access/GhafariBJPMYO20}, can help an RS to narrow down a set of selected neighbours that can better reflect a target user's preferences~\cite{DBLP:journals/eswa/BediS12}. Leveraging the trust relationship has a considerable potential to provide  trustworthy suggestions for users, which as a result can better predict the users' tastes~\cite{dong2020survey}. In addition, a combination of these features can provide more accurate recommendations; Alarcon et al.~\cite{ALARCON201869} investigated the relation between the five-factor model of personality traits and trust information, and their results demonstrated how personality can explain the processes underlying trust interactions.

\textbf{Open Issues.} Although user-related information has been incorporated into general recommender systems, more features still need to be taken into account.  

- How to exploit and integrate these important features is one of the main difficulties of RSs. Owing to the recent advancement of deep learning, which can be used for the personality detection job. For instance, Majumder et al.~\cite{DBLP:journals/expert/MajumderPGC17} reported  that the written texts can reflect the users' personality traits. Hence, in this work, five different CNN networks were trained to predict the users' personality types based on the Five Factor Model. Besides the  text-based content, CNN as one of the most commonly used techniques in object detection tasks, can also work with visual data for personality detection by analysing the users' facial features. Ventura et al.~\cite{DBLP:conf/cvpr/VenturaML17} used CNN to automatically detect the peoples' personality traits with the help of the face detection and Action Unit (AU) recognition systems.

\subsubsection{Considering More Types of Actions }
\textbf{Significance.} As we have illustrated in the action level of the  presented hierarchical structure in Chapter~\ref{chap:introduction}, we have two different types of actions: explicit actions and implicit ones. Each of these actions can create different types of data, which need to be treated differently. Sometimes, people are willing to rate the products that they have bought before or like the music they have listened or write comments for movies they have watched. These types of data are considered the explicit type of feedback, which may be more useful in the rating prediction task. However, it may be too realistic to assume that people usually like to explicitly express their level of interest in items (e.g. products, music, movies, and services) with which they have interacted. In contrast, in the implicit type of feedback people show their interests on items indirectly by clicking or viewing those items. In general, explicit type of feedback may be easier to collect and may better present users' preferences. Therefore, the explicit type of data can be modelled more accurately~\cite{10.1145/1869446.1869453}, \cite{DBLP:conf/cikm/BeheshtiBNCXZ17}, \cite{DBLP:journals/pvldb/BeheshtiBNT18} ,\cite{DBLP:journals/dpd/BeheshtiBTMBN19}, \cite{DBLP:conf/ksem/YakhchiGTF17}, \cite{DBLP:conf/ifip12/GhafariT16}.
However, modelling the implicit type of feedback may be more challenging as there are no negative instances in this type of feedback. For instance, if a user has not interacted with item $i$, it does not imply that this user is not interested in this item, but it may be possible that this user is not aware of the existence of this item. There are some collaborative filtering-based studies in the literature that have tried to learn the users' preferences from implicit data~\cite{DBLP:conf/www/AnyosaVJ18}. Therefore, consideration of  both the explicit and implicit feedback
provides different degrees of expressivity of the users’ preferences, which can result in more personalised suggestions.

\textbf{Open Issues.} Here, the main issue is how to deal with  both types of users' preferences in a model. As each type of feedback may have its own challenges, it might be difficult to incorporate both types of feedback in one model.

 - Although there may be some studies that have tried to model the implicit type of feedback, there are a few works in which both types of feedback have been integrated. Approaches based on implicit feedback (e.g. clicks) aim to formulate recommendations as a ranking problem, which highly depends on the selection of the objective loss function to optimise~\cite{DBLP:conf/recsys/KaratzoglouBS13}. For instance, Mandal et al.~\cite{DBLP:conf/complexnetworks/MandalM18} have proposed to use three types of explicit feedback, such as ratings, helpfulness score, and centrality score with one type of implicit feedback such as the view relationship with the user-item interaction in one model. Then, a probabilistic matrix factorisation (PMF)-based
model was proposed for recommendations, which considers both explicit and implicit feedback. The authors showed that fusing the two types of feedback is very effective. Generalised Probabilistic Matrix Factorisation (GPMF) is another example of combining  both the explicit and implicit types of feedback~\cite{DBLP:conf/sigir/ZhangYX17}. In this work, the `reliability score' was considered  explicit feedback and was calculated based on the review network. In this study, the extracted features from the product images could also be seen as another type of implicit feedback. Some other studies also discussed that if a user rated a particular item, it not only showed the explicit data but also was meant as the implicit data. The authors assumed that a given rate to a particular item could directly mean that this user was interested in this item while indirectly indicated the implicit data as well~\cite{DBLP:journals/kbs/ChenP18}.

\subsubsection{Modelling Cross-Level Information }
\textbf{Significance.} We have already discussed the existing levels of issues in approaches that try to model long-term users' preferences. Each level has different goals to focus on, and thus, they may address different challenges, accordingly. However, combining the benefits of some of these levels to some extent can generate a more robust recommender system. For instance, the main focus of models at the item level is to exploit the features of the items and then try to recommend those which are similar to the current ones, while some of the attribute level-based approaches may pay more attention to contextual features such as season, weather, and time, combining  these two mentioned approaches can make recommendations more personalised. For instance, assume that a user usually watches the comedy movies of a particular actor. However, currently, she/he is travelling to  Paris in autumn. In this situation, she/he may prefer to watch some romantic movies of another actor. Therefore, considering these contextual factors may increase the satisfaction's rate of this user.    

\textbf{Open Issues.} How to incorporate these parameters, how to identify a set of important contexts, and how to select a level to be better combined with another level may be challenging tasks for general recommender systems. In an early study by Yap et al.~\cite{DBLP:journals/tkde/YapTP07}, a Bayesian network was proposed  to integrate the users' contexts and items' attributes into one unique model in RSs. The authors illustrated that their proposed model could increase the performance of an RS even with noisy and incomplete contextual information. Incorporating the information from different levels could also help to mitigate the cold-start problem. For instance, Melville et al.~\cite{DBLP:conf/aaai/MelvilleMN02} used text information (e.g. movie content information) to enrich a sparse rating matrix and then applied a collaborative filtering model to their model. While different techniques for combining this information may be adopted, such as a switching mechanisms\cite{DBLP:journals/umuai/BillsusP00}, \cite{GeorgeLekakos},  a ranking methods~\cite{DBLP:journals/umuai/Burke02}, and a weighting approaches~\cite{Claypool1999CombiningCA}, the work in this area remains limited.

\subsection{Sequential Recommender Systems' Open Issues}
Similar to the previous section, here, we introduce some of the challenges of sequential recommender systems which still remain less explored. This may also point out new directions for future studies.
\subsubsection{Taking Contextual Information Into Account }
\textbf{Significance.} According to ~\cite{DBLP:conf/huc/AbowdDBDSS99}, context refers to `any information that can be used to characterise the situation of an entity, where an entity can be a person, a place, or an object relevant to the interaction between the user and application, including the user and
application themselves. A system is context-aware if it uses
context to provide relevant information and/or services to
the user, where relevancy depends on the user’s task'. The consideration of the context in recommender systems is a hot topic in RSs and has attracted much attentions from the RS community. There is a separate line of research on general recommenders, named Context-Aware RSs (CARs). CARs are well studied and broadly applied in classic types of recommenders, where the context can be considered to be a time of a day, weather, location, age, gender, mood, seasonality, nationality, budget, and expertise~\cite{DBLP:journals/kais/BagciK16}, \cite{DBLP:conf/recsys/LeviMDT12a},  \cite{DBLP:journals/hcis/BenlamriZ14}. The performance improvement of CARs as compared that of the general RS has been reported by a vast majority of the existing works in this domain~\cite{DBLP:reference/sp/AdomaviciusT15}, \cite{DBLP:conf/recsys/PaganoCLHTKQ16}.

\textbf{Open Issues.} Although the importance of incorporating contextual information into general recommender systems has been emphasised by the existing attempts in the literature~\cite{DBLP:conf/recsys/Unger15}, it is still unexplored in sequential recommender systems. In fact, selecting the main features as the context and identifying when and how to incorporate them into an SRS are the key issues that need to be addressed. CA-RNN is an example of a context-aware model in a session-based recommender system~\cite{DBLP:journals/corr/LiuWWL016}. The main difference between CA-RNN and the current RNNs-based models is that while they use the constant input matrix and transition matrix, CA-RNN uses context-specific input matrices and adaptive context-specific transition matrices. The context can be considered to be time, location, weather, and so on. Contextual Recurrent Neural Network for Recommendation (CRNN) is another example of a context-aware SRS which considers the other useful context information such as the types of user-item interactions, the exact time of the day for the occurrence of each event, and the difference between the events~\cite{DBLP:conf/recsys/SmirnovaV17}. In the work proposed by Rakkappan and Rajan~\cite{10.1145/3308558.3313567}, the importance of context was emphasised and discussed as follows: unlike most of the traditional SRSs which model the sequential information and context as two independent factors, it is better to model both of them simultaneously. The authors believed that factors such as the time difference between purchasing two items, the location, and the other contextual factors can be considered  the context. While there have been a few attempts in the literature which incorporate the context in session-based recommender systems, more efforts need to be made in this direction to fully exploit the advantages of taking the context into account.

\subsubsection{Modelling Similar Session Information}
\textbf{Significance.} One of the main characteristics of a session-based recommender system which has become considerably  popular is the capability of SRSs to work with anonymous behaviour sequential information. 
Most of the current approaches in this domain only focus on modelling the current session. While if we take a closer look at the idea behind the collaborative filtering technique which has been successful for decades in general recommender systems, we can determine the strength of these approaches in discovering a set of neighbours for a particular user. Therefore, taking the potential benefits of collaborative information can help SRSs find users who may have similar sequential behaviours. For instance, assume a user with two sessions ${a_1=\{bag 1, bag 2, bag3\}}$ and $ a_2=\{bag 1, bag 4, bag2\}$. In the context of a recommender system, this may imply that this user is looking for a bag and has thus clicked on different bags for comparison. On the basis of  this user's previous histories, we may infer that $bag 2$ is the item of interest, as this user had clicked twice on this item in two different sessions. Taking this useful information into account may increase the performance of SRSs.

\textbf{Open Issues.} How to incorporate similar session information in SRSs is a new direction that may result in recommendations improvement. For instance, Collaborative Session-based Recommendation Machine (CSRM) introduces a term called neighbourhood sessions, which can help a sequential recommender to better model the current session~\cite{DBLP:conf/sigir/WangRMCMR19}. CSRM consists of two parallel modules: an Inner Memory Encoder (IME), and an Outer Memory Encoder (OME). While IME is responsible for modelling a user’s own information in the current session, the main goal of OME is to exploit the neighbourhood sessions' information in order to discover the intent of the current sessions. Although considering this valuable information in a session-based recommender system can be a significant improvement, it is still in its early stage and more efforts need to be made this direction.

\subsubsection{Modelling Cross-Level Information}
\textbf{Significance.} As we have discussed previously in Chapter~\ref{chap:introduction}, there are different levels of issues in approaches considering the short-term users' preferences. Combining the advantages of modeling each level may result in a better recommender system. For example, in the second level of the proposed hierarchy model, the key issue is how to model a session's context, while at the third level the main focus is given to the item's dependency modelling. Assume that a user has purchased a pair of shoes, a bag , a lipstick, and a bottle of water in one transaction on Saturday because of a season sale. As this user bought different products with no particular dependency on Saturday, presumably, she/he will purchase a dress to match her new shoes in the near future. Therefore, while there might not be any dependency among the purchased items, and a combination of the mentioned key issues may suggest users with the best set of the next interesting items. 

\textbf{Open Issues.} On the one hand, how to model this information is one problem, and on the other hand, how to select which level can be better combined is another remaining problem in this domain, which needs to be addressed by the SRSs. Although the work proposed by Quadrana et al.~\cite{DBLP:journals/corr/QuadranaKHC17} have tried to model cross-session information to better capture the users' activities across sessions and the evolution of user interests over time, the very basic RNN method was applied. Hence, there is a need for further investigation in this domain.

\subsubsection{Modelling Different Heterogeneous Actions }

\textbf{Significance.} Understanding users’ interests and their behaviour is an essential task for a recommender system to be able to make the right suggestion to the users. Users can take multiple types of actions on e-commerce platforms,  each of which can show a different level of the users' interest in different items. For example, the significance of considering  different types of users' behaviour on a social media platform (i.e. posting, sharing, and commenting) in identifying their interests has been emphasised by the current studies~\cite{DBLP:conf/www/ZhaoCHC15}.  However, most of the current session-based recommender systems only model one type of users' behaviour such as clicking, or purchasing, since there are multiple types of behaviour e.g. clicking, purchasing, and adding to favourites in e-commerce, and
downloading, using, and un-installing in app usage. It has become a critical issue to model these heterogeneous actions toward items, accordingly, to better come up with personalised items for a user. 

\textbf{Open Issues.} How to model different and heterogeneous actions into a session-based recommender system is a challenging task. As stated by Le et al.~\cite{DBLP:conf/ijcai/LeLF18}, it is very common on online platforms where before an eventual purchase takes place, the user may click on several items, select some of them for adding to the cart, and put some of the clicked ones to the favourites list. The authors have believed that for predicting the best next interested items, not only are the purchased items  important, but the clicked items also need to be considered in users' the behavioural modelling. To do so, two types of sequences are introduced: a target and a support. The first one refers to high-quality and sparser interactions such as purchasing for predicting the next interaction, while the latter refers to more frequent and informative interactions such as clicking, which may be relevant for recommending the next target item. Therefore, we can infer that more investigation is required to overcome this issue. 

\subsubsection{Taking Cross-Domain
Information into Account}
\textbf{Significance.} Considering items related to movies and books belongs to the different domain, we can see cross-domain as a new line of recommender systems with the aim of exploiting knowledge
from a source domain to perform or improve recommendations in a target domain~\cite{DBLP:reference/sp/CantadorFBC15}. The success of considering cross-domain information in a recommender system has been shown by most of the approaches related to general recommender systems, where there is an enriched repository of research under the name of cross-domain recommendations~\cite{FIgnacio}. For instance, assume that a user reads the book 
`The Great Gatsby' book first, and then, she/ he  becomes interested in watching the corresponding movie. After seeing this movie, this user may become a fan of Leonardo Wilhelm DiCaprio and start watching his other movies. Moreover, probably, this user has been attracted by the special types of clothes worn by the actors in the `The Great Gatsby' movie and starts looking for such clothes online. This indicates that items from different domains can be somehow connected to each other and form a session. The strong relations between different domains have been tested and proven in marketing, behavioural, and data
mining studies as well~\cite{10.1145/3073565}, \cite{DBLP:conf/wise/BeheshtiBSS19}, \cite{DBLP:conf/wise/TabebordbarBB19}, \cite{DBLP:books/sp/BeheshtiBSGMBGR16}. According to this observation, it is more beneficial to
leverage all the available provided user data irrespective of the domains and platforms to  consider any possible overlaps or correlations between the users' preferences. Therefore, instead of treating each domain independently, the transfer of knowledge from the source domain to the target domain can help to have better recommendations and business benefits, which may trigger an increased interest for cross-domain recommendations. Furthermore, cross-domain recommendations cannot only help to make a more personalised prediction but also deal with the cold-start problem where there is a lack of sufficient information about the users or items.

\textbf{Open Issues.} One of the major challenges in cross-domain recommendations is how to exploit the knowledge from one domain to another to help RSs better understand the users' interests. Knowledge linkage and learning transfer have been widely used by models that use cross-domain information in general recommenders~\cite{DBLP:journals/tkde/PanY10}, \cite{DBLP:conf/www/ElkahkySH15}, \cite{DBLP:conf/ACMicec/ChungSS07}. However, incorporating cross-domain information in sequential recommenders may have more challenges, as the user IDs need to be extracted in order to discover and link all the sessions for the same users in different domains~\cite{DBLP:journals/expert/HuCZJCC18}, \cite{DBLP:conf/sigir/BaiNZZDW18}. In a recent study, a user's session information was represented as a graph instead of a linear structure~\cite{DBLP:journals/tois/QiuHLY20}, and a novel Mask-Readout function was proposed to construct a more expressive session embedding with the cross-session information. While this study has been effective, this area still remains a largely unexplored topic.

%
%
%
 \bibliographystyle{splncs04}
 \bibliography{Refrences}



\backmatter



\end{document}